\documentclass[mathematics,article,accept,pdftex,oneauthor]{Definitions/mdpi}

\usepackage[ruled,vlined]{algorithm2e}

\usepackage{pgfplots}

\usepackage{caption}

\firstpage{1} 
\pubvolume{13}
\issuenum{24}
\articlenumber{3981}
\pubyear{2025}
\copyrightyear{2025}
\externaleditor{Jeffery Secrest}
\datereceived{30 September 2025} 
\daterevised{4 November 2025} 
\dateaccepted{2 December 2025} 
\datepublished{13 December 2025}

\hreflink{https:// doi.org/10.3390/math13243981}

\Title{ContEvol Formalism: Numerical Methods Based on Hermite Spline Optimization}
\TitleCitation{ContEvol Formalism: Numerical Methods Based on Hermite Spline Optimization}

\Author{{Kaili Cao} $^{1,2}$\orcidA{}}
 
 \AuthorNames{Kaili Cao}
 
\AuthorCitation{Cao, K.}

\address{$^{1}$ \quad Department of Physics, The Ohio State University, 191 West Woodruff {Ave}, Columbus, OH 43210, USA; cao.1191@osu.edu\\
{$^{2}$ \quad Center for Cosmology and AstroParticle Physics (CCAPP), The Ohio State University, 191 West Woodruff Ave, Columbus, OH 43210, USA}\\
}

\abstract{We present the ContEvol (continuous evolution) formalism, a family of implicit numerical methods which only need to solve linear equations and are almost symplectic.
Combining values and derivatives of functions, ContEvol outputs allow users to recover full history and render full distributions.
Using the classic harmonic oscillator as a prototype case, we show that ContEvol methods lead to lower-order errors than two commonly used \mbox{Runge--Kutta} methods.
Applying first-order ContEvol to simple celestial mechanics problems, we demonstrate that deviation from equation(s) of motion of ContEvol tracks is still $\mathcal{O}(h^5)$ ($h$ is the step length) by our definition.
Numerical experiments with an eccentric elliptical orbit indicate that first-order ContEvol is a viable alternative to classic Runge--Kutta or the symplectic leapfrog integrator.
Solving the stationary Schr\"odinger equation in quantum mechanics, we manifest ability of ContEvol to handle boundary value or eigenvalue problems.
Important directions for future work, including mathematical foundations, higher dimensions, and technical improvements, are discussed at the end of this article.}



\begin{document}
\section{Introduction}
\label{sec:introduction}\label{sec1}

Numerical simulations are widely used in contemporary physics. For instance, famous computer codes in astrophysics include {\sc Arepo} \citep{Springel2010MNRAS} and {\sc Athena++} \citep{Jiang2014ApJS} for (magneto)hydrodynamic simulations, {\sc galpy} \citep{Bovy2015ApJS} for galactic dynamics, {\sc YREC} \citep{Demarque2008Ap&SS} and {\sc MESA} \citep{Paxton2011ApJS} for stellar evolution, {\sc mercury} \citep{Chambers1999MNRAS} and {\sc REBOUND} \citep{Rein2012A&A} for celestial mechanics, to name a few. There are certainly great works in other areas of research as well.

Because of the discreteness of the world of computers, it is common practice to convert differential equations into difference equations, so that finite difference methods can be applied.
However, at spatial scales much larger than elementary particles, the physical world is arguably continuous. Therefore, finite difference might be intrinsically limited: when we try to model the full history of a dynamic system or full details of a function of spatial location, we have to resort to spline interpolation.
Meanwhile, many physics problems are formulated as first- or second-order differential equations with analytic expressions, indicating that usage of general-purpose methods might be an overkill.
These motivate the ContEvol (continuous evolution) formalism, which we {(According to context, the pronouns ``we/us/our'' in this work may refer to: (i) the author and indirect contributors (see acknowledgements), (ii) the author and researchers with similar academic background and interests, or (iii) the author and the readers.)} present in this work.

Desire for continuity has provoked thoughts about function representation. Imaging that, in addition to {\it {values}} of a one-dimensional real function $f(x)$: $[x_{\min}, x_{\max}] \mapsto \mathbb{R}$ at a series of sampling points $\{ x_{\min}, \ldots, x_i, x_{i+1}, \ldots, x_{\max} \}$, we have its {\it {first derivative}} at the same points.
Then in each interval $x_i \leq x \leq x_{i+1}$, we can {\it {always}} find a cubic polynomial satisfying all boundary conditions at both ends, so that $f(x)$ can be represented as a piece-wise cubic function---not only is it continuous, but its first derivative is also continuous, which is favorable to some analysis in physics.
This technique is known as Hermite spline. {(Anecdote: The author ``independently'' came up with this idea about three weeks before hearing about Hermite spline. For this reason, the author feels obliged to declare the possibility that this work might be reinventing some methods.)}

It can be naturally extended to higher orders: combining values and first- to $n$th-order derivatives at both ends of an interval, we can find a $(2n+1)$st-order polynomial representation of the function.
However, it should be noted that basic calculus yields simple but powerful expressions for addition, subtraction, multiplication, division, and composition of representations with only values and first {derivatives}:
\begin{align}
    &h(x) = f(x) \pm g(x) \quad &\Rightarrow \quad &\dot h(x) = \dot f(x) \pm \dot g(x) \\
    &h(x) = f(x) \cdot g(x) \quad &\Rightarrow \quad &\dot h(x) = h(x) \left[ \frac{\dot f(x)}{f(x)} + \frac{\dot g(x)}{g(x)} \right] \\
    &h(x) = \frac{f(x)}{g(x)} \quad &\Rightarrow \quad &\dot h(x) = h(x) \left[ \frac{\dot f(x)}{f(x)} - \frac{\dot g(x)}{g(x)} \right] \\
    &h(x) = g(f(x)) \quad &\Rightarrow \quad &\dot h(x) = \dot g(f(x)) \dot f(x).
\end{align}
{Finiteness} can be a blessing and a curse---we lose some high-order information, but do not need to assume that functions are infinitely differentiable, unlike when we use spectral methods 
\citep{Grandclement2009LRR}. 

ContEvol is a family of numerical methods built on this idea. It approximates functions of space and time as polynomials and minimizes deviation from equation(s) of the problem.
While details will be presented and discussed in the rest of this work, here we briefly address how this relates to other common methods \citep{Press2007book}. 
Some of the most important dichotomies of numerical methods include: explicit or implicit, single-step or (linear) multistep, and symplectic (or in physicists' words, phase space conserving) or not.
Since ContEvol finds the optimal solution for the next step, it should be categorized as implicit. {While classic implicit methods (e.g., backward Euler) or spline collocation methods} \citep{Schiesser2017book} {usually require numerically solving non-linear equations}, ContEvol only needs to solve linear equations.  
Although this work focuses on the single-step version of ContEvol, we will argue that multistep versions are straightforward to achieve. Because of the predefined functional form, ContEvol is not strictly symplectic; however, with moderately small steps, its non-symplecticity (deviation from $1$ of determinant of Jacobian) can be rapidly below $2^{-53}$, i.e., inundated by roundoff errors of double precision.

This work is principally for illustration and discussion of general strategies. The rest of this article is structured as follows.
In Section~\ref{sec:classic}, we apply first- and second-order ContEvol methods {(An $n$th-order ContEvol method treats up to $n$th-order derivatives at sampling nodes as independent variables.)} to a prototype case, classic harmonic oscillator, and compare them to fourth- and eighth-order Runge--Kutta methods.
Then in Section~\ref{sec:celestial}, we showcase potential applications of ContEvol in celestial mechanics; examples in this work are two-body and three-body problems, in which equations of motion are non-linear and multivariate.
In Section~\ref{sec:quantum}, we use ContEvol to solve stationary Schr\"odinger equation in quantum mechanics, which is physically different from time evolution of a dynamic system.
Finally in Section~\ref{sec:discuss}, we wrap up this work by discussing important directions for future work, including mathematical foundation, higher dimensions, and technical~improvements.

\section{Prototype Case: Classic Harmonic Oscillator}
\label{sec:classic}

We start with the simplest case of a dynamical system: time evolution of a single real variable. To check results of numerical methods against exact solution, we choose the classic harmonic oscillator, for which the equation of motion (EOM) is
\vspace{-4pt}
\begin{align}
    m\ddot{x} = -kx,
\end{align}
where $m$ is the mass of the particle and $k$ is the spring constant; setting these constants to $1$, {(This is a natural choice which makes time dimensionless. A different scaling would lead to different cost functions and thus different optimization results, but is not explored in this work.)} 
the EOM becomes
\vspace{-6pt}
\begin{align}
    \label{eq:CHO_EOM} \ddot{x} = -x.
\end{align}

Without loss of generality, we are given $x(0) = x_0$, $\dot{x}(0) = v_0$ and try to solve for $x(h) = x_h$, $\dot{x}(h) = v_h$, where $h$ is the time step (usually small). The exact solution {is}
\vspace{-7pt}
\begin{adjustwidth}{-\extralength}{0cm}
\begin{align}
    \label{eq:CHO_exact} \left\{ \begin{aligned} x_{\rm exact}(t) &= x_0 \cos t + v_0 \sin t = \left[ \begin{aligned} &x_0 \left(1 - \frac{t^2}{2} + \frac{t^4}{24} - \frac{t^6}{720} + \frac{t^8}{40320} + \mathcal{O}(t^{10})\right) \\ &+ v_0 \left(t - \frac{t^3}{6} + \frac{t^5}{120} - \frac{t^7}{5040} + \frac{t^9}{362880} + \mathcal{O}(t^{11}) \right) \end{aligned} \right] \\
    v_{\rm exact}(t) &= -x_0 \sin t + v_0 \cos t = \left[ \begin{aligned} &-x_0 \left(t - \frac{t^3}{6} + \frac{t^5}{120} - \frac{t^7}{5040} + \frac{t^9}{362880} + \mathcal{O}(t^{11}) \right) \\ &+ v_0 \left(1 - \frac{t^2}{2} + \frac{t^4}{24} - \frac{t^6}{720} + \frac{t^8}{40320} + \mathcal{O}(t^{10})\right) \end{aligned} \right] \end{aligned} \right..
\end{align}
\end{adjustwidth}

Section~\ref{ss:CHO_CE1} showcases ability of the first-order ContEvol method, and Section~\ref{ss:CHO_RK} compares it to two commonly used (explicit and multistep) Runge--Kutta methods.
In Section~\ref{ss:CHO_CE2}, we explore the second-order ContEvol method, with and without strict EOM enforcement at $t=h$.

\subsection{First-Order ContEvol Method}
\label{ss:CHO_CE1}

We approximate the solution in a parametric form (subscript ``CE1'' stands for first-order ContEvol)
\vspace{-6pt}
\begin{align}
    \label{eq:CHO_CE1_xvh} x_{\rm CE1}(t) = x_0 + v_0t + Bt^2 + At^3, \quad t \in [0, h];
\end{align}
``terminal'' conditions at $t=h$ yield
\begin{align}
    &\left\{ \begin{aligned} x_{\rm CE1}(h) &= x_0 + v_0h + Bh^2 + Ah^3 = x_h \\
    \dot{x}_{\rm CE1}(h) &= v_0 + 2Bh + 3Ah^2 = v_h \end{aligned} \right. \\ \Rightarrow \quad
    &\begin{pmatrix} h^2 & h^3 \\ 2h & 3h^2 \end{pmatrix} \begin{pmatrix} B \\ A \end{pmatrix}
    = \begin{pmatrix} x_h - x_0 - v_0h \\ v_h - v_0 \end{pmatrix} \\ \Rightarrow \quad
    &\left\{ \begin{aligned} A &= 2(x_0 - x_h) h^{-3} + (v_0 + v_h) h^{-2} \\
    B &= 3(x_h - x_0) h^{-2} - (2v_0 + v_h) h^{-1} \end{aligned} \right..
\end{align}
Because of the initial conditions $(x_0, v_0)^{\rm T}$, the transformation $(x_h, v_h)^{\rm T} \to (A, B)^{\rm T}$ is affine, not linear.

We define the cost function as {the deviation from the EOM
}
\vspace{-14pt}
\begin{adjustwidth}{-\extralength}{0cm}
\begin{align}
    \label{eq:CHO_CE1_cost} \nonumber \epsilon_{\rm CE1}(A, B; h) &= \int_0^h (\ddot{x} + x)^2 \,{\rm d}t
    = \int_0^h [(2B + x_0) + (6A + v_0)t + Bt^2 + At^3]^2 \,{\rm d}t \\
    \nonumber &= \int_0^h \left[ \begin{aligned} &(4B^2+4Bx_0+x_0^2) + (24AB+12Ax_0+4Bv_0+2v_0x_0)t \\
    &+ (36A^2+12Av_0+4B^2+2Bx_0+v_0^2)t^2 + (16AB+2Ax_0+2Bv_0)t^3 \\
    &+ (12A^2+2Av_0+B^2)t^4 + 2ABt^5 + A^2 t^6 \end{aligned} \right] \,{\rm d}t \\
    &= \left[ \begin{aligned} &(4B^2+4Bx_0+x_0^2)h + (12AB+6Ax_0+2Bv_0+v_0x_0)h^2 \\
    &+ \frac{1}{3}(36A^2+12A v_0+4B^2+2Bx_0+v_0^2)h^3 + \frac{1}{2}(8AB+A _0+Bv_0)h^4 \\
    &+ \frac{1}{5}(12A^2+2Av_0+B^2)h^5 + \frac{1}{3}ABh^6 + \frac{1}{7}A^2h^7 \end{aligned} \right];
\end{align}
\end{adjustwidth}
minimizing this, we obtain
\vspace{-12pt}
\begin{adjustwidth}{-\extralength}{0cm}
\begin{align}
    &\left\{ \begin{aligned} \frac{\partial\epsilon_{\rm CE1}}{\partial A} &= (12B+6x_0)h^2 + (24A+4v_0)h^3 + \frac{1}{2}(8B+x_0)h^4 + \frac{2}{5}(12A+v_0)h^5 + \frac{1}{3}Bh^6 + \frac{2}{7}Ah^7 = 0 \\
    \frac{\partial\epsilon_{\rm CE1}}{\partial B} &= (8B+4x_0)h + (12A+2v_0)h^2 + \frac{2}{3}(4B+x_0)h^3 + \frac{1}{2}(8A+v_0)h^4 + \frac{2}{5}Bh^5 + \frac{1}{3}Ah^6 = 0 \end{aligned} \right. \\ \Rightarrow \quad
    &\begin{pmatrix} 24h^3 + \dfrac{24}{5}h^5 + \dfrac{2}{7}h^7 & 12h^2 + 4h^4 + \dfrac{1}{3}h^6 \\ 12h^2 + 4h^4 + \dfrac{1}{3}h^6 & 8h + \dfrac{8}{3}h^3 + \dfrac{2}{5}h^5 \end{pmatrix} \begin{pmatrix} A_{\rm CE1} \\ B_{\rm CE1} \end{pmatrix}
    = \begin{pmatrix} -6x_0h^2 - 4v_0h^3 - \dfrac{1}{2}x_0h^4 - \dfrac{2}{5}v_0h^5 \\ -4x_0h - 2v_0h^2 - \dfrac{2}{3}x_0h^3 - \dfrac{1}{2}v_0h^4 \end{pmatrix} \\ \Rightarrow \quad
    \label{eq:CHO_CE1_AB} &\left\{ \begin{aligned} A_{\rm CE1} &= \frac{7 (- 3600v_0 + 1800x_0h + 60v_0h^2 + 120x_0h^3 + 10x_0h^5 +3v_0h^6)}{2 (75600 + 10080h^2 + 1080h^4 + 24h^6 + 5h^8)} \\
    B_{\rm CE1} &= -\frac{15 (5040x_0 + 1092x_0h^2 + 168v_0h^3 + 72x_0h^4 + 8v_0h^5 + 5x_0h^6 + 2v_0h^7)}{2 (75600 + 10080h^2 + 1080h^4 + 24h^6 + 5h^8)} \end{aligned} \right..
\end{align}
\end{adjustwidth}

Plugging Equation~(\ref{eq:CHO_CE1_AB}) back into Equation~(\ref{eq:CHO_CE1_xvh}), our solution at $t=h$ is
\begin{align}
    \label{eq:CHO_CE1_linmap} \begin{pmatrix} x_h \\ v_h \end{pmatrix} = \begin{pmatrix} G_{{\rm CE1},00} & G_{{\rm CE1},01} \\ G_{{\rm CE1},10} & G_{{\rm CE1},11} \end{pmatrix} \begin{pmatrix} x_0 \\ v_0 \end{pmatrix}
\end{align}
with
\begin{align}
    \label{eq:CHO_CE1_evolop} \left\{ \begin{aligned} G_{{\rm CE1},00} &= \frac{151200 - 55440h^2 - 1620h^4 - 192h^6 + 5h^8}{2 (75600 + 10080h^2 + 1080h^4 + 24h^6 + 5h^8)} \\
    G_{{\rm CE1},01} &= \frac{151200h - 5040h^3 + 60h^5 - 72h^7 + h^9}{2 (75600 + 10080h^2 + 1080h^4 + 24h^6 + 5h^8)} \\
    G_{{\rm CE1},10} &= \frac{60h (-2520 + 84h^2 + 6h^4 + h^6)}{2 (75600 + 10080h^2 + 1080h^4 + 24h^6 + 5h^8)} \\
    G_{{\rm CE1},11} &= \frac{151200 - 55440h^2 - 1620h^4 - 192h^6 + 13h^8}{2 (75600 + 10080h^2 + 1080h^4 + 24h^6 + 5h^8)} \end{aligned} \right..
\end{align}

The determinant of the time evolution operator $G_{\rm CE1}$ is
\begin{align}
    \label{eq:CHO_CE1_detG} \det \begin{pmatrix} G_{{\rm CE1},00} & G_{{\rm CE1},01} \\ G_{{\rm CE1},10} & G_{{\rm CE1},11} \end{pmatrix} = 1 - \frac{19h^8}{302400 + 40320h^2 + 4320h^4 + 96h^6 + 20h^8},
\end{align}
i.e., unfortunately, ContEvol is not symplectic. However, the discrepancy $1 - \det(G_{\rm CE1}) \leq 2^{-53}$ (common double-precision floating-point format cannot tell discrepancies below this threshold) when $h \leq 0.03396$. 
Thanks to the linearity of the problem, $G$ is diagonalizable for common choices of $h$, and complexity of evolving the system for $N$ steps with fixed time step can be just $2N + \mathcal{O}(1)$.

Expanding Equations~(\ref{eq:CHO_CE1_linmap}) and (\ref{eq:CHO_CE1_evolop}), first-order ContEvol yields
\vspace{-10pt}
\begin{adjustwidth}{-\extralength}{0cm}
\begin{align}
   \label{eq:CHO_CE1_res} \left\{ \begin{aligned} x_{\rm CE1}(h) &= \left[ \begin{aligned} &x_0 \left(1 - \frac{h^2}{2} + \frac{h^4}{24} - {\color{red} 0} \cdot \frac{h^6}{720} + {\color{red} \left(-\frac{284}{15}\right)} \cdot \frac{h^8}{40320} + \mathcal{O}(h^{10})\right) \\ &+ v_0 \left(h - \frac{h^3}{6} + \frac{h^5}{120} - {\color{red} \left(-\frac{18}{7}\right)} \cdot \frac{h^7}{5040} + {\color{red} \left(-\frac{1716}{25}\right)} \cdot \frac{h^9}{362880} + \mathcal{O}(h^{11}) \right) \end{aligned} \right] \\
    v_{\rm CE1}(h) &= \left[ \begin{aligned} &-x_0 \left(h - \frac{h^3}{6} + {\color{red} \frac{2}{3}} \cdot  \frac{h^5}{120} - {\color{red} \left(-\frac{14}{3}\right)} \cdot \frac{h^7}{5040} + {\color{red} \left(-\frac{392}{5}\right)} \cdot \frac{h^9}{362880} + \mathcal{O}(h^{11}) \right) \\ &+ v_0 \left(1 - \frac{h^2}{2} + \frac{h^4}{24} - {\color{red} 0} \cdot \frac{h^6}{720} + {\color{red} (-18)} \cdot \frac{h^8}{40320} + \mathcal{O}(h^{10})\right) \end{aligned} \right] \end{aligned} \right..
\end{align}
\end{adjustwidth}
comparing to the exact solution Equation~(\ref{eq:CHO_exact}), we see that errors in $x_h$ and $v_h$ (highlighted in red) are $\mathcal{O}(h^6)$ and $\mathcal{O}(h^5)$, respectively.

According to Equation~(\ref{eq:CHO_CE1_AB}), the minimized cost function Equation~(\ref{eq:CHO_CE1_cost}) is
\vspace{-10pt}
\begin{adjustwidth}{-\extralength}{0cm}
\begin{align}
    \label{eq:CHO_CE1_mineps} \epsilon_{{\rm CE1},\min}(h) = \left[ \begin{aligned} &\frac{x_0^2}{720}h^5 + \frac{v_0x_0}{720}h^6 + \left(\frac{v_0^2}{2800}-\frac{x_0^2}{2160}\right)h^7 - \frac{v_0x_0}{2800}h^8 + \left(\frac{53x_0^2}{907200}-\frac{23v_0^2}{378000}\right)h^9 + \frac{47v_0x_0}{1512000}h^{10} \\
    &+ \left(\frac{19v_0^2}{5880000}-\frac{11x_0^2}{6804000}\right)h^{11} + \frac{41v_0x_0}{79380000}h^{12} + \left(\frac{43v_0^2}{132300000}-\frac{3223x_0^2}{5715360000}\right)h^{13} \\
    &- \frac{4681v_0x_0}{9525600000}h^{14} + \left(\frac{9461x_0^2}{85730400000}-\frac{31273v_0^2}{333396000000}\right)h^{15} + \frac{71909v_0x_0}{1000188000000}h^{16} \\
    &+ \left(\frac{18107v_0^2}{1666980000000}-\frac{360391x_0^2}{36006768000000}\right)h^{17} - \frac{287197v_0x_0}{60011280000000} h^{18} \\
    &+ \left(\frac{5933x_0^2}{135025380000000}-\frac{297667v_0^2}{700131600000000}\right)h^{19} - \frac{420823v_0x_0}{1575296100000000}h^{20} \end{aligned} \right];
\end{align}
\end{adjustwidth}
note that $\epsilon_{{\rm CE1},\min}(h) = \mathcal{O}(h^5)$ seems consistent with $x_{\rm CE1}(h) - x_{\rm exact}(h) = \mathcal{O}(h^6)$.
This minimization goal can be used to adapt step length, e.g., for $x_0 = 1$ and $v_0 = 0$ ($x_0 = 0$ and $v_0 = 1$), $\epsilon_{{\rm CE1},\min}(h) \leq 2^{-53}$ {(In this section, we use $2^{-53}$ as a general-purpose benchmark for numerical precision, although it is only a threshold for double-precision when the leading-order term is $1$.)} when $h \leq 0.002402$ ($h \leq 0.01634$). 

\subsection{Fourth- and Eighth-Order Runge--Kutta Methods}
\label{ss:CHO_RK}

To enable Runge--Kutta methods, the equation of motion Equation~(\ref{eq:CHO_EOM}) has to be written~{as} 
\vspace{-6pt}
\begin{align}
    \frac{\rm d}{{\rm d}t} \begin{pmatrix} x \\ v \end{pmatrix} = {\boldsymbol f}\begin{pmatrix} x \\ v \end{pmatrix} = \begin{pmatrix} v \\ -x \end{pmatrix}.
\end{align}
Like in many physics problems, this derivative does not have explicit time dependence.

Applying the fourth-order (i.e., classic) Runge--Kutta method, we have
\vspace{-4pt}
\begin{align}
    {\boldsymbol k}_{{\rm RK4},1} &= {\boldsymbol f}\begin{pmatrix} x_0 \\ v_0 \end{pmatrix} = \begin{pmatrix} v_0 \\ -x_0 \end{pmatrix}, \\
    {\boldsymbol k}_{{\rm RK4},2} &= {\boldsymbol f}\left(\begin{pmatrix} x_0 \\ v_0 \end{pmatrix} + \frac{{\boldsymbol k}_{{\rm RK4},1}}{2}h\right) = \left( v_0 - \frac{x_0}{2}h, -x_0 - \frac{v_0}{2}h \right)^{\rm T}, 
    \end{align}
    \begin{adjustwidth}{-\extralength}{0cm}
\begin{align}
    {\boldsymbol k}_{{\rm RK4},3} &= {\boldsymbol f}\left(\begin{pmatrix} x_0 \\ v_0 \end{pmatrix} + \frac{{\boldsymbol k}_{{\rm RK4},2}}{2}h\right) = \left( v_0 - \frac{x_0}{2}h - \frac{v_0}{4} h^2, -x_0 - \frac{v_0}{2}h + \frac{x_0}{4}h^2 \right)^{\rm T}, \\
    {\boldsymbol k}_{{\rm RK4},4} &= {\boldsymbol f}\left(\begin{pmatrix} x_0 \\ v_0 \end{pmatrix} + {\boldsymbol k}_{{\rm RK4},3}h\right) = \left( v_0 - x_0h - \frac{v_0}{2} h^2 + \frac{x_0}{4}h^3, -x_0 - v_0h + \frac{x_0}{2}h^2 + \frac{v_0}{4}h^3 \right)^{\rm T},
\end{align}
\end{adjustwidth}
and then
\begin{align}
    \nonumber \begin{pmatrix} x_h \\ v_h \end{pmatrix} &= \begin{pmatrix} x_0 \\ v_0 \end{pmatrix} + \frac{h}{6} ({\boldsymbol k}_{{\rm RK4},1} + 2{\boldsymbol k}_{{\rm RK4},2} + 2{\boldsymbol k}_{{\rm RK4},3} + {\boldsymbol k}_{{\rm RK4},4}) \\
    \nonumber &= \left( x_0 + v_0h - \frac{x_0}{2}h^2 - \frac{v_0}{6}h^3 + \frac{x_0}{24}h^4, v_0 - x_0h - \frac{v_0}{2} h^2 + \frac{x_0}{6}h^3 + \frac{v_0}{24}h^4 \right)^{\rm T} \\
    &= \begin{pmatrix} 1 - \dfrac{h^2}{2} + \dfrac{h^4}{24} & h - \dfrac{h^3}{6} \\ -h + \dfrac{h^3}{6} & 1 - \dfrac{h^2}{2} + \dfrac{h^4}{24} \end{pmatrix} \begin{pmatrix} x_0 \\ v_0 \end{pmatrix} \equiv G_{\rm RK4} \begin{pmatrix} x_0 \\ v_0 \end{pmatrix}.
\end{align}
Evidently, errors in $x_h$ and $v_h$ are both $\mathcal{O}(h^5)$.

The determinant of the time evolution operator $G_{\rm RK4}$ is
\begin{align}
    \det (G_{\rm RK4}) = 1 - \frac{h^6}{72} + \frac{h^8}{576},
\end{align}
i.e., the discrepancy $1 - \det (G_{\rm RK4})$ is two orders larger than $1 - \det (G_{\rm CE1})$; to archive \linebreak  $1 - \det (G_{\rm RK4}) \leq 2^{-53}$, one needs $h \leq 0.004472$, $7.594$ times smaller than what was required for first-order ContEvol. 
To adapt step length, the fourth-order Runge--Kutta method usually resorts to the fifth-order version, which necessitates a slight increase in computational~complexity.

Now let us try the eight-order Runge--Kutta method, {(RK8 coefficients used in this work are found on the MathWorks webpage ``Runge Kutta 8th Order Integration'':} \url{https://www.mathworks.com/matlabcentral/fileexchange/55431-runge-kutta-8th-order-integration}, {accessed on 14 April 2024}.) which gives (subscripts ``RK8'' on the right-hand side are omitted for simplicity)
 \vspace{-9pt}
\begin{align}
    {\boldsymbol k}_{{\rm RK8},0} &= {\boldsymbol f}\begin{pmatrix} x_0 \\ v_0 \end{pmatrix} = \begin{pmatrix} v_0 \\ -x_0 \end{pmatrix}, \\
    {\boldsymbol k}_{{\rm RK8},1} &= {\boldsymbol f}\left(\begin{pmatrix} x_0 \\ v_0 \end{pmatrix} + \frac{4{\boldsymbol k}_0}{27}h\right) = \left( v_0 - \frac{4x_0}{27}h, -x_0 - \frac{4v_0}{27}h \right)^{\rm T},  \\
      {\boldsymbol k}_{{\rm RK8},2} &= {\boldsymbol f}\left(\begin{pmatrix} x_0 \\ v_0 \end{pmatrix} + \frac{{\boldsymbol k}_0 + 3{\boldsymbol k}_1}{18}h\right) = \left( v_0 - \frac{2x_0}{9}h - \frac{2v_0}{81}h^2, -x_0 - \frac{2v_0}{9}h + \frac{2x_0}{81}h^2 \right)^{\rm T},\\
    {\boldsymbol k}_{{\rm RK8},3} &= {\boldsymbol f}\left(\begin{pmatrix} x_0 \\ v_0 \end{pmatrix} + \frac{{\boldsymbol k}_0 + 3{\boldsymbol k}_2}{12}h\right) = \begin{pmatrix} v_0 - \dfrac{x_0}{3}h - \dfrac{v_0}{18}h^2 + \dfrac{x_0}{162}h^3 \\  \\[0.05ex]
     -x_0 - \dfrac{v_0}{3}h + \dfrac{x_0}{18}h^2 + \dfrac{v_0}{162}h^3 \end{pmatrix}, \\
     {\boldsymbol k}_{{\rm RK8},4} &= {\boldsymbol f}\left(\begin{pmatrix} x_0 \\ v_0 \end{pmatrix} + \frac{{\boldsymbol k}_0 + 3{\boldsymbol k}_3}{8}h\right) = \begin{pmatrix} v_0 - \dfrac{x_0}{2}h - \dfrac{v_0}{8}h^2 + \dfrac{x_0}{48}h^3 + \dfrac{v_0}{432}h^4 \\ \\[0.05ex]
     -x_0 - \dfrac{v_0}{2}h + \dfrac{x_0}{8}h^2 + \dfrac{v_0}{48}h^3 - \dfrac{x_0}{432}h^4 \end{pmatrix}, 
      \end{align}
     \begin{adjustwidth}{-\extralength}{0cm}
     \small
\begin{align}
       {\boldsymbol k}_{{\rm RK8},5} &= {\boldsymbol f}\left(\begin{pmatrix} x_0 \\ v_0 \end{pmatrix} + \frac{13{\boldsymbol k}_0 - 27{\boldsymbol k}_2 + 42{\boldsymbol k}_3 + 8{\boldsymbol k}_4}{54}h\right) = \begin{pmatrix} v_0 - \dfrac{2x_0}{3}h - \dfrac{2v_0}{9}h^2 + \dfrac{4x_0}{81}h^3 + \dfrac{23v_0}{2916}h^4 - \dfrac{x_0}{2916}h^5 \\ \\[0.05ex]
       -x_0 - \dfrac{2v_0}{3}h + \dfrac{2x_0}{9}h^2 + \dfrac{4v_0}{81}h^3 - \dfrac{23x_0}{2916}h^4 - \dfrac{v_0}{2916}h^5 \end{pmatrix}, \\
    \nonumber {\boldsymbol k}_{{\rm RK8},6} &= {\boldsymbol f}\left(\begin{pmatrix} x_0 \\ v_0 \end{pmatrix} + \frac{389{\boldsymbol k}_0 - 54{\boldsymbol k}_2 + 966{\boldsymbol k}_3 - 824{\boldsymbol k}_4 + 243{\boldsymbol k}_5}{4320}h\right) \\
    &= \begin{pmatrix} v_0 - \dfrac{x_0}{6}h - \dfrac{v_0}{72}h^2 + \dfrac{x_0}{1296}h^3 + \dfrac{43v_0}{233280}h^4 - \dfrac{x_0}{466560}h^5 - \dfrac{v_0}{51840}h^6 \\ \\[0.05ex]
    -x_0 - \dfrac{v_0}{6}h + \dfrac{x_0}{72}h^2 + \dfrac{v_0}{1296}h^3 - \dfrac{43x_0}{233280}h^4 - \dfrac{v_0}{466560}h^5 + \dfrac{x_0}{51840}h^6 \end{pmatrix}, \\
    \nonumber {\boldsymbol k}_{{\rm RK8},7} &= {\boldsymbol f}\left(\begin{pmatrix} x_0 \\ v_0 \end{pmatrix} + \frac{-234{\boldsymbol k}_0 + 81{\boldsymbol k}_2 - 1164{\boldsymbol k}_3 + 656{\boldsymbol k}_4 - 122{\boldsymbol k}_5 + 800{\boldsymbol k}_6}{20}h\right) \\
    &= \begin{pmatrix} v_0 - \dfrac{17x_0}{20}h - \dfrac{v_0}{2}h^2 + \dfrac{x_0}{6}h^3 + \dfrac{29v_0}{540}h^4 - \dfrac{19x_0}{540}h^5 + \dfrac{13v_0}{6480}h^6 + \dfrac{x_0}{1296}h^7 \\ \\[0.05ex]
    -x_0 - \dfrac{17v_0}{20}h + \dfrac{x_0}{2}h^2 + \dfrac{v_0}{6}h^3 - \dfrac{29x_0}{540}h^4 - \dfrac{19v_0}{540}h^5 - \dfrac{13x_0}{6480}h^6 + \dfrac{v_0}{1296}h^7 \end{pmatrix}, \\
    \nonumber {\boldsymbol k}_{{\rm RK8},8} &= {\boldsymbol f}\left(\begin{pmatrix} x_0 \\ v_0 \end{pmatrix} + \frac{-217{\boldsymbol k}_0 + 18{\boldsymbol k}_2 - 678{\boldsymbol k}_3 + 456{\boldsymbol k}_4 - 9{\boldsymbol k}_5 + 576{\boldsymbol k}_6 + 4{\boldsymbol k}_7}{288}h\right) \\
    &= \begin{pmatrix} v_0 - \dfrac{5x_0}{6}h - \dfrac{497v_0}{1440}h^2 + \dfrac{125x_0}{1296}h^3 + \dfrac{323v_0}{15552}h^4 - \dfrac{47x_0}{10368}h^5 - \dfrac{5v_0}{10368}h^6 + \dfrac{x_0}{93312}h^7 + \dfrac{v_0}{93312}h^8 \\ \\[0.05ex]
    -x_0 - \dfrac{5v_0}{6}h + \dfrac{497x_0}{1440}h^2 + \dfrac{125v_0}{1296}h^3 - \dfrac{323x_0}{15552}h^4 - \dfrac{47v_0}{10368}h^5 + \dfrac{5x_0}{10368}h^6 + \dfrac{v_0}{93312}h^7 - \dfrac{x_0}{93312}h^8 \end{pmatrix}, \\
    \nonumber {\boldsymbol k}_{{\rm RK8},9} &= {\boldsymbol f}\left(\begin{pmatrix} x_0 \\ v_0 \end{pmatrix} + \frac{1481{\boldsymbol k}_0 - 81{\boldsymbol k}_2 + 7104{\boldsymbol k}_3 - 3376{\boldsymbol k}_4 + 72{\boldsymbol k}_5 - 5040{\boldsymbol k}_6 - 60{\boldsymbol k}_7 + 720{\boldsymbol k}_8}{820}h\right) \\
    &= \begin{pmatrix} v_0 - x_0h - \dfrac{419v_0}{820}h^2 + \dfrac{811x_0}{4920}h^3 + \dfrac{811v_0}{22140}h^4 - \dfrac{8x_0}{1845}h^5 - \dfrac{7v_0}{4920}h^6 + \dfrac{x_0}{2214}h^7 - \dfrac{5v_0}{106272}h^8 - \dfrac{x_0}{106272}h^9 \\ \\[0.05ex]
    -x_0 - v_0h + \dfrac{419x_0}{820}h^2 + \dfrac{811v_0}{4920}h^3 - \dfrac{811x_0}{22140}h^4 - \dfrac{8v_0}{1845}h^5 + \dfrac{7x_0}{4920}h^6 + \dfrac{v_0}{2214}h^7 + \dfrac{5x_0}{106272}h^8  - \dfrac{v_0}{106272}h^9 \end{pmatrix},
\end{align}
\end{adjustwidth}
and then
\vspace{-10pt}
\begin{adjustwidth}{-\extralength}{0cm}
\fontsize{8}{8}\selectfont
\begin{align}
    \nonumber \begin{pmatrix} x_h \\ v_h \end{pmatrix} &= \begin{pmatrix} x_0 \\ v_0 \end{pmatrix} + \frac{h}{840} (41{\boldsymbol k}_0 + 27{\boldsymbol k}_3 + 272{\boldsymbol k}_4 + 27{\boldsymbol k}_5 + 216{\boldsymbol k}_6+216{\boldsymbol k}_8+41{\boldsymbol k}_9) \\
    \nonumber &= \begin{pmatrix} x_0 + v_0h - \dfrac{x_0}{2}h^2 - \dfrac{v_0}{6}h^3 + \dfrac{1397x_0}{33600}h^4 + \dfrac{v_0}{120}h^5 - \dfrac{x_0}{720}h^6 - \dfrac{v_0}{5040}h^7 + \dfrac{x_0}{40320}h^8 + \dfrac{v_0}{2177280}h^9 - \dfrac{x_0}{2177280}h^{10} \\ \\[0.05ex]
    v_0 - x_0h - \dfrac{v_0}{2}h^2 + \dfrac{x_0}{6}h^3 + \dfrac{1397v_0}{33600}h^4 - \dfrac{x_0}{120}h^5 - \dfrac{v_0}{720}h^6 + \dfrac{x_0}{5040}h^7 + \dfrac{v_0}{40320}h^8 - \dfrac{x_0}{2177280}h^9 - \dfrac{v_0}{2177280}h^{10} \end{pmatrix} \\
    \nonumber &= \begin{pmatrix} 1 - \dfrac{h^2}{2} + \dfrac{1397h^4}{33600} - \dfrac{h^6}{720} + \dfrac{h^8}{40320} - \dfrac{h^{10}}{2177280} & h - \dfrac{h^3}{6} + \dfrac{h^5}{120} - \dfrac{h^7}{5040} + \dfrac{h^9}{2177280} \\ 
    \\[0.05ex]
    -h + \dfrac{h^3}{6} - \dfrac{h^5}{120} + \dfrac{h^7}{5040} - \dfrac{h^9}{2177280} & 1 - \dfrac{h^2}{2} + \dfrac{1397h^4}{33600} - \dfrac{h^6}{720} + \dfrac{h^8}{40320} - \dfrac{h^{10}}{2177280} \end{pmatrix} \begin{pmatrix} x_0 \\ v_0 \end{pmatrix} \\
    &\equiv G_{\rm RK8} \begin{pmatrix} x_0 \\ v_0 \end{pmatrix}.
\end{align}
\end{adjustwidth}
For some unknown reason, the fourth-order coefficients have a fractional error of $3/1400$, while the fifth- to eighth-order coefficients agree with the exact solution Equation~(\ref{eq:CHO_exact}).

The determinant of the time evolution operator $G_{\rm RK8}$ is
\begin{align}
    \det (G_{\rm RK8}) = \left[ \begin{aligned} &1 - \frac{h^4}{5600} + \frac{h^6}{11200} - \frac{2797h^8}{376320000} - \frac{19h^{10}}{4032000} + \frac{18119h^{12}}{18289152000} \\
    &- \frac{2197h^{14}}{36578304000} + \frac{h^{16}}{585252864} - \frac{107h^{18}}{4740548198400} + \frac{h^{20}}{4740548198400} \end{aligned} \right];
\end{align}
to archive $1 - \det (G_{\rm RK8}) \leq 2^{-53}$, one needs $h \leq 0.0008880$, $5.036$ times smaller than what was required for fourth-order Runge--Kutta. 

\subsection{Second-Order ContEvol Method}
\label{ss:CHO_CE2}

The ContEvol framework can be naturally generalized to higher orders.
Like in Section~\ref{ss:CHO_CE1}, we approximate the solution in a parametric form (subscript ``CE2'' stands for second-order ContEvol)
\begin{align}
    x_{\rm CE2}(t) = x_0 + v_0t - \frac{x_0}{2}t^2 + Ct^3 + Bt^4 + At^5, \quad t \in [0, h];
\end{align}
``terminal'' conditions at $t=h$ yield
\begin{align}
    \label{eq:CHO_CE2_xvh} &\left\{ \begin{aligned} x_{\rm CE2}(h) &= x_0 + v_0h - \frac{x_0}{2}h^2 + Ch^3 + Bh^4 + Ah^5 = x_h \\
    \dot{x}_{\rm CE2}(h) &= v_0 - x_0h + 3Ch^2 + 4Bh^3 + 5Ah^4 = v_h \\
    \ddot{x}_{\rm CE2}(h) &= - x_0 + 6Ch + 12Bh^2 + 20Ah^3 = -x_h \end{aligned} \right. \\ \Rightarrow \quad
    &\begin{pmatrix} h^3 & h^4 & h^5 \\ 3h^2 & 4h^3 & 5h^4 \\ 6h & 12h^2 & 20h^3 \end{pmatrix} \begin{pmatrix} C \\ B \\ A \end{pmatrix}
    = \begin{pmatrix} x_h - x_0 - v_0h + \dfrac{x_0}{2}h^2 \\ v_h - v_0 + x_0h \\ -x_h + x_0 \end{pmatrix} \\ \Rightarrow \quad
    \label{eq:CHO_CE2_coef} &\left\{ \begin{aligned} A &= 6(x_h-x_0)h^{-5} - 3(v_0+v_h)h^{-4} + \frac{x_0-x_h}{2} h^{-3} \\
    B &= 15(x_0-x_h)h^{-4} + (8v_0+7v_h)h^{-3} + \left(x_h-\frac{3}{2}x_0\right)h^{-2} \\
    C &= 10(x_h-x_0)h^{-3} - (6v_0+4v_h)h^{-2} + \frac{3x_0-x_h}{2}h^{-1} \end{aligned} \right..
\end{align}
Note that we have enforced the EOM at both $t=0$ and $t=h$, and the three coefficients ($A$, $B$, and $C$) are fully specified by two parameters ($x_h$ and $v_h$).

Likewise, we define the cost function as
\vspace{-12pt}
\begin{adjustwidth}{-\extralength}{0cm}
\begin{align}
    \label{eq:CHO_CE2_cost} \nonumber \epsilon_{\rm CE2}(A, B, C; h) &= \int_0^h (\ddot{x} + x)^2 \,{\rm d}t
    = \int_0^h [(6C+v_0)t + \left(12B-\frac{x_0}{2}\right)t^2 + (20A+C)t^3 + Bt^4 + At^5]^2 \,{\rm d}t \\
    \nonumber &= \int_0^h \left[ \begin{aligned} &(36C^2+12Cv_0+v_0^2)t^2 + (144BC+24Bv_0-6Cx_0-v_0x_0)t^3 \\
    &+ \left(240AC+40Av_0+144B^2-12Bx_0+12C^2+2Cv_0+\frac{x_0^2}{4}\right)t^4 \\
    &+ (480AB-20Ax_0+36BC+2Bv_0-Cx_0) t^5 \\ &+ (400A^2+52AC+2Av_0+24B^2-Bx_0+C^2)t^6 \\
    &+ (64AB-Ax_0+2BC)t^7 + (40A^2+2AC+B^2)t^8 + 2ABt^9 + A^2t^{10} \end{aligned} \right] \,{\rm d}t \\
    &= \left[ \begin{aligned} &\left(12C^2+4Cv_0+\frac{v_0^2}{3}\right) h^3 + \left(36BC+6Bv_0-\frac{3Cx_0}{2}-\frac{v_0x_0}{4}\right)h^4 \\
    &+ \frac{1}{20}\left(960AC+160Av_0+576B^2-48Bx_0+48C^2+8Cv_0+x_0^2\right) h^5 \\
    &+ \frac{1}{6}(480AB-20Ax_0+36BC+2Bv_0-Cx_0) h^6 \\
    &+ \frac{1}{7}(400A^2+52AC+2Av_0+24B^2-Bx_0+C^2) h^7 + \left(8AB-\frac{Ax_0}{8}+\frac{BC}{4}\right) h^8 \\
    &+ \frac{1}{9}(40A^2+2AC+B^2)h^9 + \frac{1}{5}AB h^{10} + \frac{1}{11}A^2 h^{11} \end{aligned} \right];
\end{align}
\end{adjustwidth}
minimizing this, we obtain
\vspace{-6pt}
\begin{align}
    &\left\{ \begin{aligned} \frac{\partial\epsilon_{\rm CE2}}{\partial A} &= \left[ \begin{aligned} &(48C+8v_0)h^5 + \left(80B-\frac{10x_0}{3}\right)h^6 + \frac{2}{7}(400A+26C+v_0)h^7 \\
    &+ \left(8B-\frac{x_0}{8}\right)h^8 +\frac{2}{9}(40A+C)h^9 + \frac{1}{5}B h^{10} + \frac{2}{11}A h^{11} \end{aligned} \right] = 0 \\
    \frac{\partial\epsilon_{\rm CE2}}{\partial B} &= \left[ \begin{aligned} &(36C+6v_0)h^4 + \frac{12}{5}\left(24B-x_0\right)h^5 + \left(80A+6C+\frac{v_0}{3}\right) h^6 \\
    &+ \frac{1}{7}(48B-x_0)h^7 + \left(8A+\frac{C}{4}\right)h^8 + \frac{2}{9}B h^9 +\frac{1}{5}A h^{10} \end{aligned} \right] = 0 \\
    \frac{\partial\epsilon_{\rm CE2}}{\partial C} &= \left[ \begin{aligned} &(24C+4v_0)h^3 + \left(36B-\frac{3x_0}{2}\right)h^4 + \frac{2}{5}(120A+12C+v_0)h^5 \\
    &+ \left(6B-\frac{x_0}{6}\right)h^6 +\frac{2}{7}(26A+C)h^7 + \frac{1}{4}B h^8 + \frac{2}{9}A h^9\end{aligned} \right] = 0 \end{aligned} \right. \\ \Rightarrow \quad
    \nonumber &\begin{pmatrix} \dfrac{800}{7}h^7 + \dfrac{80}{9}h^9 + \dfrac{2}{11}h^{11} & 80h^6 + 8h^8 + \dfrac{1}{5}h^{10} & 48h^5 + \dfrac{52}{7}h^7 + \dfrac{2}{9}h^9 \\
    80h^6 + 8h^8 + \dfrac{1}{5}h^{10} & \dfrac{288}{5}h^5 + \dfrac{48}{7}h^7 + \dfrac{2}{9}h^9 & 36h^4 + 6h^6 + \dfrac{1}{4}h^8 \\
    48h^5 + \dfrac{52}{7}h^7 + \dfrac{2}{9}h^9) & 36h^4 + 6h^6 + \dfrac{1}{4}h^8 & 24h^3 + \dfrac{24}{5}h^5 + \dfrac{2}{7}h^7 \end{pmatrix}
    \begin{pmatrix} A_{\rm CE2} \\ B_{\rm CE2} \\ C_{\rm CE2} \end{pmatrix} \\
    &= \begin{pmatrix} -8v_0h^5 + \dfrac{10}{3}x_0h^6 - \dfrac{2}{7}v_0h^7 + \dfrac{1}{8}x_0h^8 \\ -6v_0h^4 + \dfrac{12}{5}x_0h^5 - \dfrac{1}{3}v_0h^6 + \dfrac{1}{7}x_0h^7 \\ -4v_0h^3 + \dfrac{3}{2}x_0h^4 - \dfrac{2}{5}v_0h^5 + \dfrac{1}{6}x_0h^6 \end{pmatrix} 
    \end{align}
    \vspace{-20pt}
       \begin{adjustwidth}{-\extralength}{0cm}
 \begin{align}
    \label{eq:CHO_CE2_ABC} &\left\{ \begin{aligned} A_{\rm CE2} &= \frac{33 \left[ \begin{aligned} &487710720v_0 - 228614400x_0h - 26127360v_0h^2 - 4596480x_0h^3 - 1209600v_0h^4 \\ &- 42336x_0h^5 - 18240v_0h^6 - 5040x_0h^7 - 1680v_0h^8 + 175x_0h^9 \end{aligned} \right]}{16 (120708403200 + 8622028800h^2 + 337478400h^4 + 14065920h^6 + 347760h^8 + 4536h^{10} + 245h^{12})} \\
    B_{\rm CE2} &= \frac{15 \left[ \begin{aligned} &670602240x_0 + 107775360x_0h^2 + 21288960v_0h^3 + 1542240x_0h^4 + 774144v_0h^5 \\ &+ 12024x_0h^6 + 16128v_0h^7 + 1638x_0h^8 + 896v_0h^9 - 105x_0h^{10} \end{aligned} \right]}{2 (120708403200 + 8622028800h^2 + 337478400h^4 + 14065920h^6 + 347760h^8 + 4536h^{10} + 245h^{12})} \\
    C_{\rm CE2} &= \frac{-3 \left[ \begin{aligned} &26824089600v_0 + 1916006400v_0h^2 + 199584000x_0h^3 + 154828800v_0h^4 - 5322240x_0h^5 \\ &+ 5210880v_0h^6 - 312480x_0h^7 + 104160v_0h^8 + 1120x_0h^9 + 4704v_0h^{10} - 735x_0h^{11} \end{aligned} \right]}{4 (120708403200 + 8622028800h^2 + 337478400h^4 + 14065920h^6 + 347760h^8 + 4536h^{10} + 245h^{12})} \end{aligned} \right..
\end{align}
\end{adjustwidth}

Since Equation~(\ref{eq:CHO_CE2_ABC}) is inconsistent with Equation~(\ref{eq:CHO_CE2_coef}), we have two options.

\paragraph{{Option 1: With EOM enforced at $t=h$ (``direct'' solution).}
}
First, we enforce $\ddot{x}(h) = -x_h$ by rewriting the cost function Equation~(\ref{eq:CHO_CE2_cost}) as
\vspace{-8pt}
\begin{adjustwidth}{-\extralength}{0cm}
\begin{align}
    \label{eq:CHO_CE2_costp} \epsilon_{\rm CE2}(x_h, v_h; h) &= \left[ \begin{aligned} &\frac{120}{7}(x_0^2-2x_0x_h+x_h^2)h^{-3} + \frac{120}{7}(v_0x_0-v_0x_h+v_hx_0-v_hx_h)h^{-2} \\
    &+ \frac{2}{35}(96v_0^2+108v_0v_h+96v_h^2-65x_0^2+130x_0x_h-65x_h^2)h^{-1} \\
    &- \frac{2}{35}(61v_0x_0-19v_0x_h+19v_hx_0-61v_hx_h) \\
    &+ \frac{1}{2310}(-1056v_0^2+132v_0v_h-1056v_h^2+1213x_0^2+346x_0x_h+1213x_h^2)h \\
    &+ \frac{1}{154}(31v_0x_0+13v_0x_h-13v_hx_0-31v_hx_h) h^2 \\
    &+ \frac{1}{27720}(416v_0^2-532v_0v_h+416v_h^2-369x_0^2-450x_0x_h-369x_h^2)h^3 \\
    &+ \frac{1}{27720}(-69v_0x_0-52v_0x_h+52v_hx_0+69v_hx_h)h^4 + \frac{1}{27720}(3x_0^2+5x_0x_h+3x_h^2)h^5 \end{aligned} \right];
\end{align}
\end{adjustwidth}
minimizing this, we obtain (``d'' in the subscript stands for direct)
\vspace{-6pt}
\begin{adjustwidth}{-\extralength}{0cm}
\begin{align}
    &\left\{ \begin{aligned} \frac{\partial\epsilon_{\rm CE2}}{\partial x_h} &= \left[ \begin{aligned} &- \frac{240}{7}(x_0-x_h)h^{-3} - \frac{120}{7}(v_0+v_h)h^{-2} + \frac{52}{7}(x_0-x_h)h^{-1} \\
    &+ \left(\frac{38v_0}{35}+\frac{122v_h}{35}\right) + \left(\frac{173x_0}{1155}+\frac{1213x_h}{1155}\right)h + \left(\frac{13v_0}{154}-\frac{31v_h}{154}\right)h^2 \\
    &- \left(\frac{5x_0}{308}+\frac{41x_h}{1540}\right)h^3 + \left(\frac{23v_h}{9240}-\frac{13v_0}{6930}\right)h^4 + \left(\frac{x_0}{5544}+\frac{x_h}{4620}\right)h^5 \end{aligned} \right] = 0 \\
    \frac{\partial\epsilon_{\rm CE2}}{\partial v_h} &= \left[ \begin{aligned} &\frac{120}{7}(x_0-x_h)h^{-2} + \left(\frac{216v_0}{35}+\frac{384v_h}{35}\right)h^{-1} + \left(\frac{122x_h}{35}-\frac{38x_0}{35}\right) + \frac{2}{35}(v_0-16v_h)h \\
    &- \left(\frac{13x_0}{154}+\frac{31x_h}{154}\right)h^2 + \left(\frac{104v_h}{3465}-\frac{19v_0}{990}\right)h^3 + \left(\frac{13x_0}{6930}+\frac{23x_h}{9240}\right)h^4 \end{aligned} \right] = 0 \end{aligned} \right. \\ \Rightarrow \quad
    \nonumber &\begin{pmatrix} \dfrac{240}{7} - \dfrac{52}{7}h^2 + \dfrac{1213}{1155}h^4 - \dfrac{41}{1540}h^6 + \dfrac{1}{4620}h^8 & -\dfrac{120}{7}h + \dfrac{122}{35}h^3 - \dfrac{31}{154}h^5 + \dfrac{23}{9240}h^7 \\\\[0.05ex]
    -\dfrac{120}{7}h + \dfrac{122}{35}h^3 - \dfrac{31}{154}h^5 + \dfrac{23}{9240}h^7 & \dfrac{384}{35}h^2 - \dfrac{32}{35}h^4 + \dfrac{104}{3465}h^6 \end{pmatrix} \begin{pmatrix} x_{h,{\rm d}} \\ v_{h,{\rm d}} \end{pmatrix} \\
    &= \begin{pmatrix} \dfrac{240x_0}{7} + \dfrac{120v_0}{7}h - \dfrac{52x_0}{7}h^2 - \dfrac{38v_0}{35}h^3 - \dfrac{173x_0}{1155}h^4 -\dfrac{13v_0}{154} h^5 + \dfrac{5x_0}{308}h^6 + \dfrac{13v_0}{6930}h^7 - \dfrac{x_0}{5544}h^8 \\  \\[0.05ex]
    - \dfrac{120x_0}{7}h - \dfrac{216v_0}{35}h^2 + \dfrac{38x_0}{35}h^3 - \dfrac{2v_0}{35}h^4 + \dfrac{13x_0}{154}h^5 + \dfrac{19v_0}{990}h^6 -\dfrac{13x_0}{6930}h^7 \end{pmatrix} 
    \end{align}
\end{adjustwidth}
    \begin{adjustwidth}{-\extralength}{0cm}
    \small
\begin{align}
    &\left\{ \begin{aligned} x_{h,{\rm d}} &= \frac{4 \left[ \begin{aligned} &1437004800x_0 + 1437004800v_0h - 602173440x_0h^2 - 123171840v_0h^3 + 6799680x_0h^4 - 2324160v_0h^5 \\
    &+ 469872x_0h^6 + 37296v_0h^7 - 8520x_0h^8 - 4248v_0h^9 + 1125x_0h^{10} + 149v_0h^{11} - 13x_0h^{12} \end{aligned} \right]}{3 (1916006400 + 155105280h^2 + 6785280h^4 + 312576h^6 + 8464h^8 + 120h^{10} + 7h^{12})} \\
    v_{h,{\rm d}} &= \frac{\left[ \begin{aligned} 1916006400v_0 - 1916006400x_0h - 802897920v_0h^2 + 164229120x_0h^3 + 9066240v_0h^4 + 2908800x_0h^5 \\
    + 626496v_0h^6 - 31776x_0h^7 - 11360v_0h^8 + 6680x_0h^9 + 1500v_0h^{10} - 198x_0h^{11} - 12v_0h^{12} + x_0h^{13} \end{aligned} \right]}{1916006400 + 155105280h^2 + 6785280h^4 + 312576h^6 + 8464h^8 + 120h^{10} + 7h^{12}} \end{aligned} \right.,
\end{align}
\end{adjustwidth}
or equivalently
\begin{align}
    \label{eq:CHO_CE2d_linmap} \begin{pmatrix} x_{h,{\rm d}} \\ v_{h,{\rm d}} \end{pmatrix} = \begin{pmatrix} G_{{\rm CE2d},00} & G_{{\rm CE2d},01} \\ G_{{\rm CE2d},10} & G_{{\rm CE2d},11} \end{pmatrix} \begin{pmatrix} x_0 \\ v_0 \end{pmatrix}
\end{align}
with
\vspace{-10pt}
\begin{adjustwidth}{-\extralength}{0cm}
\begin{align}
    \label{eq:CHO_CE2d_evolop} \left\{ \begin{aligned} G_{{\rm CE2d},00} &= \frac{4 (1437004800 - 602173440h^2 + 6799680h^4 + 469872h^6 - 8520h^8 + 1125h^{10} - 13h^{12})}{3 (1916006400 + 155105280h^2 + 6785280h^4 + 312576h^6 + 8464h^8 + 120h^{10} + 7h^{12})} \\
    G_{{\rm CE2d},01} &= \frac{4 (1437004800h - 123171840h^3 - 2324160h^5 + 37296h^7 - 4248h^9 + 149h^{11})}{3 (1916006400 + 155105280h^2 + 6785280h^4 + 312576h^6 + 8464h^8 + 120h^{10} + 7h^{12})} \\
    G_{{\rm CE2d},10} &= \frac{- 1916006400h + 164229120h^3 + 2908800h^5 - 31776h^7 + 6680h^9 - 198h^{11} + h^{13}}{1916006400 + 155105280h^2 + 6785280h^4 + 312576h^6 + 8464h^8 + 120h^{10} + 7h^{12}} \\
    G_{{\rm CE2d},11} &= \frac{1916006400 - 802897920h^2 + 9066240h^4 + 626496h^6 - 11360h^8 + 1500h^{10} - 12h^{12}}{1916006400 + 155105280h^2 + 6785280h^4 + 312576h^6 + 8464h^8 + 120h^{10} + 7h^{12}} \end{aligned} \right..
\end{align}
\end{adjustwidth}

The determinant of the time evolution operator $G_{\rm CE2d}$ is
\vspace{-12pt}
\begin{adjustwidth}{-\extralength}{0cm}
\small
\begin{align}
    \det \begin{pmatrix} G_{{\rm CE2d},00} & G_{{\rm CE2d},01} \\ G_{{\rm CE2d},10} & G_{{\rm CE2d},11} \end{pmatrix} = 1 - \frac{17h^{12}}{3 (1916006400 + 155105280h^2 + 6785280h^4 + 312576h^6 + 8464h^8 + 120h^{10} + 7h^{12})};
\end{align}
\end{adjustwidth}
to archive $1 - \det (G_{\rm CE2d}) \leq 2^{-53}$, one only needs $h \leq 0.2406$, $7.087$ times larger than what was required for first-order ContEvol. 

Expanding Equations~(\ref{eq:CHO_CE2d_linmap}) and (\ref{eq:CHO_CE2d_evolop}), second-order ContEvol with $\ddot{x}(h) = -x_h$ enforced yields
\vspace{-6pt}
\begin{adjustwidth}{-\extralength}{0cm}
\begin{align}
     \label{eq:CHO_CE2d_res} \left\{ \begin{aligned} x_{\rm CE2d}(h) &= \left[ \begin{aligned} &x_0 \left(1 - \frac{h^2}{2} + \frac{h^4}{24} - {\color{red} \frac{29}{28}} \cdot \frac{h^6}{720} + {\color{red} \frac{1019}{630}} \cdot \frac{h^8}{40320} + \mathcal{O}(h^{10})\right) \\ &+ v_0 \left(h - \frac{h^3}{6} + \frac{h^5}{120} - {\color{red} \frac{67}{60}} \cdot \frac{h^7}{5040} + {\color{red} \frac{11513}{3850}} \cdot \frac{h^9}{362880} + \mathcal{O}(h^{11}) \right) \end{aligned} \right] \\
    v_{\rm CE2d}(h) &= \left[ \begin{aligned} &-x_0 \left(h - \frac{h^3}{6} + {\color{red} \frac{85}{84}} \cdot  \frac{h^5}{120} - {\color{red} \frac{607}{504}} \cdot \frac{h^7}{5040} + {\color{red} \frac{1559}{490}} \cdot \frac{h^9}{362880} + \mathcal{O}(h^{11}) \right) \\ &+ v_0 \left(1 - \frac{h^2}{2} + \frac{h^4}{24} - {\color{red} \frac{29}{28}} \cdot \frac{h^6}{720} + {\color{red} \frac{1019}{630}} \cdot \frac{h^8}{40320} + \mathcal{O}(h^{10})\right) \end{aligned} \right] \end{aligned} \right..
\end{align}
\end{adjustwidth}
comparing to the exact solution Equation~(\ref{eq:CHO_exact}), we see that errors in $x_h$ and $v_h$ (highlighted in red) are still $\mathcal{O}(h^6)$ and $\mathcal{O}(h^5)$, respectively, same as first-order ContEvol Equation~(\ref{eq:CHO_CE1_res}); however, the coefficients are much closer to the exact values.

The minimized cost function Equation~(\ref{eq:CHO_CE2_costp}) is
\vspace{-8pt}
\begin{adjustwidth}{-\extralength}{0cm}
\begin{align}
    \epsilon_{{\rm CE2d},\min}(h) = \frac{h^9 \left[ \begin{aligned} &1425600x_0^2 + 1425600v_0x_0h + 5616(64v_0^2-55x_0^2)h^2 - 193104v_0x_0h^3 \\ &- 36(512v_0^2-541x_0^2)h^4 + 5220v_0x_0h^5 +(256v_0^2-243x_0^2)h^6 - 31v_0x_0h^7 + x_0^2h^8 \end{aligned} \right]}{7560 (1916006400 + 155105280h^2 + 6785280h^4 + 312576h^6 + 8464h^8 + 120h^{10} + 7h^{12})};
\end{align}
\end{adjustwidth}
when $h \leq 0.1014$ ($h \leq 0.1742$), $\epsilon_{{\rm CE2d},\min}(h) \leq 2^{-53}$ for $x_0 = 1$ and $v_0 = 0$ ($x_0 = 0$ and $v_0 = 1$). 

\paragraph{{Option 2: Without EOM enforced at $t=h$ (``indirect'' solution).}}
Second, we remove the $\ddot{x}(h) = -x_h$ constraint and simply adopt Equation~(\ref{eq:CHO_CE2_ABC}); ergo (``i'' in the subscript stands for indirect)
\vspace{-12pt}
\begin{adjustwidth}{-\extralength}{0cm}
\begin{align}
    \left\{ \begin{aligned} x_{h,{\rm i}} &\equiv x_{\rm CE2i}(h) = x_0 + v_0h - \frac{x_0}{2}h^2 + C_{\rm CE2}h^3 + B_{\rm CE2}h^4 + A_{\rm CE2}h^5 \\
    &= \frac{\left[ \begin{aligned} &1931334451200x_0 + 1931334451200v_0h - 827714764800x_0h^2 - 183936614400v_0h^3 \\ &+ 16895692800x_0h^4 - 1497968640v_0h^5 + 518987520x_0h^6 + 59581440v_0h^7 - 9711360x_0h^8 \\ &- 3985920v_0h^9 + 1086048x_0h^{10} + 156096v_0h^{11} - 15568x_0h^{12} - 448v_0h^{13} + 35x_0h^{14} \end{aligned} \right]}{16 (120708403200 + 8622028800h^2 + 337478400h^4 + 14065920h^6 + 347760h^8 + 4536h^{10} + 245h^{12})} \\
    v_{h,{\rm i}} &\equiv \dot{x}_{\rm CE2i}(h) = v_0 - x_0h + 3C_{\rm CE2}h^2 + 4B_{\rm CE2}h^3 + 5A_{\rm CE2}h^4 \\
    &= \frac{\left[ \begin{aligned} &1931334451200v_0 - 1931334451200x_0h - 827714764800v_0h^2 + 183936614400x_0h^3 \\ &+ 16895692800v_0h^4 + 1426118400x_0h^5 + 558904320v_0h^6 - 51598080x_0h^7 \\ &- 10022400v_0h^8 + 4471200x_0h^9 + 1054656v_0h^{10} - 158256x_0h^{11} - 12544v_0h^{12} \end{aligned} \right]}{16 (120708403200 + 8622028800h^2 + 337478400h^4 + 14065920h^6 + 347760h^8 + 4536h^{10} + 245h^{12})} \\
    a_{h,{\rm i}} &\equiv \ddot{x}_{\rm CE2i}(h) = - x_0 + 6C_{\rm CE2}h + 12B_{\rm CE2}h^2 + 20A_{\rm CE2}h^3 \\
    &= -\frac{\left[ \begin{aligned} &482833612800x_0 + 482833612800v_0h - 206928691200x_0h^2 - 45984153600v_0h^3 \\ &+ 3864672000x_0h^4 - 566092800v_0h^5 + 163676160x_0h^6 + 14688000v_0h^7 \\ &- 1576800x_0h^8 - 921600v_0h^9 + 280224x_0h^{10} + 39312v_0h^{11} - 3325h^{12} \end{aligned} \right]}{4 (120708403200 + 8622028800h^2 + 337478400h^4 + 14065920h^6 + 347760h^8 + 4536h^{10} + 245h^{12})} \end{aligned} \right.,
\end{align}
\end{adjustwidth}
or equivalently (neglecting the acceleration; note that this ``indirect'' strategy automatically avoids accumulation of additional error in $a_h$, as it ``resets'' $a_0$ to $-x_0$ at each step)\vspace{6pt}
\begin{align}
    \label{eq:CHO_CE2i_linmap} \begin{pmatrix} x_{h,{\rm i}} \\ v_{h,{\rm i}} \end{pmatrix} = \begin{pmatrix} G_{{\rm CE2i},00} & G_{{\rm CE2i},01} \\ G_{{\rm CE2i},10} & G_{{\rm CE2i},11} \end{pmatrix} \begin{pmatrix} x_0 \\ v_0 \end{pmatrix}
\end{align}
with
\vspace{-10pt}
\begin{adjustwidth}{-\extralength}{0cm}
\small
\begin{align}
    \label{eq:CHO_CE2i_evolop} \left\{ \begin{aligned} G_{{\rm CE2i},00} &= \frac{\left[ \begin{aligned} &1931334451200 - 827714764800h^2 + 16895692800h^4 \\ &+ 518987520h^6 - 9711360h^8 + 1086048h^{10} - 15568h^{12} + 35h^{14} \end{aligned} \right]}{16 (120708403200 + 8622028800h^2 + 337478400h^4 + 14065920h^6 + 347760h^8 + 4536h^{10} + 245h^{12})} \\
    G_{{\rm CE2i},01} &= \frac{1931334451200h - 183936614400h^3 - 1497968640h^5 + 59581440h^7 - 3985920h^9 + 156096 h^{11} - 448h^{13}}{16 (120708403200 + 8622028800h^2 + 337478400h^4 + 14065920h^6 + 347760h^8 + 4536h^{10} + 245h^{12})} \\
    G_{{\rm CE2i},10} &= \frac{- 1931334451200h + 183936614400h^3 + 1426118400h^5 - 51598080h^7 + 4471200h^9 - 158256h^{11}}{16 (120708403200 + 8622028800h^2 + 337478400h^4 + 14065920h^6 + 347760h^8 + 4536h^{10} + 245h^{12})} \\
    G_{{\rm CE2i},11} &= \frac{\left[ \begin{aligned} &1931334451200 - 827714764800h^2 + 16895692800h^4 \\ &+ 558904320h^6 - 10022400h^8 + 1054656h^{10} - 12544h^{12} \end{aligned} \right]}{16 (120708403200 + 8622028800h^2 + 337478400h^4 + 14065920h^6 + 347760h^8 + 4536h^{10} + 245h^{12})} \end{aligned} \right..
\end{align} 
\end{adjustwidth}

The determinant of the time evolution operator $G_{\rm CE2i}$ is
\begin{align}
    \det \begin{pmatrix} G_{{\rm CE2i},00} & G_{{\rm CE2i},01} \\ G_{{\rm CE2i},10} & G_{{\rm CE2i},11} \end{pmatrix} = 1 - \frac{h^6 (7983360 + 77760 h^2 - 288 h^4 + 816 h^6 - h^8)}{\left[ \begin{aligned} &4 (120708403200 + 8622028800h^2 + 337478400h^4 \\ &+ 14065920h^6 + 347760h^8 + 4536h^{10} + 245h^{12}) \end{aligned} \right]};
\end{align}
to archive $1 - \det (G_{\rm CE2i}) \leq 2^{-53}$, one needs $h \leq 0.01374$, $17.52$ times smaller than what was required when we enforce $\ddot{x}(h) = -x_h$. 

Expanding Equations~(\ref{eq:CHO_CE2i_linmap}) and (\ref{eq:CHO_CE2i_evolop}), second-order ContEvol without the $\ddot{x}(h) = -x_h$ constraint yields
\vspace{-10pt}
\begin{adjustwidth}{-\extralength}{0cm}
\begin{align}
      \label{eq:CHO_CE2i_res} \left\{ \begin{aligned} x_{\rm CE2i}(h) &= \left[ \begin{aligned} &x_0 \left(1 - \frac{h^2}{2} + \frac{h^4}{24} - {\color{red} \frac{115}{112}} \cdot \frac{h^6}{720} + {\color{red} \frac{121}{84}} \cdot \frac{h^8}{40320} + \mathcal{O}(h^{10})\right) \\ &+ v_0 \left(h - \frac{h^3}{6} + \frac{h^5}{120} - {\color{red} \frac{13}{12}} \cdot \frac{h^7}{5040} + {\color{red} \frac{365}{154}} \cdot \frac{h^9}{362880} + \mathcal{O}(h^{11}) \right) \end{aligned} \right] \\
    v_{\rm CE2i}(h) &= \left[ \begin{aligned} &-x_0 \left(h - \frac{h^3}{6} + {\color{red} \frac{225}{224}} \cdot  \frac{h^5}{120} - {\color{red} \frac{751}{672}} \cdot \frac{h^7}{5040} + {\color{red} \frac{125077}{51744}} \cdot \frac{h^9}{362880} + \mathcal{O}(h^{11}) \right) \\ &+ v_0 \left(1 - \frac{h^2}{2} + \frac{h^4}{24} - {\color{red} \frac{85}{84}} \cdot \frac{h^6}{720} + {\color{red} \frac{635}{462}} \cdot \frac{h^8}{40320} + \mathcal{O}(h^{10})\right) \end{aligned} \right] \end{aligned} \right..
\end{align}
\end{adjustwidth}
comparing to the exact solution Equation~(\ref{eq:CHO_exact}), we see that errors in $x_h$ and $v_h$ (highlighted in red) are once again $\mathcal{O}(h^6)$ and $\mathcal{O}(h^5)$, respectively.
Note that the ``indirect'' coefficients Equation~(\ref{eq:CHO_CE2i_res}) are slightly closer to the exact version than their ``direct'' counterparts Equation~(\ref{eq:CHO_CE2d_res}).

According to the optimal coefficients Equation~(\ref{eq:CHO_CE2_ABC}), the minimized cost function Equation~(\ref{eq:CHO_CE2_cost}) is
\vspace{-10pt}
\begin{adjustwidth}{-\extralength}{0cm}
\small
\begin{align}
    \epsilon_{{\rm CE2i},\min}(h) = \frac{h^9 \left[ \begin{aligned} &199584000x_0^2 + 191600640v_0x_0h + (46448640v_0^2-39916800x_0^2)h^2 - 24030720(v_0x_0)h^3 \\ &- (2211840v_0^2 - 2337120x_0^2)h^4 + 604800v_0x_0h^5 + (28672v_0^2-27216x_0^2)h^6 - 3360v_0x_0h^7 + 105x_0^2h^8 \end{aligned} \right]}{26880 (120708403200 + 8622028800h^2 + 337478400h^4 + 14065920h^6 + 347760h^8 + 4536h^{10} + 245h^{12})};
\end{align}
\end{adjustwidth}
when $h \leq 0.1068$ ($h \leq 0.1832$), $\epsilon_{{\rm CE2i},\min}(h) \leq 2^{-53}$ for $x_0 = 1$ and $v_0 = 0$ ($x_0 = 0$ and $v_0 = 1$). 

To summarize, the marginal benefit of raising ContEvol to second order is moderate: this reduces the minimized cost function from $\mathcal{O}(h^5)$ to $\mathcal{O}(h^9)$---leading to a better representation of the evolutionary track---but does not reduce the order of errors in $x_h$ or $v_h$.
One is advised to enforce equation of motion at $t=h$ if symplecticity is more important, but to remove this constraint if error control takes priority. In the rest of this work, we only consider first-order ContEvol methods, as for real-world problems, second derivatives might be unavailable or unaffordable.

\section{Celestial Mechanics: Two-Body and Three-Body Problems}
\label{sec:celestial}

In this section, we extend the ContEvol framework to time evolution of multiple real variables. As astrophysicists, we choose two simplest cases from celestial mechanics, two-body and three-body problems.

The equations of motion (EOMs) for a two-body problem are
\vspace{-6pt}
\begin{align}
    \left\{ \begin{aligned} m_0\ddot{\boldsymbol r}_0 &= -Gm_1m_0 \frac{{\boldsymbol r}_0 - {\boldsymbol r}_1}{\Vert{\boldsymbol r}_0 - {\boldsymbol r}_1\Vert^3} \\
    m_1\ddot{\boldsymbol r}_1 &= -Gm_0m_1 \frac{{\boldsymbol r}_1 - {\boldsymbol r}_0}{\Vert{\boldsymbol r}_1 - {\boldsymbol r}_0\Vert^3} \end{aligned} \right.,
\end{align}
where $G$ is the gravitational constant and $m_i$ denotes masses of the two objects; setting the constant $G(m_0+m_1)$ to $1$, these can be straightforwardly reduced to
\begin{align}
    \label{eq:2BP_EOM} \ddot{\boldsymbol r} = -\frac{\boldsymbol r}{r^3},
\end{align}
with ${\boldsymbol r} \equiv {\boldsymbol r}_1 - {\boldsymbol r}_0$ and $r \equiv \Vert{\boldsymbol r}\Vert$. {(In the rest of this section, we use regular symbols to denote magnitudes of vectors without further notice.)}
This problem only needs to be solved in two dimensions, as the particle never leaves the plane spanned by initial conditions---or the line, if the initial position and velocity are collinear, but it is trivial to apply full results to the one-dimensional case.
The general solution to the above EOM can be expressed in parametric forms, which we do not include here; exact solutions to specific problems (i.e., for specific initial values) will be presented when needed.

The (unrestricted) three-body problem is more complicated, with equations of motion
\vspace{-6pt}
\begin{align}
    \left\{ \begin{aligned} m_0\ddot{\boldsymbol r}_0' &= -Gm_1m_0 \frac{{\boldsymbol r}_0' - {\boldsymbol r}_1'}{\Vert{\boldsymbol r}_0' - {\boldsymbol r}_1'\Vert^3} - Gm_2m_0 \frac{{\boldsymbol r}_0' - {\boldsymbol r}_2'}{\Vert{\boldsymbol r}_0' - {\boldsymbol r}_2'\Vert^3}\\
    m_1\ddot{\boldsymbol r}_1' &= -Gm_0m_1 \frac{{\boldsymbol r}_1' - {\boldsymbol r}_0'}{\Vert{\boldsymbol r}_1' - {\boldsymbol r}_0'\Vert^3} - Gm_2m_1 \frac{{\boldsymbol r}_1' - {\boldsymbol r}_2'}{\Vert{\boldsymbol r}_1' - {\boldsymbol r}_2'\Vert^3} \\
    m_2\ddot{\boldsymbol r}_2' &= -Gm_0m_2 \frac{{\boldsymbol r}_2' - {\boldsymbol r}_0'}{\Vert{\boldsymbol r}_2' - {\boldsymbol r}_0'\Vert^3} - Gm_1m_2 \frac{{\boldsymbol r}_2' - {\boldsymbol r}_1'}{\Vert{\boldsymbol r}_2' - {\boldsymbol r}_1'\Vert^3} \end{aligned} \right.,
\end{align}
where the prime ``$'$'' denotes inertial coordinate system; writing ${\boldsymbol r}_i \equiv {\boldsymbol r}_i' - {\boldsymbol r}_0'$, $r_i \equiv \Vert{\boldsymbol r}_i\Vert$ and $G(m_0+m_1+m_2) = 1$, $\mu_i \equiv m_i/(m_0+m_1+m_2) < 1$ for $i = 1, 2$, these equations can be reduced to
\vspace{-6pt}
\begin{align}
    \label{eq:3BP_EOM} \left\{ \begin{aligned} \ddot{\boldsymbol r}_1 &= -(1-\mu_2) \frac{{\boldsymbol r}_1}{r_1^3} - \mu_2 \left( \frac{{\boldsymbol r}_2}{r_2^3} + \frac{{\boldsymbol r}_1 - {\boldsymbol r}_2}{\Vert{\boldsymbol r}_1 - {\boldsymbol r}_2\Vert^3} \right) \\
    \ddot{\boldsymbol r}_2 &= -(1-\mu_1) \frac{{\boldsymbol r}_2}{r_2^3} - \mu_1 \left( \frac{{\boldsymbol r}_1}{r_1^3} + \frac{{\boldsymbol r}_2 - {\boldsymbol r}_1}{\Vert{\boldsymbol r}_2 - {\boldsymbol r}_1\Vert^3} \right) \end{aligned} \right..
\end{align}
The above equations do not have a closed-form solution in general.

Although $r_{x(i)}$ and $r_{y(i)}$ will be written as polynomials, Taylor expansion of \linebreak  $r_{(i)} = \sqrt{r_{x(i)}^2 + r_{y(i)}^2}$ has infinitely many terms, hence some truncation is necessary.
In Section~\ref{ss:2BP_CE1A}, we apply first-order ContEvol method to the two-body problem, keeping ``adequately'' many terms. We show that this is equivalent to linearization and Taylor expansion in Section~\ref{ss:2BP_CE1L}.
In Section~\ref{ss:2BP_CE1C}, we investigate conservation of mechanic energy and angular momentum, before moving on to numerical tests with an eccentric elliptical orbit in \mbox{Section~\ref{ss:2BP_CE1N}}.
Finally in Section~\ref{ss:3BP_CE1}, we describe how ContEvol is supposed to be applied to the three-body problem.

\subsection{Two-Body, First-Order ContEvol with ``Adequate'' Expansion}
\label{ss:2BP_CE1A}

Without loss of generality, we are given ${\boldsymbol r}(0) = {\boldsymbol r}_0$, $\dot{\boldsymbol r}(0) = {\boldsymbol v}_0$ and try to solve for ${\boldsymbol r}(h) = {\boldsymbol r}_h$, $\dot{\boldsymbol r}(h) = {\boldsymbol v}_h$, where $h$ is the time step.
Like in Section~\ref{ss:CHO_CE1}, we approximate the solution in a parametric form (subscript ``CE2'' now stands for ContEvol and two-body problem; note that we are recycling the subscripts)
\vspace{-3pt}
\begin{align}
    \label{eq:2BP_CE1_rvh} {\boldsymbol r}_{\rm CE2}(t) = {\boldsymbol r}_0 + {\boldsymbol v}_0t + {\boldsymbol B}t^2 + {\boldsymbol A}t^3, \quad t \in [0, h],
\end{align}
with coefficients ${\boldsymbol A}$ and ${\boldsymbol B}$ yielded by ``terminal'' conditions at $t=h$
\vspace{-3pt}
\begin{align}
    \label{eq:2BP_CE1_AB} \left\{ \begin{aligned} {\boldsymbol A} &= 2({\boldsymbol r}_0 - {\boldsymbol r}_h) h^{-3} + ({\boldsymbol v}_0 + {\boldsymbol v}_h) h^{-2} \\
    {\boldsymbol B} &= 3({\boldsymbol r}_h - {\boldsymbol r}_0) h^{-2} - (2{\boldsymbol v}_0 + {\boldsymbol v}_h) h^{-1} \end{aligned} \right..
\end{align}

To define the cost function $\epsilon$ as a finite polynomial of $h$, we have to truncate the Taylor expansion on the right-hand side of the EOM Equation~(\ref{eq:2BP_EOM}).
Since ${\boldsymbol r}_{\rm CE2}(t)$ traces up to the third order, we do not expect any benefit from going beyond the third order; justifying this statement is left for future work---note that non-linear coefficients in ${\boldsymbol A}$ and ${\boldsymbol B}$ start to occur at the fourth order, so one would have to solve non-linear equations to minimize the cost function.
Thus we have
\vspace{-3pt}
\begin{align}
    r_{\rm CE2}^2(t) \approx {\boldsymbol r}_0^2 + 2 {\boldsymbol r}_0\cdot{\boldsymbol v}_0 t + (2 {\boldsymbol B}\cdot{\boldsymbol r}_0 + v_0^2) t^2 + 2 ({\boldsymbol A}\cdot{\boldsymbol r}_0 + {\boldsymbol B}\cdot{\boldsymbol v}_0) t^3, \quad t \in [0, h],
\end{align}
and the cost function is defined as
\vspace{-12pt}
\begin{adjustwidth}{-\extralength}{0cm}
\begin{align}
    \label{eq:2BP_CE1_cost} \nonumber \epsilon_{\rm CE2}({\boldsymbol A}, {\boldsymbol B}; h) &= \int_0^h \left(\ddot{\boldsymbol r} + \frac{\boldsymbol r}{r^3}\right)^2 \,{\rm d}t
    = \int_0^h \left[(2{\boldsymbol B} + 6{\boldsymbol A}t) + \frac{{\boldsymbol r}_0 + {\boldsymbol v}_0t + {\boldsymbol B}t^2 + {\boldsymbol A}t^3}{\Vert{\boldsymbol r}_0 + {\boldsymbol v}_0t + {\boldsymbol B}t^2 + {\boldsymbol A}t^3\Vert^3} \right]^2 \,{\rm d}t \\
    \nonumber &\approx \int_0^h \left[ {\boldsymbol C}_0 + {\boldsymbol C}_1 t + {\boldsymbol C}_2 t^2 + {\boldsymbol C}_3 t^3 \right]^2 \,{\rm d}t \\
    \nonumber &= \int_0^h \left[ \begin{aligned} &C_0^2 + 2{\boldsymbol C}_0\cdot{\boldsymbol C}_1 t + (C_1^2 + 2{\boldsymbol C}_0\cdot{\boldsymbol C}_2) t^2 + 2({\boldsymbol C}_0\cdot{\boldsymbol C}_3 + {\boldsymbol C}_1\cdot{\boldsymbol C}_2) t^3 \\
    &+ (C_2^2 + 2{\boldsymbol C}_1\cdot{\boldsymbol C}_3) t^4 + 2{\boldsymbol C}_2\cdot{\boldsymbol C}_3 t^5 + C_3^2 t^6\end{aligned} \right] \,{\rm d}t \\
    &= \left[ \begin{aligned} &C_0^2 h + {\boldsymbol C}_0\cdot{\boldsymbol C}_1 h^2 + \frac{1}{3}(C_1^2 + 2{\boldsymbol C}_0\cdot{\boldsymbol C}_2) h^3 + \frac{1}{2}({\boldsymbol C}_0\cdot{\boldsymbol C}_3 + {\boldsymbol C}_1\cdot{\boldsymbol C}_2) h^4 \\
    &+ \frac{1}{5}(C_2^2 + 2{\boldsymbol C}_1\cdot{\boldsymbol C}_3) h^5 + \frac{1}{3}{\boldsymbol C}_2\cdot{\boldsymbol C}_3 h^6 + \frac{1}{7}C_3^2 h^7 \end{aligned} \right]
\end{align}
\end{adjustwidth}
with
\begin{align}
    \label{eq:2BP_CE1_C0123} \left\{ \begin{aligned} {\boldsymbol C}_0 &= 2{\boldsymbol B}+\frac{{\boldsymbol r}_0}{r_0^3} \\
    {\boldsymbol C}_1 &= 6{\boldsymbol A} + \frac{{\boldsymbol v}_0}{r_0^3} - \frac{3 {\boldsymbol r}_0\cdot{\boldsymbol v}_0}{r_0^5}{\boldsymbol r}_0 \\
    {\boldsymbol C}_2 &= \frac{\boldsymbol B}{r_0^3} - \frac{3{\boldsymbol r}_0\cdot{\boldsymbol v}_0}{r_0^5} {\boldsymbol v}_0 - \frac{3}{2} \left( \frac{2{\boldsymbol B}\cdot{\boldsymbol r}_0 + v_0^2}{r_0^5} - \frac{5 ({\boldsymbol r}_0\cdot{\boldsymbol v}_0)^2}{r_0^7} \right) {\boldsymbol r}_0 \\
    {\boldsymbol C}_3 &= \left[ \begin{aligned} &\frac{\boldsymbol A}{r_0^3} - \frac{3 {\boldsymbol r}_0\cdot{\boldsymbol v}_0}{r_0^5} \boldsymbol B - \frac{3}{2} \left( \frac{2{\boldsymbol B}\cdot{\boldsymbol r}_0 + v_0^2}{r_0^5} - \frac{5 ({\boldsymbol r}_0\cdot{\boldsymbol v}_0)^2}{r_0^7} \right) {\boldsymbol v}_0 \\
    &- \left( \frac{3 ({\boldsymbol A}\cdot{\boldsymbol r}_0 + {\boldsymbol B}\cdot{\boldsymbol v}_0)}{r_0^5} - \frac{15 (2 {\boldsymbol B}\cdot{\boldsymbol r}_0 + v_0^2) ({\boldsymbol r}_0\cdot{\boldsymbol v}_0)}{2 r_0^7} + \frac{35 ({\boldsymbol r}_0\cdot{\boldsymbol v}_0)^3}{2 r_0^9} \right) {\boldsymbol r}_0 \end{aligned} \right] \end{aligned} \right.;
\end{align}
because of the ${\boldsymbol B}\cdot{\boldsymbol r}_0$, ${\boldsymbol A}\cdot{\boldsymbol r}_0$, and ${\boldsymbol B}\cdot{\boldsymbol v}_0$ terms, the two components are coupled with each~other.

Minimizing this, we obtain
\begin{align}
    &\left\{ \begin{aligned} \frac{\partial\epsilon_{\rm CE2}}{\partial A_x} &= 12\left(2B_x + \frac{r_{0x}}{r_0^3}\right)h^2 + 4\left(6A_x + \frac{v_{0x}}{r_0^3} - \frac{3 {\boldsymbol r}_0\cdot{\boldsymbol v}_0}{r_0^5}r_{0x}\right)h^3 + \cdots = 0 \\
    \frac{\partial\epsilon_{\rm CE2}}{\partial B_x} &= 4\left(2B_x+\frac{r_{0x}}{r_0^3}\right)h + 2\left(6A_x + \frac{v_{0x}}{r_0^3} - \frac{3 {\boldsymbol r}_0\cdot{\boldsymbol v}_0}{r_0^5}r_{0x}\right)h^2 + \cdots = 0 \end{aligned} \right.,
\end{align}
where we have omitted some high-order terms (``$\cdots$''; up to $\mathcal{O}(h^7)$) for simplicity, and equations $\partial\epsilon_{\rm CE2}/\partial A_y = 0$ and $\partial\epsilon_{\rm CE2}/\partial B_y = 0$ as they can be easily obtained via swapping subscripts $x$ and $y$; because of the coupling mentioned above, there are cross terms in high-order coefficients.

Put in matrix form, the system of equations is
\begin{align}
    \label{eq:2BP_CE1_linsys} \begin{pmatrix} M_{11} & M_{12} & M_{13} & M_{14} \\ M_{21} & M_{22} & M_{23} & M_{24} \\ M_{31} & M_{32} & M_{33} & M_{34} \\ M_{41} & M_{42} & M_{43} & M_{44} \end{pmatrix} \begin{pmatrix} A_{x,{\rm CE2}} \\ A_{y,{\rm CE2}} \\ B_{x,{\rm CE2}} \\ B_{y,{\rm CE2}} \end{pmatrix} = \begin{pmatrix} b_1 \\ b_2 \\ b_3 \\ b_4 \end{pmatrix}
\end{align}
with
\begin{align}
    \left\{ \begin{aligned} &M_{11} = 24h^3 - \frac{24(2 r_{0x}^2-r_{0y}^2)}{5r_0^5}h^5 + \cdots \\
    &M_{22} = 24h^3 + \frac{24(r_{0x}^2-2r_{0y}^2)}{5r_0^5}h^5 + \cdots
    &&M_{12} = M_{21} = - \frac{72r_{0x}r_{0y}}{5r_0^5}h^5 + \cdots \\
    &M_{13} = M_{31} = 12h^2 - \frac{4(2r_{0x}^2-r_{0y}^2)}{r_0^5}h^4 + \cdots 
    &&M_{14} = M_{41} = - \frac{12r_{0x}r_{0y}}{r_0^5} h^4 + \cdots \\
    &M_{24} = M_{42} = 12h^2 + \frac{4(r_{0x}^2-2r_{0y}^2)}{r_0^5}h^4 + \cdots
    &&M_{23} = M_{32} = - \frac{12r_{0x}r_{0y}}{r_0^5}h^4 + \cdots \\
    &M_{33} = 8h - \frac{8(2 r_{0x}^2-r_{0y}^2)}{3r_0^5}h^3 + \cdots \\
    &M_{44} = 8h + \frac{8(r_{0x}^2-2r_{0y}^2)}{3r_0^5}h^3 + \cdots
    &&M_{34} = M_{43} = - \frac{8r_{0x}r_{0y}}{r_0^5}h^3 + \cdots \end{aligned} \right.
\end{align}
and
\begin{align}
    \left\{ \begin{aligned} b_1 &= - \frac{6r_{0x}}{r_0^3}h^2 - 4\left( \frac{v_{0x}}{r_0^3} - \frac{3 {\boldsymbol r}_0\cdot{\boldsymbol v}_0}{r_0^5}r_{0x} \right)h^3 + \cdots \\
    b_2 &= - \frac{6r_{0y}}{r_0^3}h^2 - 4\left( \frac{v_{0y}}{r_0^3} - \frac{3 {\boldsymbol r}_0\cdot{\boldsymbol v}_0}{r_0^5}r_{0y} \right)h^3 + \cdots \\
    b_3 &= - \frac{4r_{0x}}{r_0^3}h - 2\left( \frac{v_{0x}}{r_0^3} - \frac{3 {\boldsymbol r}_0\cdot{\boldsymbol v}_0}{r_0^5}r_{0x} \right)h^2 + \cdots \\
    b_4 &= - \frac{4r_{0y}}{r_0^3}h - 2\left( \frac{v_{0y}}{r_0^3} - \frac{3 {\boldsymbol r}_0\cdot{\boldsymbol v}_0}{r_0^5}r_{0y} \right)h^2 + \cdots \end{aligned} \right.;
\end{align}
the solution {(To prevent Wolfram Mathematica from taking forever, one is advised to keep only up to} $\mathcal{O}(h^7)$ {(or another desired order) terms} {\it {at each step}}. {This advice also applies to computation of determinant of the Jacobian matrix in this case.)} is 
\vspace{-6pt}
\begin{adjustwidth}{-\extralength}{0cm}
\begin{align}
    \label{eq:2BP_CE1_ABopt} &\left\{ \begin{aligned} A_{x,{\rm CE2}} &= \left[ \begin{aligned} &\frac{(2r_{0x}^2-r_{0y}^2)v_{0x} + 3r_{0x}r_{0y}v_{0y}}{6r_0^5} \\
    &- \frac{3r_{0x}^3 (2v_{0x}^2-v_{0y}^2) + r_{0x} [2r_0 + 3r_{0y}^2(4v_{0y}^2-3v_{0x}^2)] + 6(4r_{0x}^2-r_{0y}^2) r_{0y} v_{0x} v_{0y}}{12r_0^7}h + \cdots \end{aligned} \right] \\
    B_{x,{\rm CE2}} &= - \frac{r_{0x}}{2r_0^3} + \frac{3r_{0x}^3 (2v_{0x}^2-v_{0y}^2) + r_{0x} [2r_0 + 3r_{0y}^2(4v_{0y}^2-3v_{0x}^2)] + 6(4r_{0x}^2-r_{0y}^2) r_{0y} v_{0x} v_{0y}}{24r_0^7}h^2 + \cdots \end{aligned} \right.,
\end{align}
\end{adjustwidth}
where again we have omitted some high-order terms (up to $\mathcal{O}(h^7)$) and expressions for $y$~components.

Plugging Equation~(\ref{eq:2BP_CE1_ABopt}) back into Equation~(\ref{eq:2BP_CE1_rvh}), our solution at $t = h$ is
\vspace{-12pt}
\begin{adjustwidth}{-\extralength}{0cm}
\begin{align}
    \label{eq:2BP_CE1_rvh_res} \left\{ \begin{aligned} r_{hx} &= r_{0x} + v_{0x}h - \frac{r_{0x}}{2r_0^3}h^2 - \left( \frac{v_{0x}}{6r_0^3} - \frac{{\boldsymbol r}_0\cdot{\boldsymbol v}_0}{2r_0^5}r_{0x} \right)h^3 + \cdots \\
    v_{hx} &= \left[ \begin{aligned} &v_{0x} - \frac{r_{0x}}{r_0^3}h + \left( \frac{v_{0x}}{2r_0^3} - \frac{3 {\boldsymbol r}_0\cdot{\boldsymbol v}_0}{2r_0^5}r_{0x} \right)h^2 \\
    &- \frac{3r_{0x}^3 (2v_{0x}^2-v_{0y}^2) + r_{0x} [2r_0 + 3r_{0y}^2(4v_{0y}^2-3v_{0x}^2)] + 6(4r_{0x}^2-r_{0y}^2) r_{0y} v_{0x} v_{0y}}{6r_0^7}h^3 + \cdots\end{aligned} \right] \end{aligned} \right.;
\end{align}
\end{adjustwidth}
thus the Jacobian matrix is
\begin{align}
    J = \begin{pmatrix} \partial r_{hx}/\partial r_{0x} & \partial r_{hx}/\partial r_{0y} & \partial r_{hx}/\partial v_{0x} & \partial r_{hx}/\partial v_{0y} \\
    \partial r_{hy}/\partial r_{0x} & \partial r_{hy}/\partial r_{0y} & \partial r_{hy}/\partial v_{0x} & \partial r_{hy}/\partial v_{0y} \\
    \partial v_{hx}/\partial r_{0x} & \partial v_{hx}/\partial r_{0y} & \partial v_{hx}/\partial v_{0x} & \partial v_{hx}/\partial v_{0y} \\
    \partial v_{hy}/\partial r_{0x} & \partial v_{hy}/\partial r_{0y} & \partial v_{hy}/\partial v_{0x} & \partial v_{hy}/\partial v_{0y} \end{pmatrix}
    \equiv \begin{pmatrix} J_{11} & J_{12} & J_{13} & J_{14} \\ J_{21} & J_{22} & J_{23} & J_{24} \\ J_{31} & J_{32} & J_{33} & J_{34} \\ J_{41} & J_{42} & J_{43} & J_{44} \end{pmatrix}
\end{align}
with
\begin{align}
    \label{eq:2BP_CE1_Jac_rr} &\left\{ \begin{aligned} &J_{11} = 1 + \frac{2 r_{0x}^2 - r_{0y}^2}{2r_0^5}h^2 + \frac{-2r_{0x}^3v_{0x} + 3r_{0x}r_{0y}^2v_{0x} - 4r_{0x}^2r_{0y}v_{0y} + r_{0y}^3v_{0y}}{2r_0^7}h^3 + \cdots \\
    &J_{22} = 1 + \frac{-r_{0x}^2 + 2r_{0y}^2}{2r_0^5}h^2 + \frac{r_{0x}^3v_{0x} - 4r_{0x}r_{0y}^2v_{0x} + 3r_{0x}^2r_{0y}v_{0y} - 2r_{0y}^3v_{0y}}{2r_0^7}h^3 + \cdots \\
    &J_{12} = \frac{3r_{0x}r_{0y}}{2r_0^5}h^2 + \frac{-4r_{0x}^2r_{0y}v_{0x} + r_{0y}^3v_{0x} + r_{0x}^3v_{0y} - 4r_{0x}r_{0y}^2v_{0y}}{2r_0^7}h^3 + \cdots \\
    &J_{21} = \frac{3r_{0x}r_{0y}}{2r_0^5}h^2 + \frac{-4r_{0x}^2r_{0y}v_{0x} + r_{0y}^3v_{0x} + r_{0x}^3v_{0y} - 4r_{0x}r_{0y}^2v_{0y}}{2r_0^7}h^3 + \cdots \end{aligned} \right.
    \end{align}
     \vspace{-9pt}
    \begin{align}
    \label{eq:2BP_CE1_Jac_rv} &\left\{ \begin{aligned} &J_{13} = h + \frac{2 r_{0x}^2 - r_{0y}^2}{6r_0^5}h^3 + \frac{-2r_{0x}^3v_{0x} + 3r_{0x}r_{0y}^2v_{0x} - 4r_{0x}^2r_{0y}v_{0y} + r_{0y}^3v_{0y}}{4r_0^7}h^4 + \cdots \\
    &J_{24} = h + \frac{-r_{0x}^2 + 2r_{0y}^2}{6r_0^5}h^3 + \frac{r_{0x}^3v_{0x} - 4r_{0x}r_{0y}^2v_{0x} + 3r_{0x}^2r_{0y}v_{0y} - 2r_{0y}^3v_{0y}}{4r_0^7}h^4 + \cdots \\
    &J_{31} = \frac{2r_{0x}^2 - r_{0y}^2}{r_0^5}h - \frac{3 (2r_{0x}^3v_{0x} - 3r_{0x}r_{0y}^2v_{0x} + 4r_{0x}^2r_{0y}v_{0y} - r_{0y}^3v_{0y})}{2r_0^7}h^2 + \cdots \\
    &J_{42} = \frac{-r_{0x}^2 + 2r_{0y}^2}{r_0^5}h + \frac{3 (r_{0x}^3v_{0x} - 4r_{0x}r_{0y}^2v_{0x} + 3r_{0x}^2r_{0y}v_{0y} - 2r_{0y}^3v_{0y})}{2r_0^7}h^2 + \cdots \end{aligned} \right. 
     \end{align}

       \begin{align}
    \label{eq:2BP_CE1_Jac_vr} &\left\{ \begin{aligned} &J_{14} = \frac{r_{0x}r_{0y}}{2r_0^5}h^3 + \frac{-4r_{0x}^2r_{0y}v_{0x} + r_{0y}^3v_{0x} + r_{0x}^3v_{0y} - 4r_{0x}r_{0y}^2v_{0y}}{4r_0^7}h^4 + \cdots \\
    &J_{23} = \frac{r_{0x}r_{0y}}{2r_0^5}h^3 + \frac{-4r_{0x}^2r_{0y}v_{0x} + r_{0y}^3v_{0x} + r_{0x}^3v_{0y} - 4r_{0x}r_{0y}^2v_{0y}}{4r_0^7}h^4 + \cdots \\
    &J_{41} = \frac{3r_{0x}r_{0y}}{r_0^5}h + \frac{3 (-4r_{0x}^2r_{0y}v_{0x} + r_{0y}^3v_{0x} + r_{0x}^3v_{0y} - 4r_{0x}r_{0y}^2v_{0y})}{2r_0^7}h^2 + \cdots \\
    &J_{32} = \frac{3r_{0x}r_{0y}}{r_0^5}h + \frac{3 (-4r_{0x}^2r_{0y}v_{0x} + r_{0y}^3v_{0x} + r_{0x}^3v_{0y} - 4r_{0x}r_{0y}^2v_{0y})}{2r_0^7}h^2 + \cdots \end{aligned} \right.  \\
    \label{eq:2BP_CE1_Jac_vv} &\left\{ \begin{aligned} &J_{33} = 1 + \frac{2r_{0x}^2 - r_{0y}^2}{2r_0^5}h^2 + \frac{-2r_{0x}^3v_{0x} + 3r_{0x}r_{0y}^2v_{0x} - 4r_{0x}^2r_{0y}v_{0y} + r_{0y}^3v_{0y}}{r_0^7}h^3 + \cdots \\
    &J_{44} = 1 + \frac{-r_{0x}^2 + 2 r_{0y}^2}{2r_0^5}h^2 + \frac{r_{0x}^3 v_{0x} - 4 r_{0x} r_{0y}^2 v_{0x} + 3 r_{0x}^2 r_{0y} v_{0y} - 2 r_{0y}^3 v_{0y}}{r_0^7}h^3 + \cdots \\
    &J_{34} = \frac{3r_{0x}r_{0y}}{2r_0^5}h^2 + \frac{-4r_{0x}^2r_{0y}v_{0x} + r_{0y}^3v_{0x} + r_{0x}^3v_{0y} - 4r_{0x}r_{0y}^2v_{0y}}{r_0^7}h^3 + \cdots \\
    &J_{43} = \frac{3r_{0x}r_{0y}}{2r_0^5}h^2 + \frac{-4r_{0x}^2r_{0y}v_{0x} + r_{0y}^3v_{0x} + r_{0x}^3v_{0y} - 4r_{0x}r_{0y}^2v_{0y}}{r_0^7}h^3 + \cdots \end{aligned} \right.
\end{align}
Note that ``symmetries'' in the Jacobian are broken at high orders. Its determinant is
\vspace{-8pt}
\begin{adjustwidth}{-\extralength}{0cm}
\begin{align}
    \label{eq:2BP_CE1_Jac_det} \det (J) = 1 + \frac{{\boldsymbol r}_0\cdot{\boldsymbol v}_0 [119r_0 + 30r_{0x}^2(4v_{0x}^2-3v_{0y}^2) + 420r_{0x}r_{0y}v_{0x}v_{0y} - 30r_{0y}^2(3v_{0x}^2-4v_{0y}^2)]}{60r_0^9} h^5 + \cdots,
\end{align}
\end{adjustwidth}
i.e., the non-symplecticity is at the $\mathcal{O}(h^5)$ level, three orders larger than applying first-order ContEvol to classic harmonic oscillator (see Section~\ref{ss:CHO_CE1}, specifically Equation~(\ref{eq:CHO_CE1_detG})).

According to Equation~(\ref{eq:2BP_CE1_ABopt}), the minimized cost function Equation~(\ref{eq:2BP_CE1_cost}) is
\vspace{-8pt}
\begin{adjustwidth}{-\extralength}{0cm}
\begin{align}
    \epsilon_{{\rm CE2},\min}(h) = \left[ \begin{aligned} &\frac{1}{180r_0^{10}} + \frac{6r_{0x}r_{0y}v_{0x}v_{0y} + (2r_{0x}^2-r_{0y}^2)v_{0x}^2 - (r_{0x}^2-2 r_{0y}^2)v_{0y}^2}{60 r_0^{11}} \\
    &+\frac{\left\{ \begin{aligned} &(4r_{0x}^4+r_{0y}^4)v_{0x}^4 + 4(4r_{0x}^3r_{0y}-r_{0x}r_{0y}^3)v_{0x}^3v_{0y} + 30r_{0x}^2r_{0y}^2v_{0x}^2v_{0y}^2 \\ &- 4(r_{0x}^3r_{0y}-4r_{0x}r_{0y}^3)v_{0x}v_{0y}^3 + (r_{0x}^4+4r_{0y}^4)v_{0y}^4 \end{aligned} \right\}}{80 r_0^{12}} \end{aligned} \right] h^5 + \cdots;
\end{align}
\end{adjustwidth}
the order in $h$ is same as in the prototype case Equation~(\ref{eq:CHO_CE1_mineps}); however, when $r_0$ is small, i.e., when $r_0 \lesssim \sqrt{h}$, the above expression can still be large.

\paragraph{{Test case 1: Uniform circular motion.}}
Consider the initial conditions ${\boldsymbol r}_0 = (1, 0)^{\rm T}$ and ${\boldsymbol v}_0 = (0, 1)^{\rm T}$. The particle will perform a uniform circular motion along the unit circle.

The exact solution is (subscript ``UCM'' stands for uniform circular motion)\vspace{12pt}
\begin{align}
    \left\{ \begin{aligned} {\boldsymbol r}_{\rm UCM}(t) &= \begin{pmatrix} \cos t \\ \sin t \end{pmatrix} = \begin{pmatrix} 1 - \dfrac{t^2}{2} + \dfrac{t^4}{24} - \dfrac{t^6}{720} + \mathcal{O}(t^8) \\ t - \dfrac{t^3}{6} + \dfrac{t^5}{120} - \dfrac{t^7}{5040} + \mathcal{O}(t^9) \end{pmatrix} \\
    {\boldsymbol v}_{\rm UCM}(t) &= \begin{pmatrix} -\sin t \\ \cos t \end{pmatrix} = \begin{pmatrix} -\left[ t - \dfrac{t^3}{6} + \dfrac{t^5}{120} - \dfrac{t^7}{5040} + \mathcal{O}(t^9) \right] \\ 1 - \dfrac{t^2}{2} + \dfrac{t^4}{24} - \dfrac{t^6}{720} + \mathcal{O}(t^8) \end{pmatrix} \end{aligned} \right.,
\end{align}
while first-order ContEvol with ``adequate'' expansion {yields}
\begin{align}
   \label{eq:2BP_CE1_UCM} \left\{ \begin{aligned} {\boldsymbol r}_h &= \begin{pmatrix} 1 - \dfrac{h^2}{2} + \dfrac{h^4}{24} - {\color{red} 0} \cdot \dfrac{h^6}{720} + \mathcal{O}(h^8) \\ h - \dfrac{h^3}{6} + \dfrac{h^5}{120} - {\color{red} \dfrac{303}{5}} \cdot \dfrac{h^7}{5040} + \mathcal{O}(h^9) \end{pmatrix} \\
    {\boldsymbol v}_h &= \begin{pmatrix} -\left[ h - \dfrac{h^3}{6} + {\color{red} \dfrac{4}{3}} \cdot \dfrac{h^5}{120} - {\color{red} \dfrac{1486}{15}} \cdot \dfrac{h^7}{5040} + \mathcal{O}(h^9) \right] \\ 1 - \dfrac{h^2}{2} + \dfrac{h^4}{24} - {\color{red} 27} \cdot \dfrac{h^6}{720} + \mathcal{O}(h^8) \end{pmatrix} \end{aligned} \right.,
\end{align}
i.e., like in Section~\ref{ss:CHO_CE1}, errors in ${\boldsymbol r}_h$ and ${\boldsymbol v}_h$ (highlighted in red) are $\mathcal{O}(h^6)$ and $\mathcal{O}(h^5)$, respectively.

\paragraph{{Test case 2: Parabolic motion.}}
Consider the initial conditions ${\boldsymbol r}_0 = (2, 0)^{\rm T}$ and ${\boldsymbol v}_0 = (-1/\sqrt{2}, 1/\sqrt{2})^{\rm T}$. The particle will move along the parabola $r_y = 1 - r_x^2/4$.

According to conservation of angular momentum and mechanic energy (see Section~\ref{ss:2BP_CE1C} for further treatment), the exact solution is (subscript ``PBM'' stands for parabolic motion)
\vspace{-10pt}
\begin{adjustwidth}{-\extralength}{0cm}
\begin{align}
    \left\{ \begin{aligned} r_{{\rm PBM},x}(t) &= \frac{2 \cdot 2^{2/3}}{\sqrt[3]{\sqrt{80 - 48\sqrt{2}t + 18t^2} + 3\sqrt{2}t - 8}} - \sqrt[3]{2}\sqrt[3]{\sqrt{80 - 48\sqrt{2}t + 18t^2} + 3\sqrt{2}t - 8} \\
    &= 2 - \frac{t}{\sqrt{2}} - \frac{t^2}{8} - \frac{t^3}{24\sqrt{2}} - \frac{5t^4}{768} - \frac{t^5}{768\sqrt{2}} + \frac{7t^6}{36864} + \frac{13t^7}{36864\sqrt{2}} + \mathcal{O}(t^8) \\
    r_{{\rm PBM},y}(t) &= 1 - \frac{r_{{\rm PBM},x}^2(t)}{4} = \frac{t}{\sqrt{2}} - \frac{t^3}{48\sqrt{2}} - \frac{t^4}{128} - \frac{7t^5}{1536\sqrt{2}} - \frac{7t^6}{6144} - \frac{35t^7}{73728\sqrt{2}} + \mathcal{O}(t^8) \end{aligned} \right.,
\end{align}
\end{adjustwidth}
where $t$ is within the radius of convergence for the expansion, and
\vspace{-10pt}
\begin{adjustwidth}{-\extralength}{0cm}
\small
\begin{align}
    \left\{ \begin{aligned} v_{{\rm PBM},x}(t) &= - \frac{\sqrt{2/r_{\rm PBM}(t)}}{\sqrt{1+[-r_{{\rm PBM},x}(t)/2]^2}} = -\frac{1}{\sqrt{2}} - \frac{t}{4} - \frac{t^2}{8\sqrt{2}} - \frac{5t^3}{192} - \frac{5t^4}{768\sqrt{2}} + \frac{7t^5}{6144} + \frac{91t^6}{36864\sqrt{2}} + \frac{341t^7}{294912} + \mathcal{O}(t^8) \\
    v_{{\rm PBM},y}(t) &= [-r_{{\rm PBM},x}(t)/2] \cdot v_{{\rm PBM},x}(t) = \frac{1}{\sqrt{2}} - \frac{t^2}{16\sqrt{2}} - \frac{t^3}{32} - \frac{35t^4}{1536\sqrt{2}} - \frac{7t^5}{1024} - \frac{245t^6}{73728\sqrt{2}} - \frac{9t^7}{16384} + \mathcal{O}(t^8) \end{aligned} \right..
\end{align}
\end{adjustwidth}

First-order ContEvol with ``adequate'' expansion yields
\vspace{-12pt}
\begin{adjustwidth}{-\extralength}{0cm}
\small
\begin{align}
      \label{eq:2BP_CE1_PBM} \left\{ \begin{aligned} {\boldsymbol r}_h &= \begin{pmatrix} 2 - \dfrac{h}{\sqrt{2}} - \dfrac{h^2}{8} - \dfrac{h^3}{24\sqrt{2}} - \dfrac{5h^4}{768} - \dfrac{h^5}{768\sqrt{2}} + {\color{red} 0} \cdot \dfrac{7h^6}{36864} + {\color{red} \dfrac{528}{455}} \cdot \dfrac{13h^7}{36864\sqrt{2}} + \mathcal{O}(h^8) \\
    \dfrac{h}{\sqrt{2}} - \dfrac{h^3}{48\sqrt{2}} - \dfrac{h^4}{128} - \dfrac{7h^5}{1536\sqrt{2}} - {\color{red} 0} \cdot \dfrac{7h^6}{6144} - {\color{red} \dfrac{3}{25}} \cdot \dfrac{35h^7}{73728\sqrt{2}} + \mathcal{O}(h^8) \end{pmatrix} \\
    {\boldsymbol v}_h &= \begin{pmatrix} -\dfrac{1}{\sqrt{2}} - \dfrac{h}{4} - \dfrac{h^2}{8\sqrt{2}} - \dfrac{5h^3}{192} - \dfrac{5h^4}{768\sqrt{2}} + {\color{red} \left( -\dfrac{512}{2373} \right)} \cdot \dfrac{7h^5}{6144} + {\color{red} \dfrac{216}{455}} \cdot \dfrac{91h^6}{36864\sqrt{2}} + {\color{red} \dfrac{13348}{35805}} \cdot \dfrac{341h^7}{294912} + \mathcal{O}(h^8) \\
    \dfrac{1}{\sqrt{2}} - \dfrac{h^2}{16\sqrt{2}} - \dfrac{h^3}{32} - \dfrac{35h^4}{1536\sqrt{2}} - {\color{red} \left( -\dfrac{2}{105} \right)} \cdot \dfrac{7h^5}{1024} - {\color{red} \dfrac{27}{1225}} \cdot \dfrac{245h^6}{73728\sqrt{2}} - {\color{red} \dfrac{976}{2835}} \cdot \dfrac{9h^7}{16384} + \mathcal{O}(h^8) \end{pmatrix} \end{aligned} \right..
\end{align}
\end{adjustwidth}
Again, errors in ${\boldsymbol r}_h$ and ${\boldsymbol v}_h$ (highlighted in red) are $\mathcal{O}(h^6)$ and $\mathcal{O}(h^5)$, respectively.

\subsection{Two-Body, Equivalence with Linearization and Taylor Expansion}
\label{ss:2BP_CE1L}

In this section, we show that first-order ContEvol with ``adequate'' expansion is equivalent to both linearization and fifth-order Taylor expansion of the equation of motion.

\paragraph{{Equivalence with linearization.}
}
An alternative way to handle the right hand side of the EOM Equation~(\ref{eq:2BP_EOM}) is to define
\begin{align}
    \label{eq:f_rdivr3} {\boldsymbol f}(t) = {\boldsymbol f}({\boldsymbol r}(t)) = \frac{\boldsymbol r}{r^3}
\end{align}
and use its derivatives at $t=0$ and $t=h$ to approximate it as (again, subscript ``CE2'' stands for ContEvol and two-body problem)
\begin{align}
    \label{eq:2BP_CE1L_fh} {\boldsymbol f}_{\rm CE2}(t) = {\boldsymbol f}_0 + \dot{\boldsymbol f}_0t + {\boldsymbol B}_{\boldsymbol f}t^2 + {\boldsymbol A}_{\boldsymbol f}t^3, \quad t \in [0, h],
\end{align}
with coefficients ${\boldsymbol A}_{\boldsymbol f}$ and ${\boldsymbol B}_{\boldsymbol f}$ yielded by ``terminal'' conditions at $t=h$
\begin{align}
    \label{eq:2BP_CE1L_ABf} \left\{ \begin{aligned} {\boldsymbol A}_{\boldsymbol f} &= 2({\boldsymbol f}_0 - {\boldsymbol f}_h) h^{-3} + (\dot{\boldsymbol f}_0 + \dot{\boldsymbol f}_h) h^{-2} \\
    {\boldsymbol B}_{\boldsymbol f} &= 3({\boldsymbol f}_h - {\boldsymbol f}_0) h^{-2} - (2\dot{\boldsymbol f}_0 + \dot{\boldsymbol f}_h) h^{-1} \end{aligned} \right..
\end{align}
Evidently, we have ${\boldsymbol f}_0 = {\boldsymbol C}_0 - 2{\boldsymbol B}$ (see Equation~(\ref{eq:2BP_CE1_C0123}) for ${\boldsymbol C}_i$, $i = 0, 1, 2, 3$).

Since ${\boldsymbol f}(t)$ only depends on time through ${\boldsymbol r}(t)$, its derivative is
\vspace{-10pt}
\begin{adjustwidth}{-\extralength}{0cm}
\begin{align}
    \nonumber \dot{\boldsymbol f}(t) &= \dot{f}_i {\boldsymbol e}_i = v_j \frac{\partial f_i}{\partial r_j} {\boldsymbol e}_i = v_j \frac{\partial}{\partial r_j} \left[ \frac{r_i}{(r_kr_k)^{3/2}} \right] {\boldsymbol e}_i = v_j \frac{\delta_{ij} (r_kr_k)^{3/2} - r_i (3/2) (r_kr_k)^{1/2} (2r_j)}{(r_kr_k)^3} {\boldsymbol e}_i \\
    &= \frac{v_i {\boldsymbol e}_i}{(r_kr_k)^{3/2}} - \frac{3r_jv_j}{(r_kr_k)^{5/2}} r_i{\boldsymbol e}_i = \frac{{\boldsymbol v}}{r^3} - \frac{3 {\boldsymbol r}\cdot{\boldsymbol v}}{r^5}{\boldsymbol r},
\end{align}
\end{adjustwidth}
where we have used Einstein notation. Similarly, we have $\dot{\boldsymbol f}_0 = {\boldsymbol C}_1 - 6{\boldsymbol A}$.

The coefficients ${\boldsymbol A}_{\boldsymbol f}$ and ${\boldsymbol B}_{\boldsymbol f}$ can be fully specified by either ${\boldsymbol A}$ and ${\boldsymbol B}$ (see Equation~(\ref{eq:2BP_CE1_AB})) or ${\boldsymbol r}_h$ and ${\boldsymbol v}_h$.
Proceeding with ${\boldsymbol A}$ and ${\boldsymbol B}$, the function ${\boldsymbol f}(t)$ at $t=h$ is
\vspace{-12pt}
\begin{adjustwidth}{-\extralength}{0cm}
\begin{align}
    \nonumber {\boldsymbol f}_h &= {\boldsymbol f}({\boldsymbol r}_h) = \frac{{\boldsymbol r}_h}{r_h^3} = \frac{{\boldsymbol r}_0 + {\boldsymbol v}_0h + {\boldsymbol B}h^2 + {\boldsymbol A}h^3}{\Vert{\boldsymbol r}_0 + {\boldsymbol v}_0h + {\boldsymbol B}h^2 + {\boldsymbol A}h^3\Vert^3} \\
    \nonumber &= \left[ \begin{aligned} &\frac{{\boldsymbol r}_0}{r_0^3} + \left( \frac{{\boldsymbol v}_0}{r_0^3} - \frac{3 {\boldsymbol r}_0\cdot{\boldsymbol v}_0}{r_0^5}{\boldsymbol r}_0 \right)h + \left[ \frac{\boldsymbol B}{r_0^3} - \frac{3{\boldsymbol r}_0\cdot{\boldsymbol v}_0}{r_0^5} {\boldsymbol v}_0 - \frac{3}{2} \left( \frac{2{\boldsymbol B}\cdot{\boldsymbol r}_0 + v_0^2}{r_0^5} - \frac{5 ({\boldsymbol r}_0\cdot{\boldsymbol v}_0)^2}{r_0^7} \right) {\boldsymbol r}_0 \right]h^2 \\
    &+ \left\{ \begin{aligned} &\frac{\boldsymbol A}{r_0^3} - \frac{3 {\boldsymbol r}_0\cdot{\boldsymbol v}_0}{r_0^5} \boldsymbol B - \frac{3}{2} \left( \frac{2{\boldsymbol B}\cdot{\boldsymbol r}_0 + v_0^2}{r_0^5} - \frac{5 ({\boldsymbol r}_0\cdot{\boldsymbol v}_0)^2}{r_0^7} \right) {\boldsymbol v}_0 \\
    &- \left( \frac{3 ({\boldsymbol A}\cdot{\boldsymbol r}_0 + {\boldsymbol B}\cdot{\boldsymbol v}_0)}{r_0^5} - \frac{15 (2 {\boldsymbol B}\cdot{\boldsymbol r}_0 + v_0^2) ({\boldsymbol r}_0\cdot{\boldsymbol v}_0)}{2 r_0^7} + \frac{35 ({\boldsymbol r}_0\cdot{\boldsymbol v}_0)^3}{2 r_0^9} \right) {\boldsymbol r}_0 \end{aligned} \right\}h^3 + \mathcal{O}(h^4) \end{aligned} \right] \\
    &= ({\boldsymbol C}_0 - 2{\boldsymbol B}) + ({\boldsymbol C}_1 - 6{\boldsymbol A})h + {\boldsymbol C}_2h^2 + {\boldsymbol C}_3h^3 + \mathcal{O}(h^4),
\end{align}
\end{adjustwidth}
and its derivative $\dot{\boldsymbol f}(t)$ at $t=h$ is
\vspace{-12pt}
\begin{adjustwidth}{-\extralength}{0cm}
\begin{align}
    \nonumber \dot{\boldsymbol f}_h &= \dot{\boldsymbol f}({\boldsymbol r}_h, {\boldsymbol v}_h) = \frac{{\boldsymbol v}}{r^3} - \frac{3 {\boldsymbol r}\cdot{\boldsymbol v}}{r^5}{\boldsymbol r} \\
    \nonumber &= \frac{{\boldsymbol v}_0 + 2{\boldsymbol B}h + 3{\boldsymbol A}h^2}{\Vert{\boldsymbol r}_0 + {\boldsymbol v}_0h + {\boldsymbol B}h^2 + {\boldsymbol A}h^3\Vert^3} + \frac{3 ({\boldsymbol r}_0 + {\boldsymbol v}_0h + {\boldsymbol B}h^2 + {\boldsymbol A}h^3) \cdot ({\boldsymbol v}_0 + 2{\boldsymbol B}h + 3{\boldsymbol A}h^2)}{\Vert{\boldsymbol r}_0 + {\boldsymbol v}_0h + {\boldsymbol B}h^2 + {\boldsymbol A}h^3\Vert^5} ({\boldsymbol r}_0 + {\boldsymbol v}_0h + {\boldsymbol B}h^2 + {\boldsymbol A}h^3) \\
    \nonumber &= \left[ \begin{aligned} &\left( \frac{{\boldsymbol v}_0}{r_0^3} - \frac{3 {\boldsymbol r}_0\cdot{\boldsymbol v}_0}{r_0^5}{\boldsymbol r}_0 \right) + 2\left[ \frac{\boldsymbol B}{r_0^3} - \frac{3{\boldsymbol r}_0\cdot{\boldsymbol v}_0}{r_0^5} {\boldsymbol v}_0 - \frac{3}{2} \left( \frac{2{\boldsymbol B}\cdot{\boldsymbol r}_0 + v_0^2}{r_0^5} - \frac{5 ({\boldsymbol r}_0\cdot{\boldsymbol v}_0)^2}{r_0^7} \right) {\boldsymbol r}_0 \right]h^2 \\
    &+ 3\left\{ \begin{aligned} &\frac{\boldsymbol A}{r_0^3} - \frac{3 {\boldsymbol r}_0\cdot{\boldsymbol v}_0}{r_0^5} \boldsymbol B - \frac{3}{2} \left( \frac{2{\boldsymbol B}\cdot{\boldsymbol r}_0 + v_0^2}{r_0^5} - \frac{5 ({\boldsymbol r}_0\cdot{\boldsymbol v}_0)^2}{r_0^7} \right) {\boldsymbol v}_0 \\
    &- \left( \frac{3 ({\boldsymbol A}\cdot{\boldsymbol r}_0 + {\boldsymbol B}\cdot{\boldsymbol v}_0)}{r_0^5} - \frac{15 (2 {\boldsymbol B}\cdot{\boldsymbol r}_0 + v_0^2) ({\boldsymbol r}_0\cdot{\boldsymbol v}_0)}{2 r_0^7} + \frac{35 ({\boldsymbol r}_0\cdot{\boldsymbol v}_0)^3}{2 r_0^9} \right) {\boldsymbol r}_0 \end{aligned} \right\}h^2 + \mathcal{O}(h^3) \end{aligned} \right] \\
    &= ({\boldsymbol C}_1 - 6{\boldsymbol A}) + 2{\boldsymbol C}_2h + 3{\boldsymbol C}_3h^2 + \mathcal{O}(h^3),
\end{align}
\end{adjustwidth}
where we have ``adequately'' expanded ${\boldsymbol f}_h$ and $\dot{\boldsymbol f}_h$ to keep all the terms without non-linear coefficients in ${\boldsymbol A}$ and ${\boldsymbol B}$.
Plugging these into Equation~(\ref{eq:2BP_CE1L_ABf}), we obtain the simple relations ${\boldsymbol A}_{\boldsymbol f} \approx {\boldsymbol C}_3$ and ${\boldsymbol B}_{\boldsymbol f} \approx {\boldsymbol C}_2$.

With the function ${\boldsymbol f}(t)$, the cost function is defined as (here the prime ``$'$'' denotes linearization)
\vspace{-13pt}
\begin{adjustwidth}{-\extralength}{0cm}
\begin{align}
    \nonumber \epsilon_{\rm CE2}'({\boldsymbol A}, {\boldsymbol B}; h) &= \int_0^h [\ddot{\boldsymbol r} + {\boldsymbol f}({\boldsymbol r})]^2 \,{\rm d}t = \int_0^h [(2{\boldsymbol B} + 6{\boldsymbol A}t) + ({\boldsymbol f}_0 + \dot{\boldsymbol f}_0t + {\boldsymbol B}_{\boldsymbol f}t^2 + {\boldsymbol A}_{\boldsymbol f}t^3) ]^2 \,{\rm d}t \\
    \nonumber &= \int_0^h [(2{\boldsymbol B} + {\boldsymbol f}_0) + (6{\boldsymbol A} + \dot{\boldsymbol f}_0)t + {\boldsymbol B}_{\boldsymbol f}t^2 + {\boldsymbol A}_{\boldsymbol f}t^3 ]^2 \,{\rm d}t \\
    &\approx \int_0^h \left[ {\boldsymbol C}_0 + {\boldsymbol C}_1 t + {\boldsymbol C}_2 t^2 + {\boldsymbol C}_3 t^3 \right]^2 \,{\rm d}t = \epsilon_{\rm CE2}({\boldsymbol A}, {\boldsymbol B}; h).
\end{align}
\end{adjustwidth}

Therefore, linearization is equivalent to ``adequate'' expansion (see Section~\ref{ss:2BP_CE1A}); nevertheless, this approach should be more suitable when the function ${\boldsymbol f}(t)$ does not have a simple expression, e.g., when it has to be numerically computed by interpolating in lookup~tables.

\paragraph{{Equivalence with Taylor expansion.}}
By successively differentiate the equation of motion Equation~(\ref{eq:2BP_EOM}), one can attain the third derivative (jerk)
\begin{align}
    {\boldsymbol r}^{(3)} = \frac{{\rm d}}{{\rm d} t} \left( - \frac{r_j {\boldsymbol e}_j}{(r_k r_k)^{3/2}} \right) = - \frac{\dot r_j {\boldsymbol e}_j}{r^3} + \frac{3}{2} \frac{2 r_{k'} \dot r_{k'}}{(r_k r_k)^{5/2}} {\boldsymbol r} = - \frac{\dot {\boldsymbol r}}{r^3} + \frac{3 {\boldsymbol r} \cdot \dot{\boldsymbol r}}{r^5} {\boldsymbol r},
\end{align}
the fourth derivative (snap)
\begin{align}
    \nonumber {\boldsymbol r}^{(4)} &= \frac{{\rm d}}{{\rm d} t} \left( - \frac{\dot r_j {\boldsymbol e}_j}{(r_k r_k)^{3/2}} + \frac{3 r_l \dot r_l}{(r_k r_k)^{5/2}} r_j {\boldsymbol e}_j \right) \\
    \nonumber &= - \frac{\ddot r_j {\boldsymbol e}_j}{r^3} + \frac{3}{2} \frac{2 r_{k'} \dot r_{k'}}{(r_k r_k)^{5/2}} \dot{\boldsymbol r} + 3 \frac{\dot r_l \dot r_l + r_l \ddot r_l}{r^5} {\boldsymbol r} + \frac{3 {\boldsymbol r} \cdot \dot{\boldsymbol r}}{r^5} \dot r_j {\boldsymbol e}_j - \frac{5}{2} \frac{2 r_{k'} \dot r_{k'}}{(r_k r_k)^{7/2}} 3 ({\boldsymbol r} \cdot \dot{\boldsymbol r}) {\boldsymbol r} \\
    &= - \frac{\ddot{\boldsymbol r}}{r^3} + 3 \frac{2 ({\boldsymbol r} \cdot \dot{\boldsymbol r}) \dot{\boldsymbol r} + (\dot{\boldsymbol r} \cdot \dot{\boldsymbol r} + {\boldsymbol r} \cdot \ddot{\boldsymbol r}) {\boldsymbol r}}{r^5} - 15 \frac{({\boldsymbol r} \cdot \dot{\boldsymbol r})^2}{r^7} {\boldsymbol r},
\end{align}
and the fifth derivative (crackle)
\vspace{-10pt}
\begin{adjustwidth}{-\extralength}{0cm}
\begin{align}
    \nonumber {\boldsymbol r}^{(5)} &= \frac{{\rm d}}{{\rm d} t} \left( - \frac{\ddot{r}_j {\boldsymbol e}_j}{(r_k r_k)^{3/2}} + 3 \frac{2 (r_l \dot r_l) \dot r_j {\boldsymbol e}_j + (\dot r_l \dot r_l + r_l \ddot r_l) r_j {\boldsymbol e}_j}{(r_k r_k)^{5/2}} - 15 \frac{(r_l \dot r_l)^2}{(r_k r_k)^{7/2}} r_j {\boldsymbol e}_j \right) \\
    \nonumber &= \left[ \begin{aligned} &- \frac{r_j^{(3)} {\boldsymbol e}_j}{r^3} + \frac{3}{2} \frac{2 r_{k'} \dot r_{k'}}{(r_k r_k)^{5/2}} \ddot{\boldsymbol r} + 6 \frac{(\dot r_l \dot r_l + r_l \ddot r_l) \dot {\boldsymbol r} + ({\boldsymbol r} \cdot \dot{\boldsymbol r}) \ddot r_j {\boldsymbol e}_j}{r^5} \\
    &+ 3 \frac{(3 \dot r_l \ddot r_l + r_l r_l^{(3)}) {\boldsymbol r} + (\dot{\boldsymbol r} \cdot \dot{\boldsymbol r} + {\boldsymbol r} \cdot \ddot{\boldsymbol r}) \dot r_j {\boldsymbol e}_j}{r^5} - \frac{5}{2} \frac{2 r_{k'} \dot r_{k'}}{(r_k r_k)^{7/2}} 3 [2 ({\boldsymbol r} \cdot \dot{\boldsymbol r}) \dot{\boldsymbol r} + (\dot{\boldsymbol r} \cdot \dot{\boldsymbol r} + {\boldsymbol r} \cdot \ddot{\boldsymbol r}) {\boldsymbol r}] \\
    &- 15 \frac{2 (r_l \dot r_l) (\dot r_{l'} \dot r_{l'} + r_{l'} \ddot r_{l'}) {\boldsymbol r} + ({\boldsymbol r} \cdot \dot{\boldsymbol r})^2 \dot r_j {\boldsymbol e}_j}{r^7} + \frac{7}{2} \frac{2 r_{k'} \dot r_{k'}}{(r_k r_k)^{9/2}} 15 ({\boldsymbol r} \cdot \dot{\boldsymbol r})^2 {\boldsymbol r} \end{aligned} \right] \\
    &= \left[ \begin{aligned} &- \frac{{\boldsymbol r}^{(3)}}{r^3} + 3 \frac{3 ({\boldsymbol r} \cdot \dot{\boldsymbol r}) \ddot{\boldsymbol r} + 3 (\dot{\boldsymbol r} \cdot \dot{\boldsymbol r} + {\boldsymbol r} \cdot \ddot{\boldsymbol r}) \dot {\boldsymbol r} + (3 \dot{\boldsymbol r} \cdot \ddot{\boldsymbol r} + {\boldsymbol r} \cdot {\boldsymbol r}^{(3)}) {\boldsymbol r}}{r^5} \\
    &- 45 ({\boldsymbol r} \cdot \dot{\boldsymbol r}) \frac{({\boldsymbol r} \cdot \dot{\boldsymbol r}) \dot{\boldsymbol r} + (\dot{\boldsymbol r} \cdot \dot{\boldsymbol r} + {\boldsymbol r} \cdot \ddot{\boldsymbol r}) {\boldsymbol r}}{r^7} + 105 \frac{({\boldsymbol r} \cdot \dot{\boldsymbol r})^3}{r^9} {\boldsymbol r} \end{aligned} \right]
\end{align}
\end{adjustwidth}
of the position vector ${\boldsymbol r}$; using these derivatives, the Taylor expansion of the EOM is
\begin{align}
    \left\{ \begin{aligned} {\boldsymbol r}(t) &= {\boldsymbol r}_0 + \dot{\boldsymbol r}_0 t + \frac{1}{2} \ddot{\boldsymbol r}_0 t^2 + \frac{1}{6} {\boldsymbol r}^{(3)}_0 t^3 + \frac{1}{24} {\boldsymbol r}^{(4)}_0 t^4 + \frac{1}{120} {\boldsymbol r}^{(5)}_0 t^5 + \mathcal{O}(t^6) \\
    {\boldsymbol v}(t) &= \dot{\boldsymbol r}_0 + \ddot{\boldsymbol r}_0 t + \frac{1}{2} {\boldsymbol r}^{(3)}_0 t^2 + \frac{1}{6} {\boldsymbol r}^{(4)}_0 t^3 + \frac{1}{24} {\boldsymbol r}^{(5)}_0 t^4 + \mathcal{O}(t^5) \end{aligned} \right..
\end{align}

It is verified that the first-order ContEvol solution Equation~(\ref{eq:2BP_CE1_rvh_res}) is identical to
\begin{align}
    \label{eq:2BP_CE1_Taylor} \left\{ \begin{aligned} {\boldsymbol r}_{{\rm CE1},h} &= {\boldsymbol r}_0 + \dot{\boldsymbol r}_0 h + \frac{1}{2} \ddot{\boldsymbol r}_0 h^2 + \frac{1}{6} {\boldsymbol r}^{(3)}_0 h^3 + \frac{1}{24} {\boldsymbol r}^{(4)}_0 h^4 + \frac{1}{120} {\boldsymbol r}^{(5)}_0 h^5 + \mathcal{O}(h^7) \\
    {\boldsymbol v}_{{\rm CE1},h} &= \dot{\boldsymbol r}_0 + \ddot{\boldsymbol r}_0 h + \frac{1}{2} {\boldsymbol r}^{(3)}_0 h^2 + \frac{1}{6} {\boldsymbol r}^{(4)}_0 h^3 + \frac{1}{24} {\boldsymbol r}^{(5)}_0 h^4 + \mathcal{O}(h^5) \end{aligned} \right.;
\end{align}
note that the $\mathcal{O}(h^6)$ term of ${\boldsymbol r}_{{\rm CE1},h}$ is missing.
Therefore, at least for the two-body problem, first-order ContEvol is equivalent to fifth-order Taylor expansion of the EOM in terms of position, and fourth-order in terms of velocity.

For relatively simple equation(s), successive derivatives are feasible; however, when the system is complicated, ContEvol could provide a ``shortcut'' to obtain fifth/fourth-order Taylor expansion of the evolution numerically.
Specifically, one can compute counterparts of the ${\boldsymbol C}_i$ coefficients Equation~(\ref{eq:2BP_CE1_C0123}) numerically, use them to construct a linear system like Equation~(\ref{eq:2BP_CE1_linsys}), and then solve it to obtain counterparts of ${\boldsymbol A}$ and ${\boldsymbol B}$.
In Section~\ref{ss:3BP_CE1}, we will outline how this is supposed to be done for the three-body problem.

The procedure described above is not the only way to implement a ContEvol method. For relatively simple problems like the two-body problem, one can choose to directly use expressions for results at $t=h$, e.g., ${\boldsymbol r}_h$ and ${\boldsymbol v}_h$ Equation~(\ref{eq:2BP_CE1_rvh_res}).
We refer to the two strategies as implementation by optimization process and implementation by optimization results, respectively.
In Section~\ref{ss:2BP_CE1N}, while implementing first-order ContEvol for an eccentric orbit, we will adopt the second strategy, i.e., directly utilize Equation~(\ref{eq:2BP_CE1_Taylor}), truncating it at $\mathcal{O}(h^7)$ for ${\boldsymbol r}_h$ and $\mathcal{O}(h^5)$ for ${\boldsymbol v}_h$.

\subsection{Two-Body, Conservation of Mechanic Energy and Angular Momentum}
\label{ss:2BP_CE1C}

As mentioned in the second test case of Section~\ref{ss:2BP_CE1A}, two quantities should be conserved in the two body problem: mechanic energy and angular momentum. In terms of ${\boldsymbol r}$ and ${\boldsymbol v}$, these are
\vspace{-8pt}
\begin{align}
    \label{eq:2BP_EandLz} \left\{ \begin{aligned} E ({\boldsymbol r}, {\boldsymbol v}) &= -\frac{1}{r} + \frac{v^2}{2} \\
    L_z ({\boldsymbol r}, {\boldsymbol v}) &= r_x v_y - r_y v_x \end{aligned} \right.,
\end{align}
respectively; note that ${\boldsymbol L} = {\boldsymbol r} \times {\boldsymbol v} = L_z \hat{\boldsymbol z}$ in the case of a two-body problem, hence we only need to track its $z$ component. The proofs are straightforward:
\begin{align}
    \left\{ \begin{aligned} \dot E &= - \dot r_i \frac{\partial}{\partial r_i} \frac{1}{(r_k r_k)^{1/2}} + \ddot r_i \frac{\partial}{\partial \dot r_i} \frac{\dot r_k \dot r_k}{2} = \dot r_i \frac{2 r_i}{2 (r_k r_k)^{3/2}} + \ddot r_i \dot r_i = \dot{\boldsymbol r} \cdot \left( \frac{{\boldsymbol r}}{r^3} + \ddot{\boldsymbol r} \right) = 0 \\
    \dot{\boldsymbol L} &= \dot{\boldsymbol r} \times \dot{\boldsymbol r} + {\boldsymbol r} \times \ddot{\boldsymbol r} = {\boldsymbol r} \times \left( - \frac{{\boldsymbol r}}{r^3} \right) = {\boldsymbol 0} \end{aligned} \right.,
\end{align}
where we have used the equation of motion Equation~(\ref{eq:2BP_EOM}) in both.

Using these two conservation laws, we can express ${\boldsymbol v}$ in terms of ${\boldsymbol r}$ as
\begin{align}
    \label{eq:2BP_vxy_pm} \left\{ \begin{aligned} v_x &= \frac{-r_y L_z \pm |r_x| \sqrt{\Delta_{\boldsymbol r}}}{r^2} \\
    v_y &= \frac{r_x L_z \pm \operatorname{sgn}(r_x) r_y \sqrt{\Delta_{\boldsymbol r}}}{r^2} \end{aligned} \right. \quad r_x \neq 0,
\end{align}
where $\operatorname{sgn}(\cdot)$ is the sign function, or
\begin{align}
    \label{eq:2BP_vyx_pm} \left\{ \begin{aligned} v_x &= \frac{-r_y L_z \pm \operatorname{sgn}(r_y) r_x \sqrt{\Delta_{\boldsymbol r}}}{r^2} \\
    v_y &= \frac{r_x L_z \pm |r_y| \sqrt{\Delta_{\boldsymbol r}}}{r^2} \end{aligned} \right. \quad r_y \neq 0
\end{align}
with
\begin{align}
    \Delta_{\boldsymbol r} = 2(r+E r^2) - L_z^2 = (rv)^2 - (rv \sin \langle {\boldsymbol r}, {\boldsymbol v} \rangle)^2 = (rv \cos \langle {\boldsymbol r}, {\boldsymbol v} \rangle)^2 \geq 0.
\end{align}
One should not use $\sqrt{\Delta_{\boldsymbol r}} = |rv \cos \langle {\boldsymbol r}, {\boldsymbol v} \rangle| = |{\boldsymbol r}\cdot{\boldsymbol v}|$ to simplify Equation~(\ref{eq:2BP_vxy_pm}) or (\ref{eq:2BP_vyx_pm}), as ${\boldsymbol v}$ is what we are trying to derive.

To resolve the ambiguity of the $\pm$ symbols, we write ${\boldsymbol r}_h \equiv {\boldsymbol r}(h)$ and ${\boldsymbol v}_h \equiv {\boldsymbol v}(h)$ as
\vspace{-6pt}
\begin{align}
    \label{eq:2BP_rvh_vabar} \left\{ \begin{aligned} {\boldsymbol r}_h &= {\boldsymbol r}_0 + \bar{\boldsymbol v}h \\
    {\boldsymbol v}_h &= {\boldsymbol v}_0 + \bar{\boldsymbol a}h \end{aligned} \right.
\end{align}
and derive $E$ and $L_z$ from initial conditions ${\boldsymbol r}_0 \equiv {\boldsymbol r}(0)$ and ${\boldsymbol v}_0 \equiv {\boldsymbol v}(0)$
\vspace{-6pt}
\begin{align}
    \label{eq:2BP_EandLz0} \left\{ \begin{aligned} E &= -\frac{1}{r_0} + \frac{v_0^2}{2} \\
    L_z &= r_{0x} v_{0y} - r_{0y} v_{0x} \end{aligned} \right..
\end{align}
Note that for this purpose, we are treating each pair of ${\boldsymbol r}$ and ${\boldsymbol v}$ as ${\boldsymbol r}_0$ and ${\boldsymbol v}_0$; in other words, we imagine an infinitesimal next step $h \to 0$ for any given position and velocity.

Plugging these into and expanding Equation~(\ref{eq:2BP_vxy_pm}), we obtain
\begin{adjustwidth}{-\extralength}{0cm}
\begin{align}
    v_{xh} = v_{0x} + \left[ \begin{aligned} &\bar v_x \left( \frac{2 r_{0x} r_{0y} L_z}{r_0^4} \pm \frac{r_{0x} [2 r_{0x}^2 r_{0y}^2 v_{0x}^2 - 2 r_{0x} r_{0y} (r_{0x}^2 - r_{0y}^2) v_{0x} v_{0y} + (r_{0y}^4 + r_{0x}^4) v_{0y}^2 - r_0 r_{0x}^2]}{r_0^4 |r_{0x} {\boldsymbol r}_0\cdot{\boldsymbol v}_0|} \right) \\
    &- \bar v_y \left( \frac{(r_{0x}^2 - r_{0y}^2) L_z}{r_0^4} \pm \frac{r_{0x}^2 r_{0y} [(r_{0x}^2 - r_{0y}^2) (v_{0x}^2 - v_{0y}^2) + 4 r_{0x} r_{0y} v_{0x} v_{0y} + r_0]}{r_0^4 |r_{0x} {\boldsymbol r}_0\cdot{\boldsymbol v}_0|} \right) \end{aligned} \right]h + \mathcal{O}(h^2),
\end{align}
\end{adjustwidth}
where we have used $L_z$ to simplify notation, and a similar expression for $r_{hy}$; in the limit $h \to 0$, we should have $\bar{\boldsymbol v} \to {\boldsymbol v}_0$, and thus
\vspace{-12pt}
\begin{adjustwidth}{-\extralength}{0cm}
\begin{align}
    v_{xh} \to v_{0x} + \left[ \begin{aligned} &a_{0x} + (\pm \operatorname{sgn}(r_{0x} {\boldsymbol r}_0\cdot{\boldsymbol v}_0) - 1) \\ &\cdot \frac{2 r_{0x} r_{0y}^2 v_{0x}^2 - r_{0y} (3 r_{0x}^2 - r_{0y}^2) v_{0x} v_{0y} + r_{0x} (r_{0x}^2 - r_{0y}^2) v_{0y}^2 - r_0 r_{0x}}{r_0^4} \end{aligned} \right]h + \mathcal{O}(h^2).
\end{align}
\end{adjustwidth}
Since $\bar{\boldsymbol a} \to {\boldsymbol a}_0 = - {\boldsymbol r}_0/r_0^3$, the $\pm$ symbols in Equation~(\ref{eq:2BP_vxy_pm}) should take the same sign as $r_{0x} {\boldsymbol r}_0\cdot{\boldsymbol v}_0$.

The $\pm$ symbols in Equation~(\ref{eq:2BP_vyx_pm}) can be determined in the same way, and the final expressions are the same.
In conclusion, based on the conserved quantities $E$ and $L_z$, unambiguous expression for ${\boldsymbol v}$ in terms of ${\boldsymbol r}$ is
\begin{align}
    \label{eq:2BP_vxy_rxy} \left\{ \begin{aligned} v_x &= \frac{-r_y L_z + \operatorname{sgn}({\boldsymbol r} \cdot{\boldsymbol v}') r_x \sqrt{\Delta_{\boldsymbol r}}}{r^2} \\
    v_y &= \frac{r_x L_z + \operatorname{sgn}({\boldsymbol r} \cdot{\boldsymbol v}') r_y \sqrt{\Delta_{\boldsymbol r}}}{r^2} \end{aligned} \right. \quad r_x r_y \neq 0,
\end{align}
where the prime ``$'$'' for ${\boldsymbol v}'$ will be explained later in this section. Note that the condition $r_x r_y \neq 0$ is always satisfied unless $L_z = 0$, which reduces the two-body problem to its one-dimensional case and is usually not of interest, except for calculating the ``free-fall''~timescale.

Similarly, we can express ${\boldsymbol r}$ in terms of ${\boldsymbol v}$ as
\begin{align}
    \label{eq:2BP_rxy_pm} \left\{ \begin{aligned} r_x &= \frac{v_y L_z \pm |v_x| \sqrt{\Delta_{\boldsymbol v}}}{v^2} \\
    r_y &= \frac{- v_x L_z \pm \operatorname{sgn}(v_x) v_y \sqrt{\Delta_{\boldsymbol v}}}{v^2} \end{aligned} \right. \quad v_x \neq 0
\end{align}
or
\begin{align}
    \label{eq:2BP_ryx_pm} \left\{ \begin{aligned} r_x &= \frac{v_y L_z \pm \operatorname{sgn}(v_y) v_x \sqrt{\Delta_{\boldsymbol v}}}{v^2} \\
    r_y &= \frac{- v_x L_z \pm |v_y| \sqrt{\Delta_{\boldsymbol v}}}{v^2} \end{aligned} \right. \quad v_y \neq 0
\end{align}
with
\begin{align}
    \Delta_{\boldsymbol v} = \frac{v^2}{(-E+v^2/2)^2} - L_z^2 = (rv)^2 - (rv \sin \langle {\boldsymbol r}, {\boldsymbol v} \rangle)^2 = (rv \cos \langle {\boldsymbol r}, {\boldsymbol v} \rangle)^2 \geq 0.
\end{align}

Plugging Equations~(\ref{eq:2BP_rvh_vabar}) and (\ref{eq:2BP_EandLz0}) into and expanding Equation~(\ref{eq:2BP_rxy_pm}), we obtain
\vspace{-8pt}
\begin{adjustwidth}{-\extralength}{0cm}
\begin{align}
    r_{hx} = r_{0x} + \left[ \begin{aligned} &\bar a_x v_{0x} \left( -2 \frac{v_{0y} L_z \pm |v_{0x} {\boldsymbol r}_0\cdot{\boldsymbol v}_0|}{v_0^4} \pm 2 \frac{r_{0x}}{v_0^2} \operatorname{sgn}(v_{0x} {\boldsymbol r}_0\cdot{\boldsymbol v}_0) \pm \frac{r_{0y}^2 - r_0^3 v_{0x}^2}{|v_{0x} {\boldsymbol r}_0\cdot{\boldsymbol v}_0|} \right) \\
    &- \bar a_y \left( 2 v_{0y} \frac{v_{0y} L_z \pm |v_{0x} {\boldsymbol r}_0\cdot{\boldsymbol v}_0|}{v_0^4} - \frac{L_z}{v_0^2} \pm \frac{r_0^2 v_{0x}^2 v_{0y} (r_0 v_0^2 - 1)}{|v_{0x} {\boldsymbol r}_0\cdot{\boldsymbol v}_0|} \right) \end{aligned} \right]h + \mathcal{O}(h^2),
\end{align}
\end{adjustwidth}
where again we have used $L_z$ to simplify notation, and a similar expression for $r_{hy}$; in the limit $h \to 0$, we should have $\bar{\boldsymbol a} \to {\boldsymbol a}_0 = - {\boldsymbol r}_0/r_0^3$, and thus
\vspace{-12pt}
\begin{adjustwidth}{-\extralength}{0cm}
\begin{align}
    r_{hx} \to r_{0x} + \left[ \begin{aligned} &v_{0x} + (\pm \operatorname{sgn}(v_{0x} {\boldsymbol r}_0\cdot{\boldsymbol v}_0) - 1) \\
    &\cdot \left( v_{0x} - \frac{2 r_{0x}^2 v_{0x} v_{0y}^2 + r_{0x} r_{0y} v_{0y} (v_{0y}^2 - 3 v_{0x}^2) + r_{0y}^2 v_{0x} (v_{0x}^2 - v_{0y}^2)}{r_0^3 v_0^4} \right) \end{aligned} \right]h + \mathcal{O}(h^2).
\end{align}
\end{adjustwidth}
Since $\bar{\boldsymbol v} \to {\boldsymbol v}_0$, $\pm$ symbols in Equation~(\ref{eq:2BP_rxy_pm}) should take the same sign as $v_{0x} {\boldsymbol r}_0\cdot{\boldsymbol v}_0$.

The $\pm$ symbols in Equation~(\ref{eq:2BP_ryx_pm}) can be determined in the same way, and the final expressions are the same.
In conclusion, based on the conserved quantities $E$ and $L_z$, unambiguous expression for ${\boldsymbol r}$ in terms of ${\boldsymbol v}$ is
\begin{align}
    \label{eq:2BP_rxy_vxy} \left\{ \begin{aligned} r_x &= \frac{v_y L_z + \operatorname{sgn}({\boldsymbol r}' \cdot{\boldsymbol v}) v_x \sqrt{\Delta_{\boldsymbol v}}}{v^2} \\
    r_y &= \frac{- v_x L_z + \operatorname{sgn}({\boldsymbol r}' \cdot{\boldsymbol v}) v_y \sqrt{\Delta_{\boldsymbol v}}}{v^2} \end{aligned} \right. \quad v_x v_y \neq 0.
\end{align}
The condition $v_x v_y \neq 0$ is also always satisfied unless $L_z = 0$.

Admittedly, ${\boldsymbol v}$ should not appear when we express ${\boldsymbol v}$ in terms of ${\boldsymbol r}$, neither vice versa.
Fortunately, numerical methods (including Runge--Kutta, ContEvol, etc.) usually predict both ${\boldsymbol r}$ and ${\boldsymbol v}$ after each time step, hence when we use ${\boldsymbol r}$ (or ${\boldsymbol v}$) to derive ${\boldsymbol v}$ (or ${\boldsymbol r}$), ${\boldsymbol v}'$ (or ${\boldsymbol r}'$) provided by the original numerical methods can be treated as a reasonable initial guess; these are denoted with a prime ``$'$'' in Equations~(\ref{eq:2BP_vxy_rxy}) and (\ref{eq:2BP_rxy_vxy}).

Behavior of the sign function near zero is worth more discussion.
When ${\boldsymbol r} \cdot{\boldsymbol v}' \approx 0$, i.e., when ${\boldsymbol r}$ and ${\boldsymbol v}'$ are perpendicular to each other, $\Delta_{\boldsymbol r} \approx 0$, so that value of $\operatorname{sgn}({\boldsymbol r} \cdot{\boldsymbol v}')$ does not matter.
Similarly, when ${\boldsymbol r}' \cdot{\boldsymbol v} \approx 0$, $\Delta_{\boldsymbol v} \approx 0$, so that value of $\operatorname{sgn}({\boldsymbol r}' \cdot{\boldsymbol v})$ does not matter either.
In practice, neither ${\boldsymbol r} \cdot{\boldsymbol v}'$ nor ${\boldsymbol r}' \cdot{\boldsymbol v}$ can be exactly zero, except for initial conditions or very rare coincidences, yet we need to consider the cases where they are about zero, as wrong signs can change the direction of the history, which is undesirable.
To resolve this issue, one can specify a threshold $\delta$, and set the value of the sign function to $0$ when $|{\boldsymbol r} \cdot{\boldsymbol v}'| < \delta$ or $|{\boldsymbol r}' \cdot{\boldsymbol v}| < \delta$, or make a smoother transition using, e.g., a rescaled logistic function.

In the context of ContEvol, there are two approaches to make use of these conservation~laws.

\paragraph{{Approach 1: Use} ${\boldsymbol r}_h$ to correct ${\boldsymbol v}_h$.}
As shown in Section~\ref{ss:2BP_CE1A}, errors in ${\boldsymbol r}_h$ and ${\boldsymbol v}_h$ of first-order ContEvol are $\mathcal{O}(h^6)$ and $\mathcal{O}(h^5)$, respectively. Because of this difference, after each step, using ${\boldsymbol r}_h$ to correct ${\boldsymbol v}_h$ according to Equation~(\ref{eq:2BP_vxy_rxy}) could be beneficial.

To testify the usefulness of this approach, we plug ${\boldsymbol r}_h$ given by Equation~(\ref{eq:2BP_CE1_rvh_res}) into Equation~(\ref{eq:2BP_vxy_rxy}) to attain a corrected version of ${\boldsymbol v}_h$, denoted as ${\boldsymbol v}_{h,{\rm CC}}$; the discrepancy between uncorrected and corrected expressions is at the fifth order, hence we omit the latter here.
Assuming ${\boldsymbol r} \cdot {\boldsymbol v} < 0$, the corrected Jacobian matrix is (subscript ``CC'' stands for conservation~correction)
\vspace{-10pt}
\begin{adjustwidth}{-\extralength}{0cm}
\begin{align}
    J_{\rm CC} = \begin{pmatrix} \partial r_{hx}/\partial r_{0x} & \partial r_{hx}/\partial r_{0y} & \partial r_{hx}/\partial v_{0x} & \partial r_{hx}/\partial v_{0y} \\
    \partial r_{hy}/\partial r_{0x} & \partial r_{hy}/\partial r_{0y} & \partial r_{hy}/\partial v_{0x} & \partial r_{hy}/\partial v_{0y} \\
    \partial v_{hx,{\rm CC}}/\partial r_{0x} & \partial v_{hx,{\rm CC}}/\partial r_{0y} & \partial v_{hx,{\rm CC}}/\partial v_{0x} & \partial v_{hx,{\rm CC}}/\partial v_{0y} \\
    \partial v_{hy,{\rm CC}}/\partial r_{0x} & \partial v_{hy,{\rm CC}}/\partial r_{0y} & \partial v_{hy,{\rm CC}}/\partial v_{0x} & \partial v_{hy,{\rm CC}}/\partial v_{0y} \end{pmatrix}
    \equiv \begin{pmatrix} J_{11} & J_{12} & J_{13} & J_{14} \\ J_{21} & J_{22} & J_{23} & J_{24} \\ J_{31,{\rm CC}} & J_{32,{\rm CC}} & J_{33,{\rm CC}} & J_{34,{\rm CC}} \\ J_{41,{\rm CC}} & J_{42,{\rm CC}} & J_{43,{\rm CC}} & J_{44,{\rm CC}} \end{pmatrix},
\end{align}
\end{adjustwidth}
where the matrix elements are the same as those given by {Equations}~(\ref{eq:2BP_CE1_Jac_rr})--(\ref{eq:2BP_CE1_Jac_vr}) for the first two rows, since we are using the same expressions for ${\boldsymbol r}_h$; for the last two rows, they are different from those in {Equations}~(\ref{eq:2BP_CE1_Jac_rv})--(\ref{eq:2BP_CE1_Jac_vv}), but again, the leading orders are not affected, so we refrain from showing them here.
Most importantly, the determinant of the corrected Jacobian is
\vspace{-12pt}
\begin{adjustwidth}{-\extralength}{0cm}
\small
\begin{align}
    \det (J_{\rm CC}) = 1 + \frac{\left[ \begin{aligned} &22 r_0^3 -4 r_0^2 [r_{0x}^2 (95 v_{0x}^2 + 22 v_{0y}^2) + r_{0y}^2 (22 v_{0x}^2 + 95 v_{0y}^2) + 146 r_{0x} r_{0y} v_{0x} v_{0y}] \\
    &-3 r_0 \left\{ \begin{aligned} &r_{0x}^4 (596 v_{0x}^4 - 386 v_{0x}^2 v_{0y}^2 - 37 v_{0y}^4) + r_{0y}^4 (-37 v_{0x}^4 - 386 v_{0x}^2 v_{0y}^2 + 596 v_{0y}^4) \\
    &- 2 r_{0x}^2 r_{0y}^2 (193 v_{0x}^4 - 2449 v_{0x}^2 v_{0y}^2 + 193 v_{0y}^4) \\
    &+ 12 r_{0x} r_{0y} v_{0x} v_{0y} [r_{0y}^2 (-52 v_{0x}^2 + 263 v_{0y}^2) + r_{0x}^2 (263 v_{0x}^2 - 52 v_{0y}^2)] \end{aligned} \right\} \\
    &-45 \left\{ \begin{aligned} &r_{0x}^6 (16 v_{0x}^6 - 72 v_{0x}^4 v_{0y}^2 + 18 v_{0x}^2 v_{0y}^4 + v_{0y}^6) + r_{0y}^6 (v_{0x}^6 + 18 v_{0x}^4 v_{0y}^2 - 72 v_{0x}^2 v_{0y}^4 + 16 v_{0y}^6) \\
    &+ 30 r_{0x} r_{0y} v_{0x} v_{0y} \left[ \begin{aligned} &r_{0x}^4 (8 v_{0x}^4 - 12 v_{0x}^2 v_{0y}^2 + v_{0y}^4) + r_{0y}^4 (v_{0x}^4 - 12 v_{0x}^2 v_{0y}^2 + 8 v_{0y}^4) \\ &- 2 r_{0x}^2 r_{0y}^2 (6 v_{0x}^4 - 23 v_{0x}^2 v_{0y}^2 + 6 v_{0y}^4)\end{aligned} \right] \\
    &- 3 r_{0x}^2 r_{0y}^2 \left[ \begin{aligned} &r_{0x}^2 (24 v_{0x}^6 - 308 v_{0x}^4 v_{0y}^2 + 187 v_{0x}^2 v_{0y}^4 - 6 v_{0y}^6) \\ &+ r_{0y}^2 (-6 v_{0x}^6 + 187 v_{0x}^4 v_{0y}^2 - 308 v_{0x}^2 v_{0y}^4 + 24 v_{0y}^6) \end{aligned} \right] \end{aligned} \right\} \end{aligned} \right]}{720 r_0^{11} ({\boldsymbol r}_0 \cdot{\boldsymbol v}_0)^2}h^6 + \cdots,
\end{align}
\end{adjustwidth}
i.e., the non-simplecticity has been reduced from $\mathcal{O}(h^5)$ (see Equation~(\ref{eq:2BP_CE1_Jac_det})) to $\mathcal{O}(h^6)$. However, it blows up when ${\boldsymbol r}_0 \cdot{\boldsymbol v}_0 = 0$, in other words, when ${\boldsymbol r}_0$ and ${\boldsymbol v}_0$ are perpendicular to each other.

Then we take another look at test case 1 (of Section~\ref{ss:2BP_CE1A}): uniform circular motion. Plugging ${\boldsymbol r}_h$ given by Equation~(\ref{eq:2BP_CE1_UCM}) into Equation~(\ref{eq:2BP_vxy_rxy}), we obtain
\begin{align}
  {\boldsymbol v}_{h,{\rm CC}} &= \begin{pmatrix} -\left[ h - \dfrac{h^3}{6} + \dfrac{h^5}{120} - {\color{red} \dfrac{1119}{14}} \cdot \dfrac{h^7}{5040} + \mathcal{O}(h^9) \right] \\ 1 - \dfrac{h^2}{2} + \dfrac{h^4}{24} - {\color{red} 2} \cdot \dfrac{h^6}{720} + \mathcal{O}(h^8) \end{pmatrix}.
\end{align}
The fractional error of fifth-order coefficient has been eliminated, as expected; those of sixth and seventh (highlighted in red) have been reduced as well.
Equation~(\ref{eq:2BP_EandLz}) tells us that the discrepancies in conserved quantities are ameliorated as well
\vspace{-10pt}
\begin{adjustwidth}{-\extralength}{0cm}
\begin{align}
    \left\{ \begin{aligned} E_h &= - \frac{1}{2} - \frac{13}{240}h^6 + \frac{2857}{100800}h^8 + \mathcal{O}(h^9) \\ L_{zh} &= 1 - \frac{13}{240}h^6 + \frac{2857}{100800}h^8 + \mathcal{O}(h^9) \end{aligned} \right. \quad \Rightarrow \quad
    \left\{ \begin{aligned} E_{h,{\rm CC}} &= - \frac{1}{2} - \frac{2669}{100800}h^8 + \mathcal{O}(h^9) \\ L_{zh,{\rm CC}} &= 1 - \frac{2669}{100800}h^8 + \mathcal{O}(h^9) \end{aligned} \right.,
\end{align}
\end{adjustwidth}
i.e., deviations from conservation laws have been reduced by two orders. Note that the $\mathcal{O}(h^8)$ errors may arise from truncation, since we only kept up to $\mathcal{O}(h^7)$ terms in Section~\ref{ss:2BP_CE1A}.
Interestingly, errors in these two quantities are the same in both cases (before and after correction). We emphasize that, since $E$ and $L_z$ are derived from initial conditions, errors of the ``CC'' version do not accumulate.

In test case 2: parabolic motion, the corrected velocity vector {is}
\vspace{-12pt}
\begin{adjustwidth}{-\extralength}{0cm}
\begin{align}
    {\boldsymbol v}_{h,{\rm CC}} &= \begin{pmatrix} -\dfrac{1}{\sqrt{2}} - \dfrac{h}{4} - \dfrac{h^2}{8\sqrt{2}} - \dfrac{5h^3}{192} - \dfrac{5h^4}{768\sqrt{2}} + \dfrac{7h^5}{6144} + {\color{red} \dfrac{13}{10}} \cdot \dfrac{91h^6}{36864\sqrt{2}} + {\color{red} \dfrac{1306}{1705}} \cdot \dfrac{341h^7}{294912} + \mathcal{O}(h^8) \\
    \dfrac{1}{\sqrt{2}} - \dfrac{h^2}{16\sqrt{2}} - \dfrac{h^3}{32} - \dfrac{35h^4}{1536\sqrt{2}} - \dfrac{7h^5}{1024} - {\color{red} \dfrac{8}{7}} \cdot \dfrac{245h^6}{73728\sqrt{2}} - {\color{red} \dfrac{11059}{5670}} \cdot \dfrac{9h^7}{16384} + \mathcal{O}(h^8) \end{pmatrix};
\end{align}
\end{adjustwidth}
the mechanic energy and the angular momentum before and after conservation correction~are
\vspace{-12pt}
\begin{adjustwidth}{-\extralength}{0cm}
\small
\begin{align}
    \left\{ \begin{aligned} E_h &= \frac{767 h^5}{92160 \sqrt{2}} + \frac{1891 h^6}{737280} + \frac{7759 h^7}{6193152 \sqrt{2}} + \frac{25337 h^8}{55050240} + \mathcal{O}(h^9) \\ L_{zh} &= \sqrt{2} + \frac{107 h^5}{7680} + \frac{113 h^6}{61440 \sqrt{2}} - \frac{223 h^7}{368640} - \frac{1961 h^8}{8257536 \sqrt{2}} + \mathcal{O}(h^9) \end{aligned} \right. \quad \Rightarrow \quad
    \left\{ \begin{aligned} E_{h,{\rm CC}} &= \frac{99077 h^8}{165150720} + \mathcal{O}(h^9) \\ L_{zh,{\rm CC}} &= \sqrt{2} + \frac{8045 h^8}{8257536 \sqrt{2}} + \mathcal{O}(h^9) \end{aligned} \right..
\end{align}
\end{adjustwidth}
respectively. The situation is basically the same as test case 1, except that the reduction in $E$ and $L_z$ errors is three orders in this case.

As indicated by Section~\ref{ss:CHO_RK}, for Runge--Kutta methods, errors in ${\boldsymbol r}_h$ and ${\boldsymbol v}_h$ have the same order, hence it is probably not well-motivated to use one to correct another; nevertheless, the correction described in this section should still be able to produce better~conservation.

\paragraph{{Approach 2: Enforce conservation laws in the formalism.}}
Alternatively, we can try to enforce conservation of machanic energy and angular momentum in the ContEvol formalism.

Plugging our polynomial approximation Equation~(\ref{eq:2BP_CE1_rvh}) into and expanding Equation~(\ref{eq:2BP_vxy_rxy}), we obtain
\begin{align}
    {\boldsymbol v}(t) = {\boldsymbol v}_0 - \frac{{\boldsymbol r}_0}{r_0^3}t + \mathcal{O}(t^2) 
    = {\boldsymbol v}_0 + 2{\boldsymbol B}t + \mathcal{O}(t^2) \quad \Rightarrow \quad
    {\boldsymbol B} = - \frac{{\boldsymbol r}_0}{2 r_0^3};
\end{align}
further expansion (based on ${\boldsymbol B}$ found above) yields
\begin{align}
    \nonumber {\boldsymbol v}(t) &= {\boldsymbol v}_0 - \frac{{\boldsymbol r}_0}{r_0^3}t + \frac{1}{2r_0^5} \begin{pmatrix} (2r_{0x}^2-r_{0y}^2)v_{0x} + 3r_{0x}r_{0y}v_{0y} \\ (2r_{0y}^2-r_{0x}^2)v_{0y} + 3r_{0x}r_{0y}v_{0x} \end{pmatrix}t^2 + \mathcal{O}(t^3) \\
    &= {\boldsymbol v}_0 + 2{\boldsymbol B}t + 3{\boldsymbol A}t^2 + \mathcal{O}(t^3) \quad \Rightarrow \quad
    {\boldsymbol A} = \frac{1}{6r_0^5} \begin{pmatrix} (2r_{0x}^2-r_{0y}^2)v_{0x} + 3r_{0x}r_{0y}v_{0y} \\ (2r_{0y}^2-r_{0x}^2)v_{0y} + 3r_{0x}r_{0y}v_{0x} \end{pmatrix}.
\end{align}
In short, the ${\boldsymbol A}$ and ${\boldsymbol B}$ coefficients determined in this way are simply zeroth-order terms of Equation~(\ref{eq:2BP_CE1_ABopt}), which are not very useful.
Therefore, in the context of ContEvol, conservation laws are better used for correction purposes.

To conclude this section, we briefly comment on how conservation of mechanic energy and angular momentum can be used in more realistic cases.
\begin{itemize}
    \item In galactic dynamics, when the matter distribution is axisymmetric, e.g., in the cases of some disk or elliptical galaxies, the situation is very similar to the two-body problem we consider here.
    Although both position and velocity of the particle are three-dimensional vectors now, mechanic energy and $z$ component of angular momentum are still conserved.
    Hence we can use these two constraints to correct $v_x$ and $v_y$ using all components of ${\boldsymbol r}$ and $v_z$---note that $r_z$ and $v_z$ do not appear in the expression of $L_z$, and are usually significantly smaller (in terms of absolute values) than their counterparts in $x$ and $y$ directions.

    \item In general relativity, mechanic energy and angular momentum are conserved at the $\mathcal{O}(c^{-4})$ level ($c$ is the speed of light in vacuum), before gravitational waves enter the scene.
    Thenceforth, while studying orbital motion of a planet around a star (e.g., Mercury around the Sun) or a star around a supermassive black hole using ContEvol (or another method which lead to different orders in position and velocity), conservation correction may also be useful.

    \item Back to Newtonian gravity. For a general three-body problem (see Section~\ref{ss:3BP_CE1} for further discussion), there are twelve components in total (two particles, positions and velocities, three directions) and four conserved quantities (total mechanic energy and three components of total angular momentum).
    Therefore, especially in almost coplanar cases, we can use $\{ {\boldsymbol r}_i \}$ and $\{ v_{iz} \}$ to correct $\{ v_{ix} \}$ and $\{ v_{iy} \}$, where $i = 1, 2$ is the index of particle; in a restricted three-body problem, where one of the particles is much less massive than the others, we can choose a different set of four velocity~components.

    \item For a general $n$-body problem, there are $6(n-1)$ components in total, but the number of conserved quantities are still four. Consequently, conservation laws become less and less useful as the number of particles increases.
    However, they are probably useful in hierarchical systems where we can still identify ``important'' velocity components. Further discussion on this topic is beyond the scope of this work.
\end{itemize}

The above discussion is only about the conservation laws per se.
Since the ContEvol formalism promises to ``recover'' full evolutionary histories, when mechanic energy and angular momentum are not conserved for individual objects, in principle it allows users to perform corrections using energy-work and angular impulse-momentum theorems.
To go one step further, if global sums of $E$ or ${\boldsymbol L}$ components (all of which should be conserved) obtained via these theorems deviate from the initial values, it is reasonable to globally rescale such sums before correcting individual quantities.
However, such corrections are computationally expensive, and are only recommended when conservation laws are crucial.

\subsection{Two-Body, Numerical Tests with an Eccentric Elliptical Orbit}
\label{ss:2BP_CE1N}

In this section, we conduct numerical experiments to compare first-order ContEvol with some other low-order methods for celestial mechanics. We choose a highly eccentric elliptical orbit for testing purposes.

Specifically, this elliptical orbit has eccentricity $e$, semi-major axis $a$, semi-minor axis $b = a \sqrt{1-e^2}$, and focal distance $c = a e$. We write the equation of this ellipse as 
\begin{align}
    \frac{(x-c)^2}{a^2} + \frac{y^2}{b^2} = 0,
\end{align}
so that location of the ``central object,'' i.e., origin of our coordinate system $(0, 0)^{\rm T}$, is at the right focus. The mechanic energy of this orbit is (subscript ``M'' stands for mechanic and is added to distinguish energy from eccentric anomaly)
\vspace{-8pt}
\begin{align}
    E_{\rm M} = - \frac{1}{2a},
\end{align}
while the orbital period is given by Kepler's third law
\begin{align}
    T = 2\pi a^{3/2}.
\end{align}

We let the particle start at the pericenter $(a(1-e), 0)^{\rm T}$ and move counter-clockwise. The vis-viva equation tells us the initial speed
\begin{align}
    v_{0} = \sqrt{\frac{2}{a(1-e)} - \frac{1}{a}} = \sqrt{\frac{1}{a} \frac{1+e}{1-e}},
\end{align}
so that the initial velocity is $(0, v_0)^{\rm T}$, and thus ($z$ component of) the angular momentum is
\begin{align}
    L_z = r_0 v_0 = \sqrt{a (1-e^2)}.
\end{align}

At time $t$, the position of our particle is given by
\begin{align}
    \label{eq:eccentric_exact} {\boldsymbol r}(t) = \begin{pmatrix} a (\cos E - e) \\ b \sin E \end{pmatrix},
\end{align}
where the eccentric anomaly $E$ is related to the mean anomaly
\begin{align}
    M = \frac{2\pi}{T} t = \frac{t}{a^{3/2}}
\end{align}
by Kepler's equation
\begin{align}
    \label{eq:Kepler} M = E - e \sin E,
\end{align}
which is a transcendental equation and has to be solved numerically.

The velocity can be obtained via Equation~(\ref{eq:2BP_vxy_rxy}) or expressed as
\begin{align}
    {\boldsymbol v} = \dot{\boldsymbol r} = \begin{pmatrix} - a \sin E \\ b \cos E \end{pmatrix} \dot E.
\end{align}
Ergo we have
\vspace{-8pt}
\begin{align}
    \nonumber {\boldsymbol r} \cdot {\boldsymbol v} &= \begin{pmatrix} a (\cos E - e) \\ b \sin E \end{pmatrix}^{\rm T} \begin{pmatrix} - a \sin E \\ b \cos E \end{pmatrix} \dot E \\
    \nonumber &= a^2 [- (\cos E - e) \sin E + (1-e^2) \cos E \sin E] \dot E \\
    &= a^2 e (1 - e \cos E) \sin E \dot E;
\end{align}
since $1 - e \cos E \geq 1 - e > 0$ and $\dot E > 0$, ${\boldsymbol r} \cdot {\boldsymbol v}$ always has the same sign as $\sin E$, or equivalently $r_y$.
This relation will be used for conservation correction (see Section~\ref{ss:2BP_CE1C}), since this section is dedicated to testing numerical methods, not the sign determination strategy.

For numerical tests in this work, we choose the following orbital parameters:
\begin{itemize}
    \item Eccentricity $e = 63/64 \approx 0.9844$, semi-major axis $a = 16$, semi-minor axis \linebreak  $b = \sqrt{127}/4 \approx 2.817$, and focal distance $c = 63/4 = 15.75$.
    \item Orbital period $T = 128\pi \approx 402.1$, mechanic energy $E_{\rm M} = -1/32 = -0.03125$, and angular momentum $L_z = \sqrt{127}/16 \approx 0.7043$.
    \item Pericenter at ${\boldsymbol r}_p = (1/4, 0)^{\rm T}$, where the velocity is ${\boldsymbol v}_p = (0, \sqrt{127}/4)^{\rm T} \approx (0, 2.817)^{\rm T}$; apocenter at ${\boldsymbol r}_a = (-127/4, 0)^{\rm T} = (-31.75, 0)^{\rm T}$, where the velocity is \linebreak  ${\boldsymbol v}_a = (0, 1/(4\sqrt{127}))^{\rm T} \approx (0, -0.02218)^{\rm T}$.
\end{itemize}

Meanwhile, technical choices include:
\begin{itemize}
    \item Numerical methods: leapfrog integrator (which is simple but simplectic), fourth-order Runge--Kutta, and first-order ContEvol methods, without and with conservation correction. Note that all these methods have higher-order counterparts.
    \item Total duration $t_{\max} = 432$; four fixed time steps: $h = 1/16 = 0.0625$, $h = 1/64 \approx 0.0156$, $h = 1/256 \approx 0.0039$, and $h = 1/1024 \approx 0.0010$. For the two $h < 1/64$ cases, we only record position and velocity every $\Delta t = 1/64$.
    \item Programming language: Python with just-in-time compilation (see data availability).
    Processor information: 11th Gen Intel(R) Core(TM) i7-1165G7@2.80 GHz, 2803 Mhz, 4~Core(s), 8 Logical Processor(s). We do not use multiprocessing explicitly.
\end{itemize}

\begin{figure}[H]
       \includegraphics[width=\textwidth]{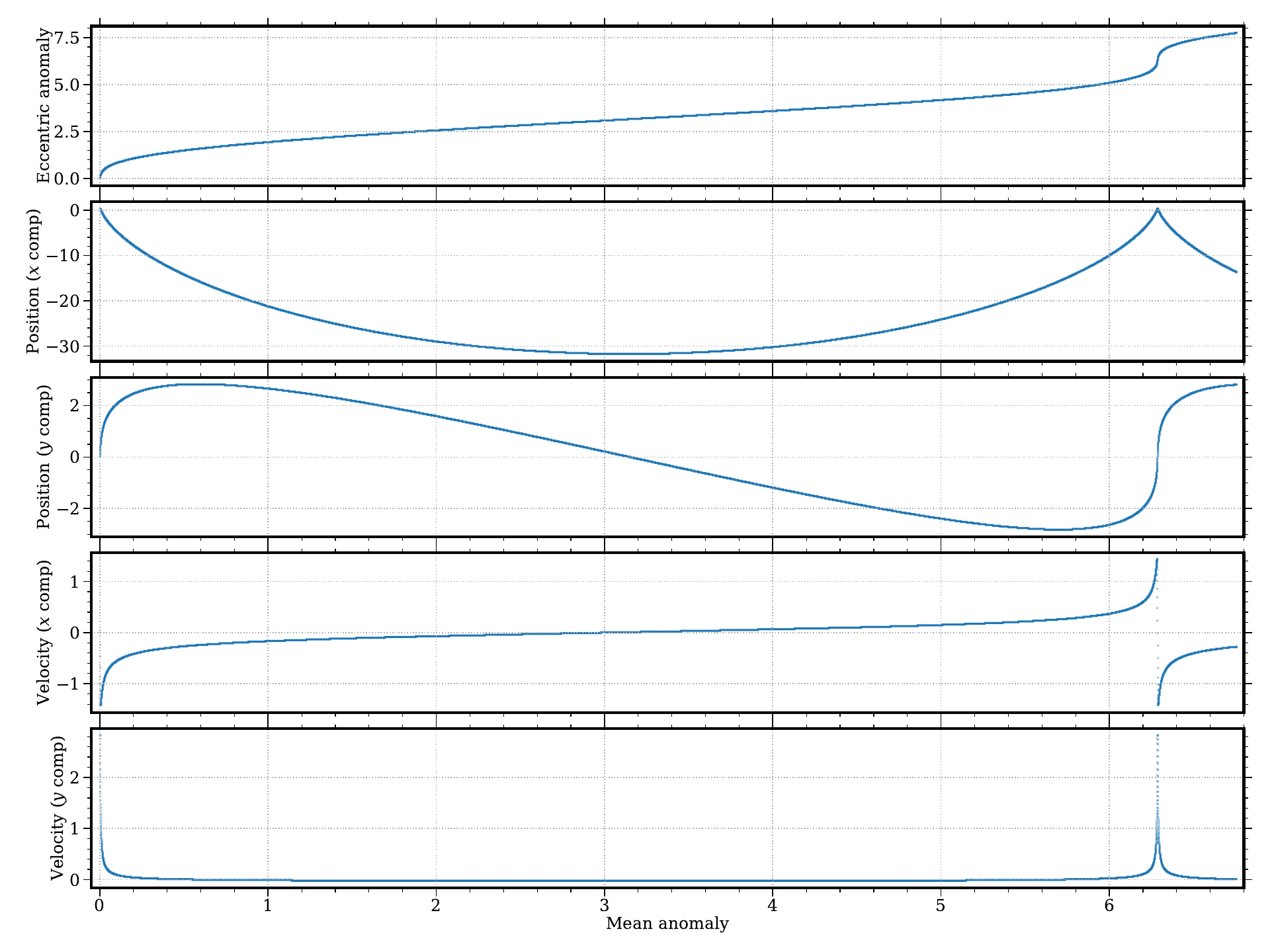}
    \caption{{Exact} solution to the eccentric orbit specified in Section~\ref{ss:2BP_CE1N}. From (\textbf{{top}}) to (\textbf{{bottom}}), plotted versus mean anomaly $M$ are eccentric anomaly $E$, position ${\boldsymbol r}$ ($x$ and $y$ components) and velocity ${\boldsymbol v}$ ($x$ and $y$ components).}
    \label{fig:eccentric_exact}
\end{figure}

Figure~\ref{fig:eccentric_exact} displays the exact solution of this scenario.
Since the particle reaches its maximum speed at the pericenter, both its position and velocity change rapidly near $M = 0$ and $M = 2\pi$.
In the case of velocity near $M = 2\pi$, the $x$ component reaches its maximum and quickly flips its sign, while the $y$ component reaches a larger maximum and quickly falls back.
These rapid changes constitute a ``stress test'' for the numerical methods.
\vspace{-8pt}

Table~\ref{tab:eccentric} presents the time consumption of each configuration (integrator, conservation correction, and time step) tested in this work.
Since the time step is fixed in each case, the time consumption is roughly inversely proportional to the time step, as expected.
As a second-order method, leapfrog integrator is $\sim$3 times faster than fourth-order Runge--Kutta; for these two methods, conservation correction increases the time consumption by a significant fraction---despite the simplicity of Equation~(\ref{eq:2BP_vxy_rxy}), it still takes time to perform floating point operations.
Without conservation correction, first-order ContEvol costs about one half more time than fourth-order Runge--Kutta; with correction, it becomes slightly faster, since calculating ${\boldsymbol v}_h$ from Equation~(\ref{eq:2BP_vxy_rxy}) is simpler than from Equation~(\ref{eq:2BP_CE1_Taylor}).
In principle, this trick can be applied to Runge--Kutta as well, but we have not explored this possibility in this work, since it would encounter more overhead and an acceleration is not~guaranteed.

\begin{table}[H]
  \caption{Time consumption of leapfrog (``LF''), fourth-order Runge--Kutta (``RK4''), and first-order ContEvol (``CE1'') integrators, without and with conservation correction (``CC''), all for the configuration specified in Section~\ref{ss:2BP_CE1N}. All quotes are obtained using the {\sc timeit} standard library of {Python~3.11}.}
    \label{tab:eccentric}
   \begin{adjustwidth}{-\extralength}{0cm}
  \begin{tabularx}{\fulllength }{m{2cm}<{\centering}m{3.5cm}<{\centering}m{3.5cm}<{\centering}m{3.5cm}<{\centering}m{3.5cm}<{\centering}}
   \toprule
        \textbf{Integrator} & \boldmath{$h = 1/16$} & \boldmath{$h = 1/64$} & \boldmath{$h = 1/256$} & \boldmath{$h = 1/1024$} \\
 \midrule
        LF & 2.58 ms $\pm$ 62.2 {{$\upmu$}s }
& 11.3 ms $\pm$ 999 {$\upmu$}s & 32.8 ms $\pm$ 1.65 ms & 133 ms $\pm$ 1.10 ms \\
        LFCC & 3.16 ms $\pm$ 53.6 {$\upmu$}s & 13.2 ms $\pm$ 2.07 ms & 47.3 ms $\pm$ 688 {$\upmu$}s & 210 ms $\pm$ 9.66 ms \\
        RK4 & 7.99 ms $\pm$ 397 {$\upmu$}s & 31.7 ms $\pm$ 1.37 ms & 131 ms $\pm$ 5.65 ms & 514 ms $\pm$ 29.2 ms \\
        RK4CC & 8.97 ms $\pm$ 467 {$\upmu$}s & 39.2 ms $\pm$ 1.49 ms & 146 ms $\pm$ 9.60 ms & 631 ms $\pm$ 26.7 ms \\
        CE1 & 11.6 ms $\pm$ 443 {$\upmu$}s & 45.5 ms $\pm$ 365 {$\upmu$}s & 193 ms $\pm$ 8.94 ms & 734 ms $\pm$ 27.4 ms \\
        CE1CC & 11.3 ms $\pm$ 636 {$\upmu$}s & 45.0 ms $\pm$ 2.19 ms & 181 ms $\pm$ 9.56 ms & 718 ms $\pm$ 36.0 ms \\
  \bottomrule
    \end{tabularx}
\end{adjustwidth}
  \end{table}

\begin{figure}[H]

\begin{adjustwidth}{-\extralength}{0cm}
\centering 
    \includegraphics[width=1.2\textwidth]{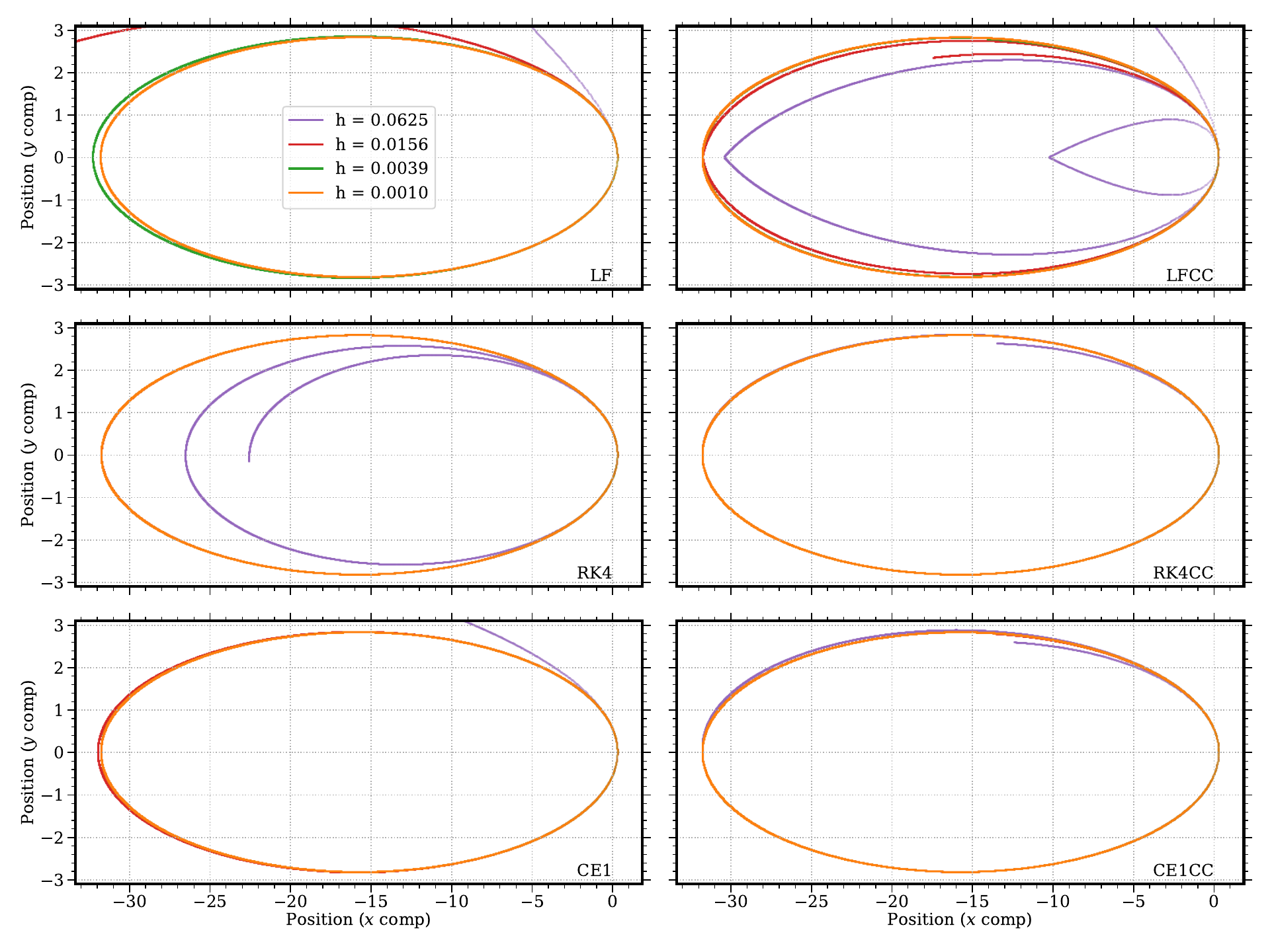}
\end{adjustwidth}
    \caption{{Orbit} predicted by leapfrog (``LF''; {\bf top row}), fourth-order Runge--Kutta (``RK4''; {\bf middle row}), and first-order ContEvol (``CE1''; {\bf bottom row}) integrators, without ({\bf left column}) and with ({\bf right column}) conservation correction (``CC''), all based on initial conditions specified in Section~\ref{ss:2BP_CE1N}.
    For each integrator, $h=1/16$ (``tab:purple''), $h=1/64$ (``tab:red''), $h=1/256$ (``tab:green''), and $h=1/1024$ (``tab:orange'') results are shown in different colors.}
    \label{fig:eccentric_orbits}
\end{figure}

\textls[-10]{Figure~\ref{fig:eccentric_orbits} shows orbits predicted by configurations tested in this work. Those close to the exact solution Equation~(\ref{eq:eccentric_exact}), e.g., $h=1/1024$ ellipses, will be further investigated in the next few paragraphs; here we comment on significantly deviatory ones.
Without conservation correction, leapfrog integrator produces a hyperbolic trajectory with $h=1/16$, and a significantly larger and incomplete ellipse with $h=1/64$---it only finishes slightly over half a cycle at our terminal time, $t_{\max} = 432$.
With conservation correction, the $h=1/16$ leapfrog orbit involves more artifacts, featuring two teardrop-shaped laps with different size, and then a segment of probably the third one---apparently, the correction permanently alters the history by suddenly changing the sign of $v_x$; however, the $h=1/64$ did become more reasonable.
Because of their higher-order precision, fourth-order \mbox{Runge--Kutta} and first-order ContEvol integrators only show substantial deviations when $h=1/16$.
Without conservation correction, the Runge--Kutta orbit ``loses'' energy and shrinks, while its ContEvol counterpart ``gains'' energy and leaves the ``central object''.
With conservation correction, both orbits slightly flattens in the second lap, possibly due to artifacts induced by the correction, although these artifacts are less noticeable than in the case of leapfrog.}
\vspace{-6pt}

\paragraph{{Method 1: Leapfrog integrator.}
}

\begin{figure}[H]

\begin{adjustwidth}{-\extralength}{0cm}
\centering 
    \includegraphics[width=1.2\textwidth]{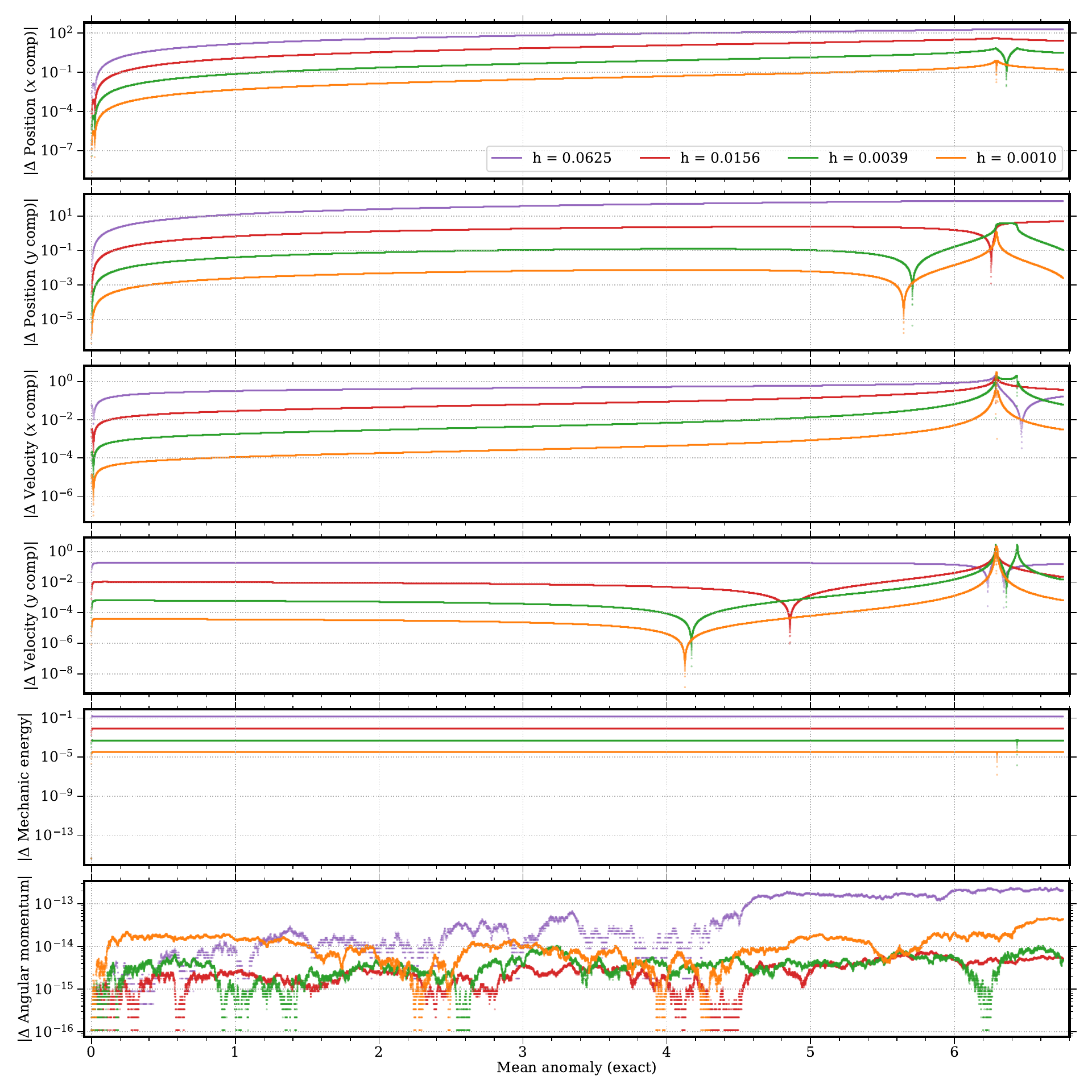}
\end{adjustwidth}
    \caption{Deviation from exact solution to the eccentric orbit specified in Section~\ref{ss:2BP_CE1N} of prediction by leapfrog (``LF'') integrator without conservation correction (``CC'').
    From ({\bf top}) to ({\bf bottom}), plotted versus mean anomaly $M$ (exact, proportional to time) are absolute errors in position ${\boldsymbol r}$ ($x$ and $y$ components), velocity ${\boldsymbol v}$ ($x$ and $y$ components), mechanic energy $E_{\rm M}$, and angular momentum $L_z$.
    In each panel, different time steps are shown in different colors; the mapping is the same as in Figure~\ref{fig:eccentric_orbits}.}
    \label{fig:eccentric_LF}
\end{figure}

\begin{figure}[H]

\begin{adjustwidth}{-\extralength}{0cm}
\centering 
    \includegraphics[width=1.2\textwidth]{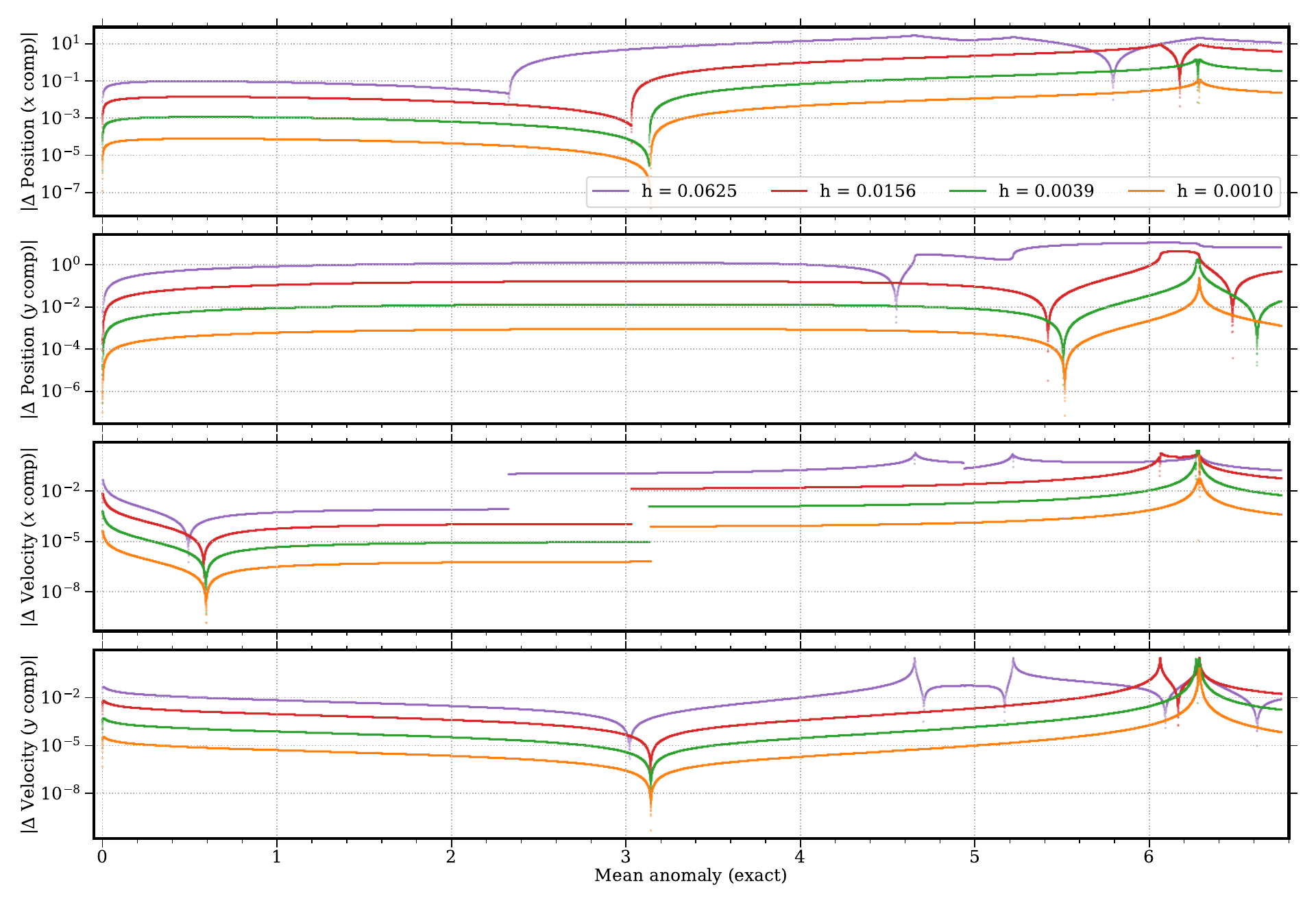}
\end{adjustwidth}
    \caption{Deviation from exact solution to the eccentric orbit specified in Section~\ref{ss:2BP_CE1N} of prediction by leapfrog (``LF'') integrator with conservation correction (``CC'').
    From ({\bf top}) to ({\bf bottom}), plotted versus mean anomaly $M$ (exact, proportional to time) are absolute errors in position ${\boldsymbol r}$ ($x$ and $y$ components) and velocity ${\boldsymbol v}$ ($x$ and $y$ components).
    In each panel, different time steps are shown in different colors; the mapping is the same as in Figure~\ref{fig:eccentric_orbits}.}
    \label{fig:eccentric_LFCC}
\end{figure}

Figures~\ref{fig:eccentric_LF} and \ref{fig:eccentric_LFCC} display deviations from exact solution of predictions by leapfrog (``LF'') integrator without and with conservation correction (``CC''), respectively.
Thanks to its symplectic nature, leapfrog (without conservation correction) conserves angular momentum remarkably well---better than both ``higher-order'' methods tested in this work---regardless of the time step.
The mechanic energy is also well-conserved, except at the beginning $M = 0$, where the particle gets an ``initial kick,'' of which the magnitude seems proportional to the time step; nevertheless, near $M = 2\pi$, none of the leapfrog orbits gets a ``second kick,'' making leapfrog eligible for studies of long-term (or secular) behaviors of the particle, if the energy discrepancy is acceptable.
Without or with conservation correction, shrinking the time step by a factor of $4$ reduces errors in position and velocity by about an order of magnitude.
However, since the correction breaks simplecticity and causes artifacts when $r_y$ reaches $0$ (most noticeable in the $v_x$ panel of Figure~\ref{fig:eccentric_LFCC}), it only improves leapfrog in the first half of the first lap.

\vspace{-6pt}

\paragraph{{Method 2: Fourth-order Runge--Kutta.}}

\begin{figure}[H]
    \includegraphics[width=0.9\textwidth]{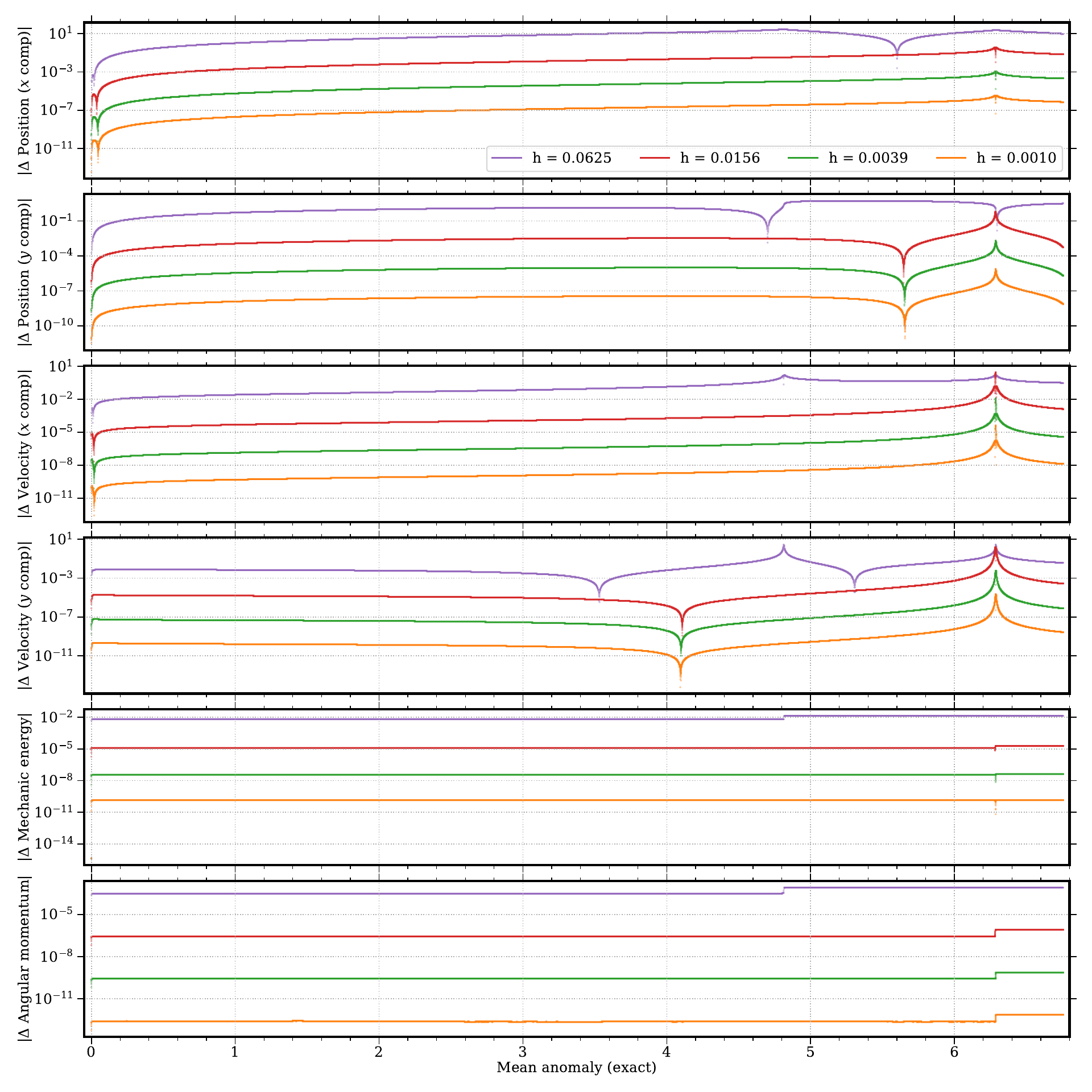}
    \caption{Deviation from exact solution to the eccentric orbit specified in Section~\ref{ss:2BP_CE1N} of prediction by fourth-order Runge--Kutta (``RK4'') integrator without conservation correction (``CC'').
    Panels and colors are same as in Figure~\ref{fig:eccentric_LF}.}
    \label{fig:eccentric_RK4}
\end{figure}

\begin{figure}[H]
    \includegraphics[width=0.9\textwidth]{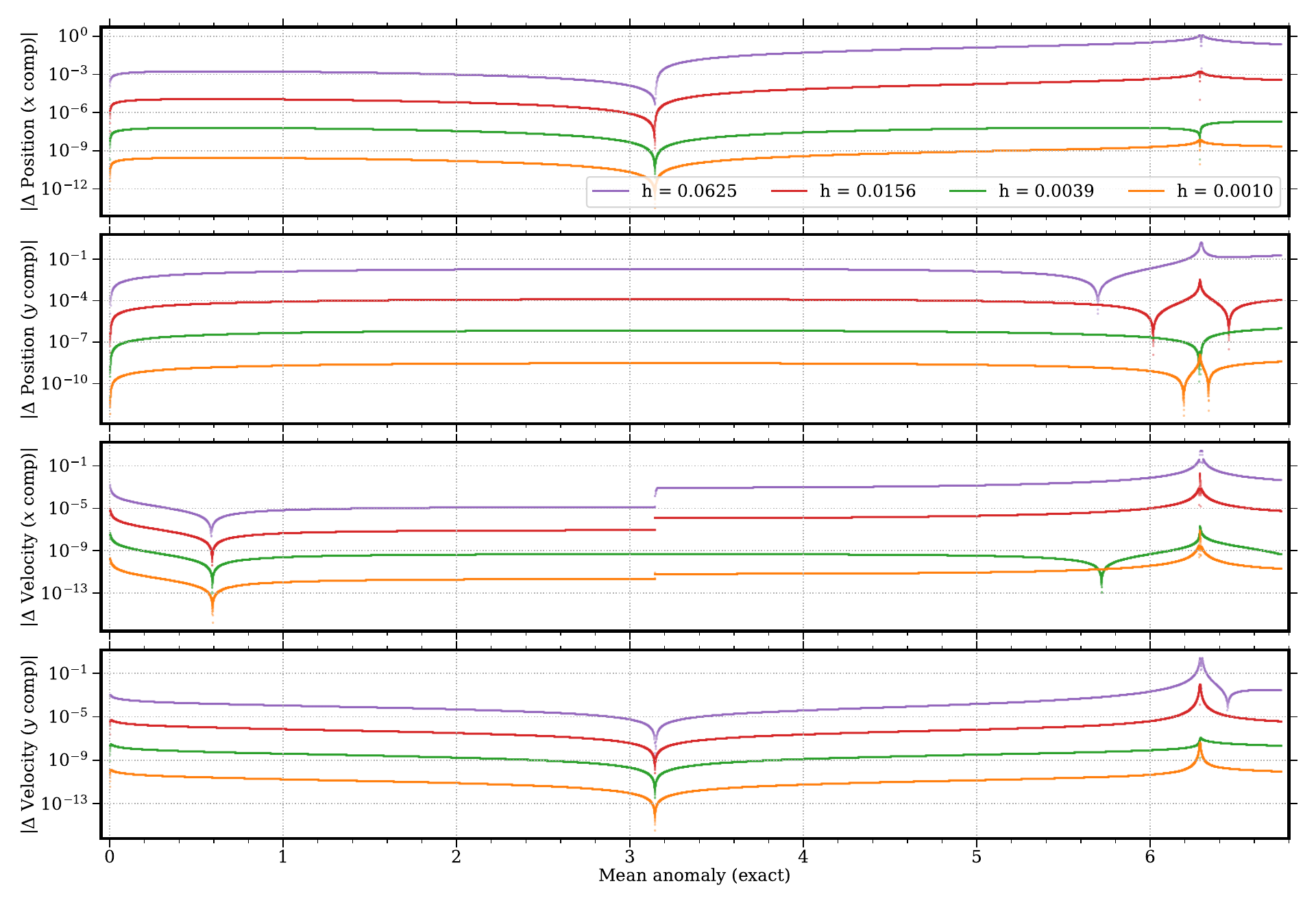}
    \caption{Deviation from exact solution to the eccentric orbit specified in Section~\ref{ss:2BP_CE1N} of prediction by fourth-order Runge--Kutta (``RK4'') integrator with conservation correction (``CC'').
    Panels and colors are same as in Figure~\ref{fig:eccentric_LFCC}.}
    \label{fig:eccentric_RK4CC}
\end{figure}

Figures~\ref{fig:eccentric_RK4} and \ref{fig:eccentric_RK4CC} display deviations from exact solution of predictions by fourth-order Runge--Kutta (``RK4'') integrator without and with conservation correction (``CC''), respectively.
As a higher-order method, Runge--Kutta (without conservation correction) significantly reduces the ``initial kick'' (in terms of mechanic energy and angular momentum) the particle gets at $M = 0$; however, the particle does get a ``second kick'' near $M = 2\pi$, of which the amplitude shrinks with time step for mechanic energy, but is constantly about half an order of magnitude for angular momentum regardless of the time step.
Therefore, quality of Runge--Kutta predictions possibly deteriorates after several laps; yet for the first lap, shrinking the time step by $4$ reduces errors by almost three (two and a half) orders of magnitude without (with) conservation correction, which is much better than leapfrog.
In the first half of the first lap, with $h=1/1024$, conservation correction improves Runge--Kutta by nearly three orders of magnitude in terms of $x$ components, and almost an order of magnitude in terms of $y$ components. Because of different scaling relations described above, these improvements are slightly more significant for larger time steps; due to roundoff errors, time steps smaller than $1/1024$ probably do not make much sense.
However, a closer look at the $v_x$ panel of Figure~\ref{fig:eccentric_RK4CC} would reveal a slight jump near $M = \pi$, which is an artifact of the correction.

\vspace{-12pt}

\paragraph{{Method 3: First-order ContEvol.}}

\begin{figure}[H]
    \includegraphics[width=0.95\textwidth]{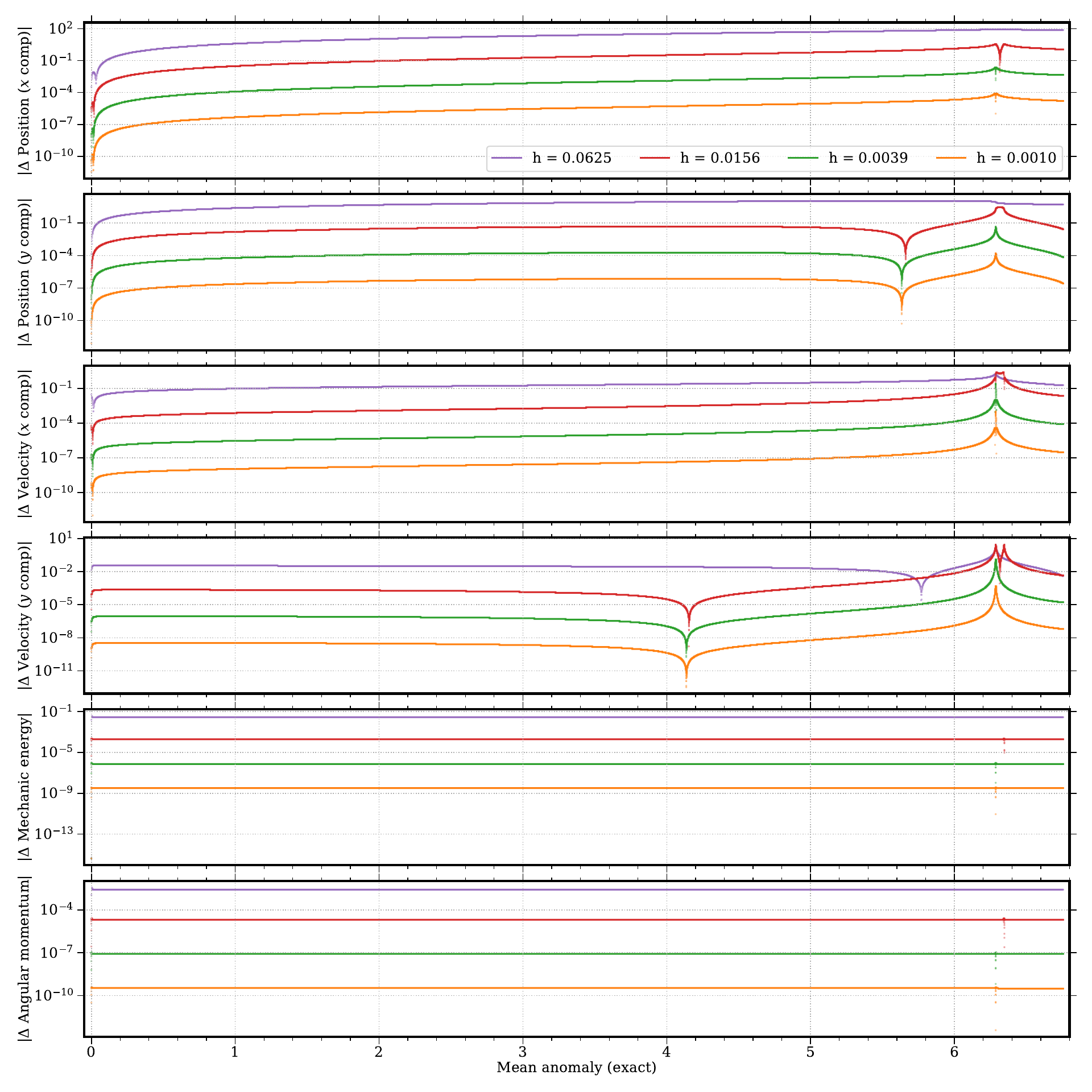}
    \caption{Deviation from exact solution to the eccentric orbit specified in Section~\ref{ss:2BP_CE1N} of prediction by first-order ContEvol (``CE1'') integrator without conservation correction (``CC'').
    Panels and colors are same as in Figure~\ref{fig:eccentric_LF}.}
    \label{fig:eccentric_CE1}
\end{figure}

\begin{figure}[H]
    \includegraphics[width=\textwidth]{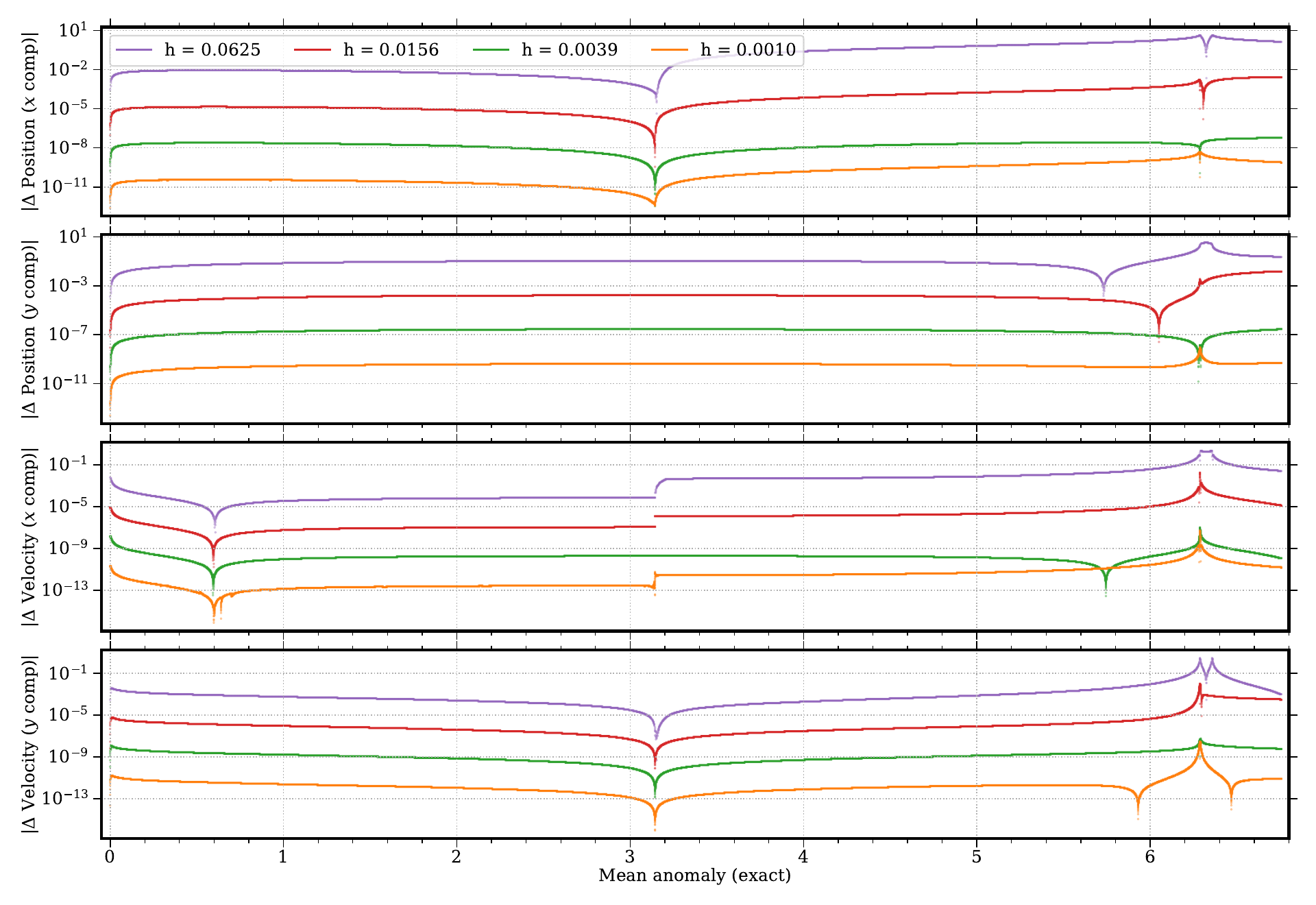}
    \caption{Deviation from exact solution to the eccentric orbit specified in Section~\ref{ss:2BP_CE1N} of prediction by first-order ContEvol (``CE1'') integrator with conservation correction (``CC'').
    Panels and colors are same as in Figure~\ref{fig:eccentric_LFCC}.}
    \label{fig:eccentric_CE1CC}
\end{figure}

Figures~\ref{fig:eccentric_CE1} and \ref{fig:eccentric_CE1CC} display deviations from exact solution of predictions by first-order ContEvol (``CE1'') integrator without and with conservation correction (``CC''), respectively.
Without conservation correction, ContEvol does not perform as well as Runge--Kutta for the first lap---the ``initial kick'' is almost an order of magnitude larger in terms of mechanic energy, and up to three orders of magnitude in term of angular momentum; errors in position and velocity are also about an order of magnitude larger.
This is not unexpected, because although ContEvol (as implemented for these tests, see Section~\ref{ss:2BP_CE1L}) accurately traces ${\boldsymbol r}_h$ to $\mathcal{O}(h^5)$, it only traces ${\boldsymbol v}_h$ to $\mathcal{O}(h^4)$, and the higher-order terms are just zero; meanwhile, Runge--Kutta accurately traces both ${\boldsymbol r}_h$ and ${\boldsymbol v}_h$ to $\mathcal{O}(h^4)$, but the $\mathcal{O}(h^5)$ terms could be partially right, hence it performs better when errors accumulate.
Nonetheless, a comparison between $E_{\rm M}$ and $L_z$ panels of Figures~\ref{fig:eccentric_RK4} and \ref{fig:eccentric_CE1} tells us that, thanks to its closeness to simplecticity, ContEvol errors in these two quantities are not amplified at all near $M = 2\pi$, thus it could win out after several laps.
Such possibility is not explored in this work, but we note that the $\mathcal{O}(h^5)$ term of the determinant of the first order ContEvol Jacobian Equation~(\ref{eq:2BP_CE1_Jac_det}) vanishes when ${\boldsymbol r}_0 \cdot {\boldsymbol v}_0 = 0$, which might not have been affected by our truncation (see Section~\ref{ss:2BP_CE1L}).
With conservation correction, ContEvol accurately traces ${\boldsymbol v}_h$ to $\mathcal{O}(h^5)$ as well, therefore it becomes more accurate than its Runge--Kutta counterpart by up to an order of magnitude, especially with smaller time steps.

\vspace{-6pt}

To summarize, with different pros and cons, first-order ContEvol is a viable alternative to classic Runge--Kutta or the symplectic leapfrog integrator, especially for some specific situations or after some further developments.

\subsection{Three-Body, First-Order ContEvol (Description)}
\label{ss:3BP_CE1}

To simplify notation, we follow Equation~(\ref{eq:2BP_CE1_C0123}) to generalize Equation~(\ref{eq:f_rdivr3}) as a series of~functionals
\vspace{-12pt}
\begin{adjustwidth}{-\extralength}{0cm}
\begin{align}
    \label{eq:3BP_CE1_f0123} \left\{ \begin{aligned} {\boldsymbol f}_0 [{\boldsymbol r}(t)] &= \frac{{\boldsymbol r}_0}{r_0^3} \\
    {\boldsymbol f}_1 [{\boldsymbol r}(t)] &= \frac{{\boldsymbol v}_0}{r_0^3} - \frac{3 {\boldsymbol r}_0\cdot{\boldsymbol v}_0}{r_0^5}{\boldsymbol r}_0 \\
    {\boldsymbol f}_2 [{\boldsymbol r}(t)] &= \frac{\boldsymbol B}{r_0^3} - \frac{3{\boldsymbol r}_0\cdot{\boldsymbol v}_0}{r_0^5} {\boldsymbol v}_0 - \frac{3}{2} \left( \frac{2{\boldsymbol B}\cdot{\boldsymbol r}_0 + v_0^2}{r_0^5} - \frac{5 ({\boldsymbol r}_0\cdot{\boldsymbol v}_0)^2}{r_0^7} \right) {\boldsymbol r}_0 \\
    {\boldsymbol f}_3 [{\boldsymbol r}(t)] &= \left[ \begin{aligned} &\frac{\boldsymbol A}{r_0^3} - \frac{3 {\boldsymbol r}_0\cdot{\boldsymbol v}_0}{r_0^5} \boldsymbol B - \frac{3}{2} \left( \frac{2{\boldsymbol B}\cdot{\boldsymbol r}_0 + v_0^2}{r_0^5} - \frac{5 ({\boldsymbol r}_0\cdot{\boldsymbol v}_0)^2}{r_0^7} \right) {\boldsymbol v}_0 \\
    &- \left( \frac{3 ({\boldsymbol A}\cdot{\boldsymbol r}_0 + {\boldsymbol B}\cdot{\boldsymbol v}_0)}{r_0^5} - \frac{15 (2 {\boldsymbol B}\cdot{\boldsymbol r}_0 + v_0^2) ({\boldsymbol r}_0\cdot{\boldsymbol v}_0)}{2 r_0^7} + \frac{35 ({\boldsymbol r}_0\cdot{\boldsymbol v}_0)^3}{2 r_0^9} \right) {\boldsymbol r}_0 \end{aligned} \right] \end{aligned} \right.
\end{align}
\end{adjustwidth}
for any ${\boldsymbol r}(t)$ given or approximated by Equations~(\ref{eq:2BP_CE1_rvh}) and (\ref{eq:2BP_CE1_AB}), so that the (reduced) equations of motion for the three body problem Equation~(\ref{eq:3BP_EOM}) can be written as
\begin{align}
    \left\{ \begin{aligned} \ddot{\boldsymbol r}_1 &= -(1-\mu_2) \left( \sum_{i=0}^3 {\boldsymbol f}_i [{\boldsymbol r}_1] t^i \right) - \mu_2 \left[ \left( \sum_{i=0}^3 {\boldsymbol f}_i [{\boldsymbol r}_2] t^i \right) + \left( \sum_{i=0}^3 {\boldsymbol f}_i [{\boldsymbol r}_1-{\boldsymbol r}_2] t^i \right) \right] \\
    \ddot{\boldsymbol r}_2 &= -(1-\mu_1) \left( \sum_{i=0}^3 {\boldsymbol f}_i [{\boldsymbol r}_2] t^i \right) - \mu_1 \left[ \left( \sum_{i=0}^3 {\boldsymbol f}_i [{\boldsymbol r}_1] t^i \right) + \left( \sum_{i=0}^3 {\boldsymbol f}_i [{\boldsymbol r}_2-{\boldsymbol r}_1] t^i \right) \right] \end{aligned} \right.,
\end{align}
and the cost function can be defined as (subscript ``CE3'' stands for ContEvol and three-body~problem)
\vspace{-6pt}
\begin{align}
    \label{eq:3BP_CE1_cost} \epsilon_{\rm CE3}(\{{\boldsymbol A}_i\}, \{{\boldsymbol B}_i\}; h) = \int_0^h \left[ \begin{aligned} &\left\Vert (2{\boldsymbol B}_1 + 6{\boldsymbol A}_1t) + (1-\mu_2) \left( \sum_{i=0}^3 {\boldsymbol f}_i [{\boldsymbol r}_1] t^i \right) \right. \\ &+ \left. \mu_2 \left[ \left( \sum_{i=0}^3 {\boldsymbol f}_i [{\boldsymbol r}_2] t^i \right) + \left( \sum_{i=0}^3 {\boldsymbol f}_i [{\boldsymbol r}_1-{\boldsymbol r}_2] t^i \right) \right] \right\Vert^2 \\
    &+ \left\Vert (2{\boldsymbol B}_2 + 6{\boldsymbol A}_2t) + (1-\mu_1) \left( \sum_{i=0}^3 {\boldsymbol f}_i [{\boldsymbol r}_2] t^i \right) \right. \\ &+ \left. \mu_1 \left[ \left( \sum_{i=0}^3 {\boldsymbol f}_i [{\boldsymbol r}_1] t^i \right) + \left( \sum_{i=0}^3 {\boldsymbol f}_i [{\boldsymbol r}_2-{\boldsymbol r}_1] t^i \right) \right] \right\Vert^2 \end{aligned} \right] \,{\rm d}t.
\end{align}

We refrain from proceeding with a symbolic analysis of the above cost function in this work, as orders of the discrepancy between determinant of Jacobian and $1$ (which mirrors non-symplecticity), the minimized cost function, and the errors in results at $t=h$ are not expected to be different from those in Sections~\ref{ss:2BP_CE1A} and \ref{ss:2BP_CE1C}.

From a perspective of numerical implementation, we can ``flatten'' the combination of $1$ and all the coefficients to be determined as
\vspace{-12pt}
\begin{adjustwidth}{-\extralength}{0cm}
\begin{align}
    {\boldsymbol x} = (x_0, x_1, x_2, \ldots, x_{12})^{\rm T} \equiv (1, A_{1x}, A_{1y}, A_{1z}, B_{1x}, B_{1y}, B_{1z}, A_{2x}, A_{2y}, A_{2z}, B_{2x}, B_{2y}, B_{2z})^{\rm T},
\end{align}
\end{adjustwidth}
so that the cost function can be succinctly expressed as
\begin{align}
    \epsilon_{\rm CE3}({\boldsymbol x}; h) = \int_0^h w_\alpha \Vert{\mathcal E}_{\alpha ijk} h^i {\boldsymbol e_j} x_k\Vert^2 \,{\rm d}t
\end{align}
with weights $w_\alpha$ (see below for discussion) and the fourth-order tensor ${\mathcal E}_{\alpha ijk}$, wherein $\alpha = 1, 2$ is the index of equation, $i = 0, 1, 2, 3$ is the index of order, $j = x, y, z$ is the index of direction, and $k = 0, 1, 2, \ldots, 12$ is the index of location in the ${\boldsymbol x}$ vector; note that Einstein summation is assumed for all four indices, including $i$ in $h^i$.
All its $2 \times 4 \times 3 \times 13 = 312$ elements can be numerically evaluated using initial conditions and information about the dynamic system, e.g., Equation~(\ref{eq:3BP_CE1_f0123}); many intermediate quantities can be shared between~elements.

Then to minimize the cost function, we have
\begin{align}
    \frac{\partial \epsilon_{\rm CE3}}{\partial x_k} = \frac{\partial^2 \epsilon_{\rm CE3}}{\partial x_{k'} \partial x_k} x_{k'} \equiv {\boldsymbol M}_k \cdot {\boldsymbol x} = 0, \quad k = 1, 2, \ldots, 12,
\end{align}
where the vectors ${\boldsymbol M}_k$ can be derived from the tensor ${\mathcal E}_{\alpha ijk}$, and all their elements are guaranteed to be constants; put in a matrix form, this system of equations is
\begin{align}
    \begin{pmatrix} M_{11} & M_{12} & \cdots & M_{1,12} \\ M_{21} & M_{22} & \cdots & M_{2,12} \\ \vdots & \vdots & \ddots & \vdots \\ M_{12,1} & M_{12,2} & \cdots & M_{12,12} \end{pmatrix}
    \begin{pmatrix} x_1 \\ x_2 \\ \vdots \\ x_{12} \end{pmatrix} = \begin{pmatrix} b_1 \\ b_2 \\ \vdots \\ b_{12} \end{pmatrix},
\end{align}
where $b_k = -M_{k0}$. Intuitively, the Hessian matrix $M$ should be positive semidefinite, since the cost function $\epsilon_{\rm CE3}$ is by definition non-negative; yet because of the {difference} 
 between affine and linear transformations, such intuition requires further justification.
If it is indeed positive semidefinite, then efficient linear algebra solvers, e.g., Cholesky decomposition, can be used to solve the above linear system; if it is not, more general solvers must be used. Either way, this produces optimal coefficients $\{{\boldsymbol A}_i\}$ and $\{{\boldsymbol B}_i\}$, which tell us the position and velocity of each particle at $t=h$.
As advertied in Section~\ref{sec1}, ContEvol methods are implicit but only need to solve linear equations. 

Here we conclude Section~\ref{sec:celestial} with several remarks.
\begin{itemize}
    \item First, the framework described above can be naturally extended to more particles and more interactions.
    Equation~(\ref{eq:3BP_CE1_f0123}) is general for many-body problem in celestial mechanics, and should facilitate programming for both symbolic derivation and numerical implementation.
    The functionals ${\boldsymbol f}_i$, $i = 0, 1, 2, 3$ are also applicable to some electromagnetic problems, since Coulomb's law has the same form as Newton's law of universal gravitation.

    \item Second, whenever we have multiple equations (e.g., $2$ in the case of three-body problem), it is possible and sometimes natural to assign different weights to them while defining the cost function.
    Equation~(\ref{eq:3BP_CE1_cost}) does not do so because the two EOMs are symmetric, and thanks to $\mu_1$ and $\mu_2$, more weights are automatically assigned to more massive objects.
    While different equations describe different quantities, one is advised to rescale the equations and use the dimensionless version to define the cost function, and assign $\mathcal{O}(1)$ weights to them if necessary.

    \item Third, in principle, one can combine Sections~\ref{ss:CHO_CE2} and \ref{ss:3BP_CE1} to study celestial mechanics with second- (or even higher-) order ContEvol method.
    Since the cost function, which describes the discrepancy between approximated and ``true'' histories of the dynamic system, gets much better with higher order, results like Poincar\'e sections based on post hoc analysis (instead of combining tiny time steps and backwards evolution with traditional methods) should be more accurate than those based on lower-order~ContEvol.
\end{itemize}

\section{Quantum Mechanics: Stationary Schr\"odinger Equation}
\label{sec:quantum}

Now we switch topic from initial value problems (IVPs) to boundary value problems (BVPs). Again as physicists, we choose two simplest cases from quantum mechanics, infinite potential well and (quantum) harmonic oscillator, and then a more realistic case, Coulomb potential.

In one dimension, the stationary Schr\"odinger equation is
\vspace{-6pt}
\begin{align}
    H'\psi = -\frac{\hbar^2}{2m} \ddot{\psi} + V'\psi = E'\psi,
\end{align}
where $H'$ is the Hamiltonian (an operator), $\hbar$ is the reduced Planck constant, $m$ is the mass of the particle, $V'$ is the potential energy (a function), and $E'$ is the energy of the particle (a scalar); setting $\hbar^2/2m$ to $1$, this becomes
\vspace{-6pt}
\begin{align}
    \label{eq:Schrodinger} H\psi = -\ddot{\psi} + V\psi = E\psi.
\end{align}
In this work, we require the wavefunction $\psi$ to be a real function.

To solve this eigenvalue problem, the general strategy of ContEvol is:
\begin{enumerate}
    \item Represent the wavefunction $\psi$ as two series, $\{ \psi_i \equiv \psi(x_i) \}$ and $\{ \dot\psi_i \equiv \dot\psi(x_i) \}$, where $\{ x_i \}$ is a finite sampling of the real axis.
    \item Find the optimal approximation $\phi \equiv H\psi$, represented as $\{ \phi_i \equiv \phi(x_i) \}$ and \linebreak  $\{ \dot\phi_i \equiv \dot\phi(x_i) \}$, by minimizing a cost function. We treat the wavefunction $\psi$ as ``known'' for this purpose.
    \item Formulate the Hamiltonian $H$ as a linear transformation, and solve for the eigenvalues and eigenvectors of the matrix.
    \item Normalize, orthogonalize (not implemented in this work), and ``render'' the eigenvectors as continuous wavefunctions.
\end{enumerate}

To set a benchmark, we start by solving the infinite potential well using simple discretization in Section~\ref{ss:IPW_simple}, before addressing the same problem with first-order ContEvol method in Section~\ref{ss:IPW_CE1}.
Then in Section~\ref{ss:QHO_CE1}, we describe how ContEvol is supposed to be applied to a slightly trickier problem, quantum harmonic oscillator.
In Section~\ref{ss:QCP_CE1}, we try to solve a more realistic problem, one-dimensional Coulomb potential.

\subsection{Infinite Potential Well, Simple Discretization}
\label{ss:IPW_simple}

In this section and the next, we study the infinite potential well
\begin{align}
    \label{eq:IPW_potential} V(x) = \left\{ \begin{aligned} &0 &&0 \leq x \leq 1 \\
    &+\infty &&{\rm otherwise} \end{aligned} \right.,
\end{align}
for which the exact solution is
\begin{align}
    \label{eq:IPW_exact} \psi^{(n)}(x) = \left\{ \begin{aligned} &\sqrt{2} \sin (n\pi x) &&0 \leq x \leq 1 \\
    &0 &&{\rm otherwise} \end{aligned} \right. \quad {\rm and} \quad
    E_n = (n\pi)^2, \quad n \in {\mathbb N}^+.
\end{align}

We divide the interval $[0, 1]$ into $N+1$ equal parts with $N+2$ nodes
\begin{align}
    x_i = \frac{i}{N+1}, \quad i = 0, 1, \ldots, N+1.
\end{align}
With $\{ \psi_i \}$ and linear spline interpolation, the wavefunction is sampled as
\begin{align}
    \psi(x) = \left\{ \begin{aligned} &\psi_i + \frac{\psi_{i+1} - \psi_i}{h} (x-x_i) &&x_i \leq x \leq x_{i+1} \\
    &0 &&x < 0 \ {\rm or}\  x > 1 \end{aligned} \right.,
\end{align}
where $h \equiv 1/(N+1)$ is now the length of each sub-interval. Boundary conditions at $x_0=0$ and $x_{N+1}=1$ indicate that $\psi_0 = \psi_{N+1} = 0$.

At each sampling node, the second-order derivative $\ddot\psi$ is approximated as
\begin{align}
    \ddot\psi_i \approx \frac{\dot\psi_{i+1/2} - \dot\psi_{i-1/2}}{h} \approx \frac{1}{h} \left( \frac{\psi_{i+1} - \psi_i}{h} - \frac{\psi_i - \psi_{i-1}}{h} \right) = \frac{\psi_{i+1} - 2\psi_i + \psi_{i-1}}{h^2},
\end{align}
and thus the $N \times N$ (for $i = 1, 2, \ldots, N$) Hamiltonian $H$ is simply
\begin{align}
    H = h^{-2} \begin{pmatrix} 2 & -1 & 0 & \cdots & 0 & 0 \\
    -1 & 2 & -1 & \ddots & 0 & 0 \\
    0 & -1 & 2 & \ddots & 0 & 0 \\
    \vdots & \ddots & \ddots & \ddots & \ddots & \vdots \\
    0 & 0 & 0 & \ddots & 2 & -1 \\
    0 & 0 & 0 & \cdots & -1 & 2 \end{pmatrix},
\end{align}
where the minus sign comes from Equation~(\ref{eq:Schrodinger}). This Hamiltonian matrix is Hermitian, as it should.

Before moving on to examples, we note that the eigenvectors need to be ``renormalized'' (even if they have already been normalized as usual vectors) as\vspace{12pt} 
\begin{align}
    \label{eq:IPW_simple_normal} \nonumber 1 &= \int_0^1 [{\mathcal N}\psi(x)]^2 \,{\rm d}x = {\mathcal N}^2 \sum_{i=0}^N \int_{x_i}^{x_{i+1}} \left[ \psi_i + \frac{\psi_{i+1} - \psi_i}{h} (x-x_i) \right]^2 \,{\rm d}x \\
    &= {\mathcal N}^2 \sum_{i=0}^N \int_0^h \left( \psi_i + \frac{\psi_{i+1} - \psi_i}{h} x \right)^2 \,{\rm d}x = {\mathcal N}^2 \sum_{i=0}^N \frac{h}{3} (\psi_i^2 + \psi_i \psi_{i+1} + \psi_{i+1}^2),
\end{align}
where ${\mathcal N}$ is the normalization factor; similarly, in principle, they may need to be ``reorthogonalized'' according to the ``inner product'' defined as follows
\vspace{-10pt}
\begin{adjustwidth}{-\extralength}{0cm}
\small
\begin{align}
    \label{eq:IPW_simple_ortho} \nonumber \langle \psi^{(k)} | \psi^{(l)} \rangle &= \langle \{ \psi_i^{(k)} \} | \{ \psi_i^{(l)} \} \rangle
    = \sum_{i=0}^{N} \int_{x_i}^{x_{i+1}} \left[ \left( \psi_i^{(k)} + \frac{\psi_{i+1}^{(k)} - \psi_i^{(k)}}{h} (x-x_i) \right) \cdot \left( \psi_i^{(l)} + \frac{\psi_{i+1}^{(l)} - \psi_i^{(l)}}{h} (x-x_i) \right) \right] \,{\rm d}x \\
    \nonumber &= \sum_{i=0}^{N} \int_0^h \left[ \left( \psi_i^{(k)} + \frac{\psi_{i+1}^{(k)} - \psi_i^{(k)}}{h} x \right) \cdot \left( \psi_i^{(l)} + \frac{\psi_{i+1}^{(l)} - \psi_i^{(l)}}{h} x \right) \right] \,{\rm d}x \\
    &= \sum_{i=0}^{N} \frac{h}{6} [\psi_i^{(k)} (2\psi_i^{(l)} + \psi_{i+1}^{(l)}) + \psi_{i+1}^{(k)} (\psi_i^{(l)} + 2\psi_{i+1}^{(l)})],
\end{align}
\end{adjustwidth}
where we have not written the complex conjugate symbol ``$*$'' as our wavefunctions are real.
Yet intuitively, the eigenvectors should be orthogonal to each other, as they correspond to different eigenvalues of a Hermitian operator. Since this work is principally for illustration purposes, we simply present the normalized wavefunctions, and leave investigation of orthogonality for future work.

\begin{figure}[H]
    \subfloat{
        \includegraphics[width=0.48\textwidth]{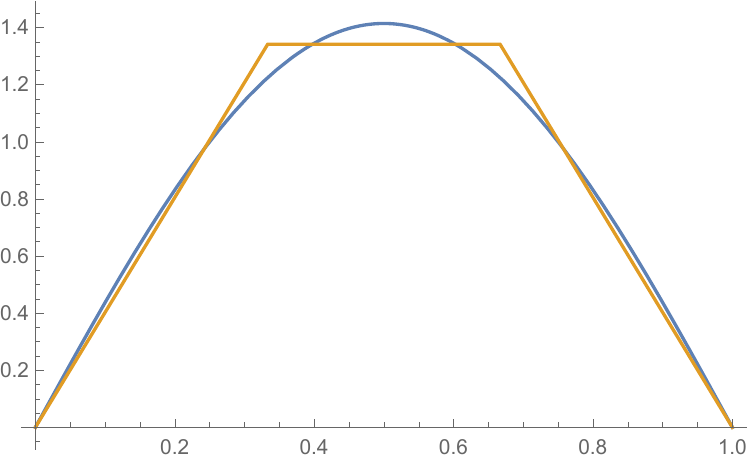}
    }
    \subfloat{
        \includegraphics[width=0.48\textwidth]{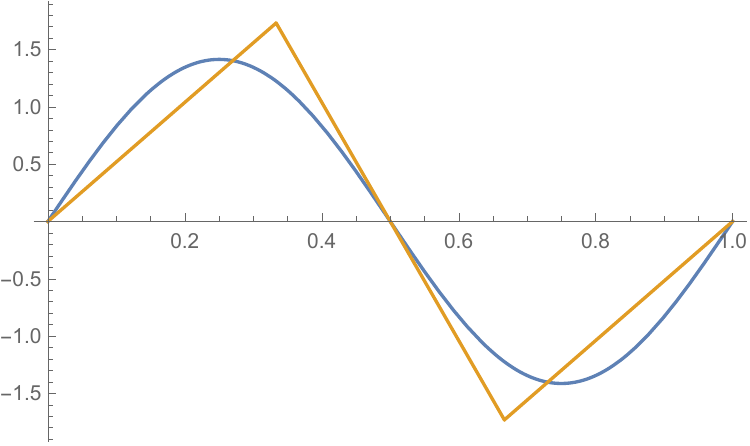}
    }
    \caption{{Infinite} potential well, comparisons between simple discretization $N=2$ results (orange) with exact solution Equation~(\ref{eq:IPW_exact}) (blue) for $n=1$ ({\bf left}) and $n=2$ ({\bf right}).}
    \label{fig:IPW_simple_N=2_vec}
\end{figure}

\begin{figure}[H]
    \subfloat{
        \includegraphics[width=0.32\textwidth]{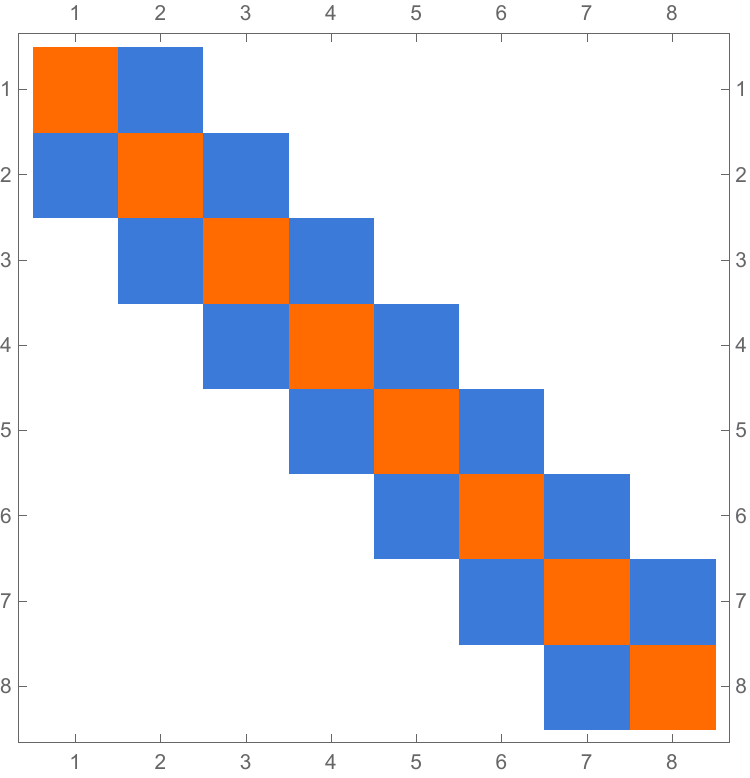}
    }
    \subfloat{
        \includegraphics[width=0.32\textwidth]{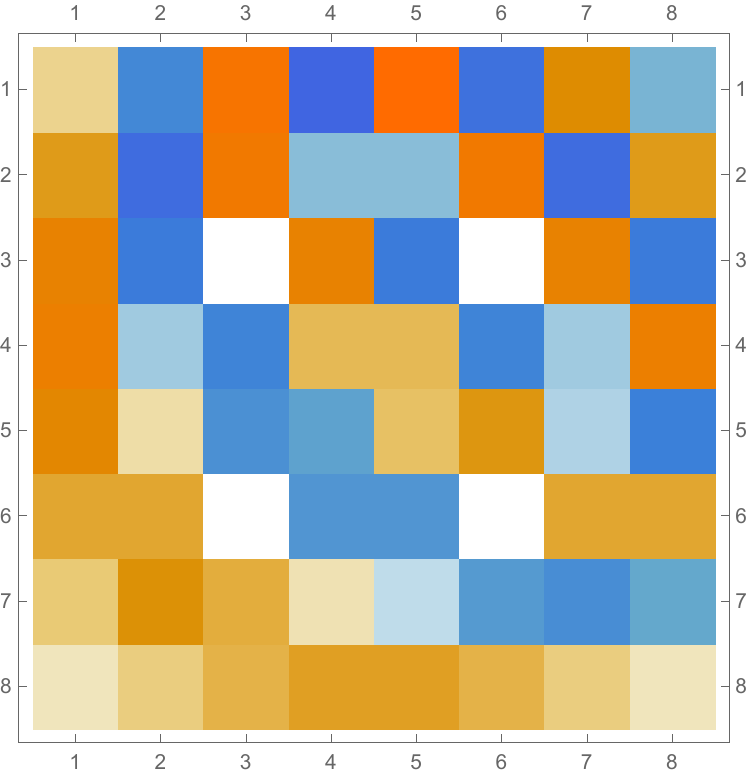}
    }
    \subfloat{
        \includegraphics[width=0.32\textwidth]{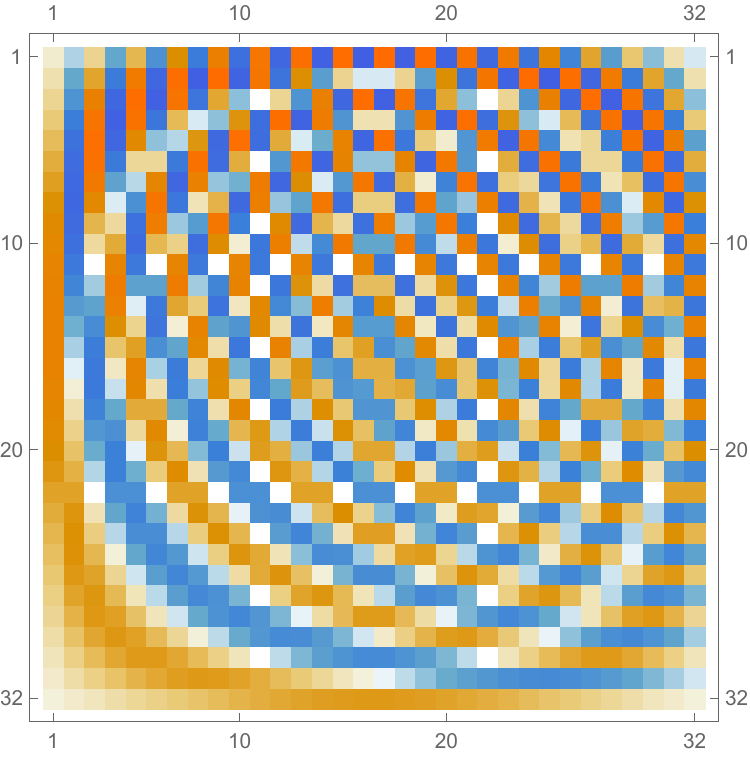}
    }

    \subfloat{
        \includegraphics[width=0.32\textwidth]{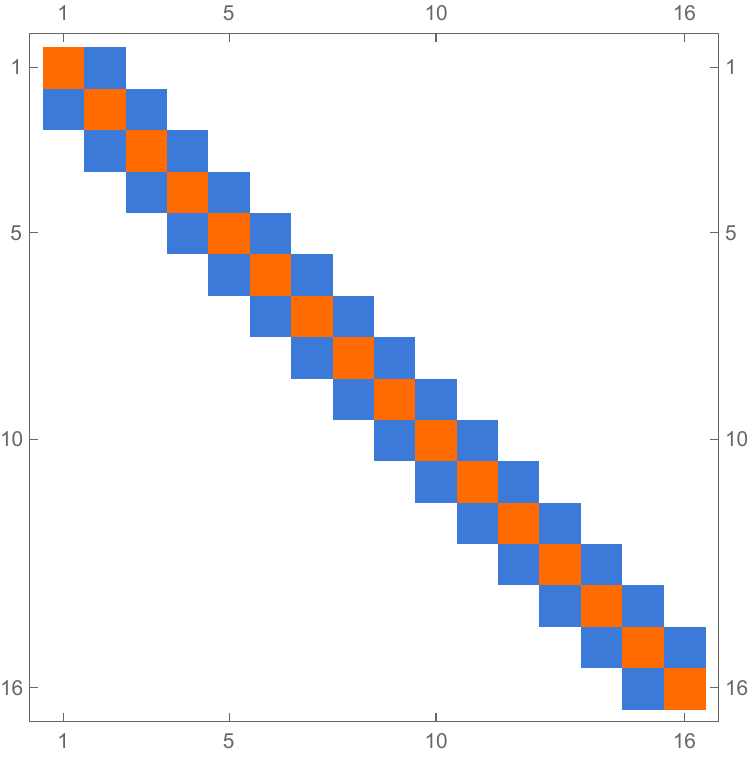}
    }
    \subfloat{
        \includegraphics[width=0.32\textwidth]{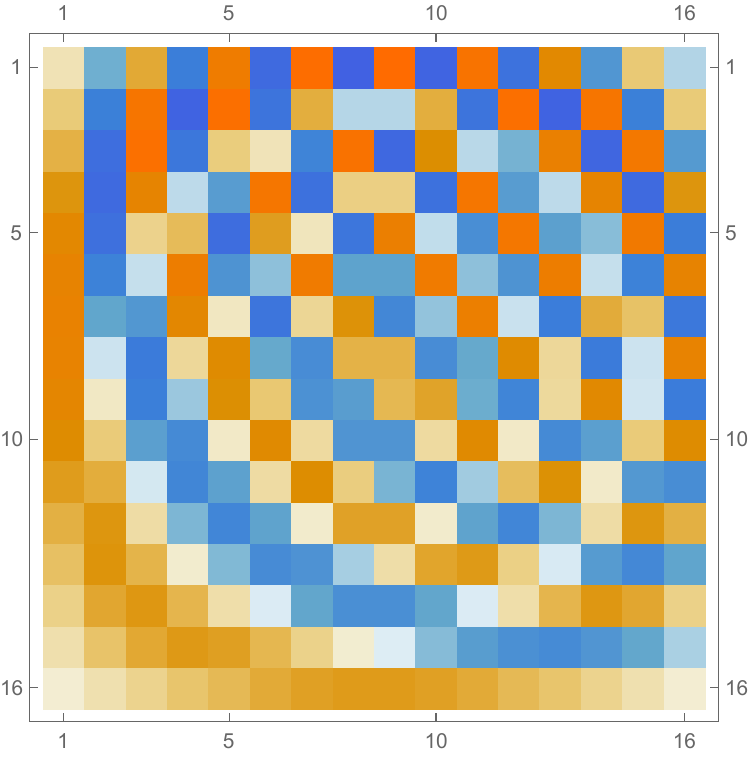}
    }
    \subfloat{
        \includegraphics[width=0.32\textwidth]{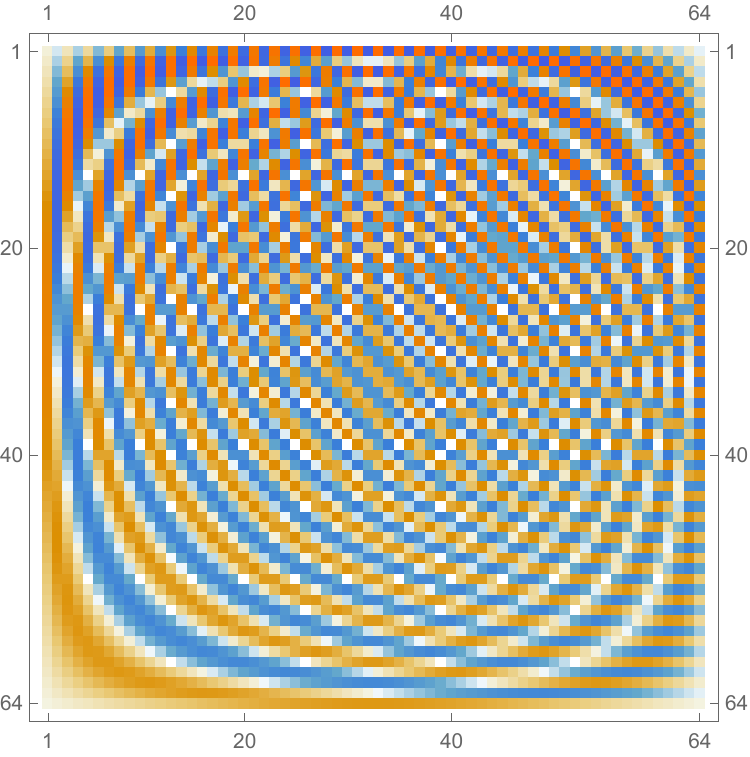}
    }
    \caption{{Infinite} potential well, $H$ matrices ({\bf first column}) for $N=8$ and $N=16$, and eigenvectors ({\bf second} and {\bf third columns}) for $N=8$, $N=16$, $N=32$, and $N=64$ versions of simple discretization.
    Following Mathematica convention, the eigenvectors are presented horizontally and ordered by decreasing eigenvalues (i.e., first row is $\psi^{(N)}$, last row is $\psi^{(1)}$).
    They are normalized in terms of Equation~(\ref{eq:IPW_simple_normal}), but not deliberately orthogonalized in terms of Equation~(\ref{eq:IPW_simple_ortho}); their signs are set so that $\psi_1$ (the first component) is positive in all cases.}
    \label{fig:IPW_simple_mats_vecs}
\end{figure}

Figure~\ref{fig:IPW_simple_N=2_vec} compares simple discretization with $N=2$ and exact solution Equation~(\ref{eq:IPW_exact}) for $n=1$ and $n=2$. Note that these two wavefunctions are automatically orthogonal to each other.
Figure~\ref{fig:IPW_simple_mats_vecs} shows two $H$ matrices ($N=8$ and $N=16$) and normalized but not necessarily orthogonal eigenvectors produced by $N=8$, $N=16$, $N=32$, and $N=64$ versions of simple discretization; the other two $H$ matrices ($N=32$ and $N=64$) are omitted as the tridiagonal structure is the same.
With increasing $n$ (note that $\psi^{(n)}$ has $n-1$ zero points between the two end points), the eigenvectors become less and less smooth.

\vspace{-6pt}

\begin{figure}[H]
    \includegraphics[width=.8\textwidth]{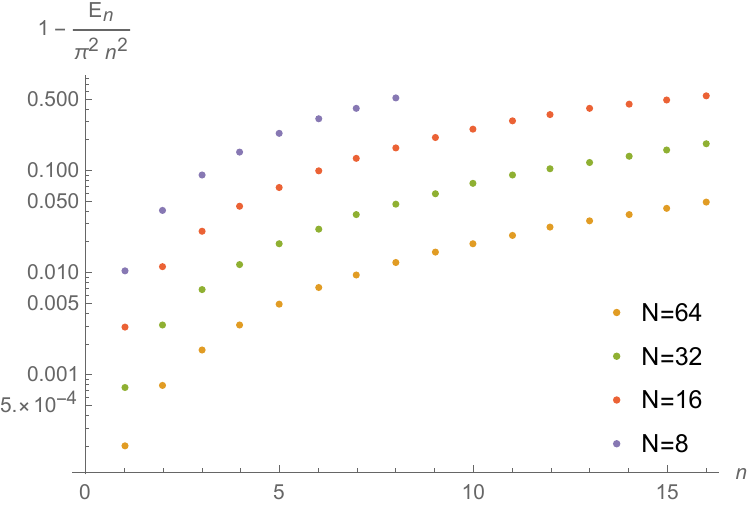}
    \caption{{Infinite} potential well, $1$ minus $n$th eigenvalue $E_n$ divided by its exact counterpart Equation~(\ref{eq:IPW_exact}) versus quantum number $n$ for $n=1, 2, \ldots, 16$. $N=64$ (orange), $N=32$ (green), $N=16$ (red), and $N=8$ (purple) results of simple discretization are shown in different colors.}
    \label{fig:IPW_simple_eigenvalues}
\end{figure}

\begin{figure}[H]
    \subfloat{
        \includegraphics[width=0.32\textwidth]{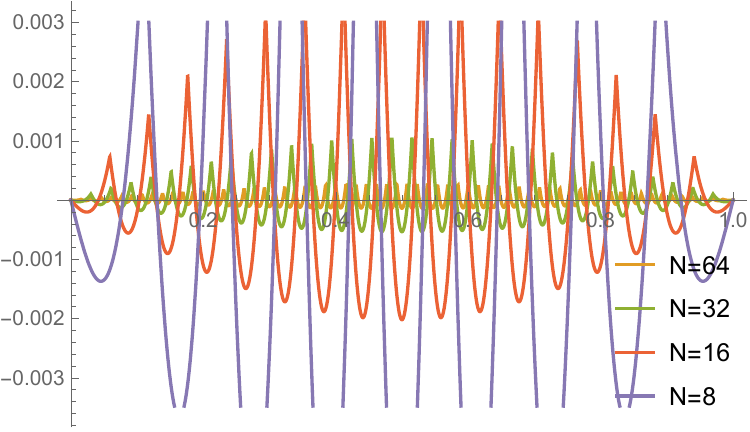}
    }
    \subfloat{
        \includegraphics[width=0.32\textwidth]{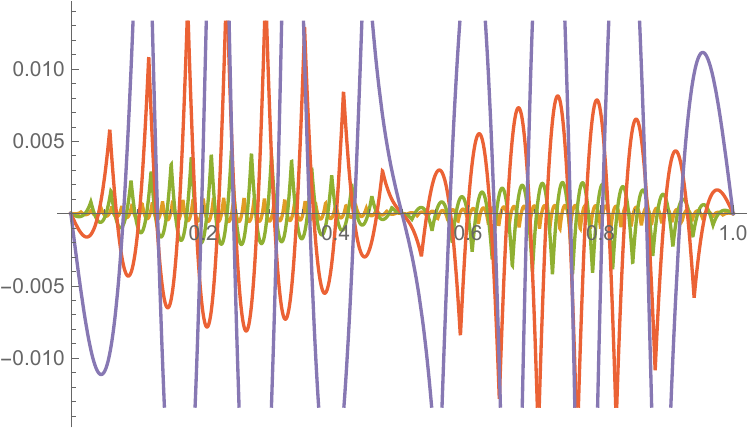}
    }
    \subfloat{
        \includegraphics[width=0.32\textwidth]{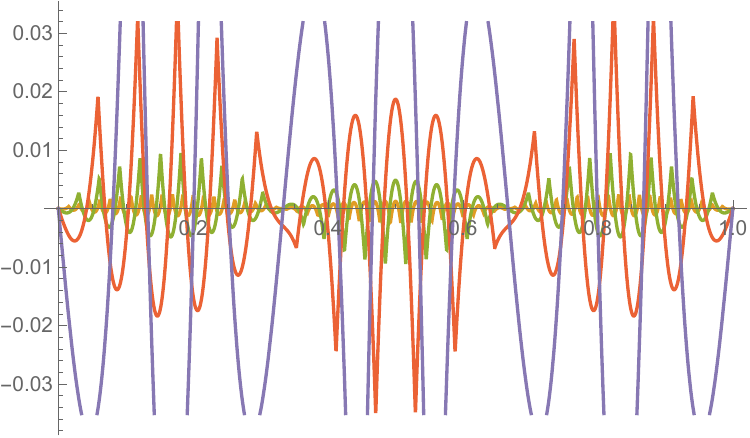}
    }

    \subfloat{
        \includegraphics[width=0.32\textwidth]{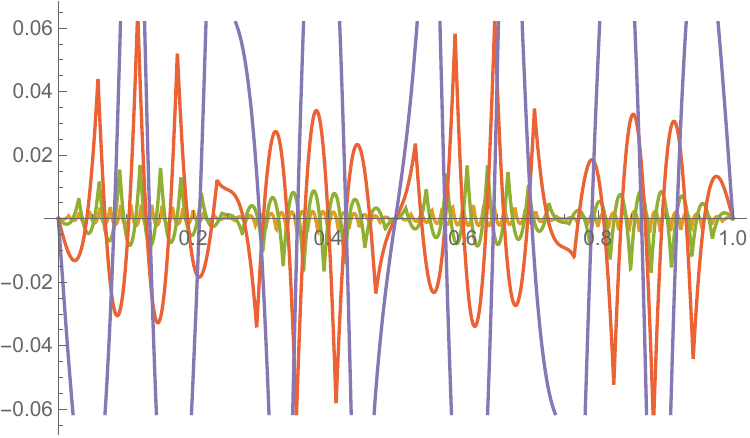}
    }
    \subfloat{
        \includegraphics[width=0.32\textwidth]{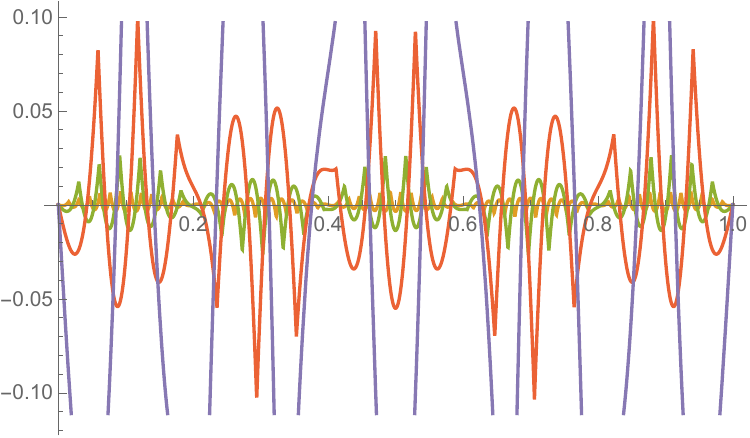}
    }
    \subfloat{
        \includegraphics[width=0.32\textwidth]{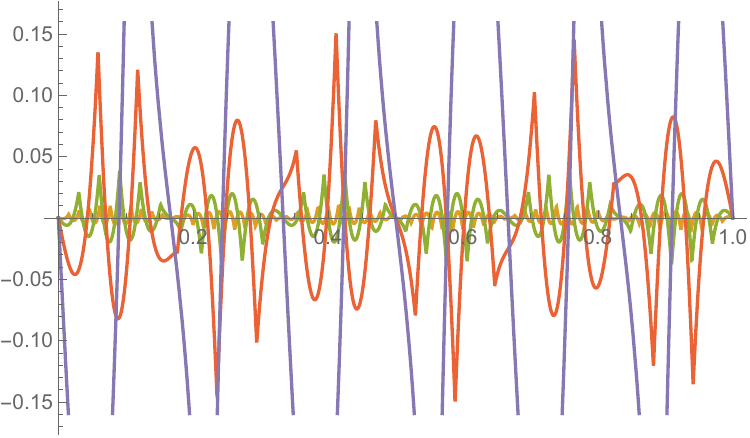}
    }

   \centering \subfloat{
        \includegraphics[width=0.32\textwidth]{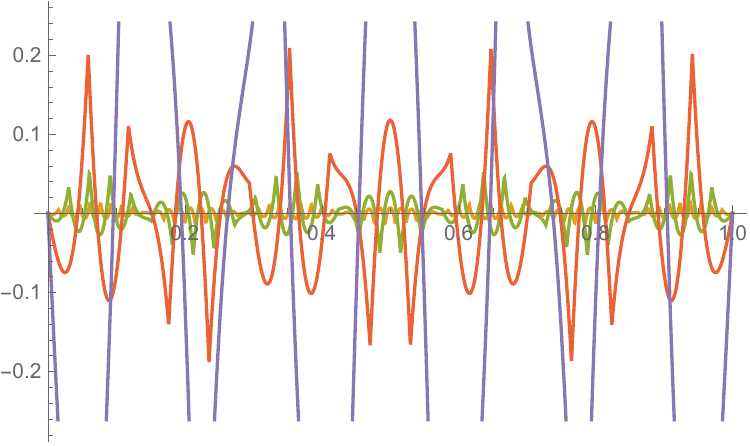}
    }
     \centering \subfloat{
        \includegraphics[width=0.32\textwidth]{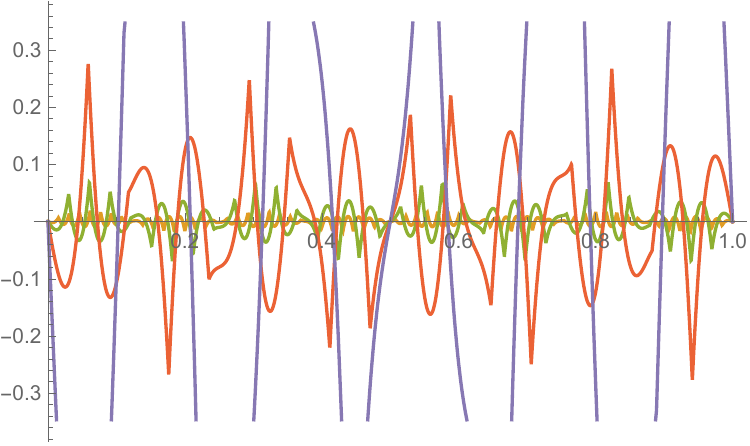}
    }

    \subfloat{
        \includegraphics[width=0.32\textwidth]{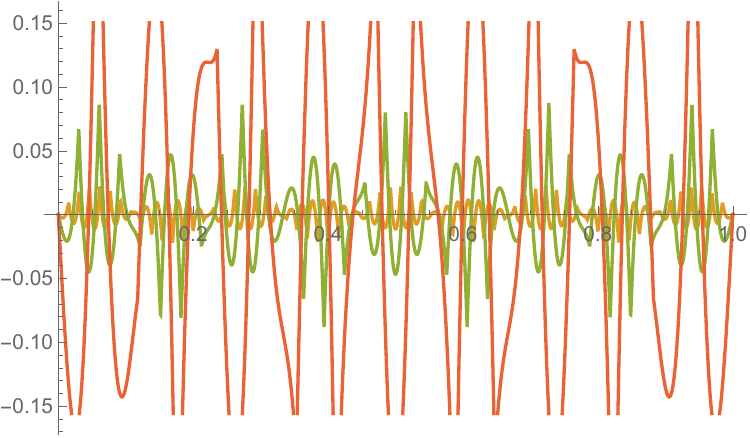}
    }
    \subfloat{
        \includegraphics[width=0.32\textwidth]{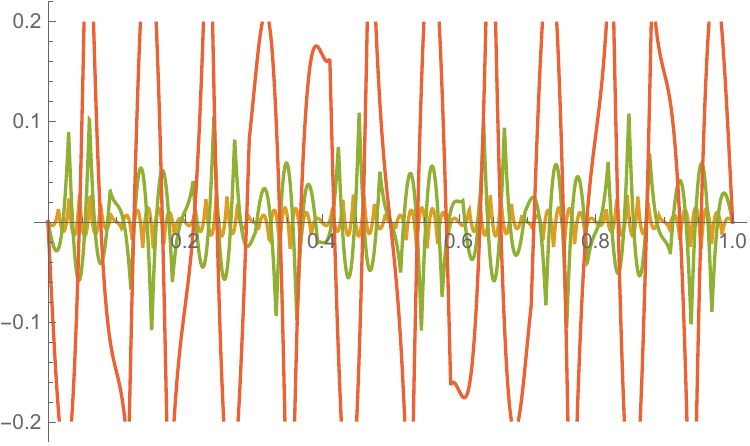}
    }
    \subfloat{
        \includegraphics[width=0.32\textwidth]{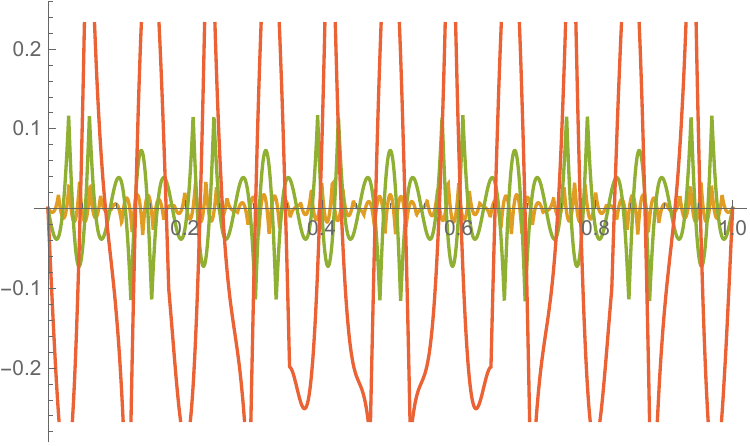}
    }

    \subfloat{
        \includegraphics[width=0.32\textwidth]{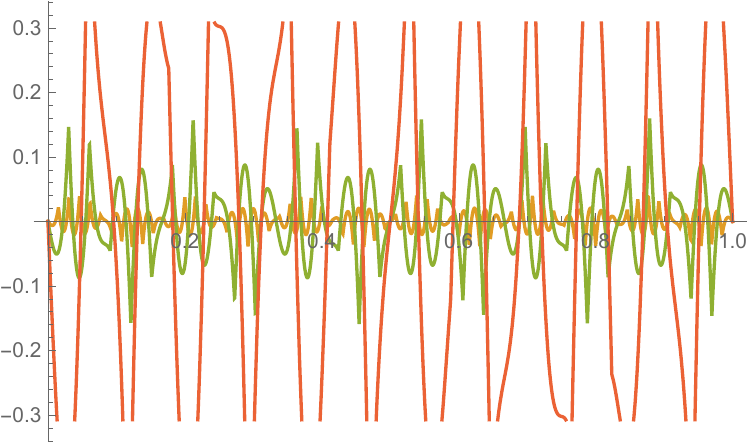}
    }
    \subfloat{
        \includegraphics[width=0.32\textwidth]{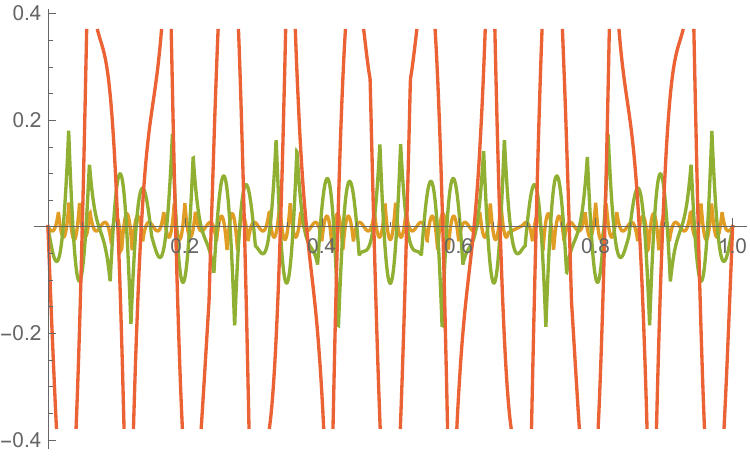}
    }
    \subfloat{
        \includegraphics[width=0.32\textwidth]{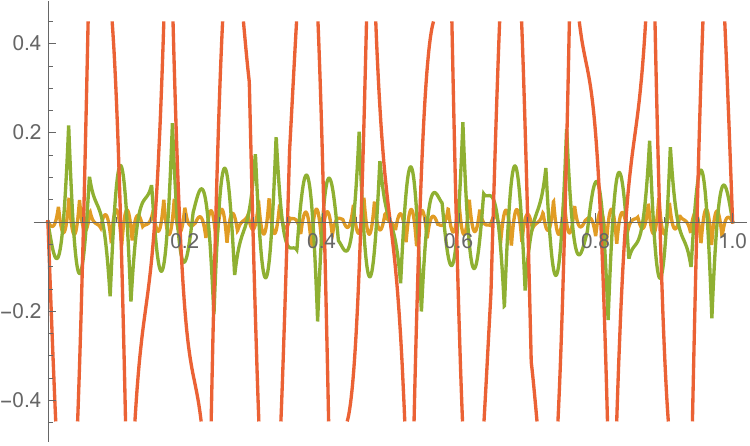}
    }

    \subfloat{
        \includegraphics[width=0.32\textwidth]{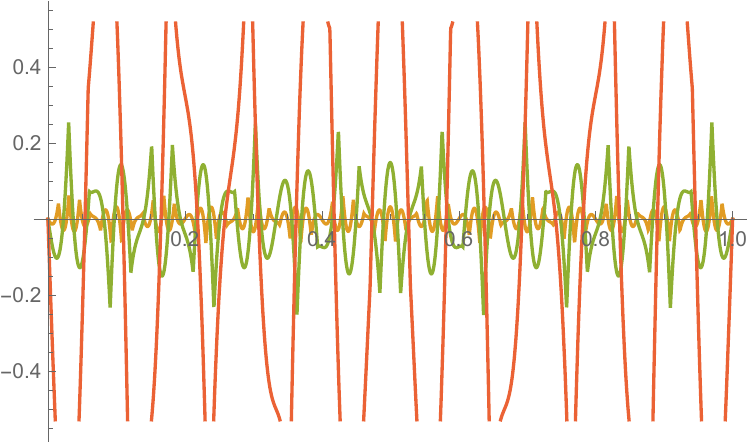}
    }
    \subfloat{
        \includegraphics[width=0.32\textwidth]{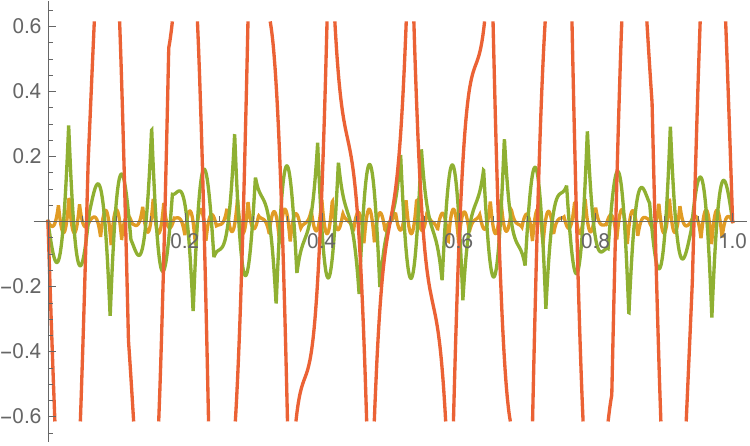}
    }
    \caption{{Infinite} potential well, errors in rendered wavefunctions of $N=64$ (orange), $N=32$ (green), $N=16$ (red), and $N=8$ (purple) results of simple discretization. Note that magnitude of exact wavefunctions is $\sqrt{2}$.}
    \label{fig:IPW_simple_eigenvectors}
\end{figure}

Figures~\ref{fig:IPW_simple_eigenvalues} and \ref{fig:IPW_simple_eigenvectors} display errors in eigenvalues and rendered eigenvectors of $N=8$, $N=16$, $N=32$, and $N=64$ Hamiltonians, respectively.
Although a $N \times N$ Hermitian matrix has $N$ eigenpairs, $E_n$ and $\psi^{(n)}$ with $n \geq 17$ are not shown in these figures.
At small $n$, the approximated wavefunctions are reasonably smooth; however, as $n$ approaches $N/2$, the broken features become much more noticeable.
It should be noted that all the eigenvalues produced by simple discretization are smaller than their exact counterparts, unlike those yielded by first ContEvol method, as we will show in the next section.

\subsection{Infinite Potential Well, First-Order ContEvol}
\label{ss:IPW_CE1}

Now we present the ContEvol treatment of the same problem.
We divide the interval $[0, 1]$ into $N$ equal parts with $N+1$ nodes
\vspace{-4pt}
\begin{align}
    x_i = \frac{i}{N}, \quad i = 0, 1, \ldots, N;
\end{align}
investigating if an unequal partition leads to better results is left for future work.
With $\{ \psi_i \}$ and $\{ \dot\psi_i \}$, the wavefunction is sampled as
\begin{align}
    \psi(x) = \left\{ \begin{aligned} &\psi_i + \dot\psi_i(x-x_i) + B_{\psi i}(x-x_i)^2 + A_{\psi i}(x-x_i)^3 &&x_i \leq x \leq x_{i+1} \\
    &0 &&x < 0 \ {\rm or}\  x > 1 \end{aligned} \right.
\end{align}
with 
\begin{align}
    &\left\{ \begin{aligned} A_{\psi i} &= 2(\psi_i - \psi_{i+1}) h^{-3} + (\dot\psi_i + \dot\psi_{i+1}) h^{-2} \\
    B_{\psi i} &= 3(\psi_{i+1} - \psi_i) h^{-2} - (2\dot\psi_i + \dot\psi_{i+1}) h^{-1} \end{aligned} \right.,
\end{align}
where $h \equiv 1/N$ is the length of each sub-interval. Boundary conditions at $x_0=0$ and $x_N=1$ indicate that $\psi_0 = \psi_N = 0$. The desired approximation $\phi \equiv H\psi$ is represented in the same way.

We are supposed to have $\phi \approx -\ddot{\psi}$. Note that $\psi(x)$ and $\phi(x)$ are both piecewise cubic functions with continuous first derivatives, while $\ddot{\psi}$ is a piecewise linear function which is not necessarily continuous at sampling nodes.
The cost function is defined as (subscript ``IPW'' stands for infinite potential well)
\vspace{-12pt}
\begin{adjustwidth}{-\extralength}{0cm}
\begin{align}
    \label{eq:IPW_CE1_cost} \epsilon_{\rm IPW}(\{\psi_i\}, \{\dot\psi_i\}; \{\phi_i\}, \{\dot\phi_i\}; \{x_i\}) &= \sum_{i=0}^{N-1} \epsilon_{{\rm IPW},i} (\psi_i, \dot\psi_i, \psi_{i+1}, \dot\psi_{i+1}; \phi_i, \dot\phi_i, \phi_{i+1}, \dot\phi_{i+1}; x_i, x_{i+1});
\end{align}
\end{adjustwidth}
for simplicity, in the following text we omit parameters of $\epsilon_{{\rm IPW},i}$, which is
\vspace{-10pt}
\begin{adjustwidth}{-\extralength}{0cm}
\begin{align}
    \nonumber \epsilon_{{\rm IPW},i} &= \int_{x_i}^{x_{i+1}} (\ddot{\psi} + \phi)^2 \,{\rm d}x = \int_{x_i}^{x_{i+1}} [(2B_{\psi i} + \phi_i) + (6A_{\psi i} + \dot\phi_i)(x-x_i) + B_{\phi i}(x-x_i)^2 + A_{\phi i}(x-x_i)^3]^2 \,{\rm d}x \\
    \nonumber &= \int_0^h [(2B_{\psi i} + \phi_i) + (6A_{\psi i} + \dot\phi_i)x + B_{\phi i}x^2 + A_{\phi i}x^3]^2 \,{\rm d}x \\
    \nonumber &= \int_0^h \left[ \begin{aligned} &(4B_{\psi i}^2 + 4B_{\psi i}\phi_i + \phi_i^2) + (24A_{\psi i}B_{\psi i} + 12A_{\psi i}\phi_i + 4B_{\psi i}\dot\phi_i + 2\phi_i\dot\phi_i)x \\
    &+ (36A_{\psi i}^2 + 12A_{\psi i}\dot\phi_i + 4B_{\phi i}B_{\psi i} + 2B_{\phi i}\phi_i+\dot\phi_i^2)x^2 \\ &+ (12A_{\psi i}B_{\phi i} + 4A_{\phi i}B_{\psi i} + 2A_{\phi i}\phi_i + 2B_{\phi i}\dot\phi_i)x^3 \\
    &+ (12A_{\phi i}A_{\psi i} + 2A_{\phi i}\dot\phi_i + B_{\phi i}^2)x^4 + 2A_{\phi i}B_{\phi i}x^5 + A_{\phi i}^2 x^6 \end{aligned} \right] \,{\rm d}x \\
    &= \left[ \begin{aligned} &(4B_{\psi i}^2 + 4B_{\psi i}\phi_i + \phi_i^2)h + (12A_{\psi i}B_{\psi i} + 6A_{\psi i}\phi_i + 2B_{\psi i}\dot\phi_i + \phi_i\dot\phi_i)h^2 \\
    &+ \frac{1}{3}(36A_{\psi i}^2 + 12A_{\psi i}\dot\phi_i + 4B_{\phi i}B_{\psi i} + 2B_{\phi i}\phi_i+\dot\phi_i^2)h^3 \\ &+ \frac{1}{2}(6A_{\psi i}B_{\phi i} + 2A_{\phi i}B_{\psi i} + A_{\phi i}\phi_i + B_{\phi i}\dot\phi_i)h^4 \\
    &+ \frac{1}{5}(12A_{\phi i}A_{\psi i} + 2A_{\phi i}\dot\phi_i + B_{\phi i}^2)h^5 + \frac{1}{3}A_{\phi i}B_{\phi i}h^6 + \frac{1}{7}A_{\phi i}^2h^7 \end{aligned} \right],
\end{align}
\end{adjustwidth}
for $i = 0, 1, \ldots, N-1$; plugging in expressions of $A_{\psi i}$, $B_{\psi i}$, $A_{\phi i}$, and $B_{\phi i}$, this becomes
\vspace{-10pt}
\begin{adjustwidth}{-\extralength}{0cm}
\begin{align}
    \label{eq:IPW_CE1_costi} \epsilon_{{\rm IPW},i} = \left[ \begin{aligned} &12 (\psi_i-\psi_{i+1})^2 h^{-3} + 12 (\dot\psi_i+\dot\psi_{i+1}) (\psi_i-\psi_{i+1}) h^{-2} \\
    &+ \left\{ 4 (\dot\psi_i^2+\dot\psi_i\dot\psi_{i+1}+\dot\psi_{i+1}^2) - \frac{12}{5} (\psi_i-\psi_{i+1}) (\phi_i-\phi_{i+1}) \right\} h^{-1} \\
    &- \left\{\frac{12}{5} (\dot\psi_i\phi_i-\dot\psi_{i+1}\phi_{i+1}) - \frac{1}{5}(\dot\psi_i-\dot\psi_{i+1}) (\phi_i+\phi_{i+1}) + \frac{1}{5}(\psi_i-\psi_{i+1}) (\dot\phi_i+\dot\phi_{i+1}) \right\} \\
    &+ \left\{ \frac{1}{105} (39\phi_i^2+27\phi_i\phi_{i+1}+39\phi_{i+1}^2) - \frac{1}{15} (4\dot\psi_i\dot\phi_i-\dot\psi_{i+1}\dot\phi_i-\dot\psi_i\dot\phi_{i+1}+4\dot\psi_{i+1}\dot\phi_{i+1}) \right\} h \\
    &+\frac{1}{210} (22\phi_i\dot\phi_i-13\phi_i\dot\phi_{i+1}+13\dot\phi_i\phi_{i+1}-22\dot\phi_{i+1}\phi_{i+1}) h^2 + \frac{1}{210} (2\dot\phi_i^2-3\dot\phi_i\dot\phi_{i+1}+2\dot\phi_{i+1}^2) h^3 \end{aligned} \right];
\end{align}
\end{adjustwidth}
for convenience, we define $\epsilon_{{\rm IPW},-1} = \epsilon_{{\rm IPW},N} = 0$.

Partial derivatives of $\epsilon_{{\rm IPW},i}$ with respect to $\phi_i$, $\phi_{i+1}$, $\dot\phi_i$, and $\dot\phi_{i+1}$ are
\vspace{-10pt}
\begin{adjustwidth}{-\extralength}{0cm}
\begin{align}
    &\left\{ \begin{aligned} \frac{\partial\epsilon_{{\rm IPW},i}}{\partial\phi_i} &= - \frac{12(\psi_i-\psi_{i+1})}{5}h^{-1} - \frac{11\dot\psi_i+\dot\psi_{i+1}}{5} + \frac{26\phi_i+9\phi_{i+1}}{35}h + \frac{22\dot\phi_i-13\dot\phi_{i+1}}{210}h^2 \\
    \frac{\partial\epsilon_{{\rm IPW},i}}{\partial\phi_{i+1}} &= \frac{12(\psi_i-\psi_{i+1})}{5}h^{-1} + \frac{\dot\psi_i+11\dot\psi_{i+1}}{5} + \frac{9\phi_i+26\phi_{i+1}}{35}h + \frac{13\dot\phi_i-22\dot\phi_{i+1}}{210}h^2 \\
    \frac{\partial\epsilon_{{\rm IPW},i}}{\partial\dot\phi_i} &= - \frac{\psi_i-\psi_{i+1}}{5} - \frac{4\dot\psi_i-\dot\psi_{i+1}}{15}h + \frac{22\phi_i+13\phi_{i+1}}{210}h^2 + \frac{4\dot\phi_i-3\dot\phi_{i+1}}{210}h^3 \\
    \frac{\partial\epsilon_{{\rm IPW},i}}{\partial\dot\phi_{i+1}} &= - \frac{\psi_i-\psi_{i+1}}{5} + \frac{\dot\psi_i-4\dot\psi_{i+1}}{15}h - \frac{13\phi_i+22\phi_{i+1}}{210}h^2 - \frac{3\dot\phi_i-4\dot\phi_{i+1}}{210}h^3 \end{aligned} \right.,
\end{align}
\end{adjustwidth}
respectively; note that one should not set these to zero, as a node is coupled with two adjacent intervals, unless it is $x_0$ or $x_N$.
Put in matrix form, these are
\vspace{-12pt}
\begin{adjustwidth}{-\extralength}{0cm}
\begin{align}
    \label{eq:IPW_CE1_PiQi} \nonumber \begin{pmatrix} \partial/\partial\phi_i \\ \partial/\partial\phi_{i+1} \\ \partial/\partial\dot\phi_i \\ \partial/\partial\dot\phi_{i+1} \end{pmatrix} \epsilon_{{\rm IPW},i}
    &= \left[ \begin{aligned} &\begin{pmatrix} 26h/35 & 9h/35 & 11h^2/105 & -13h^2/210 \\ 9h/35 & 26h/35 & 13h^2/210 & -11h^2/105 \\ 11h^2/105 & 13h^2/210 & 2h^3/105 & -h^3/70 \\ -13h^2/210 & -11h^2/105 & -h^3/70 & 2h^3/105 \end{pmatrix} \begin{pmatrix} \phi_i \\ \phi_{i+1} \\ \dot\phi_i \\ \dot\phi_{i+1} \end{pmatrix} \\
    &+ \begin{pmatrix} -12h^{-1}/5 & 12h^{-1}/5 & -11/5 & -1/5 \\ 12h^{-1}/5 & -12h^{-1}/5 & 1/5 & 11/5 \\ -1/5 & 1/5 & -4h/15 & h/15 \\ -1/5 & 1/5 & h/15 & -4h/15 \end{pmatrix} \begin{pmatrix} \psi_i \\ \psi_{i+1} \\ \dot\psi_i \\ \dot\psi_{i+1} \end{pmatrix} \end{aligned} \right] \\
    &\equiv P^{(i)} \begin{pmatrix} \phi_i \\ \phi_{i+1} \\ \dot\phi_i \\ \dot\phi_{i+1} \end{pmatrix} + Q^{(i)} \begin{pmatrix} \psi_i \\ \psi_{i+1} \\ \dot\psi_i \\ \dot\psi_{i+1} \end{pmatrix};
\end{align}
\end{adjustwidth}
again for convenience, we define $P^{(-1)} = Q^{(-1)} = P^{(N+1)} = Q^{(N+1)} = \begin{pmatrix} 0 & 0 \\ 0 & 0 \end{pmatrix}$.

To minimize the cost function Equation~(\ref{eq:IPW_CE1_cost}), we have
\vspace{-10pt}
\begin{adjustwidth}{-\extralength}{0cm}
\begin{align}
    \label{eq:IPW_CE1_PandQ} \begin{pmatrix} \partial/\partial\phi_0 \\ \vdots \\ \partial/\partial\phi_N \\ \partial/\partial\dot\phi_0 \\ \vdots \\ \partial/\partial\dot\phi_N \end{pmatrix} \epsilon_{\rm IPW}
    = \left[ \begin{aligned} &\begin{pmatrix} P_{00} & \cdots & P_{0N} & P_{0,N+1} & \cdots & P_{0,2N+1} \\ \vdots & \ddots & \vdots & \vdots & \ddots & \vdots \\ P_{N0} & \cdots & P_{NN} & P_{N,N+1} & \cdots & P_{N,2N+1} \\ P_{N+1,0} & \cdots & P_{N+1,N} & P_{N+1,N+1} & \cdots & P_{N+1,2N+1} \\ \vdots & \ddots & \vdots & \vdots & \ddots & \vdots \\ P_{2N+1,0} & \cdots & P_{2N+1,N} & P_{2N+1,N+1} & \cdots & P_{2N+1,2N+1} \end{pmatrix} \begin{pmatrix} \phi_0 \\ \vdots \\ \phi_N \\ \dot\phi_0 \\ \vdots \\ \dot\phi_N \end{pmatrix} \\
    &+ \begin{pmatrix} Q_{00} & \cdots & Q_{0N} & Q_{0,N+1} & \cdots & Q_{0,2N+1} \\ \vdots & \ddots & \vdots & \vdots & \ddots & \vdots \\ Q_{N0} & \cdots & Q_{NN} & Q_{N,N+1} & \cdots & Q_{N,2N+1} \\ Q_{N+1,0} & \cdots & Q_{N+1,N} & Q_{N+1,N+1} & \cdots & Q_{N+1,2N+1} \\ \vdots & \ddots & \vdots & \vdots & \ddots & \vdots \\ Q_{2N+1,0} & \cdots & Q_{2N+1,N} & Q_{2N+1,N+1} & \cdots & Q_{2N+1,2N+1} \end{pmatrix} \begin{pmatrix} \psi_0 \\ \vdots \\ \psi_N \\ \dot\psi_0 \\ \vdots \\ \dot\psi_N \end{pmatrix} \end{aligned} \right]
    = \begin{pmatrix} 0 \\ \vdots \\ 0 \\ 0 \\ \vdots \\ 0 \end{pmatrix};
\end{align}
\end{adjustwidth}
since
\begin{align}
    \left\{ \begin{aligned} \frac{\partial\epsilon_{\rm IPW}}{\partial\phi_i} = \frac{\partial\epsilon_{{\rm IPW},i-1}}{\partial\phi_i} + \frac{\partial\epsilon_{{\rm IPW},i}}{\partial\phi_i} \\
    \frac{\partial\epsilon_{\rm IPW}}{\partial\dot\phi_i} = \frac{\partial\epsilon_{{\rm IPW},i-1}}{\partial\dot\phi_i} + \frac{\partial\epsilon_{{\rm IPW},i}}{\partial\dot\phi_i} \end{aligned} \right.,
\end{align}
the $(2N+2) \times (2N+2)$ $P$ and $Q$ matrices can be constructed from scratch (zero matrix) by doing
\vspace{-10pt}
\begin{adjustwidth}{-\extralength}{0cm}
\begin{align}
    \left\{ \begin{aligned} \begin{pmatrix} P_{i,i} & P_{i,i+1} & P_{i,(N+1)+i} & P_{i,(N+1)+i+1} \\ P_{i+1,i} & P_{i+1,i+1} & P_{i+1,(N+1)+i} & P_{i+1,(N+1)+i+1} \\ P_{(N+1)+i,i} & P_{(N+1)+i,i+1} & P_{(N+1)+i,(N+1)+i} & P_{(N+1)+i,(N+1)+i+1} \\ P_{(N+1)+i+1,i} & P_{(N+1)+i+1,i+1} & P_{(N+1)+i+1,(N+1)+i} & P_{(N+1)+i+1,(N+1)+i+1} \end{pmatrix} +\!\!= P^{(i)} \\
    \begin{pmatrix} Q_{i,i} & Q_{i,i+1} & Q_{i,(N+1)+i} & Q_{i,(N+1)+i+1} \\ Q_{i+1,i} & Q_{i+1,i+1} & Q_{i+1,(N+1)+i} & Q_{i+1,(N+1)+i+1} \\ Q_{(N+1)+i,i} & Q_{(N+1)+i,i+1} & Q_{(N+1)+i,(N+1)+i} & Q_{(N+1)+i,(N+1)+i+1} \\ Q_{(N+1)+i+1,i} & Q_{(N+1)+i+1,i+1} & Q_{(N+1)+i+1,(N+1)+i} & Q_{(N+1)+i+1,(N+1)+i+1} \end{pmatrix} +\!\!= Q^{(i)} \end{aligned} \right.,
\end{align}
\end{adjustwidth}
where {$+\!\!=$} denotes the addition assignment operator in common programming languages like C or Python, for $i = 0, 1, \ldots, N$.
To enforce the $\psi_0 = \psi_N = 0$ constraints, one simply needs to remove the corresponding rows and columns.

Our desired Hamiltonian is thus simply $H = -P^{-1}Q$. Eigendecomposition of $H$ should yield $2N+2$ (or $2N$) eigenpairs, $\{ \psi_i^{(k)}, \dot\psi_i^{(k)} \}$ and $E^{(k)}$, without (with) those two constraints.
With or without the $\psi_0 = \psi_N = 0$ enforcement, $P$ and $Q$ matrices are always symmetric; however, this does not guarantee that the resulting $H$ matrix is also symmetric, and thus Hermitian.

Like in Section~\ref{ss:IPW_simple}, the eigenvectors need to be ``renormalized'' as
\vspace{-14pt}
\begin{adjustwidth}{-\extralength}{0cm}
\begin{align}
    \label{eq:IPW_CE1_normal} \nonumber 1 &= \int_0^1 [{\mathcal N}\psi(x)]^2 \,{\rm d}x = {\mathcal N}^2 \sum_{i=0}^{N-1} \int_{x_i}^{x_{i+1}} [\psi_i + \dot\psi_i(x-x_i) + B_{\psi i}(x-x_i)^2 + A_{\psi i}(x-x_i)^3]^2 \,{\rm d}x \\
    \nonumber &= {\mathcal N}^2 \sum_{i=0}^{N-1} \int_0^h [\psi_i + \dot\psi_i x + B_{\psi i}x^2 + A_{\psi i}x^3]^2 \,{\rm d}x \\
    \nonumber &= {\mathcal N}^2 \sum_{i=0}^{N-1} \int_0^h \left[ \begin{aligned} &\psi_i^2 + 2 \psi_i \dot\psi_i x + (2 B_{\psi i} \psi_i + \dot\psi_i^2) x^2 + (2 A_{\psi i} \psi_i + 2 B_{\psi i} \dot\psi_i) x^3 \\ &+ (B_{\psi i}^2 + 2 A_{\psi i} \dot\psi_i) x^4 + 2 A_{\psi i} B_{\psi i} x^5 + A_{\psi i}^2 x^6 \end{aligned} \right] \,{\rm d}x \\
    \nonumber &= {\mathcal N}^2 \sum_{i=0}^{N-1} \left[ \begin{aligned} &\psi_i^2 h + \psi_i \dot\psi_i h^2 + \frac{2 B_{\psi i} \psi_i + \dot\psi_i^2}{3} h^3 + \frac{A_{\psi i} \psi_i + B_{\psi i} \dot\psi_i}{2} h^4 \\ &+ \frac{B_{\psi i}^2 + 2 A_{\psi i} \dot\psi_i}{5} h^5 + \frac{A_{\psi i} B_{\psi i}}{3} h^6 + \frac{A_{\psi i}^2}{7}h^7 \end{aligned} \right] \\
    &= {\mathcal N}^2 \sum_{i=0}^{N-1} \left[ \begin{aligned} &\frac{1}{35} (13 \psi_i^2 + 9 \psi_i \psi_{i+1} + 13 \psi_{i+1}^2) h + \frac{1}{210} (22 \psi_i \dot\psi_i - 13 \psi_i \dot\psi_{i+1} + 13 \dot\psi_i \psi_{i+1} - 22 \dot\psi_{i+1} \psi_{i+1}) h^2 \\ &+ \frac{1}{210} (2 \dot\psi_i^2 - 3 \dot\psi_i \dot\psi_{i+1} + 2 \dot\psi_{i+1}^2) h^3 \end{aligned} \right];
\end{align}
\end{adjustwidth}
they may need to be ``reorthogonalized'' according to the ``inner product'' defined as follows
\vspace{-10pt}
\begin{adjustwidth}{-\extralength}{0cm}
\small
\begin{align}
    \label{eq:IPW_CE1_ortho} \nonumber \langle \psi^{(k)} | \psi^{(l)} \rangle &= \langle \{ \psi_i^{(k)}, \dot\psi_i^{(k)} \} | \{ \psi_i^{(l)}, \dot\psi_i^{(l)} \} \rangle
    = \sum_{i=0}^{N-1} \int_{x_i}^{x_{i+1}} \left[ \begin{aligned} &\{ \psi_i^{(k)} + \dot\psi_i^{(k)}(x-x_i) + B_{\psi i}^{(k)}(x-x_i)^2 + A_{\psi i}^{(k)}(x-x_i)^3 \} \\ &\cdot \{ \psi_i^{(l)} + \dot\psi_i^{(l)}(x-x_i) + B_{\psi i}^{(l)}(x-x_i)^2 + A_{\psi i}^{(l)}(x-x_i)^3 \} \end{aligned} \right] \,{\rm d}x \\
    \nonumber &= \sum_{i=0}^{N-1} \int_0^h [ \{ \psi_i^{(k)} + \dot\psi_i^{(k)} x + B_{\psi i}^{(k)}x^2 + A_{\psi i}^{(k)}x^3 \} \cdot \{ \psi_i^{(l)} + \dot\psi_i^{(l)} x + B_{\psi i}^{(l)}x^2 + A_{\psi i}^{(l)}x^3 \} ] \,{\rm d}x \\
    \nonumber &= \sum_{i=0}^{N-1} \int_0^h \left[ \begin{aligned} &\psi_i^{(k)} \psi_i^{(l)} + (\dot\psi_i^{(k)} \psi_i^{(l)} + \psi_i^{(k)} \dot\psi_i^{(l)})x + (B_{\psi i}^{(k)} \psi_i^{(l)} + \dot\psi_i^{(k)} \dot\psi_i^{(l)} + \psi_i^{(k)} B_{\psi i}^{(l)})x^2 \\
    &+ (A_{\psi i}^{(k)} \psi_i^{(l)} + B_{\psi i}^{(k)} \dot\psi_i^{(l)} + \dot\psi_i^{(k)} B_{\psi i}^{(l)} + \psi_i^{(k)} A_{\psi i}^{(l)})x^3 \\
    &+ (A_{\psi i}^{(k)} \dot\psi_i^{(l)} + B_{\psi i}^{(k)} B_{\psi i}^{(l)} + \dot\psi_i^{(k)} A_{\psi i}^{(l)})x^4 + (A_{\psi i}^{(k)} B_{\psi i}^{(l)} + B_{\psi i}^{(k)} A_{\psi i}^{(l)})x^5 + A_{\psi i}^{(k)} A_{\psi i}^{(l)}x^6 \end{aligned} \right] \,{\rm d}x \\
    \nonumber &= \sum_{i=0}^{N-1} \left[ \begin{aligned} &\psi_i^{(k)} \psi_i^{(l)}h + \frac{1}{2}(\dot\psi_i^{(k)} \psi_i^{(l)} + \psi_i^{(k)} \dot\psi_i^{(l)})h^2 + \frac{1}{3}(B_{\psi i}^{(k)} \psi_i^{(l)} + \dot\psi_i^{(k)} \dot\psi_i^{(l)} + \psi_i^{(k)} B_{\psi i}^{(l)})h^3 \\
    &+ \frac{1}{4}(A_{\psi i}^{(k)} \psi_i^{(l)} + B_{\psi i}^{(k)} \dot\psi_i^{(l)} + \dot\psi_i^{(k)} B_{\psi i}^{(l)} + \psi_i^{(k)} A_{\psi i}^{(l)})h^4 \\
    &+ \frac{1}{5}(A_{\psi i}^{(k)} \dot\psi_i^{(l)} + B_{\psi i}^{(k)} B_{\psi i}^{(l)} + \dot\psi_i^{(k)} A_{\psi i}^{(l)})h^5 + \frac{1}{6}(A_{\psi i}^{(k)} B_{\psi i}^{(l)} + B_{\psi i}^{(k)} A_{\psi i}^{(l)})h^6 + \frac{1}{7} A_{\psi i}^{(k)} A_{\psi i}^{(l)}h^7 \end{aligned} \right] \\
    &= \sum_{i=0}^{N-1} \left[ \begin{aligned} &\frac{1}{70} (26 \psi_i^{(k)} \psi_i^{(l)} + 9 \psi_i^{(k)} \psi_{i+1}^{(l)} + 9 \psi_{i+1}^{(k)} \psi_i^{(l)} + 26 \psi_{i+1}^{(k)} \psi_{i+1}^{(l)}) h \\ &+ \frac{11}{210} (\dot\psi_i^{(k)} \psi_i^{(l)} - \dot\psi_{i+1}^{(k)} \psi_{i+1}^{(l)} + \psi_i^{(k)} \dot\psi_i^{(l)} - \psi_{i+1}^{(k)} \dot\psi_{i+1}^{(l)}) h^2 \\
    &+ \frac{13}{420} (\dot\psi_i^{(k)} \psi_{i+1}^{(l)} - \psi_i^{(k)} \dot\psi_{i+1}^{(l)} + \psi_{i+1}^{(k)} \dot\psi_i^{(l)} - \dot\psi_{i+1}^{(k)} \psi_i^{(l)}) h^2 \\ &+ \frac{1}{420} (4 \dot\psi_i^{(k)} \dot\psi_i^{(l)} - 3 \dot\psi_i^{(k)} \dot\psi_{i+1}^{(l)} - 3 \dot\psi_{i+1}^{(k)} \dot\psi_i^{(l)} + 4 \dot\psi_{i+1}^{(k)} \dot\psi_{i+1}^{(l)}) h^3 \end{aligned} \right].
\end{align}
\end{adjustwidth}

\begin{figure}[H]
    \subfloat{
        \includegraphics[width=0.48\textwidth]{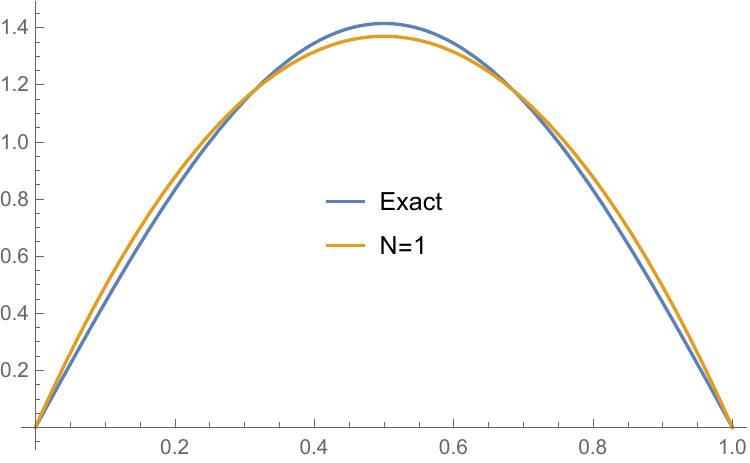}
    }
    \subfloat{
        \includegraphics[width=0.48\textwidth]{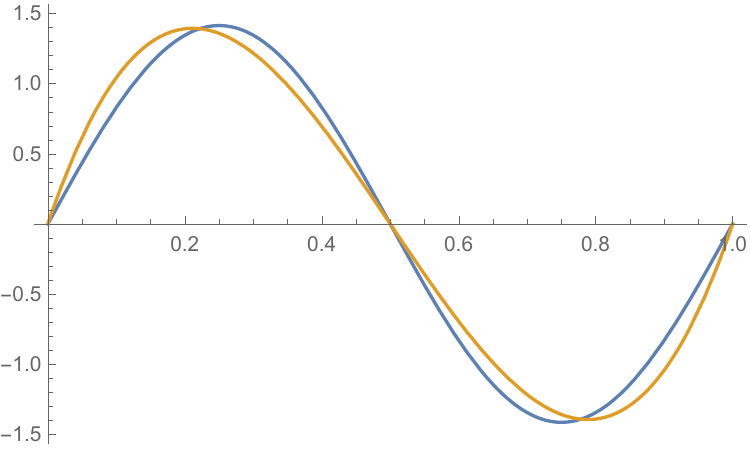}
    }
    \caption{Infinite potential well, comparisons between first-order ContEvol toy version ($N=1$) results (orange) with exact solution Equation~(\ref{eq:IPW_exact}) (blue) for $N=1$ ({\bf left}) and $N=2$ ({\bf right}).}
    \label{fig:IPW_N=1_vec}
\end{figure}

\paragraph{{Toy version: $N=1$.}
}
While $N=1$ and $h = 1/N = 1$, with $\psi_0 = \psi_N = 0$ enforced, the $P$ and $Q$ matrices are~simply 
\begin{align}
    P = \begin{pmatrix} 2/105 & -1/70 \\ -1/70 & 2/105 \end{pmatrix} \quad {\rm and} \quad
    Q = \begin{pmatrix} -4/15 & 1/15 \\ 1/15 & -4/15 \end{pmatrix},
\end{align}
and the Harmiltonian is
\begin{align}
    H = -P^{-1}Q = \begin{pmatrix} 26 & 16 \\ 16 & 26 \end{pmatrix}.
\end{align}
Eigendecomposition and normalization yield
\begin{align}
    \left\{ \begin{aligned} &\psi_1(x) = \sqrt{30}(x-x^2) \quad 0 \leq x \leq 1 \quad &&E_1 = 10 \approx 1.0132 \pi^2 \\ 
    &\psi_2(x) = \sqrt{210}(x-3x^2+2x^3) \quad 0 \leq x \leq 1 \quad &&E_2 = 42 \approx 1.0639 (2\pi)^2 \end{aligned} \right.; 
\end{align}
see Figure~\ref{fig:IPW_N=1_vec} for comparisons between these results and exact solution Equation~(\ref{eq:IPW_exact}) for $n=1$ and $n=2$. Like in Section~\ref{ss:IPW_simple}, these two wavefunctions are automatically orthogonal to each other.

\paragraph{{Realistic versions: $N=2$, $N=4$, and $N=8$.}}
Although the toy version results seem promising, one needs to use a larger $N$ for more accurate results and larger quantum numbers.

Figure~\ref{fig:IPW_mats_vecs} shows $P$, $Q$, and $H$ matrices, as well as normalized but not necessarily orthogonal eigenvectors produced by $N=1$, $N=2$, $N=4$, and $N=8$ versions of first-order ContEvol.
$P$, $Q$, and $H$ are all $2N \times 2N$ matrices. In each of them, the upper left $(N-1) \times (N-1)$ block (absent in the $N=1$ case) describes coupling between $\psi_i$ and $\psi_{i+1}$, the lower right $(N+1) \times (N+1)$ block describes coupling between $\dot\psi_i$ and $\dot\psi_{i+1}$, and {the} 
 other two blocks (both absent in the $N=1$ case) describe coupling between values and derivatives.
All these blocks are tridiagonal; because of the special form of $P^{(i)}$ and $Q^{(i)}$ submatrices Equation~(\ref{eq:IPW_CE1_PiQi}), the central diagonals of the cross blocks are uniformly zero.
From the t{hird column} 
, it is clear that the Hamiltonians are not symmetric; nevertheless, the upper left $(N-1) \times (N-1)$ blocks (absent in the $N=1$ case) and the lower right $(N+1) \times (N+1)$ blocks are symmetric.
Intuitively, the Hamiltonians should still be Hermitian if we consider them as operators on function representations $\{ \psi_i, \dot\psi_i \}$.
Shown in the {last column} are the eigenvectors: the first $N-1$ components of each row (absent in the $N=1$ case) are $\psi_i$ for $i = 1, 2, \ldots, N-1$, while the last $N+1$ components are $\dot\psi_i$ for $i = 0, 1, \ldots, N$.
Similar patterns can be seen from eigenvectors with different $N$ values. For example, both $\psi^{(2N)}$ and $\psi^{(N)}$ are zero or almost zero at nodes (not shown in the $N=1$ case), but the former has the same first derivatives, while the latter has alternating first~derivatives.

\vspace{-16pt}

\begin{figure}[H]
    \subfloat{
        \includegraphics[width=0.23\textwidth]{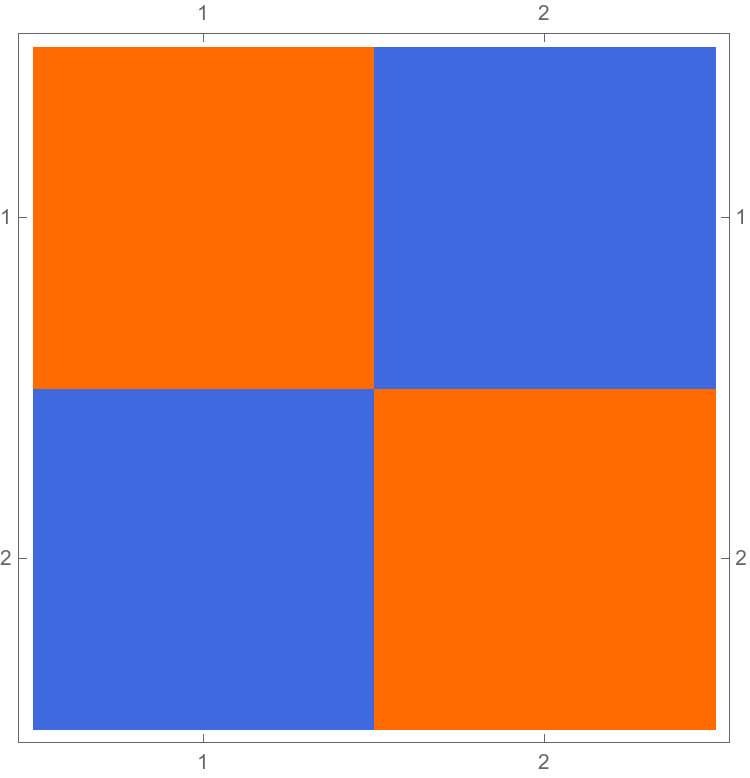}
    }
    \subfloat{
        \includegraphics[width=0.23\textwidth]{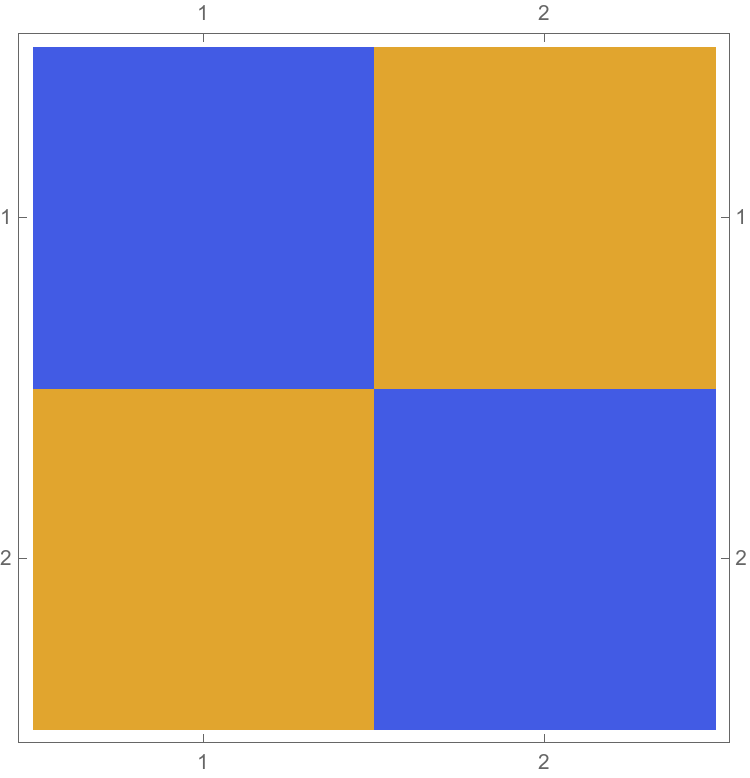}
    }
    \subfloat{
        \includegraphics[width=0.23\textwidth]{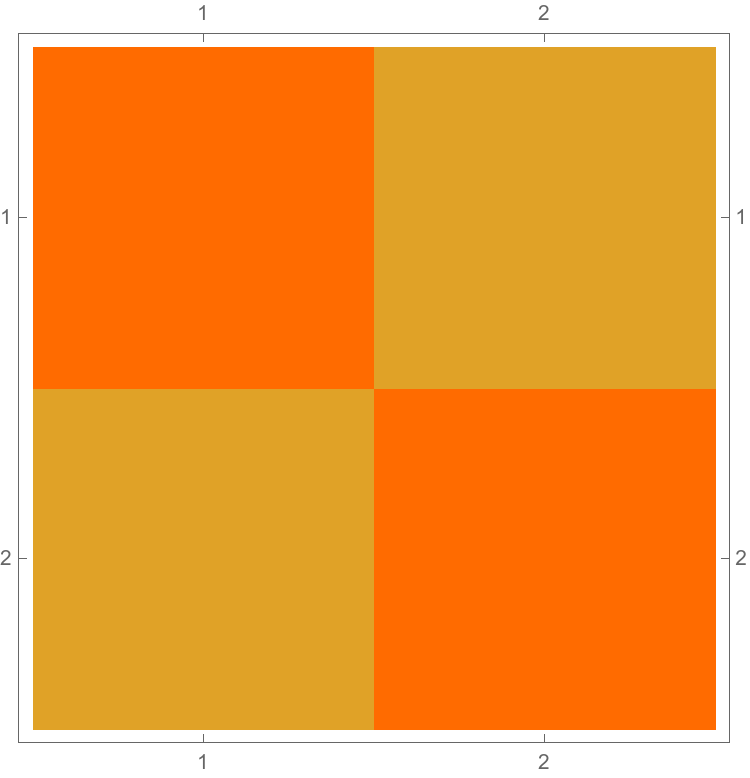}
    }
    \subfloat{
        \includegraphics[width=0.23\textwidth]{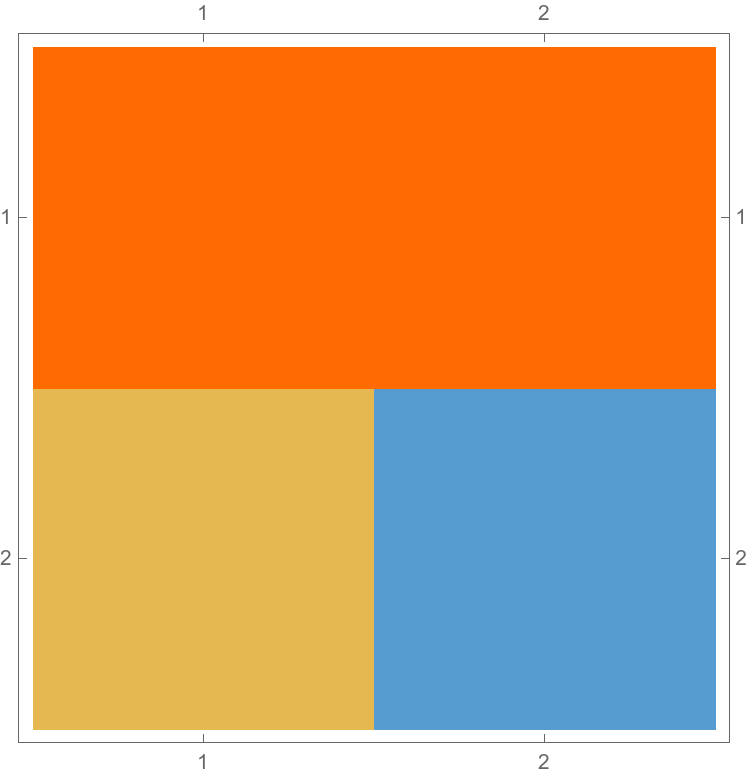}
    }

    \subfloat{
        \includegraphics[width=0.23\textwidth]{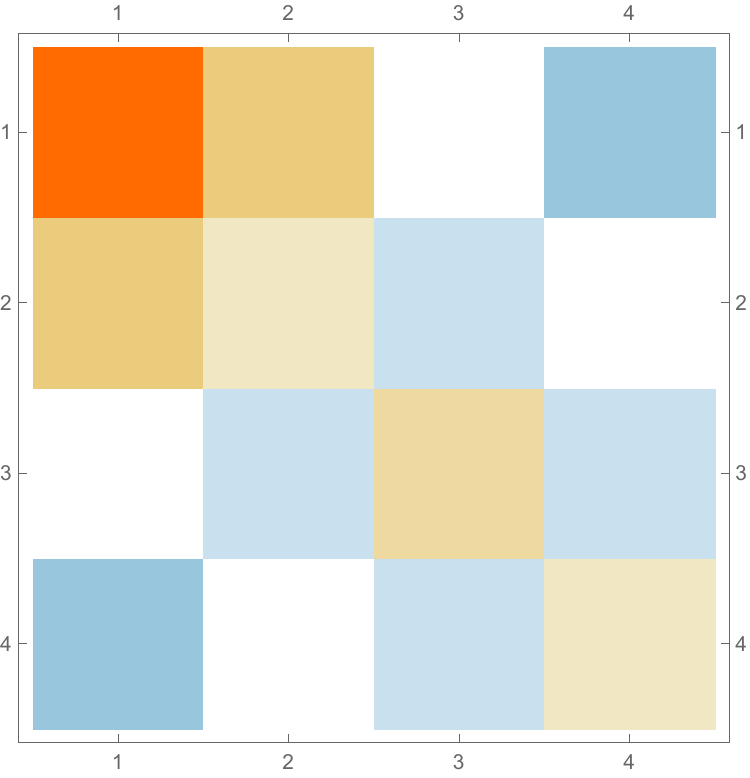}
    }
    \subfloat{
        \includegraphics[width=0.23\textwidth]{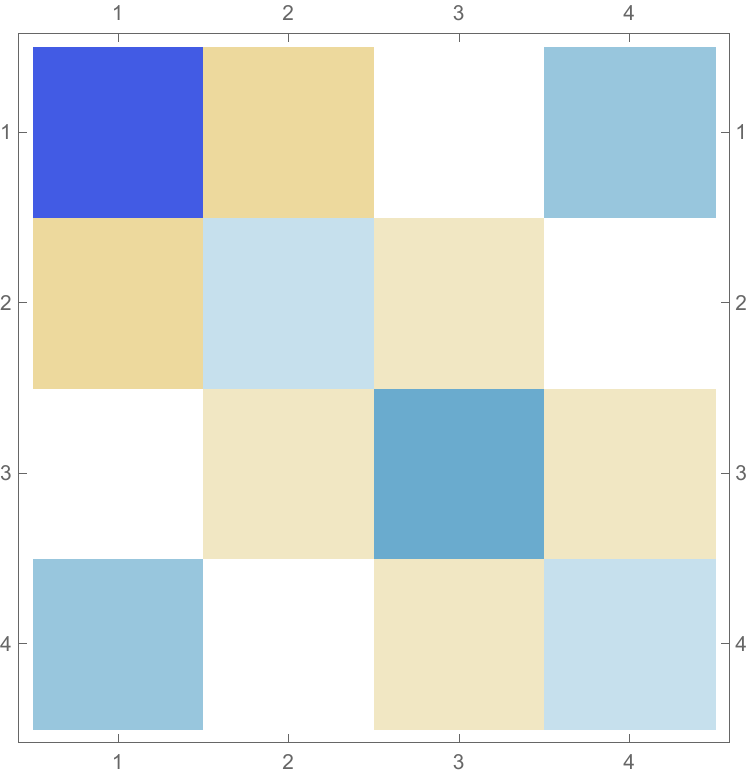}
    }
    \subfloat{
        \includegraphics[width=0.23\textwidth]{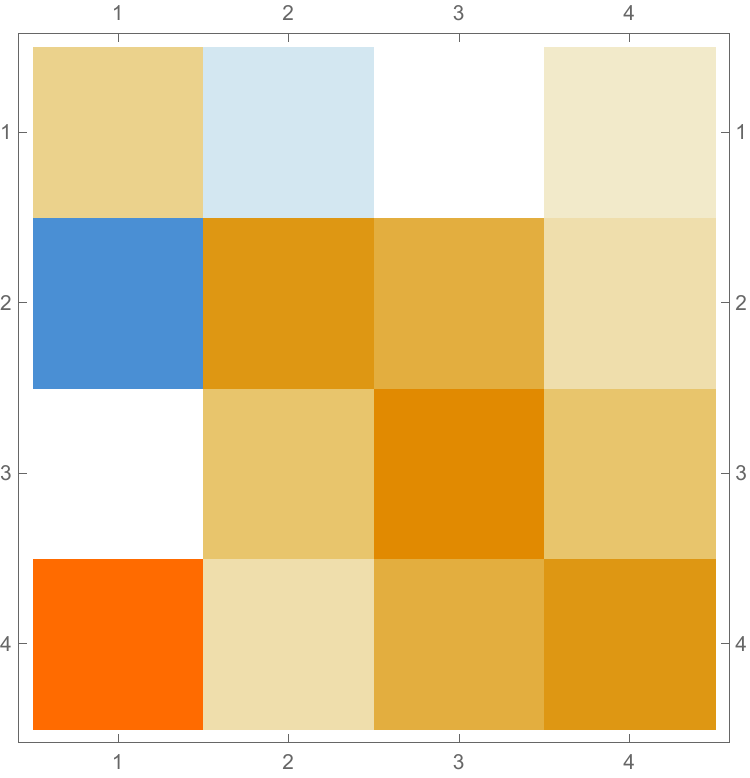}
    }
    \subfloat{
        \includegraphics[width=0.23\textwidth]{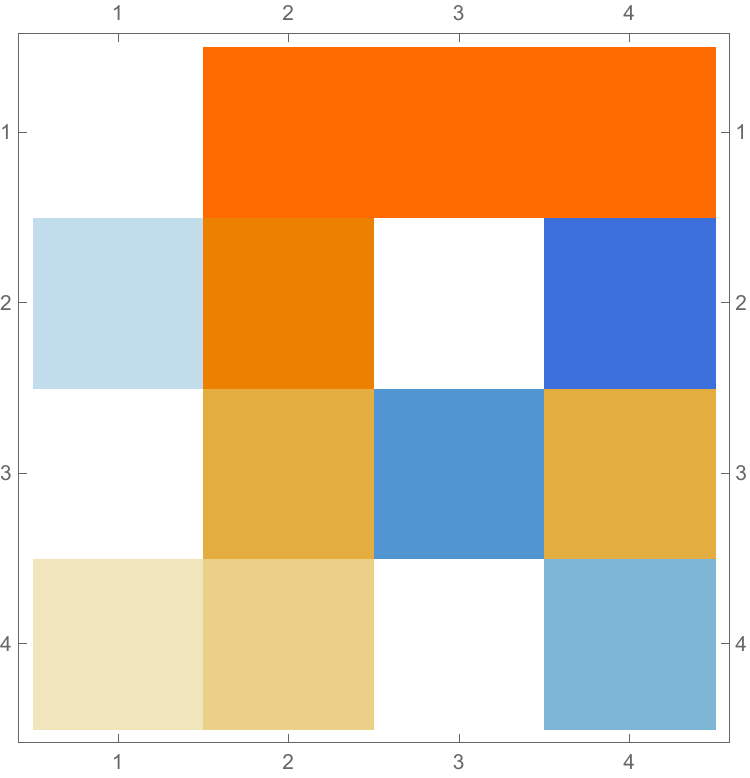}
    }
        
        \subfloat{
        \includegraphics[width=0.23\textwidth]{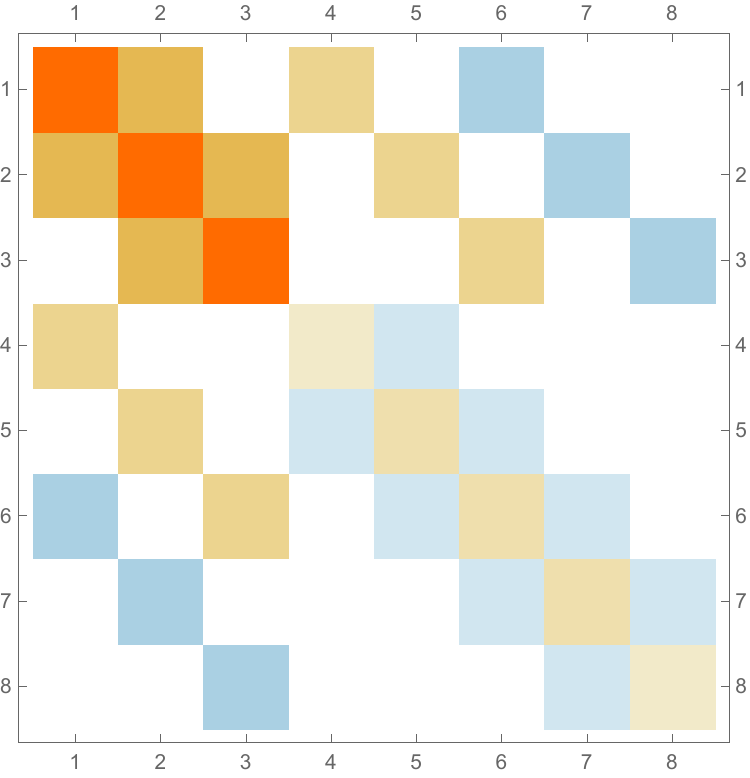}
    }
    \subfloat{
        \includegraphics[width=0.23\textwidth]{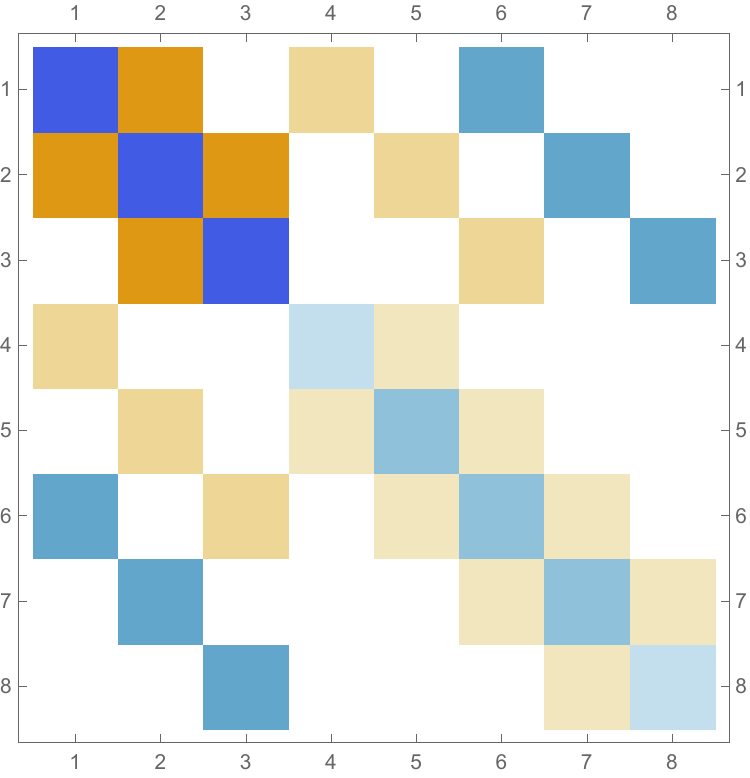}
    }
    \subfloat{
        \includegraphics[width=0.23\textwidth]{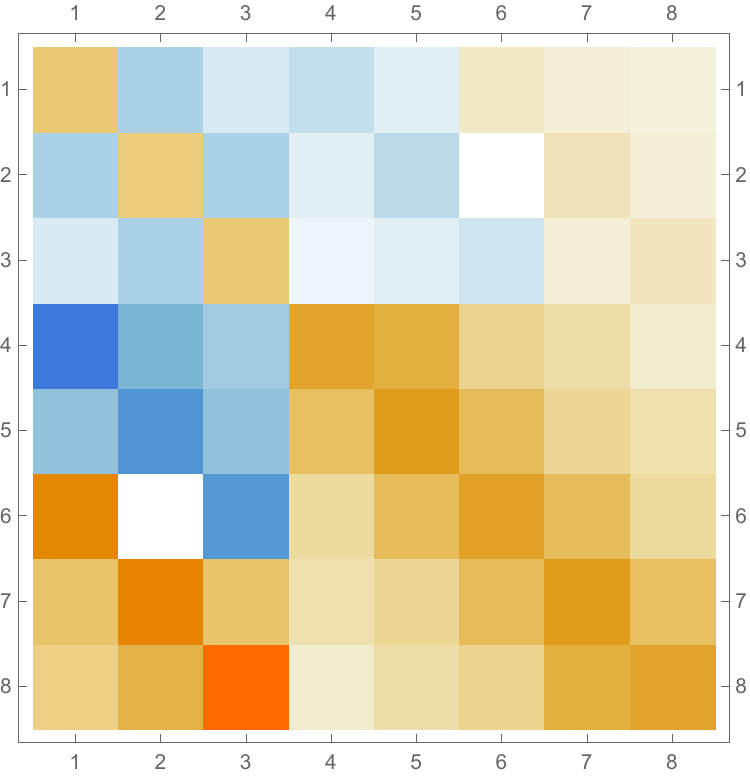}
    }
    \subfloat{
        \includegraphics[width=0.23\textwidth]{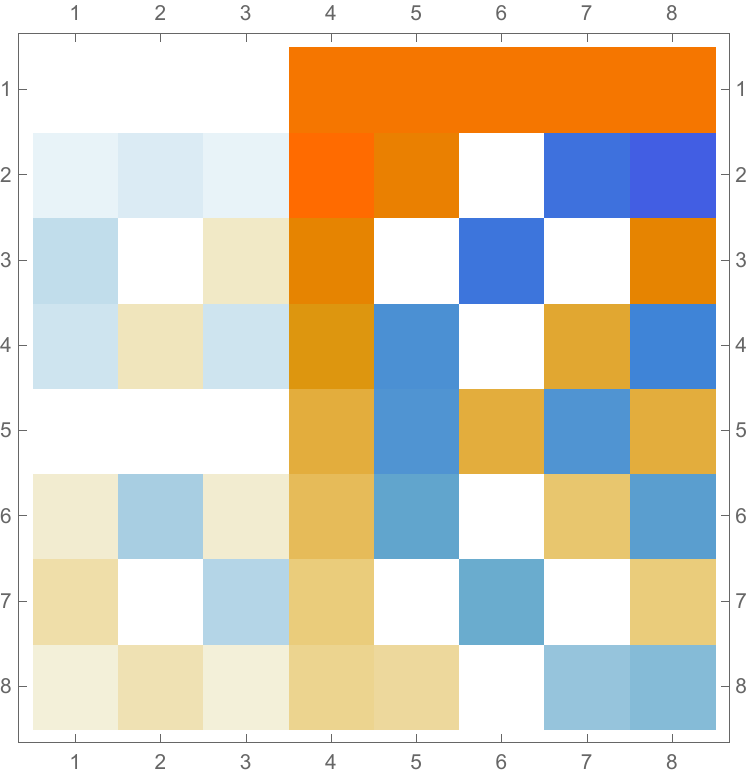}
    }
    
    \caption{\textit{Cont}.}
    \label{fig:IPW_mats_vecs}
\end{figure}

\begin{figure}[H]\ContinuedFloat

    \subfloat{
        \includegraphics[width=0.23\textwidth]{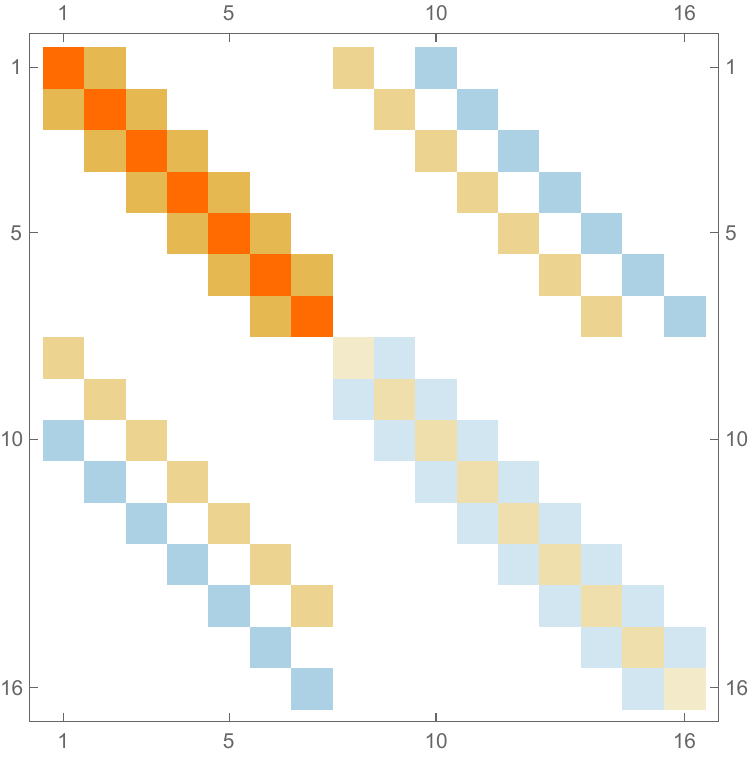}
    }
    \subfloat{
        \includegraphics[width=0.23\textwidth]{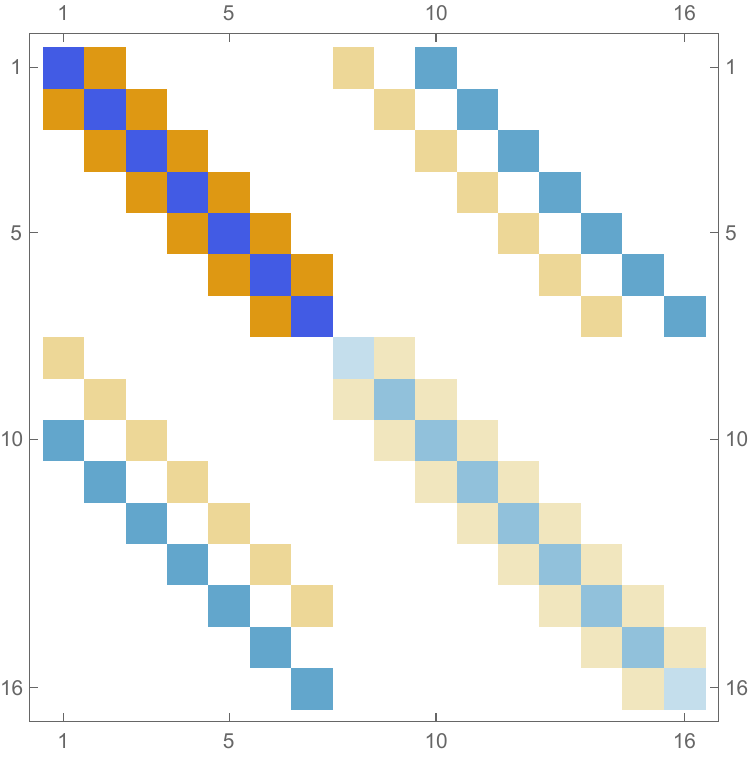}
    }
    \subfloat{
        \includegraphics[width=0.23\textwidth]{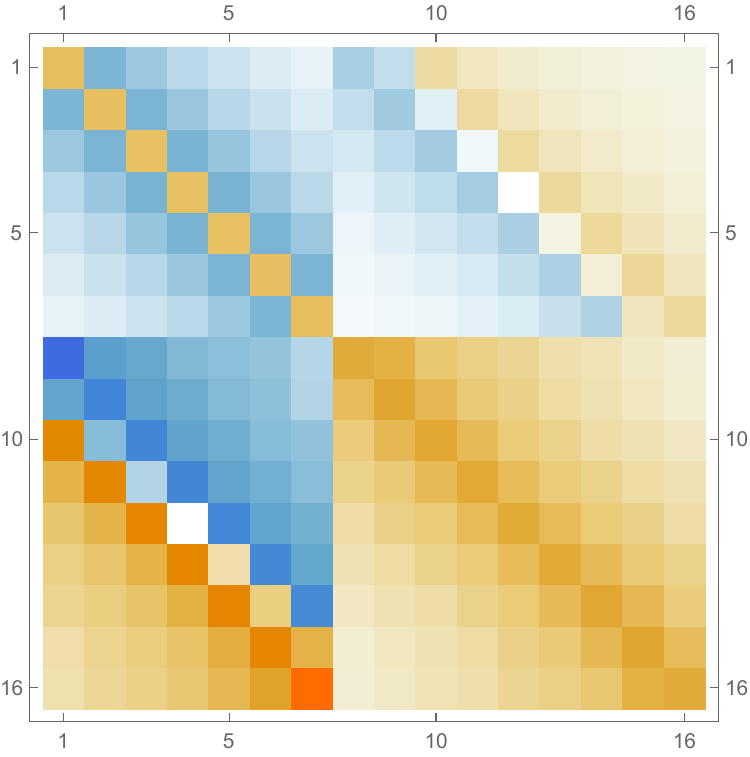}
    }
    \subfloat{
        \includegraphics[width=0.23\textwidth]{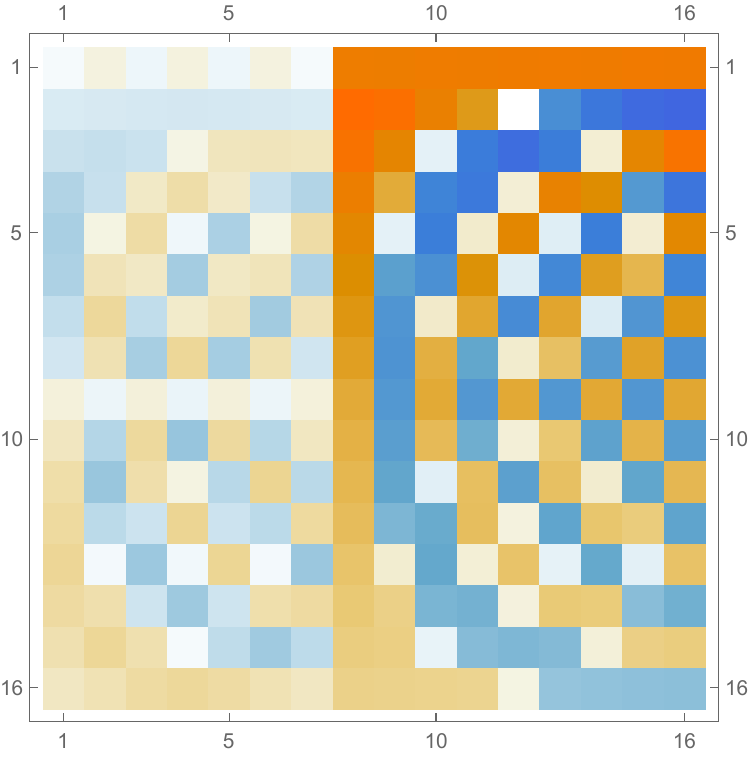}
    }
    \caption{{Infinite} potential well, $P$ ({\bf first column}), $Q$ ({\bf second column}), and $H$ matrices ({\bf third column}), together with eigenvectors ({\bf last column}) for $N=1$ ({\bf first row}), $N=2$ ({\bf second row}), $N=4$ ({\bf third row}), and $N=8$ ({\bf last row}) versions of first-order ContEvol.
    Following Mathematica convention, the eigenvectors are presented horizontally and ordered by decreasing eigenvalues (i.e., first row is $\psi^{(2N)}$, last row is $\psi^{(1)}$).
    They are normalized in terms of Equation~(\ref{eq:IPW_CE1_normal}), but not deliberately orthogonalized in terms of Equation~(\ref{eq:IPW_CE1_ortho}); their signs are set so that $\dot\psi_0$ (the $N$th component) is positive in all cases.}
    \label{fig:IPW_mats_vecs}
\end{figure}

\begin{figure}[H]
    \includegraphics[width=0.48\textwidth]{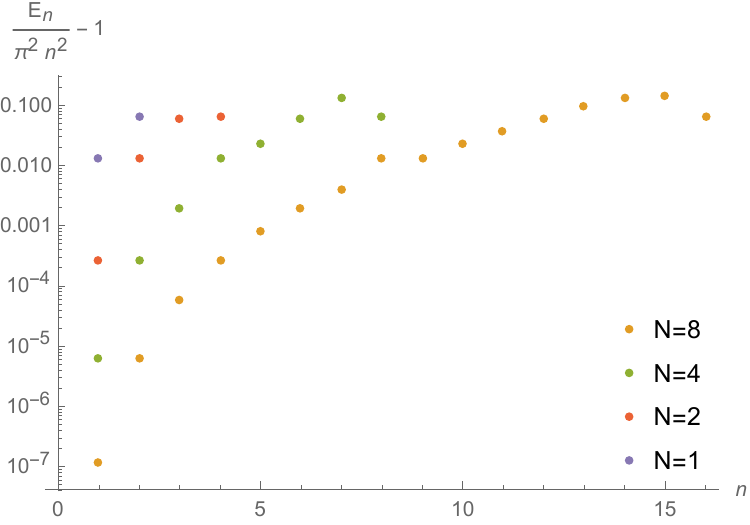}
    \caption{{Infinite} potential well, $n$th eigenvalue $E_n$ divided by its exact counterpart Equation~(\ref{eq:IPW_exact}) minus $1$ versus quantum number $n$. $N=8$ (orange), $N=4$ (green), $N=2$ (red), and $N=1$ (purple) results of first-order ContEvol are shown in different colors.}
    \label{fig:IPW_eigenvalues}
\end{figure}

Figures~\ref{fig:IPW_eigenvalues} and \ref{fig:IPW_eigenvectors} display errors in eigenvalues and rendered eigenvectors of $N=1$, $N=2$, $N=4$, and $N=8$ Hamiltonians, respectively.
With only a quarter of the number of parameters used in simple discretization (see Section~\ref{ss:IPW_simple}), ContEvol results are arguably better, especially for the ground state energy $E_1$.
Since a $2N \times 2N$ matrix only has (at most) $2N$ eigenpairs, $E_n$ and $\psi^{(n)}$ with large $n$ are only available with large $N$.
The quality of the results significantly deteriorates as $n$ approaches $2N$; it reaches the worst case at $2N-1$, and becomes reasonably good at $2N$, when our sampling nodes coincide with zero points of the wavefunctions.
Based on these two figures, a rule of thumb would be to only trust $n \leq N$ results, so that errors in eigenvalues are below or at the $\sim$1\% level.

\vspace{-13pt}

\begin{figure}[H]
        \subfloat{
        \includegraphics[width=0.32\textwidth]{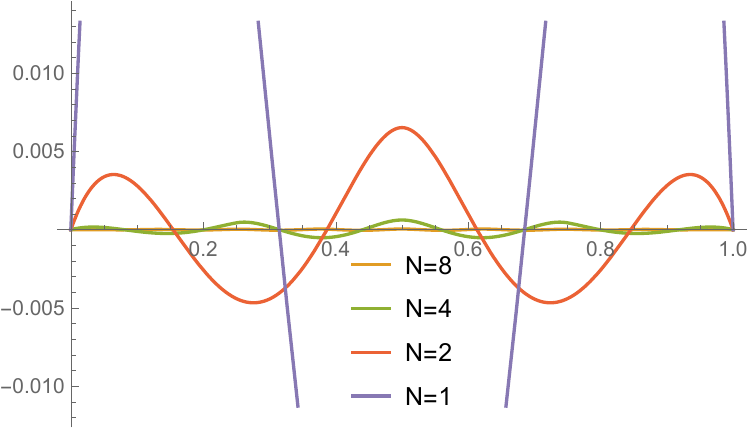}
    }
    \subfloat{
        \includegraphics[width=0.32\textwidth]{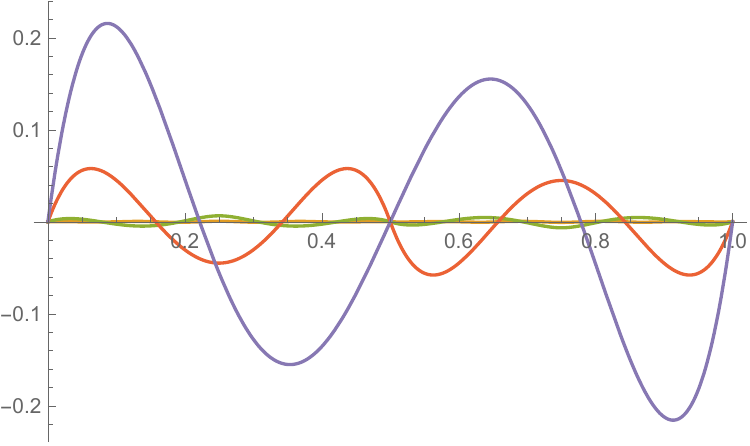}
    }
    \subfloat{
        \includegraphics[width=0.32\textwidth]{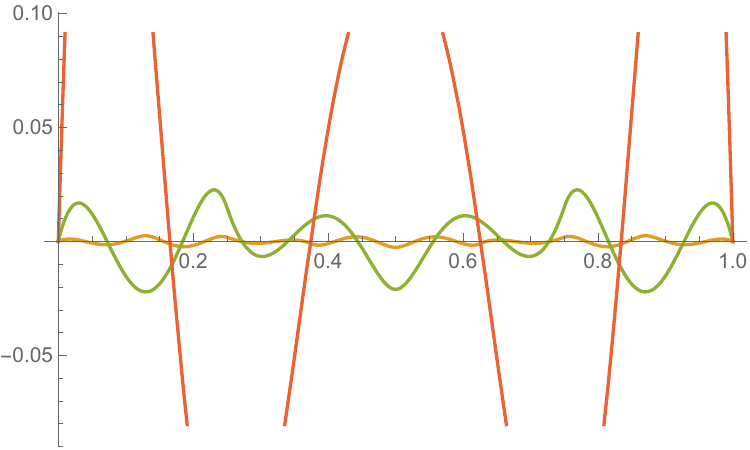}
    }
    
     \subfloat{
        \includegraphics[width=0.32\textwidth]{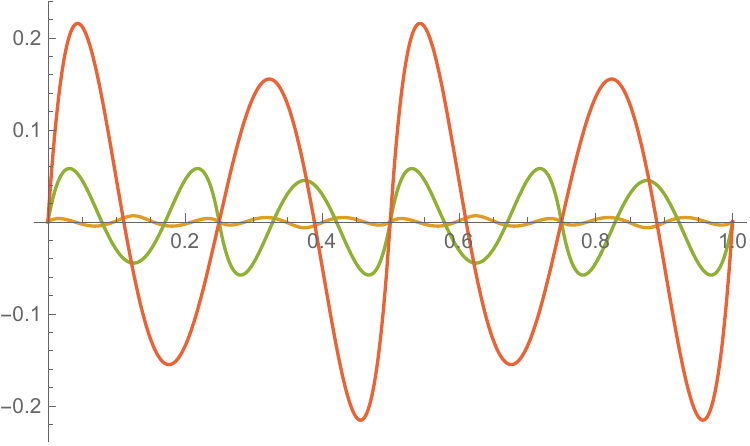}
    }
    \subfloat{
        \includegraphics[width=0.32\textwidth]{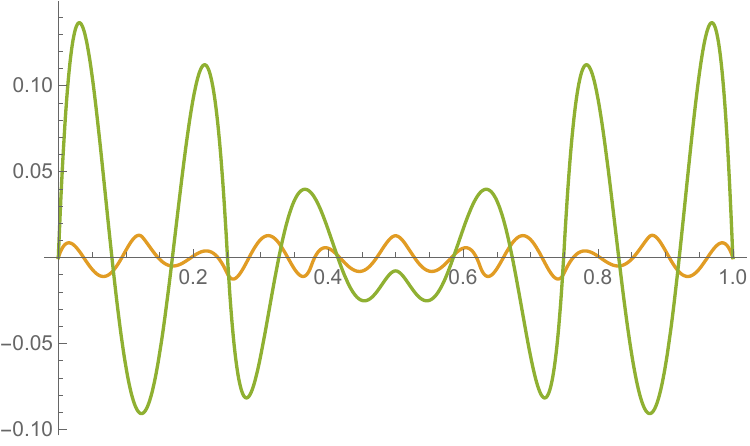}
    }
    \subfloat{
        \includegraphics[width=0.32\textwidth]{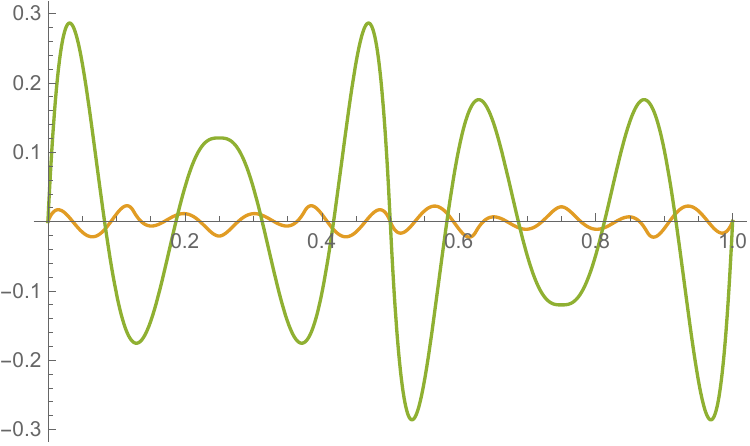}
    }
    
    \caption{\textit{Cont}.}
     \label{fig:IPW_eigenvectors}
\end{figure}

\begin{figure}[H]\ContinuedFloat

    \centering \subfloat{
        \includegraphics[width=0.32\textwidth]{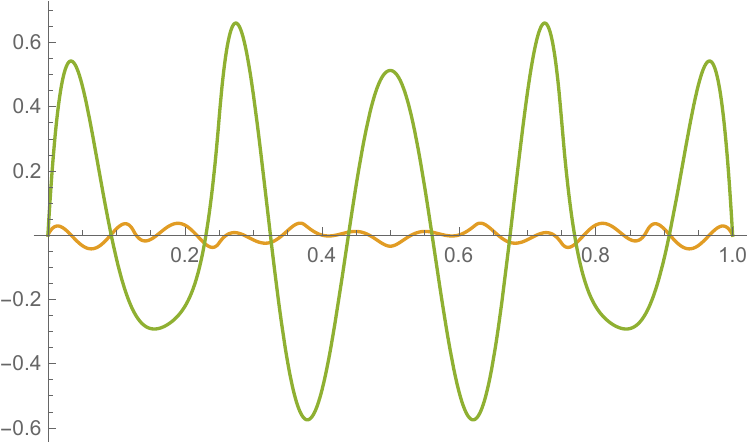}
    }
   \centering  \subfloat{
        \includegraphics[width=0.32\textwidth]{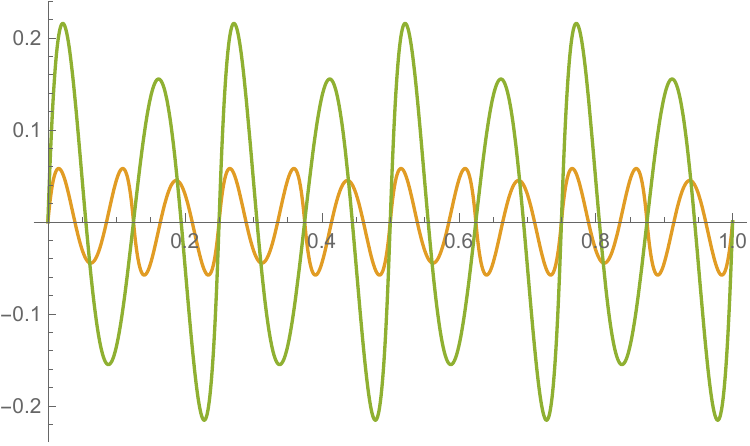}
    }

    \subfloat{
        \includegraphics[width=0.32\textwidth]{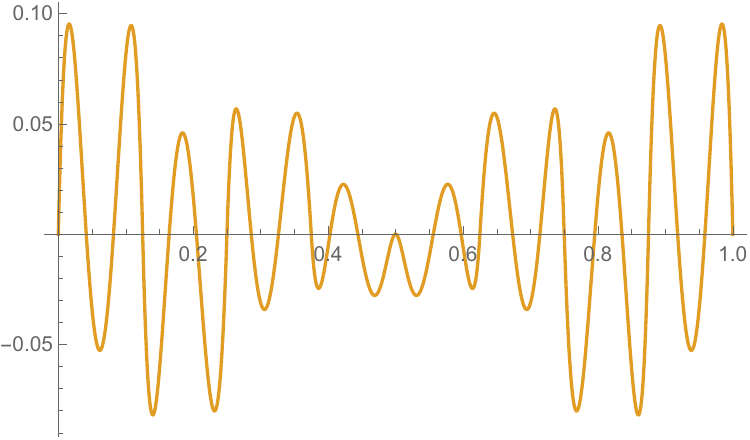}
    }
    \subfloat{
        \includegraphics[width=0.32\textwidth]{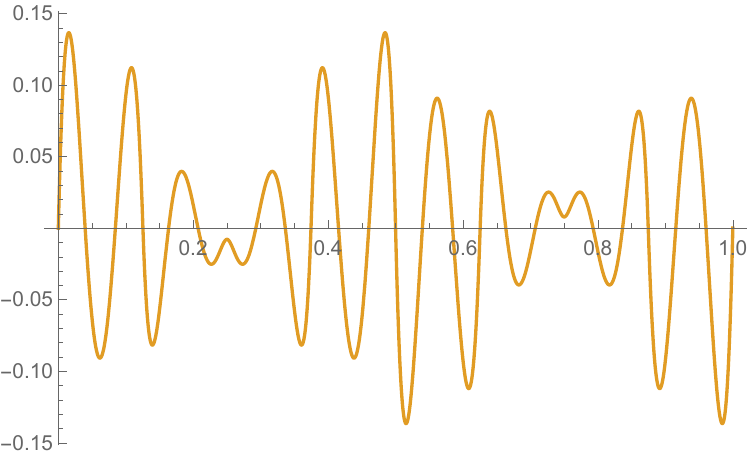}
    }
    \subfloat{
        \includegraphics[width=0.32\textwidth]{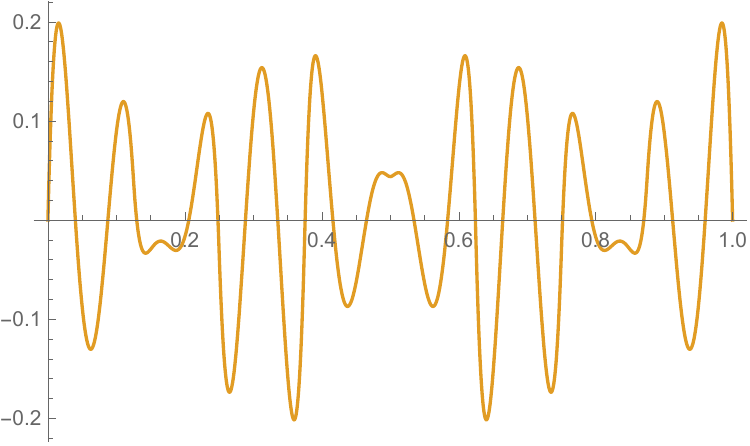}
    }

    \subfloat{
        \includegraphics[width=0.32\textwidth]{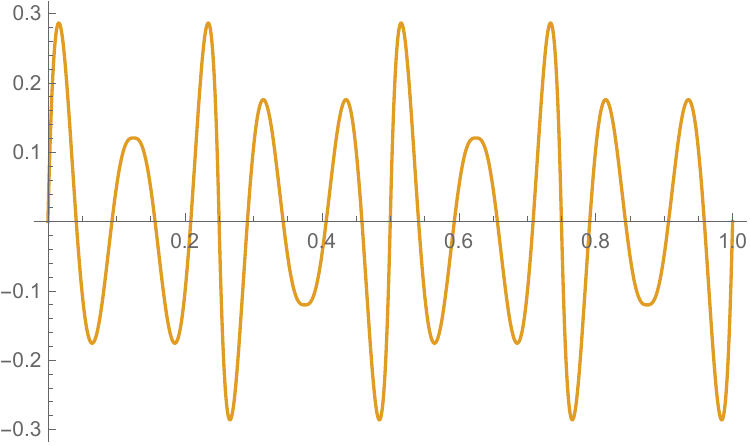}
    }
    \subfloat{
        \includegraphics[width=0.32\textwidth]{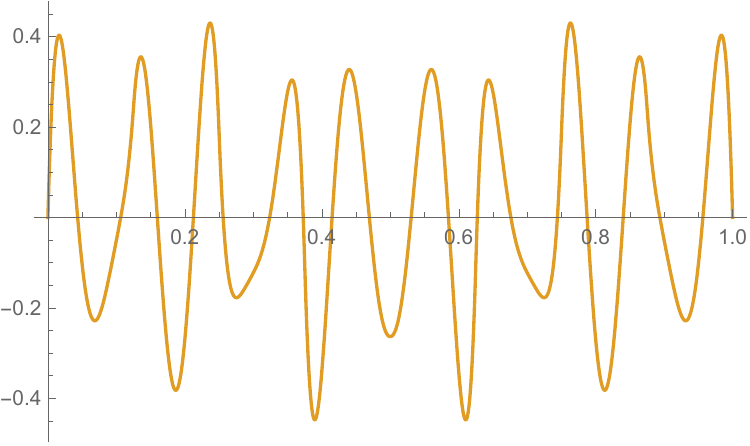}
    }
    \subfloat{
        \includegraphics[width=0.32\textwidth]{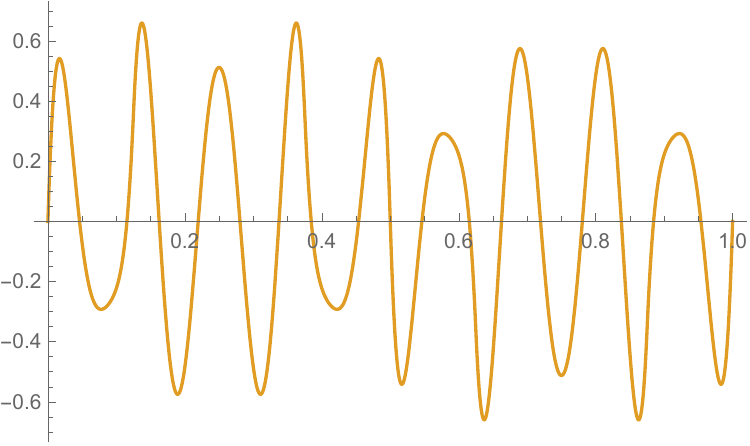}
    }

    \subfloat{
        \includegraphics[width=0.32\textwidth]{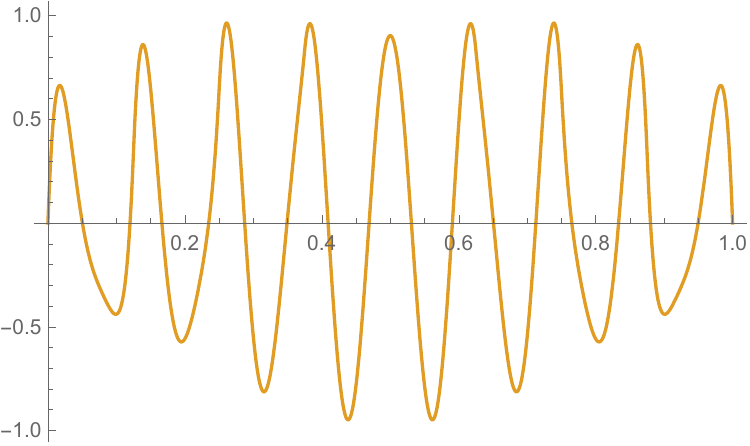}
    }
    \subfloat{
        \includegraphics[width=0.32\textwidth]{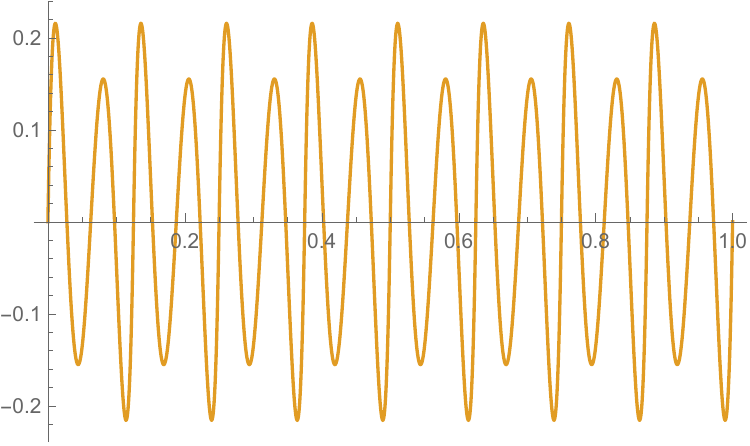}
    }
    \caption{{Infinite} potential well, errors in rendered wavefunctions of $N=8$ (orange), $N=4$ (green), $N=2$ (red), and $N=1$ (purple) results of first-order ContEvol. Note that magnitude of exact wavefunctions is $\sqrt{2}$.}
    \label{fig:IPW_eigenvectors}
\end{figure}

\subsection{Harmonic Oscillator, First-Order ContEvol (Description)}
\label{ss:QHO_CE1}

In this section, we consider (quantum) harmonic oscillator with potential
\begin{align}
    \label{eq:QHO_potential} V(x) = x^2, \quad x \in \mathbb{R},
\end{align}
where we have set the constant $k/2$ to $1$; note that this only affects the scaling of $x$. The exact wavefunctions can be expressed using Hermite polynomials; we do not include them here as no comparisons will be made.

As for application of the ContEvol method, there are three major differences between harmonic oscillator and infinite potential well, which we describe one by one.

\paragraph{{Difference 1: Position-dependent potential.}
}
In the case of infinite potential well, the potential $V(x)$ is uniformly zero in the interval of interest; the case of harmonic oscillator is different.
Consequently, each piece of the cost function needs to be written as (subscript ``QHO'' stands for quantum harmonic oscillator)
\begin{adjustwidth}{-\extralength}{0cm}
\begin{align}
    \nonumber \epsilon_{{\rm QHO},i} &= \int_{x_i}^{x_{i+1}} (\ddot{\psi} - V\psi + \phi)^2 \,{\rm d}x = \int_{x_i}^{x_{i+1}} \left[ \begin{aligned} &\{2B_{\psi i} + 6A_{\psi i}(x-x_i)\} - x^2 \\ &\cdot \{\psi_i + \dot\psi_i(x-x_i) + B_{\psi i}(x-x_i)^2 + A_{\psi i}(x-x_i)^3\} \\ &+ \{\phi_i + \dot\phi_i(x-x_i) + B_{\phi i}(x-x_i)^2 + A_{\phi i}(x-x_i)^3\} \end{aligned} \right]^2 \,{\rm d}x \\
    \nonumber &= \int_{x_i}^{x_{i+1}} \left[ \begin{aligned} &\{2B_{\psi i} + 6A_{\psi i}(x-x_i)\} - \{ x_i^2 + 2 x_i (x - x_i) + (x - x_i)^2 \} \\ &\cdot \{\psi_i + \dot\psi_i(x-x_i) + B_{\psi i}(x-x_i)^2 + A_{\psi i}(x-x_i)^3\} \\ &+ \{\phi_i + \dot\phi_i(x-x_i) + B_{\phi i}(x-x_i)^2 + A_{\phi i}(x-x_i)^3\} \end{aligned} \right]^2 \,{\rm d}x \\
    &= \int_0^h \left[ \begin{aligned} &(2B_{\psi i} + 6A_{\psi i}x) - (x_i^2 + 2 x_i x + x^2) \\ &\cdot (\psi_i + \dot\psi_i x + B_{\psi i}x^2 + A_{\psi i}x^3) + (\phi_i + \dot\phi_i x + B_{\phi i}x^2 + A_{\phi i}x^3) \end{aligned} \right]^2 \,{\rm d}x = \cdots,
\end{align}
\end{adjustwidth}
where we have omitted results of the expansion, squaring, integral, and substitution steps (``$\cdots$'').
It should be noted that, fortunately, ContEvol is robust against complications induced by the position-dependent potential function, because $\epsilon_{{\rm QHO},i}$ is still a finite polynomial of $h$, of which all coefficients are linear combinations of $\{ \psi_i, \dot\psi_i, \psi_{i+1}, \dot\psi_{i+1} \}$, $\{ \phi_i, \dot\phi_i, \phi_{i+1}, \dot\phi_{i+1} \}$, and $\{ x_i, x_{i+1} \}$.

In general, a potential function $V(x)$ can be represented as $\{ V_i \equiv V(x_i) \}$ and \linebreak  $\{ \dot V_i \equiv \dot V(x_i) \}$, even if it is a hard-to-integrate transcendental function or does not have an analytic form.
In the regime of first-order ContEvol, each piece of $V(x)$ possesses up to the third order in $x$, ergo the resulting expression of each piece of the cost function has up to the thirteenth order in $h$; when $h$ is small, it is reasonable to truncate the expansion of the square root of the integrand at the third order in $x$, so that the final expression has up to the seventh order in $h$, like in Sections~\ref{ss:CHO_CE1} or \ref{ss:2BP_CE1A}.
Note that when $h$ denotes the length of each sub-interval, it is not necessarily small, specifically not necessarily smaller than $1$, hence higher-order terms may be more important than lower-order ones.

\paragraph{{Difference 2: Lack of sharp edges.}}
Unlike Equation~(\ref{eq:IPW_potential}), Equation~(\ref{eq:QHO_potential}) does not require wavefunctions to vanish at specific, finitely distant positions; figuratively speaking, wavefunctions are allowed to (and actually should) have tails.
Therefore, we need to define $\epsilon_{{\rm QHO},-1}$ and $\epsilon_{{\rm QHO},N}$---not just for convenience, but also for accuracy.

Wavefunctions are supposed to vanish at infinity, i.e., satisfy $\psi(-\infty) = \psi(+\infty) = 0$ and $\dot\psi(-\infty) = \dot\psi(+\infty) = 0$.
Given $\psi_0$ and $\dot\psi_0$ or $\psi_N$ and $\dot\psi_N$, it is impossible to find a cubic representation of $\psi(x)$ in the interval $(-\infty, x_0]$ or $[x_N, +\infty)$; however, assuming that $\psi_0$ and $\dot\psi_0$ have same signs while $\psi_N$ and $\dot\psi_N$ have opposite signs, there is always a pair of exponential tails
\vspace{-4pt}
\begin{align}
    \psi(x) = \left\{ \begin{aligned} &\psi_0 \exp \left[ \frac{\dot\psi_0}{\psi_0} (x-x_0) \right] \quad &&x \leq x_0 \\
    &\psi_N \exp \left[ \frac{\dot\psi_N}{\psi_N} (x-x_N) \right] \quad &&x \geq x_N \end{aligned} \right.
\end{align}
satisfying all these boundary conditions.
Expressing tails of $\phi(x)$ in the same way, tails of the cost function could be defined as
\vspace{-6pt}
\begin{adjustwidth}{-\extralength}{0cm}
\begin{align}
    \left\{ \begin{aligned} \epsilon_{{\rm QHO},-1} &= \int_{-\infty}^{x_0} (\ddot{\psi} - V\psi + \phi)^2 \,{\rm d}x \\ &= \int_{-\infty}^{x_0} \left[ \frac{\dot\psi_0^2}{\psi_0} \exp \left( \frac{\dot\psi_0}{\psi_0} (x-x_0) \right) - x^2 \psi_0 \exp \left( \frac{\dot\psi_0}{\psi_0} (x-x_0) \right) + \phi_0 \exp \left( \frac{\dot\phi_0}{\phi_0} (x-x_0) \right) \right]^2 \,{\rm d}x \\
    \epsilon_{{\rm QHO},N} &= \int_{x_N}^{+\infty} (\ddot{\psi} - V\psi + \phi)^2 \,{\rm d}x \\ &= \int_{x_N}^{+\infty} \left[ \frac{\dot\psi_N^2}{\psi_N} \exp \left( \frac{\dot\psi_N}{\psi_N} (x-x_N) \right) - x^2 \psi_N \exp \left( \frac{\dot\psi_N}{\psi_N} (x-x_N) \right) + \phi_N \exp \left( \frac{\dot\phi_N}{\phi_N} (x-x_N) \right) \right]^2 \,{\rm d}x\end{aligned} \right.,
\end{align}
\end{adjustwidth}
where integrals of exponential tails multiplied by $x^2$ (actually polynomial potential functions in general) can be expressed using gamma function.
Yet unfortunately, with $\psi_0$ and $\psi_N$, $\phi_0$ and $\phi_N$ as denominators, such tails break the linearity of our ContEvol formalism.
A natural solution would be to treat $\dot\psi_0/\psi_0$ and $\dot\psi_N/\psi_N$, $\dot\phi_0/\phi_0$ and $\dot\phi_N/\phi_N$ as fixed values in the tails; as a price, one would need to fine-tune $x_0$ and $x_N$, so that these ratios are indeed close to the corresponding fixed values. A related example will be presented in the next~section.

As for second-order ContEvol, the tails could be similarly written as
\begin{align}
    \psi(x) = \left\{ \begin{aligned} &\psi_0 \exp \left[ \frac{\dot\psi_0}{\psi_0} (x-x_0) + \frac{\ddot\psi_0 \psi_0 - \dot\psi_0^2}{2\psi_0^2} (x-x_0)^2 \right] \quad &&x \leq x_0 \\
    &\psi_N \exp \left[ \frac{\dot\psi_N}{\psi_N} (x-x_N) + \frac{\ddot\psi_N \psi_N - \dot\psi_N^2}{2\psi_N^2} (x-x_N)^2 \right] \quad &&x \geq x_N \end{aligned} \right.;
\end{align}
however, even {if} 
 one is willing to deal with non-linearity, since the error function does not have an analytic form, one may need to build numerical lookup tables for $\epsilon_{{\rm QHO},-1}$ and $\epsilon_{{\rm QHO},N}$.
In the linear regime, we can treat ratios like $\ddot\psi_0/\psi_0$ and $\ddot\psi_N/\psi_N$ to zeros as well, but it is not common for first and second derivatives to simultaneously satisfy constraints, hence we can only aim for having sensible $\dot\psi_0$ and $\dot\psi_N$ values.
Better and possibly intricate circumvention is beyond the scope of this work.

\paragraph{{Difference 3: Increasing ``sizes'' of wavefunctions.}}
For scenarios like (quantum) harmonic oscillator, the ``sizes'' of wavefunctions (which can be strictly quantified using percentiles of the probability distribution) increase with larger quantum numbers.
Meanwhile, with $N$ nodes, first-order ContEvol is supposed to yield $2N$ eigenvectors. Therefore, for similar problems, the spread of nodes probably needs to be adjusted according to test results.
Since the fine-tuning may require several iterations, objective evaluation criteria can be designed to automate this process; such efforts are left for future work, and probably for specific situations.

To summarize, harmonic oscillator manifests some of the difficulties encountered in real-world problems, but ContEvol methods should be able to handle them reasonably~well.

\subsection{Coulomb Potential, First-Order ContEvol}
\label{ss:QCP_CE1}

In this final section on quantum mechanics, we look at a more realistic case, one-dimensional Coulomb potential. Following Section \ref{ss:CHO_CE1} of \citet{Pradhan2011book}, the radial part of the stationary Schr\"odinger equation for a hydrogen atom can be written as
\begin{align}
    \left[ \frac{{\rm d}^2}{{\rm d}r^2} - V(r) - \frac{l(l+1)}{r^2} + E \right] P(r) = 0, \quad r \geq 0,
\end{align}
where we have used atomic units, the potential $V(r) = -2/r$, $l$ is the angular quantum number, and $P(r) \equiv r \cdot R(r)$ is a modified version of the radial wavefunction $R(r)$.
This work focuses on the ground state $n = 1$, hence we set $l = 0$; in our notation, the equation~becomes
\begin{align}
    -\ddot{\psi} - \frac{2}{r}\psi = E\psi, \quad r \geq 0,
\end{align}
and the exact solution is
\begin{align}
    \label{eq:H_exact} \psi^{(1)}(r) = 2r e^{-r}, \quad r \geq 0 \quad {\rm and} \quad E_1 = -1.
\end{align}

For simplicity, we sample the non-negative half of the real axis with $N + 1$ nodes
\begin{align}
    r_i = i \cdot h, \quad i = 0, 1, \ldots, N,
\end{align}
where $h$ is the width of each interval; {(Caution: In this section, $i$ is always a non-negative integer and never the imaginary unit.)} 
non-uniform sampling is left for future work. To handle the $1/r$ factor in the equation, we require each piece of the wavefunction $\psi(r)$ to be proportional to $r$; note that this strategy can be applied to Yukawa potential as well. Therefore the wavefunction is written as
\begin{align}
    \label{eq:H_CE1_psi} \psi(r) = \left\{ \begin{aligned} & r (D_{\psi i} + C_{\psi i} r + B_{\psi i} r^2 + A_{\psi i} r^3) && r_i \leq r \leq r_{i+1} \\
    &\psi_N \frac{r}{r_N} \exp \left( 1 - \frac{r}{r_N} \right) && r \geq r_N \end{aligned} \right.,
\end{align}
where we exclude $\dot\psi_N$ from the tail to maintain linearity of our framework.

The coefficients $D_{\psi i}$ through $A_{\psi i}$ are yielded by terminal conditions at $r = r_i$ and $r_{i+1}$
\begin{align}
    \left\{ \begin{aligned} \psi(r_i) &= r_i (D_{\psi i} + C_{\psi i} r_i + B_{\psi i} r_i^2 + A_{\psi i} r_i^3) = \psi_i \\
    \dot\psi(r_i) &= D_{\psi i} + 2C_{\psi i} r_i + 3B_{\psi i} r_i^2 + 4A_{\psi i} r_i^3 = \dot\psi_i \\
    \psi(r_{i+1}) &= r_{i+1} (D_{\psi i} + C_{\psi i} r_{i+1} + B_{\psi i} r_{i+1}^2 + A_{\psi i} r_{i+1}^3) = \psi_{i+1} \\
    \dot\psi(r_{i+1}) &= D_{\psi i} + 2C_{\psi i} r_{i+1} + 3B_{\psi i} r_{i+1}^2 + 4A_{\psi i} r_{i+1}^3 = \dot\psi_{i+1} \end{aligned} \right.;
\end{align}
since $r_i = i \cdot h$ and $r_{i+1} = (i+1) \cdot h$, for $i > 0$ we have
\begin{align}
    &\begin{pmatrix} ih & (ih)^2 & (ih)^3 & (ih)^4 \\
    1 & 2ih & 3(ih)^2 & 4(ih)^3 \\ 
    (i+1)h & [(i+1)h]^2 & [(i+1)h]^3 & [(i+1)h]^4 \\
    1 & 2(i+1)h & 3[(i+1)h]^2 & 4[(i+1)h]^3 \end{pmatrix}
    \begin{pmatrix} D_{\psi i} \\ C_{\psi i} \\ B_{\psi i} \\ A_{\psi i} \end{pmatrix}
    = \begin{pmatrix} \psi_i \\ \dot\psi_i \\ \psi_{i+1} \\ \dot\psi_{i+1} \end{pmatrix} 
    \end{align}
    \vspace{-22pt}
  \begin{adjustwidth}{-\extralength}{0cm}
  \begin{align}
    &\left\{ \begin{aligned} A_{\psi i} &= \left[ \frac{(2i-1) \psi_i}{i^2} - \frac{(2i+3) \psi_{i+1}}{(i+1)^2} \right] h^{-4} + \left[ \frac{\dot\psi_i}{i} + \frac{\dot\psi_{i+1}}{i+1} \right] h^{-3} \\
    B_{\psi i} &= -2 \left[ \frac{(3i^2-1) \psi_i}{i^2} - \frac{(3i^2+6i+2) \psi_{i+1}}{(i+1)^2} \right] h^{-3} - \left[ \frac{(3i+2) \dot\psi_i}{i} + \frac{(3i+1) \dot\psi_{i+1}}{i+1} \right] h^{-2} \\
    C_{\psi i} &= \left[ \frac{(i+1) (6i^2-3i-1) \psi_i}{i^2} - \frac{i (6i^2+15i+8) \psi_{i+1}}{(i+1)^2} \right] h^{-2} + \left[ \frac{(i+1) (3i+1) \dot\psi_i}{i} + \frac{i (3i+2) \dot\psi_{i+1}}{i+1} \right] h^{-1} \\
    D_{\psi i} &= -2 \left[ \frac{(i+1)^2 (i-1) \psi_i}{i} - \frac{i^2 (i+2) \psi_{i+1}}{i+1} \right] h^{-1} - [(i+1)^2 \dot\psi_i + i^2 \dot\psi_{i+1}] \end{aligned} \right..
\end{align}
\end{adjustwidth}
Like in Section~\ref{ss:IPW_CE1}, the desired approximation $\phi \equiv H \psi$ is represented in the same way with $\{ \phi_i \}$ and $\{ \dot\phi_i \}$.
For convenience, we put this linear transformation in matrix form
\begin{align}
    \bar{\boldsymbol \psi}^{(i)} \equiv \begin{pmatrix} A_{\psi i} \\ B_{\psi i} \\ C_{\psi i} \\ D_{\psi i} \end{pmatrix}
    = T^{(i)} \begin{pmatrix} \psi_i \\ \psi_{i+1} \\ \dot\psi_i \\ \dot\psi_{i+1} \end{pmatrix} \equiv T^{(i)} {\boldsymbol \psi}^{(i)}
\end{align}
with the transformation matrix
\vspace{-6pt}
\begin{adjustwidth}{-\extralength}{0cm}
\begin{align}
    T^{(i)} = \begin{pmatrix} \dfrac{2i-1}{i^2}h^{-4} & - \dfrac{2i+3}{(i+1)^2}h^{-4} & \dfrac{1}{i}h^{-3} & \dfrac{1}{i+1}h^{-3} \\
    -2 \dfrac{3i^2-1}{i^2}h^{-3} & 2 \dfrac{3i^2+6i+2}{(i+1)^2}h^{-3} & -\dfrac{3i+2}{i}h^{-2} & -\dfrac{3i+1}{i+1}h^{-2} \\
    \dfrac{(i+1) (6i^2-3i-1)}{i^2}h^{-2} & - \dfrac{i (6i^2+15i+8)}{(i+1)^2}h^{-2} & \dfrac{(i+1) (3i+1)}{i}h^{-1} & \dfrac{i (3i+2)}{i+1}h^{-1} \\
    -2 \dfrac{(i+1)^2 (i-1)}{i}h^{-1} & 2 \dfrac{i^2 (i+2)}{i+1}h^{-1} & -(i+1)^2 & -i^2 \end{pmatrix},
\end{align}
\end{adjustwidth}
which is the same for $\psi(r)$ and $\phi(r)$.
Boundary condition at $r_0 = 0$ indicates that $\psi_0 = 0$. In the special case of $i = 0$, we set $A_{\psi 0} = 0$ to get
\begin{align}
    &\left\{ \begin{aligned} B_{\psi 0} &= -2 \psi_1 h^{-3} + (\dot\psi_0 + \dot\psi_1) h^{-2} \\
    C_{\psi 0} &= 3 \psi_1 h^{-2} - (2\dot\psi_0 + \dot\psi_1) h^{-1} \\
    D_{\psi 0} &= \dot\psi_0 \end{aligned} \right.
\end{align}
or
\begin{align}
    \label{eq:H_CE1_T0mat} T^{(0)} = \begin{pmatrix} -2 h^{-3} & h^{-2} & h^{-2} \\
    3 h^{-2} & -2 h^{-1} & -h^{-1} \\
    0 & 1 & 0 \end{pmatrix},
\end{align}
so that $(B_{\psi 0}, C_{\psi 0}, D_{\psi 0})^{\rm T} = T^{(0)} (\psi_1, \dot\psi_0, \dot\psi_1)^{\rm T}$.

The cost function is defined as (subscript ``H'' stands for hydrogen atom)
\vspace{-12pt}
\begin{adjustwidth}{-\extralength}{0cm}
\begin{align}
    \label{eq:H_CE1_cost} \epsilon_{\rm H}(\{\psi_i\}, \{\dot\psi_i\}; \{\phi_i\}, \{\dot\phi_i\}; h) &= \sum_{i=0}^{N-1} \epsilon_{{\rm H},i} (\psi_i, \dot\psi_i, \psi_{i+1}, \dot\psi_{i+1}; \phi_i, \dot\phi_i, \phi_{i+1}, \dot\phi_{i+1}; r_i, r_{i+1}) + \epsilon_{{\rm H},N} (\psi_N; \phi_N; r_N);
\end{align}
\end{adjustwidth}
for simplicity, in the following text we omit parameters of $\epsilon_{{\rm H},i}$, which is
\vspace{-6pt}
\begin{adjustwidth}{-\extralength}{0cm}
\small
\begin{align}
    \label{eq:H_CE1_costi} \nonumber \epsilon_{{\rm H},i} &= \int_{r_i}^{r_{i+1}} (\ddot{\psi} + \frac{2}{r} \psi + \phi)^2 \,{\rm d}r = \int_{r_i}^{r_{i+1}} \left[ \begin{aligned} &(2C_{\psi i} + 2D_{\psi i}) + (6B_{\psi i} + 2C_{\psi i} + D_{\phi i}) r \\ &+ (12A_{\psi i} + 2B_{\psi i} + C_{\phi i}) r^2 + (2A_{\psi i} + B_{\phi i}) r^3 + A_{\phi i} r^4 \end{aligned} \right]^2 \,{\rm d}r \\
    \nonumber &= \int_{ih}^{(i+1)h} \left[ \begin{aligned} &4 (C_{\psi i} + D_{\psi i})^2 + 4 (6 B_{\psi i} + 2 C_{\psi i} + D_{\phi i}) (C_{\psi i} + D_{\psi i}) r \\
    &+[(6 B_{\psi i} + 2 C_{\psi i} + D_{\phi i})^2 + 4 (12 A_{\psi i} + 2 B_{\psi i} + C_{\phi i}) (C_{\psi i} + D_{\psi i})] r^2 \\
    &+[2 (12 A_{\psi i} + 2 B_{\psi i} + C_{\phi i}) (6 B_{\psi i} + 2 C_{\psi i} + D_{\phi i}) + 4 (2 A_{\psi i} + B_{\phi i}) (C_{\psi i} + D_{\psi i})] r^3 \\
    &+ [(12 A_{\psi i} + 2 B_{\psi i} + C_{\phi i})^2 + 2 (2 A_{\psi i} + B_{\phi i}) (6 B_{\psi i} + 2 C_{\psi i} + D_{\phi i}) + 4 A_{\phi i} (C_{\psi i} + D_{\psi i})] r^4 \\
    &+ [2 (2 A_{\psi i} + B_{\phi i}) (12 A_{\psi i} + 2 B_{\psi i} + C_{\phi i}) + 2 A_{\phi i} (6 B_{\psi i} + 2 C_{\psi i} + D_{\phi i})] r^5 \\
    &+ [(2 A_{\psi i} + B_{\phi i})^2 + 2 A_{\phi i} (12 A_{\psi i} + 2 B_{\psi i} + C_{\phi i})] r^6 + 2 A_{\phi i} (2 A_{\psi i} + B_{\phi i}) r^7 + A_{\phi i}^2 r^8 \end{aligned} \right]^2 \,{\rm d}r \end{align}
    \begin{align}
    &= \left[ \begin{aligned} &4 (C_{\psi i} + D_{\psi i})^2 h + 2 (6 B_{\psi i} + 2 C_{\psi i} + D_{\phi i}) (C_{\psi i} + D_{\psi i}) d(i, 2) h^2 \\
    &+ \frac{1}{3} [(6 B_{\psi i} + 2 C_{\psi i} + D_{\phi i})^2 + 4 (12 A_{\psi i} + 2 B_{\psi i} + C_{\phi i}) (C_{\psi i} + D_{\psi i})] d(i, 3) h^3 \\
    &+ \frac{1}{4} [2 (12 A_{\psi i} + 2 B_{\psi i} + C_{\phi i}) (6 B_{\psi i} + 2 C_{\psi i} + D_{\phi i}) + 4 (2 A_{\psi i} + B_{\phi i}) (C_{\psi i} + D_{\psi i})] d(i, 4) h^4 \\
    &+ \frac{1}{5} [(12 A_{\psi i} + 2 B_{\psi i} + C_{\phi i})^2 + 2 (2 A_{\psi i} + B_{\phi i}) (6 B_{\psi i} + 2 C_{\psi i} + D_{\phi i}) + 4 A_{\phi i} (C_{\psi i} + D_{\psi i})] d(i, 5) h^5 \\
    &+ \frac{1}{6} [2 (2 A_{\psi i} + B_{\phi i}) (12 A_{\psi i} + 2 B_{\psi i} + C_{\phi i}) + 2 A_{\phi i} (6 B_{\psi i} + 2 C_{\psi i} + D_{\phi i})] d(i, 6) h^6 \\
    &+ \frac{1}{7} [(2 A_{\psi i} + B_{\phi i})^2 + 2 A_{\phi i} (12 A_{\psi i} + 2 B_{\psi i} + C_{\phi i})] d(i, 7) h^7 + \frac{1}{4} (2 A_{\phi i} A_{\psi i} + A_{\phi i} B_{\phi i}) d(i, 8) h^8 + \frac{1}{9} A_{\phi i}^2 d(i, 9) h^9 \end{aligned} \right],
\end{align}
\end{adjustwidth}
where $d(i, n) \equiv (i+1)^n - i^n$, for $i = 0, 1, \ldots, N-1$; $\epsilon_{{\rm H},N}$ will be addressed later. Again for convenience, we define $\epsilon_{{\rm H},-1} \equiv 0$.

Partial derivatives of $\epsilon_{{\rm H},i}$ with respect to $A_{\phi i}$, $B_{\phi i}$, $C_{\phi i}$, and $D_{\phi i}$ are
\vspace{-10pt}
\begin{adjustwidth}{-\extralength}{0cm}
\begin{align}
    &\left\{ \begin{aligned} \frac{\partial\epsilon_{{\rm H},i}}{\partial A_{\phi i}} &= \left[ \begin{aligned} &\frac{4}{5} (C_{\psi i} + D_{\psi i}) d(i, 5) h^5 + \frac{1}{3} (6 B_{\psi i} + 2 C_{\psi i} + D_{\phi i}) d(i, 6) h^6 \\ &+ \frac{2}{7} (12 A_{\psi i} + 2 B_{\psi i} + C_{\phi i}) d(i, 7) h^7 + \frac{1}{2} A_{\psi i} d(i, 8) h^8 + \frac{1}{4} B_{\phi i} d(i, 8) h^8 + \frac{2}{9} A_{\phi i} d(i, 9) h^9 \end{aligned} \right] \\
    \frac{\partial\epsilon_{{\rm H},i}}{\partial B_{\phi i}} &= \left[ \begin{aligned} &(C_{\psi i} + D_{\psi i}) d(i, 4) h^4 + \frac{2}{5} (6 B_{\psi i} + 2 C_{\psi i} + D_{\phi i}) d(i, 5) h^5 \\ &+ \frac{1}{3} (12 A_{\psi i} + 2 B_{\psi i} + C_{\phi i}) d(i, 6) h^6 + \frac{2}{7} (2 A_{\psi i} + B_{\phi i}) d(i, 7) h^7 + \frac{1}{4} A_{\phi i} d(i, 8) h^8 \end{aligned} \right] \\
    \frac{\partial\epsilon_{{\rm H},i}}{\partial C_{\phi i}} &= \left[ \begin{aligned} &\frac{4}{3} (C_{\psi i} + D_{\psi i}) d(i, 3) h^3 + \frac{1}{2} (6 B_{\psi i} + 2 C_{\psi i} + D_{\phi i}) d(i, 4) h^4 \\ &+ \frac{2}{5} (12 A_{\psi i} + 2 B_{\psi i} + C_{\phi i}) d(i, 5) h^5 + \frac{1}{3} (2 A_{\psi i} + B_{\phi i}) d(i, 6) h^6 + \frac{2}{7} A_{\phi i} d(i, 7) h^7) \end{aligned} \right] \\
    \frac{\partial\epsilon_{{\rm H},i}}{\partial D_{\phi i}} &= \left[ \begin{aligned} &2 (C_{\psi i} + D_{\psi i}) d(i, 2) h^2 + \frac{2}{3} (6 B_{\psi i} + 2 C_{\psi i} + D_{\phi i}) d(i, 3) h^3 \\ &+ \frac{1}{2} (12 A_{\psi i} + 2 B_{\psi i} + C_{\phi i}) d(i, 4) h^4 + \frac{2}{5} (2 A_{\psi i} + B_{\phi i}) d(i, 5) h^5 + \frac{1}{3} A_{\phi i} d(i, 6) h^6 \end{aligned} \right] \end{aligned} \right.,
\end{align}
\end{adjustwidth}
respectively; put in matrix form, these are
\begin{align}
    \begin{pmatrix} \partial/\partial A_{\phi i} \\ \partial/\partial B_{\phi i} \\ \partial/\partial C_{\phi i} \\ \partial/\partial D_{\phi i} \end{pmatrix} \epsilon_{{\rm H},i}
    \equiv \bar P^{(i)} \bar{\boldsymbol \phi}^{(i)} + \bar Q^{(i)} \bar{\boldsymbol \psi}^{(i)} = \bar P^{(i)} \begin{pmatrix} A_{\phi i} \\ B_{\phi i} \\ C_{\phi i} \\ D_{\phi i} \end{pmatrix} + \bar Q^{(i)} \begin{pmatrix} A_{\psi i} \\ B_{\psi i} \\ C_{\psi i} \\ D_{\psi i} \end{pmatrix}
\end{align}
with
\begin{adjustwidth}{-\extralength}{0cm}
\begin{align}
    \label{eq:H_CE1_PiQi} \left\{ \begin{aligned} \bar P^{(i)} &= \begin{pmatrix} 2d(i, 9) h^9/9 & d(i, 8) h^8/4 & 2d(i, 7) h^7/7 & d(i, 6) h^6/3 \\ d(i, 8) h^8/4 & 2d(i, 7) h^7/7 & d(i, 6) h^6/3 & 2d(i, 5) h^5/5 \\ 2d(i, 7) h^7/7 & d(i, 6) h^6/3 & 2d(i, 5) h^5/5 & d(i, 4) h^4/2 \\ d(i, 6) h^6/3 & 2d(i, 5) h^5/5 & d(i, 4) h^4/2 & 2d(i, 3) h^3/3 \end{pmatrix} \\
    \bar Q^{(i)} &= \begin{pmatrix} \dfrac{24}{7}d(i, 7) h^7 + \dfrac{1}{2}d(i, 8) h^8 & 2d(i, 6) h^6 + \dfrac{4}{7}d(i, 7) h^7 & \dfrac{4}{5}d(i, 5) h^5 + \dfrac{2}{3}d(i, 6) h^6 & \dfrac{4}{5}d(i, 5) h^5 \\
    4d(i, 6) h^6 + \dfrac{4}{7}d(i, 7) h^7 & \dfrac{12}{5}d(i, 5) h^5 + \dfrac{2}{3}d(i, 6) h^6 & d(i, 4) h^4 + \dfrac{4}{5}d(i, 5) h^5 & d(i, 4) h^4 \\
    \dfrac{24}{5}d(i, 5) h^5 + \dfrac{2}{3}d(i, 6) h^6 & 3d(i, 4) h^4 + \dfrac{4}{5}d(i, 5) h^5 & \dfrac{4}{3}d(i, 3) h^3 + d(i, 4) h^4 & \dfrac{4}{3}d(i, 3) h^3 \\
    6d(i, 4) h^4 + \dfrac{4}{5}d(i, 5) h^5 & 4d(i, 3) h^3 + d(i, 4) h^4 & 2d(i, 2) h^2 + \dfrac{4}{3}d(i, 3) h^3 & 2d(i, 2) h^2 \end{pmatrix} \end{aligned} \right..
\end{align}
\end{adjustwidth}
For the special case of $i = 0$, we simply need to drop the first rows and first columns of $\bar P^{(0)}$ and $\bar Q^{(0)}$.
Then partial derivatives of $\epsilon_{{\rm H},i}$ with respect to $\phi_i$, $\phi_{i+1}$, $\dot\phi_i$, and $\dot\phi_{i+1}$ can be succinctly expressed as
\begin{align}
    \nonumber \begin{pmatrix} \partial/\partial \phi_i \\ \partial/\partial \phi_{i+1} \\ \partial/\partial \dot\phi_i \\ \partial/\partial \dot\phi_{i+1} \end{pmatrix} \epsilon_{{\rm H},i}
    &= [T^{(i)}]^{\rm T} \begin{pmatrix} \partial/\partial A_{\phi i} \\ \partial/\partial B_{\phi i} \\ \partial/\partial C_{\phi i} \\ \partial/\partial D_{\phi i} \end{pmatrix} \epsilon_{{\rm H},i} = [T^{(i)}]^{\rm T} (\bar P^{(i)} \bar{\boldsymbol \phi}^{(i)} + \bar Q^{(i)} \bar{\boldsymbol \psi}^{(i)}) \\
    &= ([T^{(i)}]^{\rm T} \bar P^{(i)} T^{(i)}) {\boldsymbol \phi}^{(i)} + ([T^{(i)}]^{\rm T} \bar Q^{(i)} T^{(i)}) {\boldsymbol \psi}^{(i)} \equiv P^{(i)} {\boldsymbol \phi}^{(i)} + Q^{(i)} {\boldsymbol \psi}^{(i)}.
\end{align}

As promised, we now address $\epsilon_{{\rm H},N}$, which corresponds to the tail.
Given our assumed functional form Equation~(\ref{eq:H_CE1_psi}), this should be
\vspace{-12pt}
\begin{adjustwidth}{-\extralength}{0cm}
\begin{align}
    \nonumber \epsilon_{{\rm H},N} (\psi_N; \phi_N; r_N) &= \int_{r_N}^{\infty} (\ddot{\psi} + \frac{2}{r} \psi + \phi)^2 \,{\rm d}r = \int_{r_N}^{\infty} \left[ \left( \psi_N \frac{r-2r_N}{r_N^3} + \psi_N \frac{2}{r_N} + \phi_N \frac{r}{r_N} \right) \exp \left( 1 - \frac{r}{r_N} \right) \right]^2 \,{\rm d}r \\
    \nonumber &= \int_{r_N}^{\infty} \left[ \left\{ 2 \frac{r_N-1}{r_N^2} \psi_N + \left( \frac{\phi_N}{r_N} + \frac{\psi_N}{r_N^3} \right) r \right\} \exp \left( 1 - \frac{r}{r_N} \right) \right]^2 \,{\rm d}r \\
    \nonumber &= \frac{r_N}{4} \left[ 2 \left( 2 \frac{r_N-1}{r_N^2} \psi_N \right)^2 + 6 \left( 2 \frac{r_N-1}{r_N^2} \psi_N \right) \left( \frac{\phi_N}{r_N} + \frac{\psi_N}{r_N^3} \right) r_N + 5 \left( \frac{\phi_N}{r_N} + \frac{\psi_N}{r_N^3} \right)^2 r_N^2 \right] \\
    &= \frac{5 r_N}{4} \phi_N^2 + \left( 3 - \frac{1}{2 r_N} \right) \phi_N \psi_N + \left( \frac{2}{r_N} - \frac{1}{r_N^2} + \frac{1}{4 r_N^3} \right) \psi_N^2,
\end{align}
\end{adjustwidth}
and its partial derivative with respect to $\phi_N$ is
\begin{align}
    \frac{\partial \epsilon_{{\rm H},N}}{\partial \phi_N} = \frac{5 r_N}{2} \phi_N + \left( 3 - \frac{1}{2 r_N} \right) \psi_N \equiv P^{(N)} \phi_N + Q^{(N)} \psi_N,
\end{align}
where $P^{(N)}$ and $Q^{(N)}$ are both $1 \times 1$ matrices.

To minimize the cost function Equation~(\ref{eq:H_CE1_cost}), we have
\vspace{-12pt}
\begin{adjustwidth}{-\extralength}{0cm}
\begin{align}
    \label{eq:H_CE1_PandQ} \begin{pmatrix} \partial/\partial\phi_1 \\ \vdots \\ \partial/\partial\phi_N \\ \partial/\partial\dot\phi_0 \\ \vdots \\ \partial/\partial\dot\phi_N \end{pmatrix} \epsilon_{\rm H}
    = \left[ \begin{aligned} &\begin{pmatrix} P_{11} & \cdots & P_{1N} & P_{1,N+1} & \cdots & P_{1,2N+1} \\ \vdots & \ddots & \vdots & \vdots & \ddots & \vdots \\ P_{N1} & \cdots & P_{NN} & P_{N,N+1} & \cdots & P_{N,2N+1} \\ P_{N+1,1} & \cdots & P_{N+1,N} & P_{N+1,N+1} & \cdots & P_{N+1,2N+1} \\ \vdots & \ddots & \vdots & \vdots & \ddots & \vdots \\ P_{2N+1,1} & \cdots & P_{2N+1,N} & P_{2N+1,N+1} & \cdots & P_{2N+1,2N+1} \end{pmatrix} \begin{pmatrix} \phi_1 \\ \vdots \\ \phi_N \\ \dot\phi_1 \\ \vdots \\ \dot\phi_N \end{pmatrix} \\
    &+ \begin{pmatrix} Q_{11} & \cdots & Q_{1N} & Q_{1,N+1} & \cdots & Q_{1,2N+1} \\ \vdots & \ddots & \vdots & \vdots & \ddots & \vdots \\ Q_{N1} & \cdots & Q_{NN} & Q_{N,N+1} & \cdots & Q_{N,2N+1} \\ Q_{N+1,1} & \cdots & Q_{N+1,N} & Q_{N+1,N+1} & \cdots & Q_{N+1,2N+1} \\ \vdots & \ddots & \vdots & \vdots & \ddots & \vdots \\ Q_{2N+1,1} & \cdots & Q_{2N+1,N} & Q_{2N+1,N+1} & \cdots & Q_{2N+1,2N+1} \end{pmatrix} \begin{pmatrix} \psi_1 \\ \vdots \\ \psi_N \\ \dot\psi_1 \\ \vdots \\ \dot\psi_N \end{pmatrix} \end{aligned} \right]
    = \begin{pmatrix} 0 \\ \vdots \\ 0 \\ 0 \\ \vdots \\ 0 \end{pmatrix};
\end{align}
\end{adjustwidth}
since
\begin{align}
    \left\{ \begin{aligned} \frac{\partial\epsilon_{\rm IPW}}{\partial\phi_i} = \frac{\partial\epsilon_{{\rm IPW},i-1}}{\partial\phi_i} + \frac{\partial\epsilon_{{\rm IPW},i}}{\partial\phi_i} \\
    \frac{\partial\epsilon_{\rm IPW}}{\partial\dot\phi_i} = \frac{\partial\epsilon_{{\rm IPW},i-1}}{\partial\dot\phi_i} + \frac{\partial\epsilon_{{\rm IPW},i}}{\partial\dot\phi_i} \end{aligned} \right.,
\end{align}
the $(2N+1) \times (2N+1)$ $P$ and $Q$ matrices can be constructed from scratch (zero matrix) by doing
\begin{align}
    \left\{ \begin{aligned} \begin{pmatrix} P_{1,1} & P_{1,N+1} & P_{i,N+2} \\ P_{N+1,1} & P_{N+1,N+1} & P_{N+1,N+2} \\ P_{N+2,1} & P_{N+2,N+1} & P_{N+2,N+2} \end{pmatrix} +\!\!= P^{(0)} \\
    \begin{pmatrix} Q_{1,1} & Q_{1,N+1} & Q_{i,N+2} \\ Q_{N+1,1} & Q_{N+1,N+1} & Q_{N+1,N+2} \\ Q_{N+2,1} & Q_{N+2,N+1} & Q_{N+2,N+2} \end{pmatrix} +\!\!= Q^{(0)} \end{aligned} \right.
\end{align}
for $i=0$, and then
\vspace{-10pt}
\begin{adjustwidth}{-\extralength}{0cm}
\begin{align}
    \left\{ \begin{aligned} \begin{pmatrix} P_{i,i} & P_{i,i+1} & P_{i,(N+1)+i} & P_{i,(N+1)+i+1} \\ P_{i+1,i} & P_{i+1,i+1} & P_{i+1,(N+1)+i} & P_{i+1,(N+1)+i+1} \\ P_{(N+1)+i,i} & P_{(N+1)+i,i+1} & P_{(N+1)+i,(N+1)+i} & P_{(N+1)+i,(N+1)+i+1} \\ P_{(N+1)+i+1,i} & P_{(N+1)+i+1,i+1} & P_{(N+1)+i+1,(N+1)+i} & P_{(N+1)+i+1,(N+1)+i+1} \end{pmatrix} +\!\!= P^{(i)} \\
    \begin{pmatrix} Q_{i,i} & Q_{i,i+1} & Q_{i,(N+1)+i} & Q_{i,(N+1)+i+1} \\ Q_{i+1,i} & Q_{i+1,i+1} & Q_{i+1,(N+1)+i} & Q_{i+1,(N+1)+i+1} \\ Q_{(N+1)+i,i} & Q_{(N+1)+i,i+1} & Q_{(N+1)+i,(N+1)+i} & Q_{(N+1)+i,(N+1)+i+1} \\ Q_{(N+1)+i+1,i} & Q_{(N+1)+i+1,i+1} & Q_{(N+1)+i+1,(N+1)+i} & Q_{(N+1)+i+1,(N+1)+i+1} \end{pmatrix} +\!\!= Q^{(i)} \end{aligned} \right.,
\end{align}
\end{adjustwidth}
for $i = 1, 2, \ldots, N-1$, and finally
\begin{align}
    \left\{ \begin{aligned} \begin{pmatrix} P_{N,N} \end{pmatrix} +\!\!= P^{(N)} \\
    \begin{pmatrix} Q_{N,N} \end{pmatrix} +\!\!= Q^{(N)} \end{aligned} \right.
\end{align}
for $i=N$.

Because of our definition of the tail, we need to enforce the $\dot\psi_N = 0$ constraint if we want to maintain the continuity of first derivative at $r_N$.
In this case, simply removing the corresponding rows and columns from $P$ and $Q$ matrices constructed above would lead to erroneous results, as when four coefficients ($A_{\psi,N-1}$, $B_{\psi,N-1}$, $C_{\psi,N-1}$, and $D_{\psi,N-1}$; similar for $\phi$) are fully specified by three parameters ($\psi_{N-1}$, $\psi_N$, and $\dot\psi_{N-1}$; similar for $\phi$), the inverse transformation may not be well defined---the situation is basically the same as in Section~\ref{ss:CHO_CE2}.

Therefore, when $\dot\psi_{i+1} = 0$ and $\dot\phi_{i+1} = 0$, we have to plug the two sets of three parameters into $\epsilon_{{\rm H},i}$ Equation~(\ref{eq:H_CE1_costi}) to obtain (here the prime ``$'$'' {means} 
 with the constraints mentioned above)  
\vspace{-4pt}
\begin{align}
    \left\{ \begin{aligned} P'{}^{(i)} &= \begin{pmatrix} P'{}^{(i)}_{11} & P'{}^{(i)}_{12} & P'{}^{(i)}_{13} \\ P'{}^{(i)}_{21} & P'{}^{(i)}_{22} & P'{}^{(i)}_{23} \\ P'{}^{(i)}_{31} & P'{}^{(i)}_{32} & P'{}^{(i)}_{33} \end{pmatrix} \\
    Q'{}^{(i)} &= \begin{pmatrix} Q'{}^{(i)}_{11} & Q'{}^{(i)}_{12} & Q'{}^{(i)}_{13} \\ Q'{}^{(i)}_{21} & Q'{}^{(i)}_{22} & Q'{}^{(i)}_{23} \\ Q'{}^{(i)}_{31} & Q'{}^{(i)}_{32} & Q'{}^{(i)}_{33} \end{pmatrix} \end{aligned} \right.
\end{align}
with
\begin{align}
    &\left\{ \begin{aligned} P'{}^{(i)}_{11} &= \frac{234 i^4 + 42 i^3 - 17 i^2 - 4 i + 1}{315 i^4}h \\
    P'{}^{(i)}_{22} &= \frac{468 i^7 + 1788 i^6 + 2522 i^5 + 1560 i^4 + 360 i^3}{630 i^3 (i+1)^4}h \\
    P'{}^{(i)}_{12} = P'{}^{(i)}_{21} &= \frac{162 i^5 + 324 i^4 + 154 i^3 - 8 i^2 - 15 i}{630 i^3 (i+1)^2}h \\
    P'{}^{(i)}_{13} = P'{}^{(i)}_{31} &= \frac{132 i^3 + 60 i^2 - i - 4}{1260 i^3}h^2 \\
    P'{}^{(i)}_{23} = P'{}^{(i)}_{32} &= \frac{78 i^4 + 180 i^3 + 127 i^2 + 30 i}{1260 i^2 (i+1)^2}h^2 \\
    P'{}^{(i)}_{33} &= \frac{12 i^2 + 9 i + 2}{630 i^2}h^3 \end{aligned} \right.
\end{align}
and
\vspace{-12pt}
\begin{adjustwidth}{-\extralength}{0cm}
\begin{align}
    &\left\{ \begin{aligned} Q'{}^{(i)}_{11} &= \frac{- 8 (63 i^4 + 4 i^2 - 4 i + 1)h^{-1} + (312 i^3 - 16 i^2 - 20 i + 3)}{210 i^4} \\
    Q'{}^{(i)}_{22} &= \frac{- 8 (63 i^4 + 252 i^3 + 382 i^2 + 264 i + 72)h^{-1} + (312 i^3 + 952 i^2 + 948 i + 305)}{210 (i+1)^4} \\
    Q'{}^{(i)}_{12} = Q'{}^{(i)}_{21} &= \frac{8 (63 i^4 + 126 i^3 + 67 i^2 + 4 i - 3)h^{-1} + (108 i^3 + 162 i^2 + 20 i - 17)}{210 i^2 (i+1)^2} \\
    Q'{}^{(i)}_{13} &= \frac{- 2 (231 i^3 + 14 i^2 + i - 4) + (44 i^2 + 6 i - 3)h}{210 i^3} \\
    Q'{}^{(i)}_{31} &= \frac{- 2 (21 i^3 + 14 i^2 + i - 4) + (44 i^2 + 6 i - 3)h}{210 i^3} \\
    Q'{}^{(i)}_{23} = Q'{}^{(i)}_{32} &= \frac{2 (21 i^3 + 56 i^2 + 50 i + 12) + (26 i^2 + 46 i + 17)h}{210 i (i+1)^2} \\
    Q'{}^{(i)}_{33} &= \frac{- 4 (14 i^2 + 7 i + 2)h + (8 i + 3)h^2}{210 i^2} \end{aligned} \right..
\end{align}
\end{adjustwidth}
Note that although only the $i = N-1$ version of the above expressions is used in this work, we have written the general version for $i \neq 0$.

To construct the $2N \times 2N$ $P$ and $Q$ matrices from scratch, the procedure is the same as when we do not enforce $\dot\psi_{i+1} = 0$ and $\dot\phi_{i+1} = 0$, except for the $(N-1)$st step, which needs to be substituted by
\begin{align}
    \left\{ \begin{aligned} \begin{pmatrix} P_{N-1,N-1} & P_{N-1,N} & P_{N-1,2N} \\ P_{N,N-1} & P_{N,N} & P_{N,2N} \\ P_{2N,N-1} & P_{2N,N} & P_{2N,2N} \end{pmatrix} +\!\!= P'{}^{(N-1)} \\
    \begin{pmatrix} Q_{N-1,N-1} & Q_{N-1,N} & Q_{N-1,2N} \\ Q_{N,N-1} & Q_{N,N} & Q_{N,2N} \\ Q_{2N,N-1} & Q_{2N,N} & Q_{2N,2N} \end{pmatrix} +\!\!= Q'{}^{(N-1)} \end{aligned} \right..
\end{align}

Our desired Hamiltonian is thus simply $H = -P^{-1}Q$. Eigendecomposition of $H$ should yield $2N+1$ (or $2N$) eigenpairs, $\{ \psi_i^{(k)}, \dot\psi_i^{(k)} \}$ and $E^{(k)}$, without (with) the constraint.
With or without the $\dot\psi_N = 0$ enforcement, $\bar P^{(i)}$ ($\bar Q^{(i)}$) matrices are always (never) symmetric; consequently, $P$ ($Q$) matrices are also always (never) symmetric.

Like in Sections~\ref{ss:IPW_simple} and \ref{ss:IPW_CE1}, the eigenvectors need to be ``renormalized'' as
\vspace{-12pt}
\begin{adjustwidth}{-\extralength}{0cm}
\begin{align}
    \label{eq:H_CE1_normal} \nonumber 1 &= \int_0^\infty [{\mathcal N}\psi(r)]^2 \,{\rm d}r = {\mathcal N}^2 \left\{ \sum_{i=0}^{N-1} \int_{r_i}^{r_{i+1}} [r (D_{\psi i} + C_{\psi i} r + B_{\psi i} r^2 + A_{\psi i} r^3)]^2 \,{\rm d}r + \int_{r_N}^{\infty} \left[ \psi_N \frac{r}{r_N} \exp \left( 1 - \frac{r}{r_N} \right) \right]^2 \,{\rm d}r \right\} \\
    \nonumber &= {\mathcal N}^2 \left\{ \sum_{i=0}^{N-1} \int_{i h}^{(i+1) h} \left[ \begin{aligned} &D_{\psi i}^2 r^2 + 2 C_{\psi i} D_{\psi i} r^3 + (C_{\psi i}^2 + 2 B_{\psi i} D_{\psi i}) r^4 + (2 B_{\psi i} C_{\psi i} + 2 A_{\psi i} D_{\psi i}) r^5 \\ &+ (B_{\psi i}^2 + 2 A_{\psi i} C_{\psi i}) r^6 + 2 A_{\psi i} B_{\psi i} r^7 + A_{\psi i}^2 r^8 \end{aligned} \right] \,{\rm d}r + \frac{5}{4} r_N \psi_N^2 \right\} \\
    &= {\mathcal N}^2 \Bigg\{ \sum_{i=0}^{N-1} \left[ \begin{aligned} &\frac{1}{3} D_{\psi i}^2 d(i, 3) h^3 + \frac{1}{2} C_{\psi i} D_{\psi i} d(i, 4) h^4 + \frac{1}{5} (C_{\psi i}^2 + 2 B_{\psi i} D_{\psi i}) d(i, 5) h^5 \\ &+ \frac{1}{3} (B_{\psi i} C_{\psi i} + A_{\psi i} D_{\psi i}) d(i, 6) h^6 + \frac{1}{7} (B_{\psi i}^2 + 2 A_{\psi i} C_{\psi i}) d(i, 7) h^7 \\ &+ \frac{1}{4} A_{\psi i} B_{\psi i} d(i, 8) h^8 + \frac{1}{9} A_{\psi i}^2 d(i, 9) h^9 \end{aligned} \right] + \frac{5}{4} r_N \psi_N^2 \Bigg\};
\end{align}
\end{adjustwidth}
we omit the ``inner product'' definition here as this section focuses on the ground state.

\paragraph{{Special version: $N=0$.}
}
Because of the tail, it is possible to study the $N=0$ case, for which our wavefunction is simply
\begin{align}
    \psi(r) = \dot\psi_0 r e^{-r}, \quad r \geq 0,
\end{align}
which, after normalization, coincides with the exact solution Equation~(\ref{eq:H_exact}). Nevertheless, we still need to study the energy predicted by ContEvol.

In this special case, the cost function is
\begin{align}
    \nonumber \epsilon_{{\rm H},N=0} &= \int_0^\infty (\ddot{\psi} + \frac{2}{r} \psi + \phi)^2 \,{\rm d}r = \int_0^\infty [\dot\psi_0 (r-2) e^{-r} + 2 \dot\psi_0 e^{-r} + \dot\phi_0 r e^{-r}]^2 \,{\rm d}r \\
    &= \int_0^\infty [(\dot\psi_0 + \dot\phi_0) r e^{-r}]^2 \,{\rm d}r = (\dot\psi_0 + \dot\phi_0)^2 \int_0^\infty (r e^{-r})^2 \,{\rm d}r.
\end{align}
Evidently, minimizing this would yield $\dot\phi_0 = -\dot\psi_0$, i.e., the Hamiltonian $H = \begin{pmatrix} -1 \end{pmatrix}$, and the ground state energy also coincides with the exact solution.
Of course, such coincidence should not be relied upon, hence we move on to more realistic $N$ values.

\paragraph{{Toy version: $N=1$.}}

\begin{figure}[H]
    \subfloat{
        \includegraphics[width=0.149\textwidth]{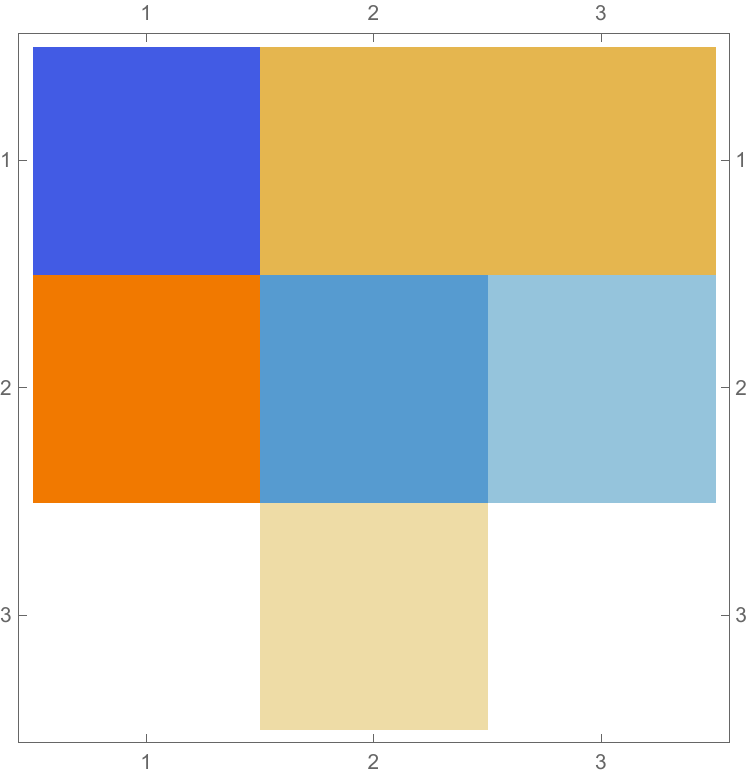}
    }
    \subfloat{
        \includegraphics[width=0.149\textwidth]{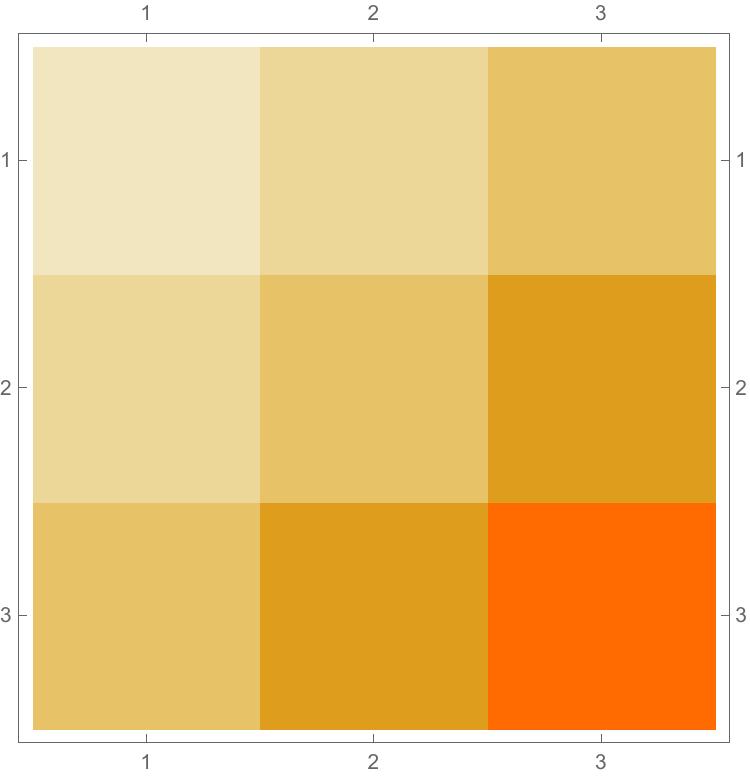}
    }
    \subfloat{
        \includegraphics[width=0.149\textwidth]{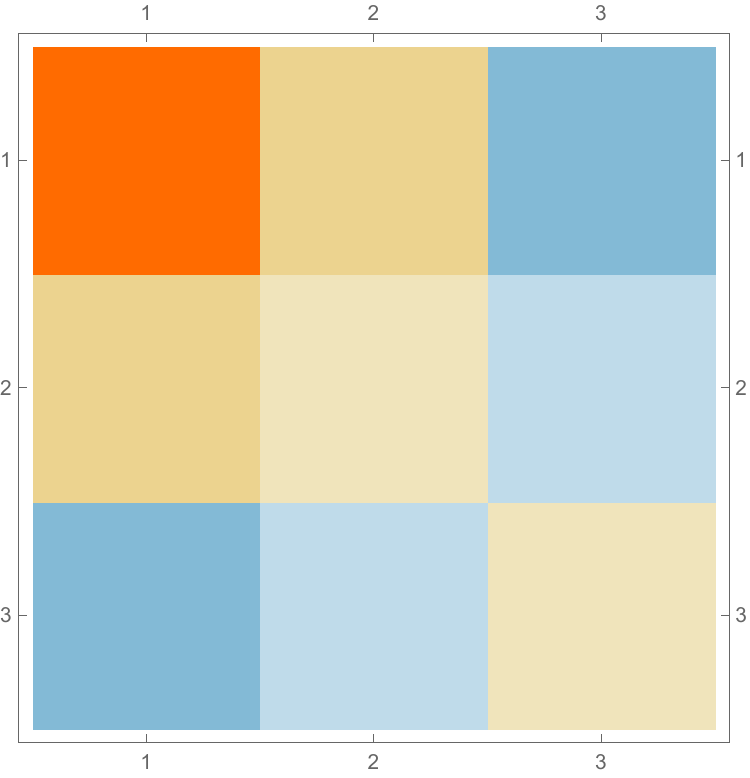}
    }
    \subfloat{
        \includegraphics[width=0.149\textwidth]{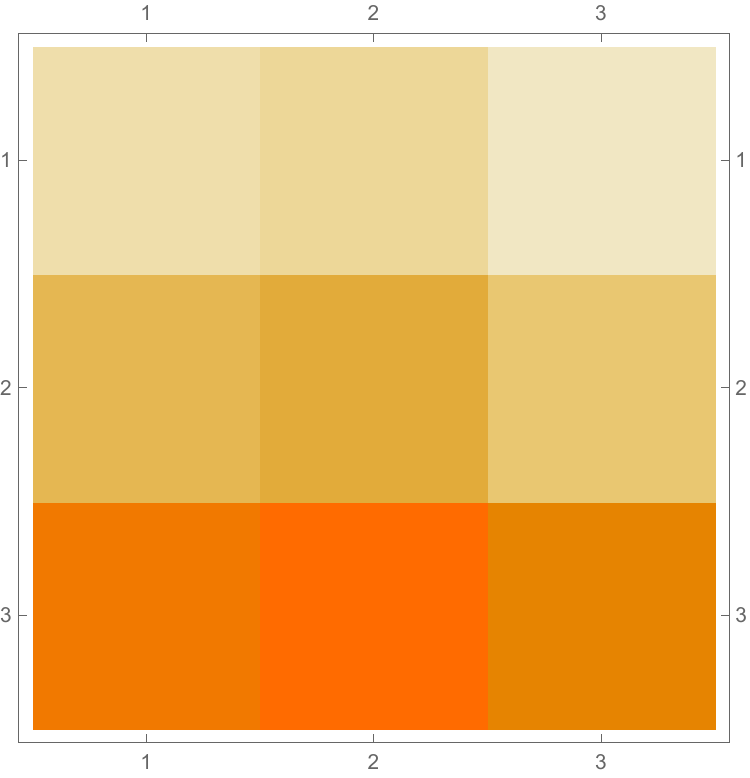}
    }
    \subfloat{
        \includegraphics[width=0.149\textwidth]{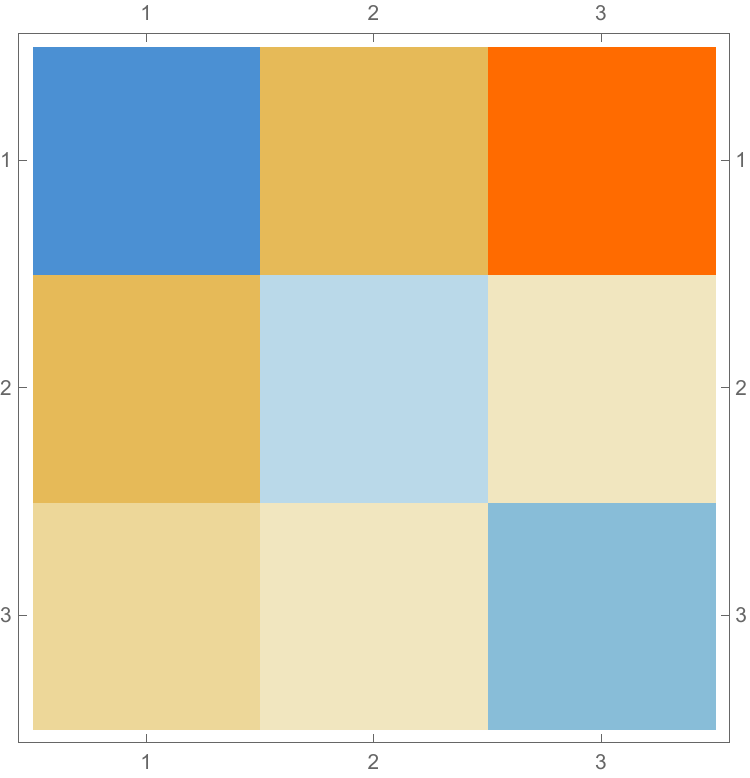}
    }
    \subfloat{
        \includegraphics[width=0.149\textwidth]{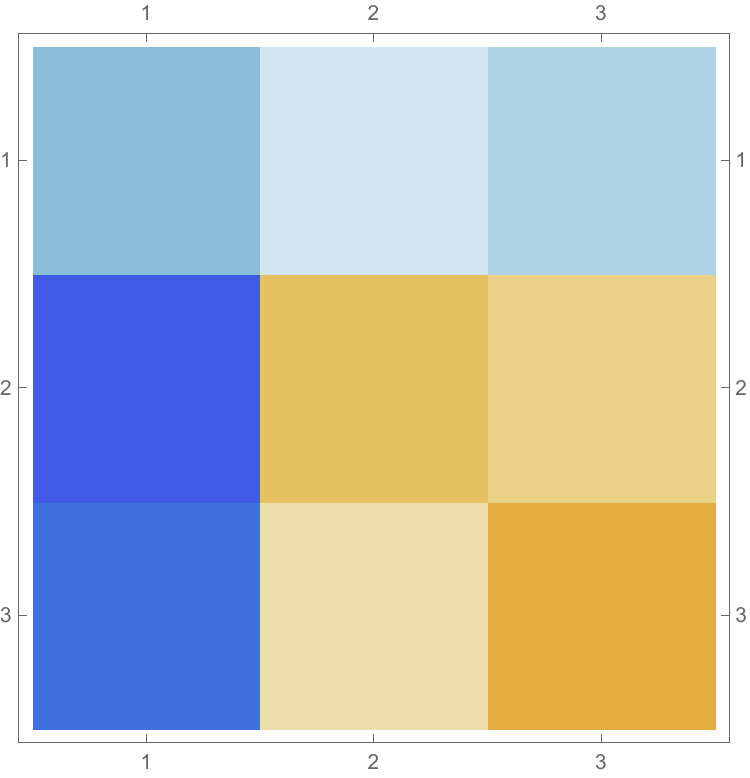}
    }

    \subfloat{
        \includegraphics[width=0.149\textwidth]{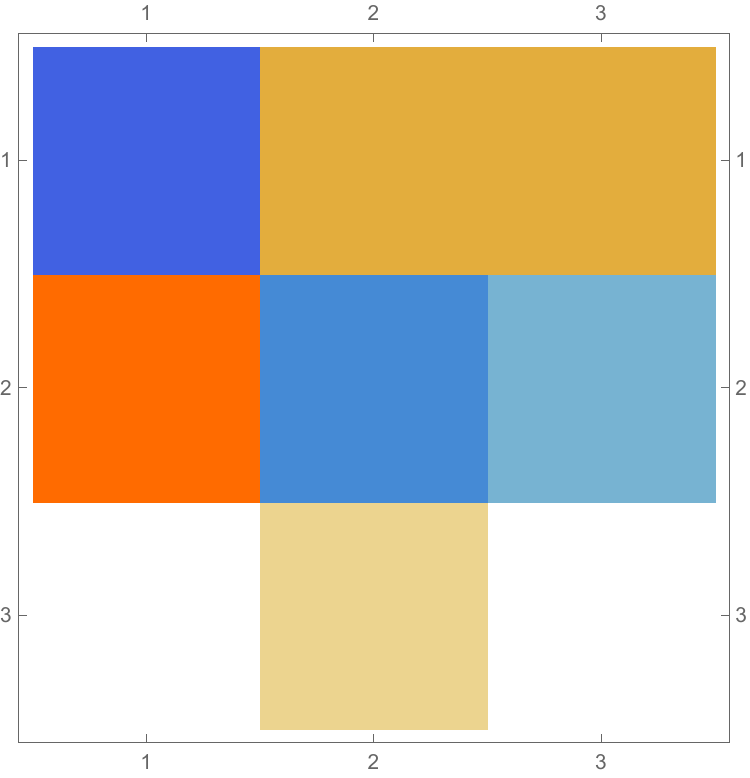}
    }
    \subfloat{
        \includegraphics[width=0.149\textwidth]{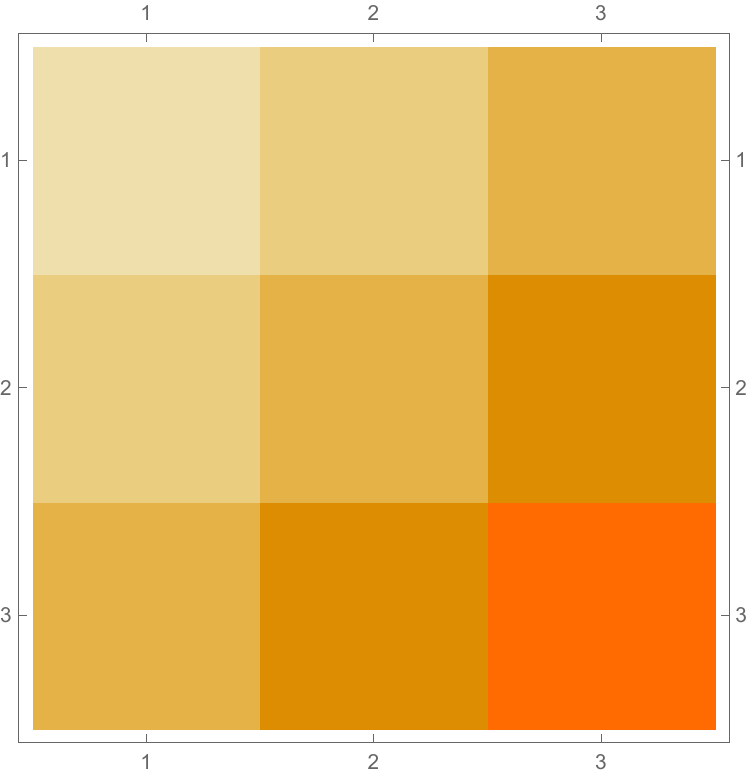}
    }
    \subfloat{
        \includegraphics[width=0.149\textwidth]{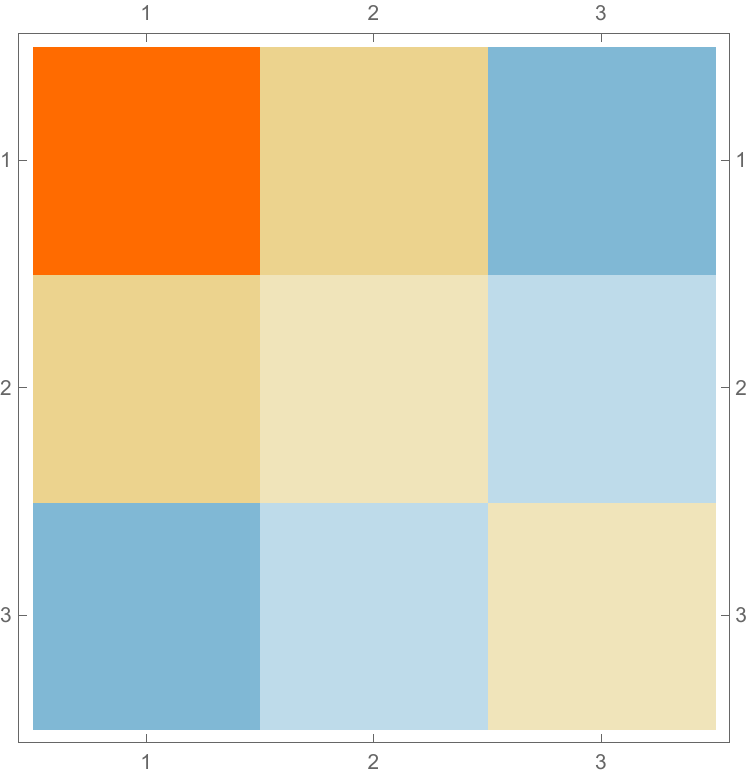}
    }
    \subfloat{
        \includegraphics[width=0.149\textwidth]{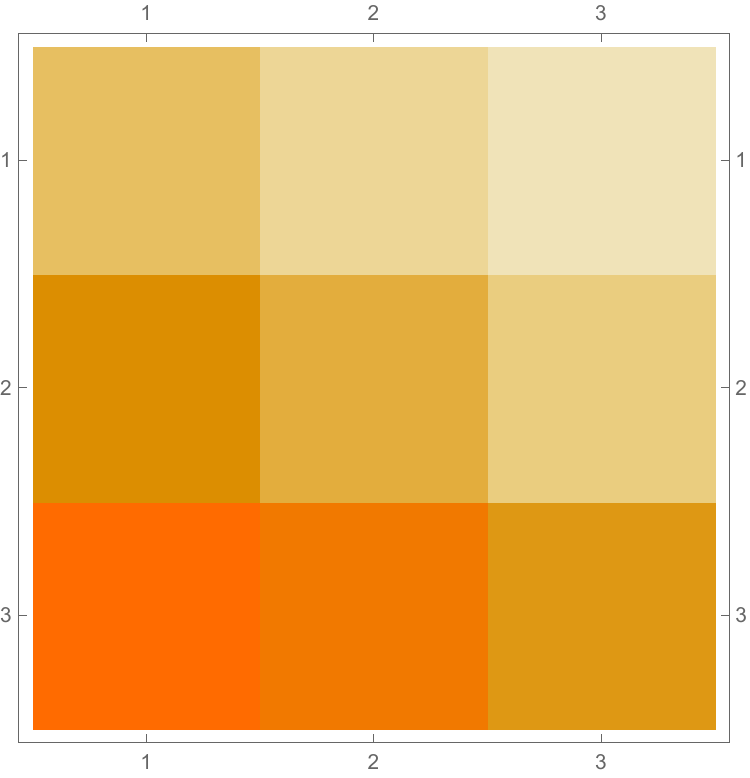}
    }
    \subfloat{
        \includegraphics[width=0.149\textwidth]{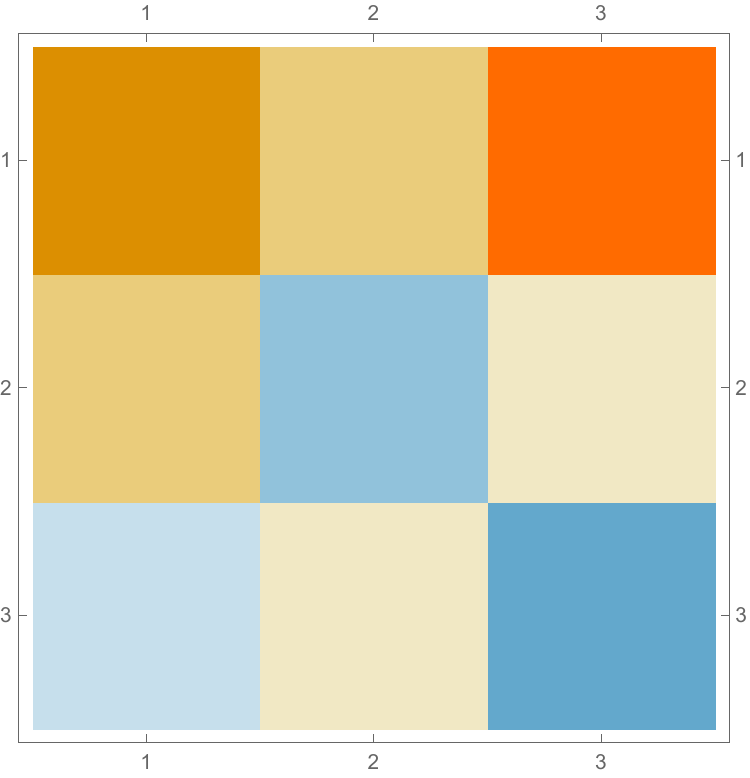}
    }
    \subfloat{
        \includegraphics[width=0.149\textwidth]{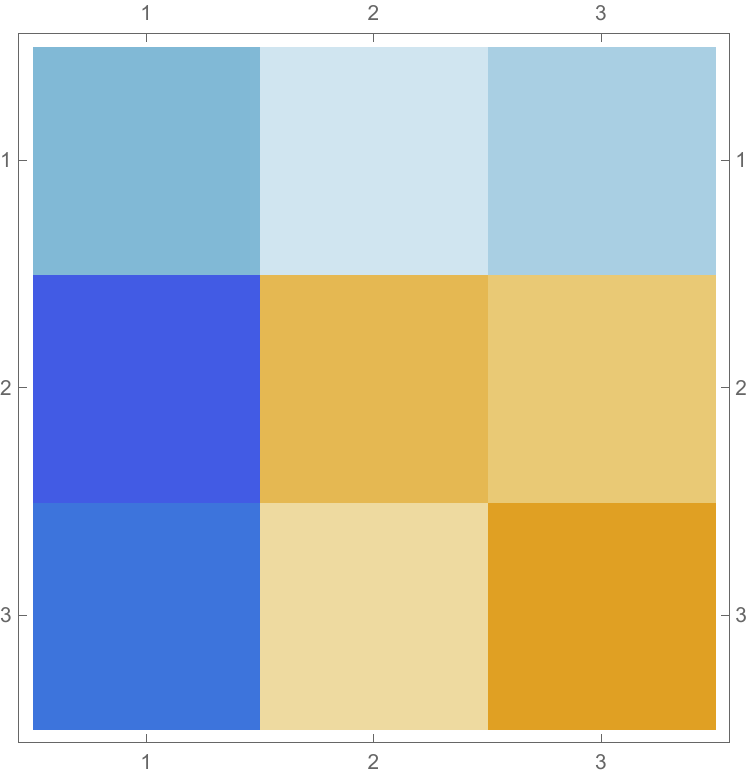}
    }

    \subfloat{
        \includegraphics[width=0.149\textwidth]{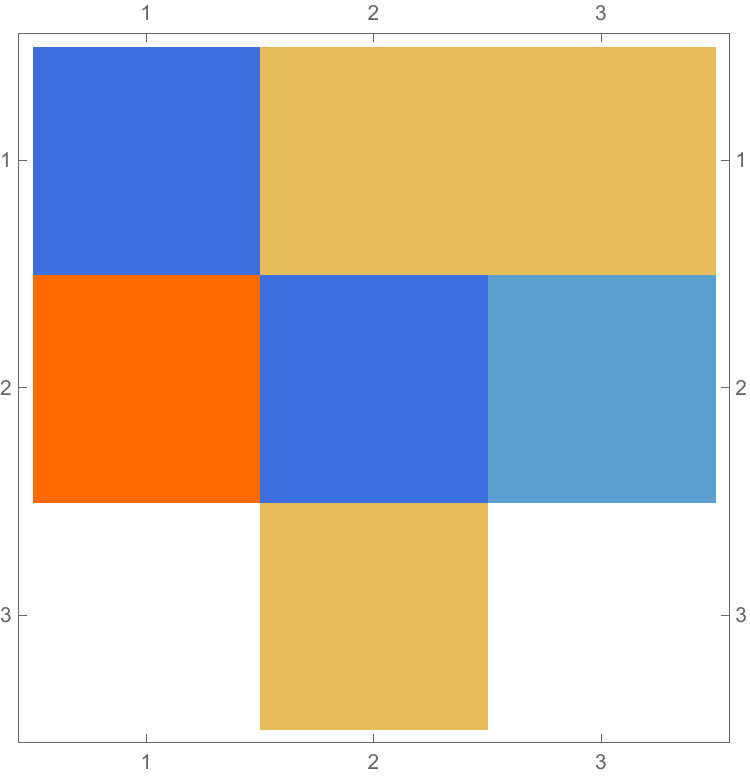}
    }
    \subfloat{
        \includegraphics[width=0.149\textwidth]{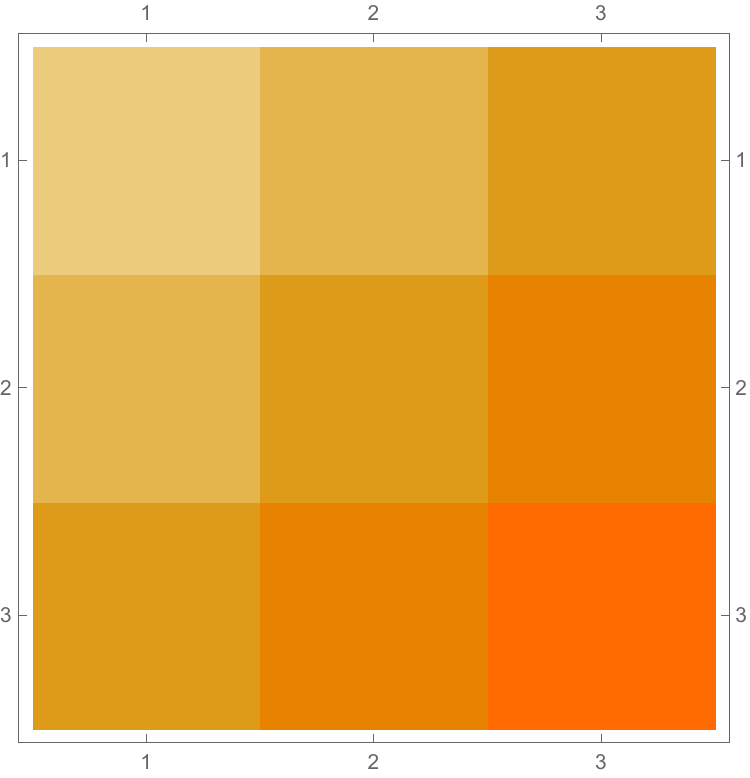}
    }
    \subfloat{
        \includegraphics[width=0.149\textwidth]{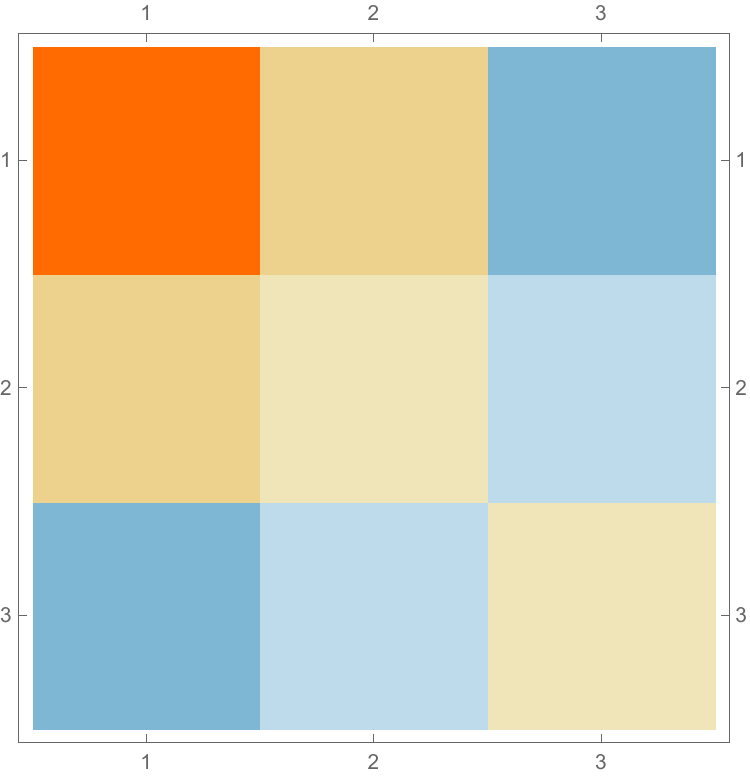}
    }
    \subfloat{
        \includegraphics[width=0.149\textwidth]{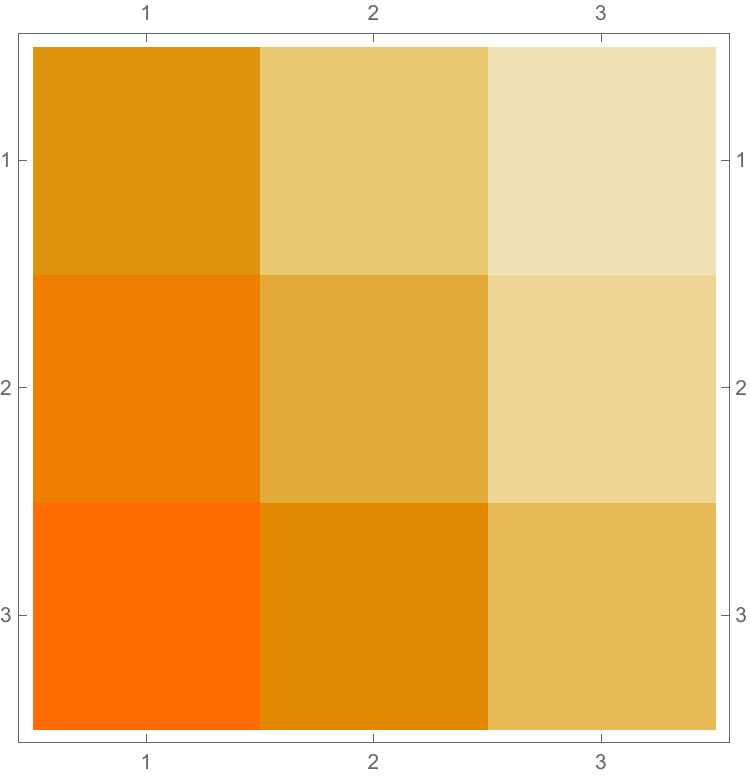}
    }
    \subfloat{
        \includegraphics[width=0.149\textwidth]{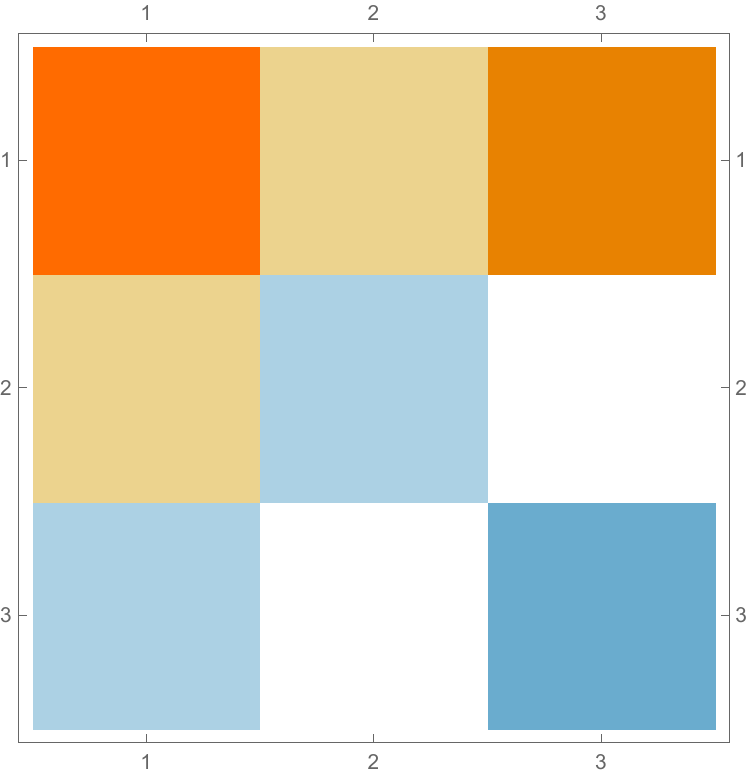}
    }
    \subfloat{
        \includegraphics[width=0.149\textwidth]{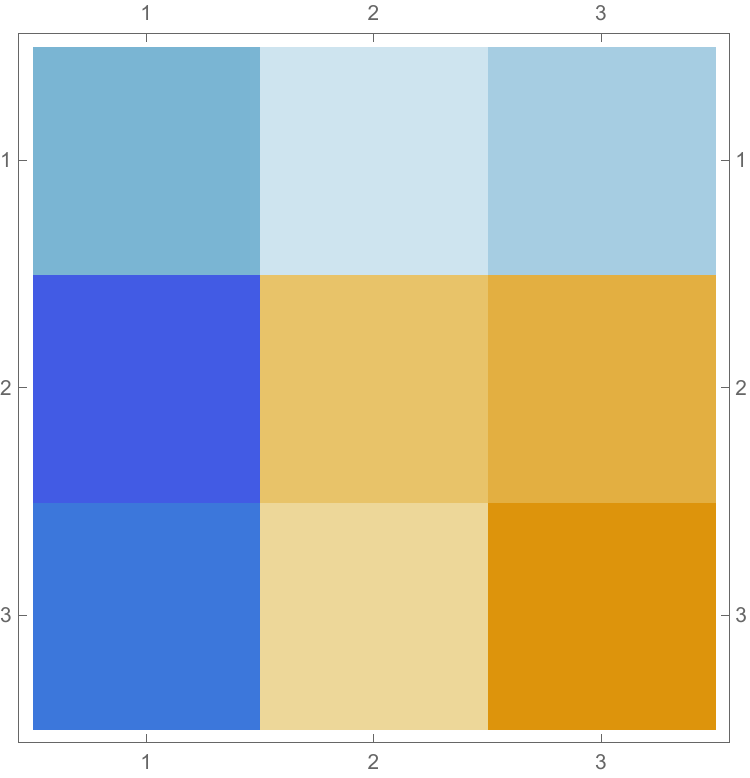}
    }

    \subfloat{
        \includegraphics[width=0.149\textwidth]{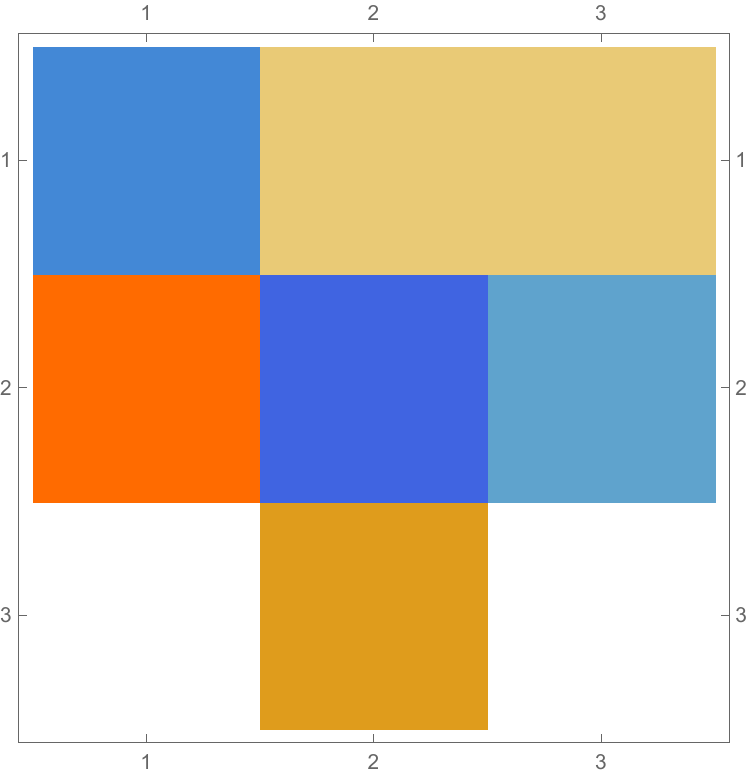}
    }
    \subfloat{
        \includegraphics[width=0.149\textwidth]{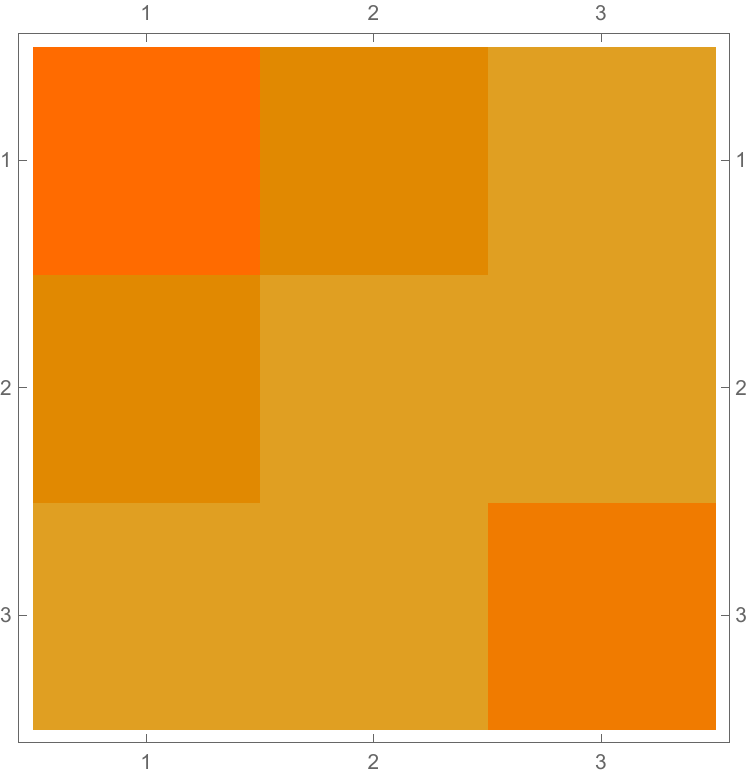}
    }
    \subfloat{
        \includegraphics[width=0.149\textwidth]{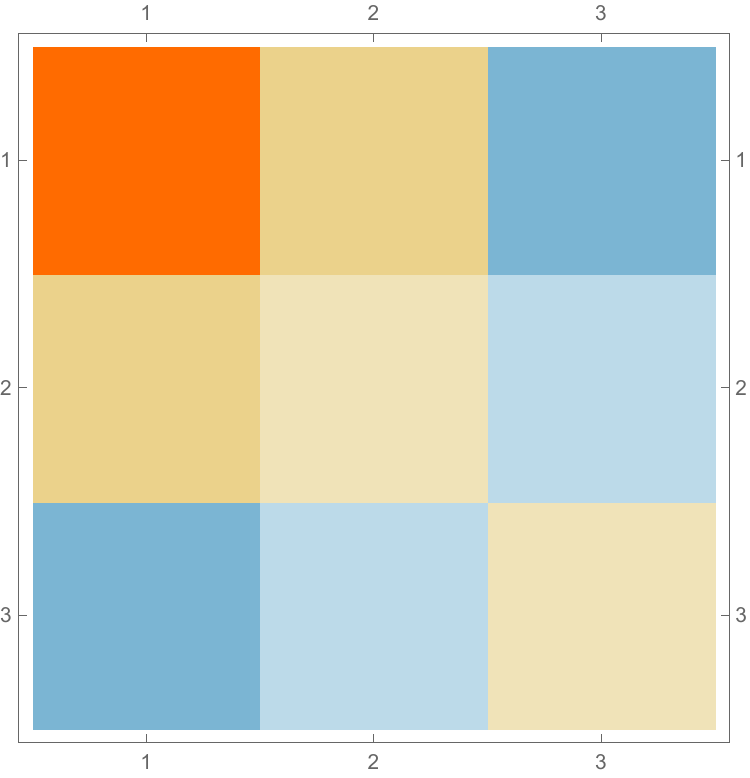}
    }
    \subfloat{
        \includegraphics[width=0.149\textwidth]{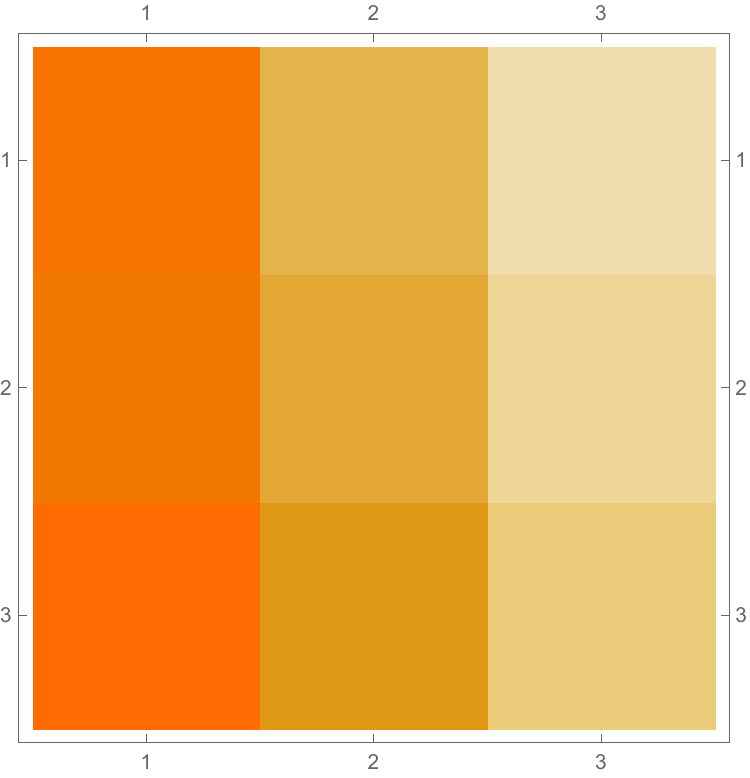}
    }
    \subfloat{
        \includegraphics[width=0.149\textwidth]{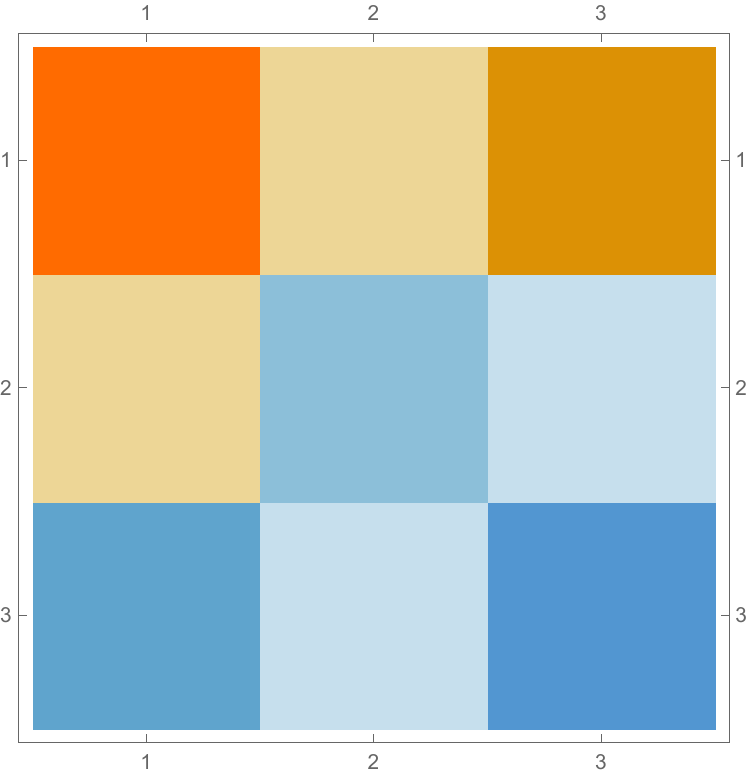}
    }
    \subfloat{
        \includegraphics[width=0.149\textwidth]{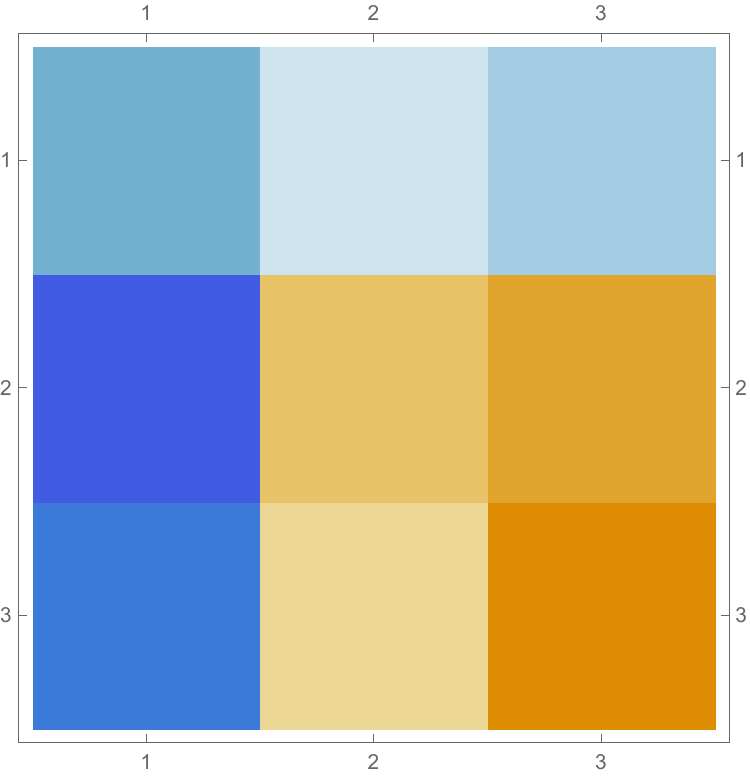}
    }

    \subfloat{
        \includegraphics[width=0.149\textwidth]{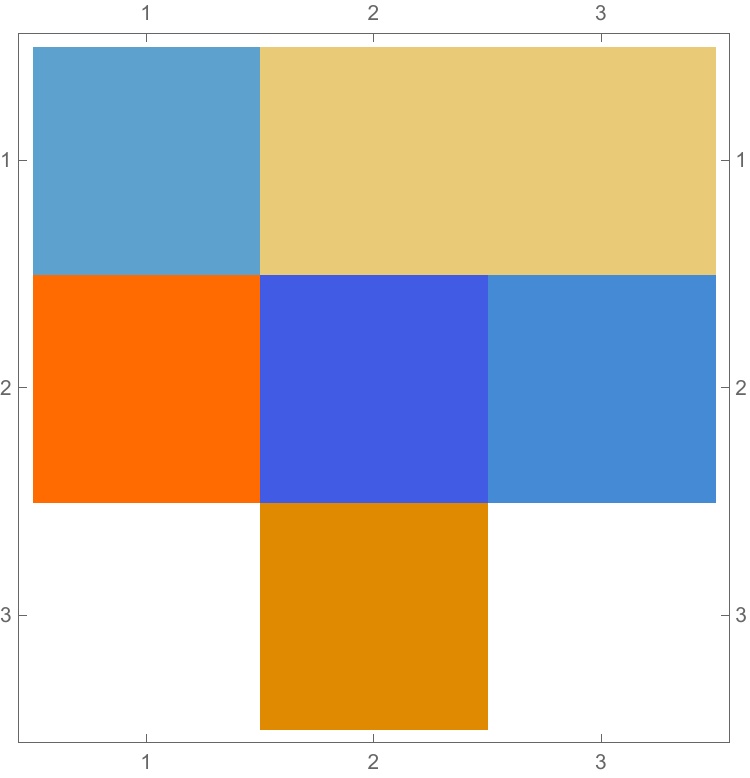}
    }
    \subfloat{
        \includegraphics[width=0.149\textwidth]{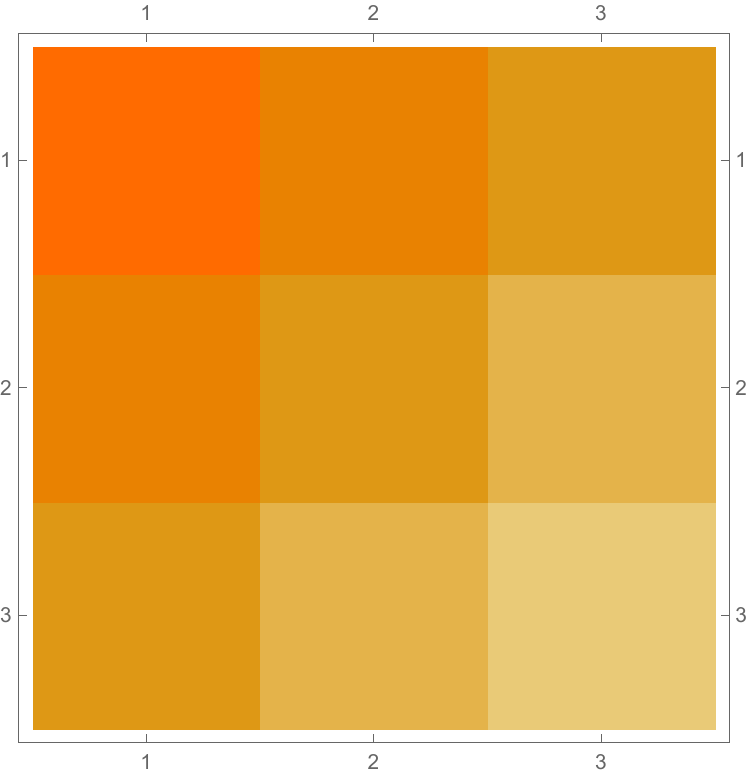}
    }
    \subfloat{
        \includegraphics[width=0.149\textwidth]{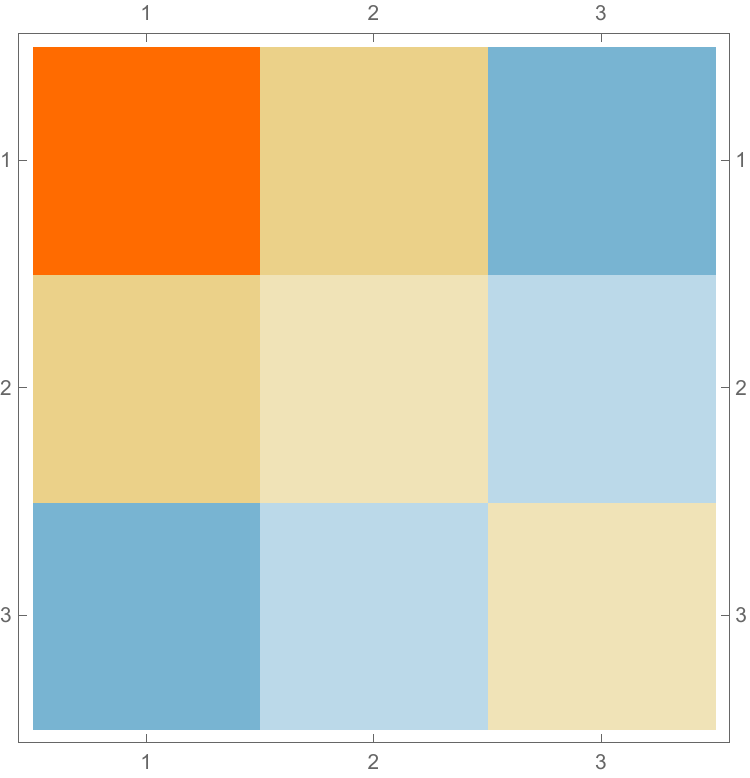}
    }
    \subfloat{
        \includegraphics[width=0.149\textwidth]{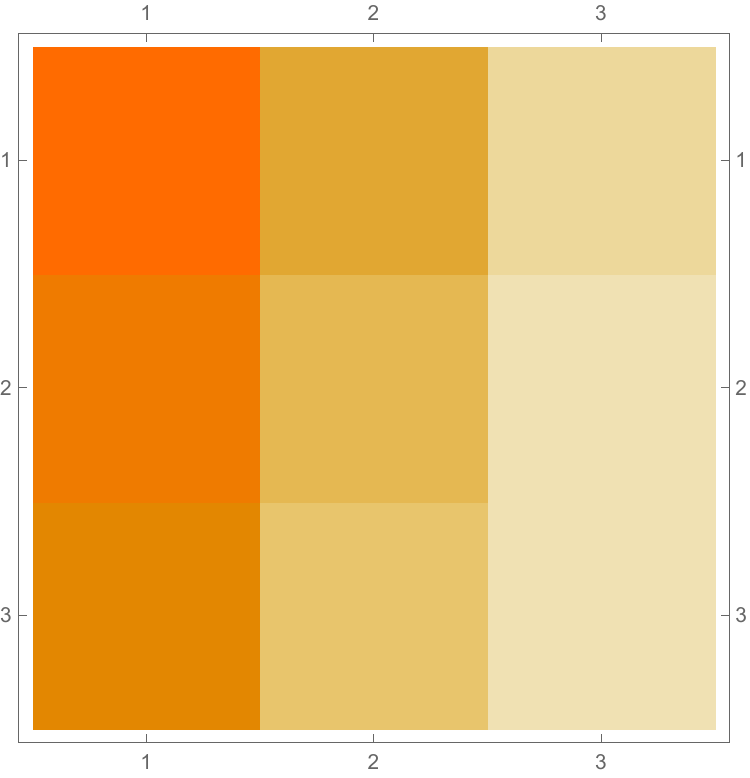}
    }
    \subfloat{
        \includegraphics[width=0.149\textwidth]{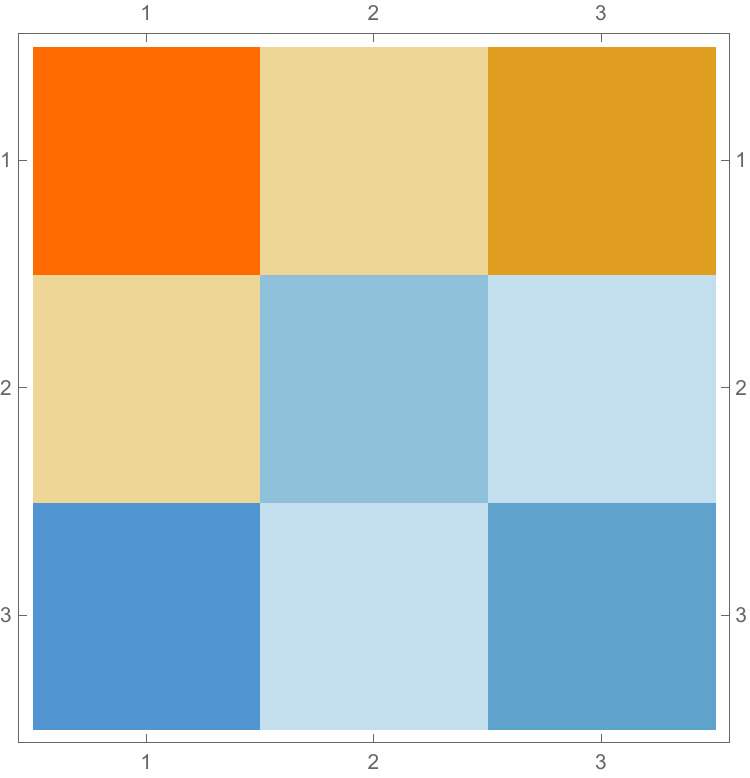}
    }
    \subfloat{
        \includegraphics[width=0.149\textwidth]{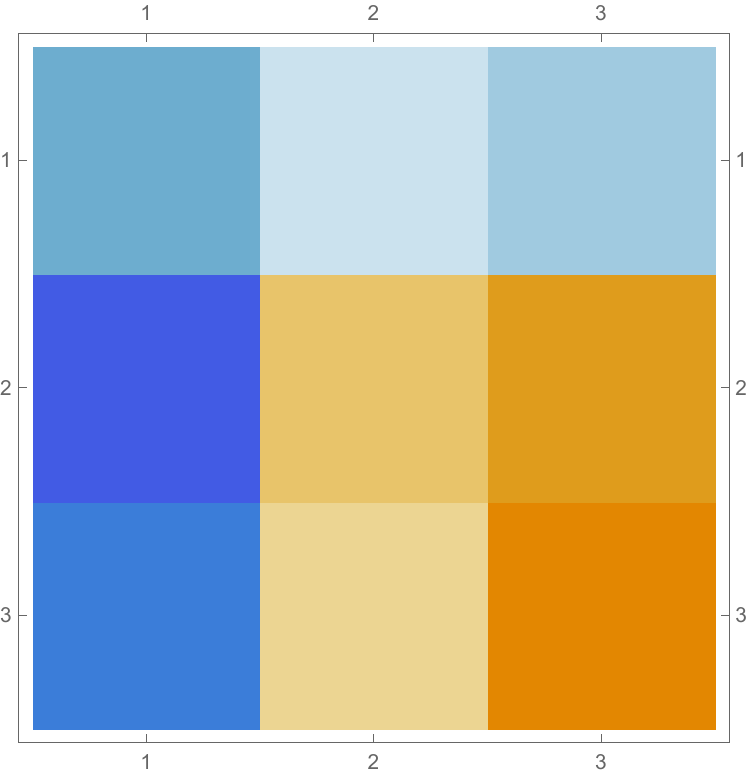}
    }
    \caption{{Coulomb} potential, $T^{(0)}$, $\bar P^{(0)}$ (with first rows and first columns dropped), $P$, $\bar Q^{(0)}$ (with first rows and first columns dropped), $Q$, and $H$ matrices (from ({\bf first column}) to ({\bf last column})) of $N=1$ version of first-order ContEvol with $h=1/2$, $h=3/4$, $h=1$, $h=5/4$, and $h=3/2$ (from ({\bf first row}) to ({\bf last row})).}
    \label{fig:H_CE1_varh_mats}
\end{figure}

Then we explore the $N=1$ case, which only has one single interval $[0, h]$ in addition to the tail.
Figure~\ref{fig:H_CE1_varh_mats} presents five sets of six $3 \times 3$ matrices based on different values of $h$.
All non-zero elements of $T^{(0)}$ matrices are shown in gradually varying colors, illustrating how $T^{(0)}$ changes with $h$; note that Equation~(\ref{eq:H_CE1_T0mat}) tells us that the matrix element $T^{(0)}_{32}$ is always $1$ regardless of $h$.
The symmetric $\bar P^{(0)}$ matrices (with first rows and first columns dropped) manifest similar gradual variation, with largest element ``migrating'' from lower-right corner to upper-left corner; however, combining variations of $T^{(0)}$ and $\bar P^{(0)}$, as well as $P'{}^{(0)}$ added for the tail, the $P$ matrices seem very similar to each other, although the color scales (not shown in Figure~\ref{fig:H_CE1_varh_mats}) are different.
The $\bar Q^{(0)}$ matrices (also with first rows and first columns dropped) are intrinsically asymmetric, and the largest element ``migrates'' from lower-center to lower-left; the resulting $Q$ matrices seem quite different with different values of $h$, yet gradual variation can still be revealed if we examine the elements one at a time.
Finally, the $H$ matrices also look similar to each other, although slightly variation can still be noticed; their eigenvectors are not shown as a matrix, since this section focuses on the ground state.
Here we comment that the other two eigenvalues are positive, and the corresponding wavefunctions are quasi-sinusoidal in the interval $[0, h]$ and almost zero in the tail; to study the actual excited states, one needs to repeat the fine-tuning exercise described below.

\vspace{-12pt}

\begin{figure}[H]
    \subfloat{
        \includegraphics[width=0.48\textwidth]{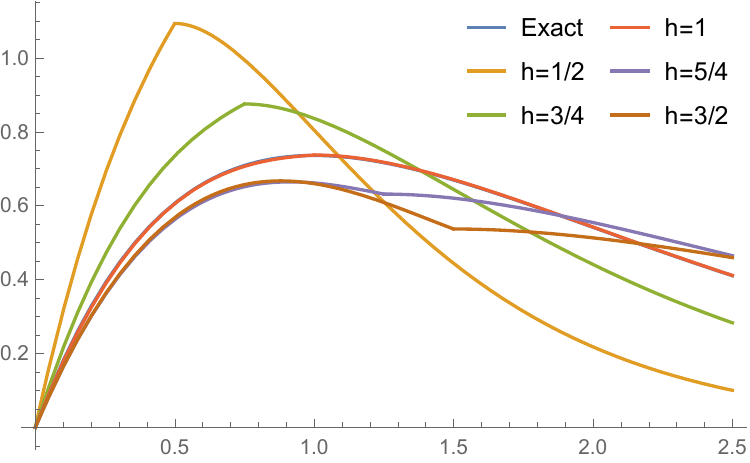}
    }
    \subfloat{
        \includegraphics[width=0.48\textwidth]{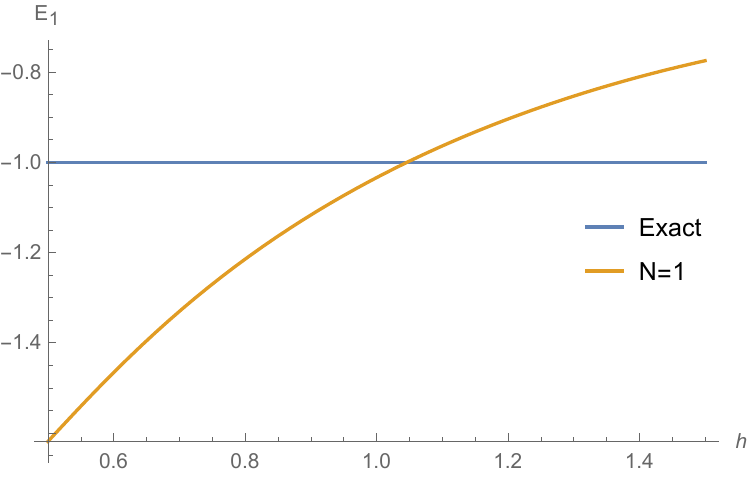}
    }
    \caption{Coulomb potential. ({\bf Left}): exact (blue) ground state wavefunction and rendered counterparts produced by $N=1$ version of first-order ContEvol with $h=1/2$ (orange), $h=3/4$ (green), $h=1$ (red), $h=5/4$ (purple), and $h=3/2$ (brown), which are shown in different colors; the exact solution is largely behind the $h=1$ version. ({\bf Right}): ground state energy produced by $N=1$ version of first-order ContEvol with varying $h$; the exact value $-1$ is shown as a horizontal line.}
    \label{fig:H_CE1_N=1_varh}
\end{figure}

\begin{figure}[H]
    \includegraphics[width=0.8\textwidth]{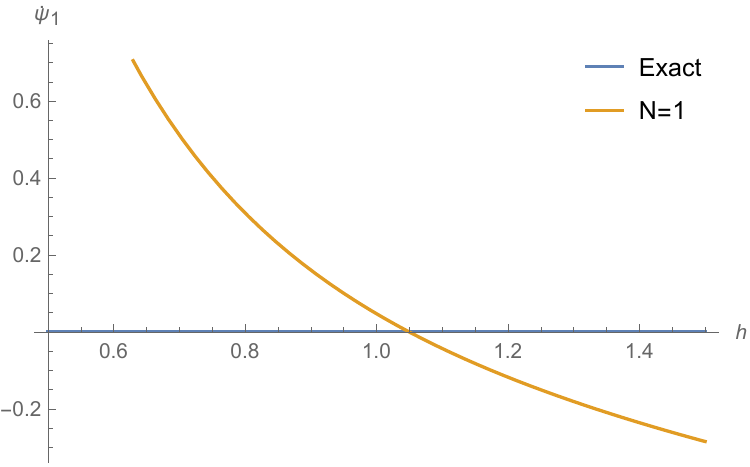}
    \caption{Coulomb potential, derivative at the first node $\dot\psi_1$ predicted by $N=1$ version of first-order ContEvol with varying $h$; the exact value $0$ is shown as a horizontal line.}
    \label{fig:H_CE1_N=1_dotpsi1}
\end{figure}

Rendered ground state wavefunctions based on the $H$ matrices in Figure~\ref{fig:H_CE1_varh_mats} are shown in the {left} 
 panel of Figure~\ref{fig:H_CE1_N=1_varh}.
The $h=1$ version agrees with the exact solution Equation~(\ref{eq:H_exact}) remarkably well, while other values of $h$ are limited by not-so-good predefined functional forms.
The {right} panel of Figure~\ref{fig:H_CE1_N=1_varh} plots the ground state energy as a function of $h$. The ContEvol solution coincides with the exact value at $h \approx 1.0469$.
However, how shall we determine the optimal value of $h$ when we have no idea about the exact solution? Similar to an argument in Section~\ref{ss:QHO_CE1}, we can fine-tune $h$ so that $\dot\psi_N$, in this case $\dot\psi_1$, is close to zero.
Figure~\ref{fig:H_CE1_N=1_dotpsi1} plots $\dot\psi_1$ as a function of $h$. It is exactly zero at $h \approx 1.0493$, which is close but not identical to the value quoted above.
In practice, we can adjust values of $h$ and $N$ in turn: for example, we explore a small interval around $h \approx 1.0493$ with $N=2$, get a better estimate of $h$, and explore a smaller interval around the updated $h$ with a larger $N$, etc., until the errors are below some threshold.
Such iterative process is not implemented for this work. In the following, we simply adopt $r_N = Nh = 1$, and enforce the $\dot\psi_N = 0$ constraint; investigating how $h$ affects the accuracy of $N > 1$ results is left for future work.

\vspace{-10pt}

\vspace{-10pt}

\paragraph{{Realistic versions: $N=2$ to $N=8$.}}

\begin{figure}[H]
    \subfloat{
        \includegraphics[width=0.32\textwidth]{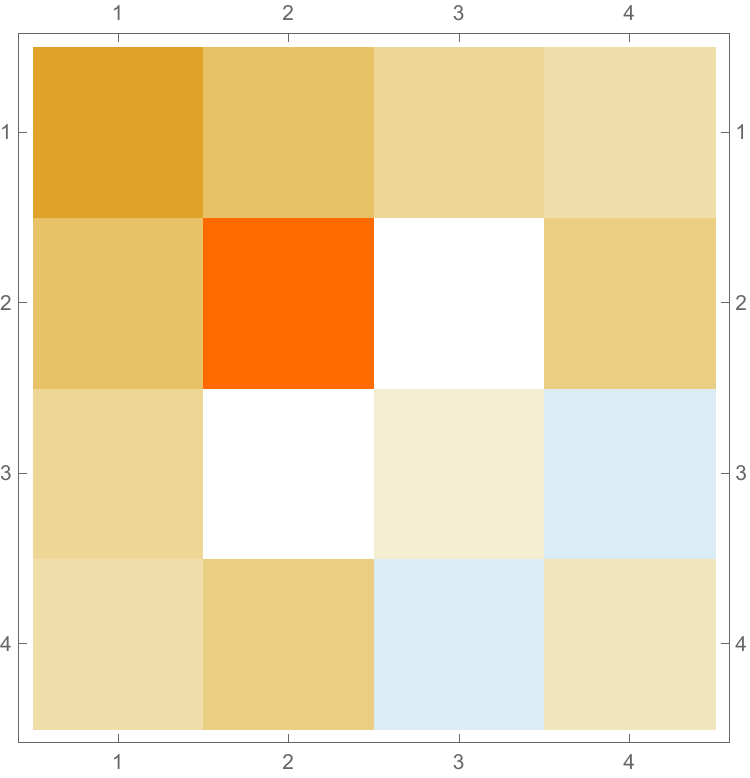}
    }
    \subfloat{
        \includegraphics[width=0.32\textwidth]{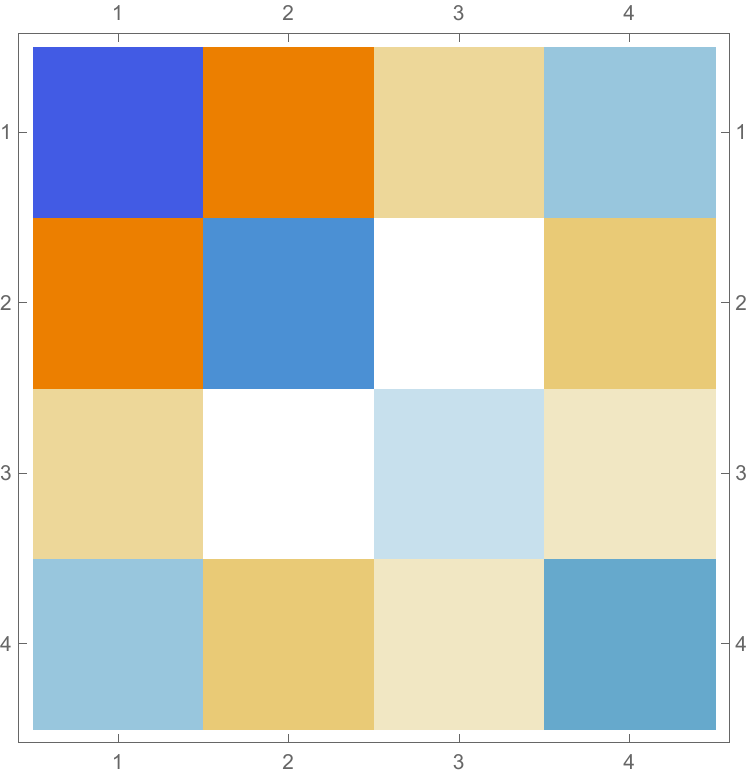}
    }
    \subfloat{
        \includegraphics[width=0.32\textwidth]{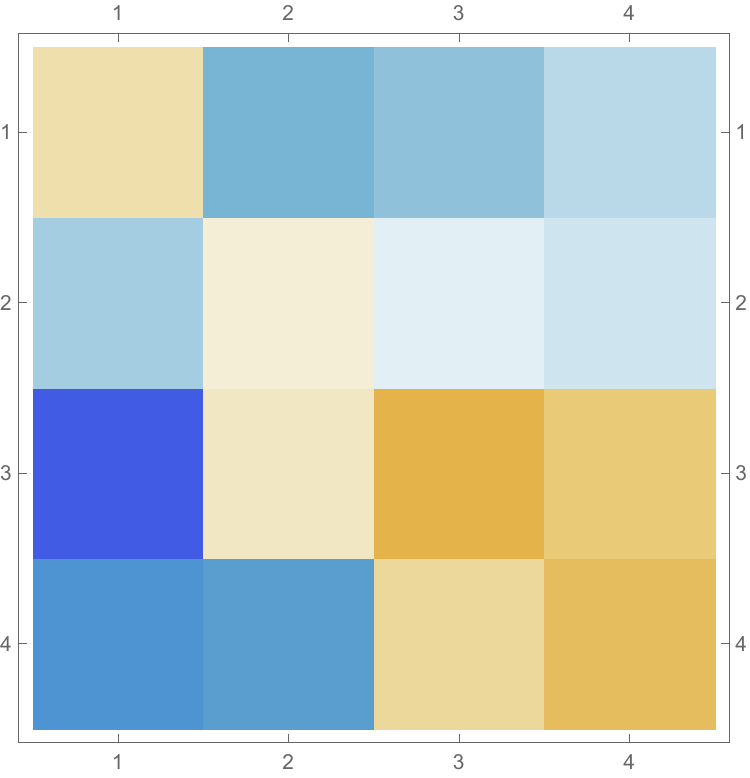}
    }

    \subfloat{
        \includegraphics[width=0.32\textwidth]{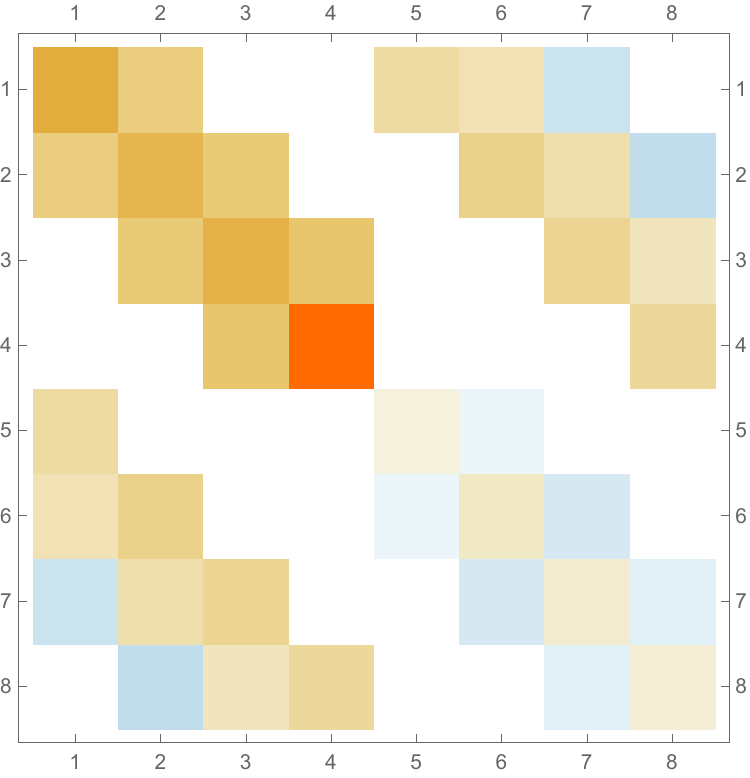}
    }
    \subfloat{
        \includegraphics[width=0.32\textwidth]{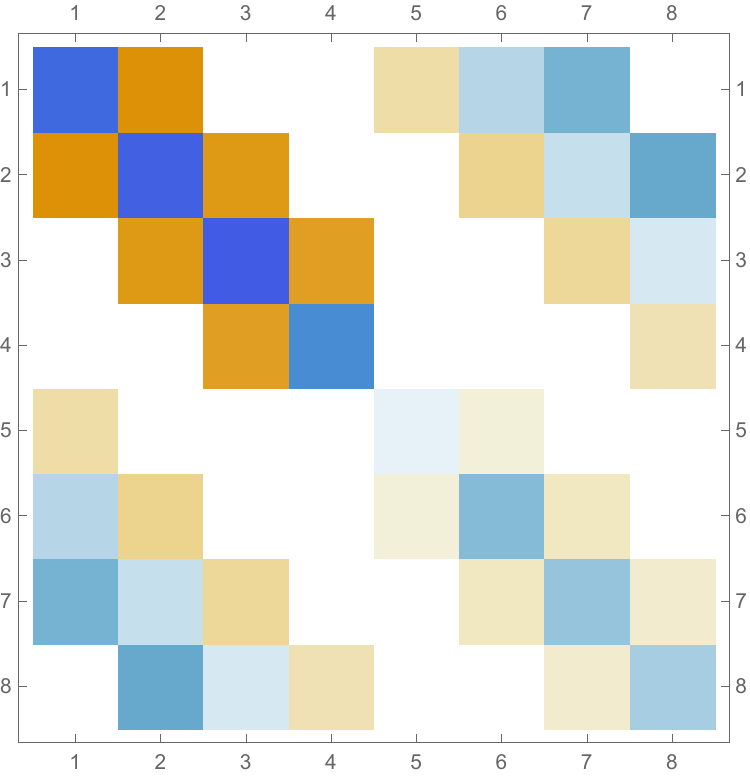}
    }
    \subfloat{
        \includegraphics[width=0.32\textwidth]{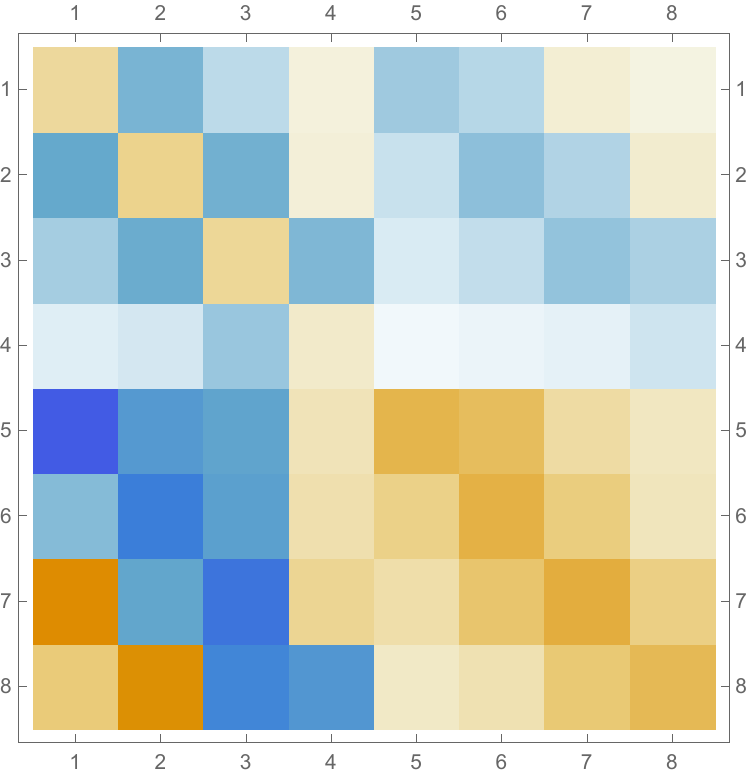}
    }

    \subfloat{
        \includegraphics[width=0.32\textwidth]{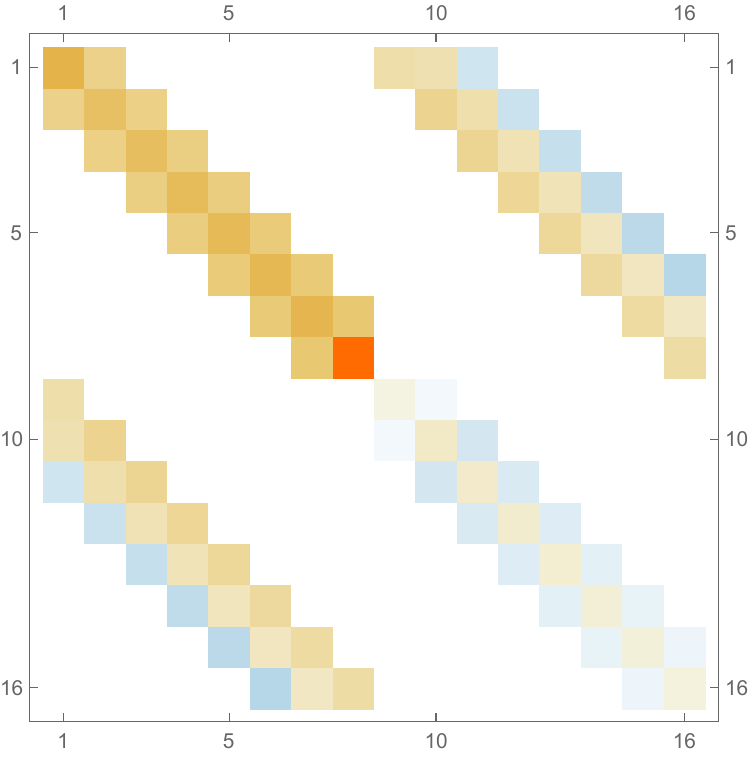}
    }
    \subfloat{
        \includegraphics[width=0.32\textwidth]{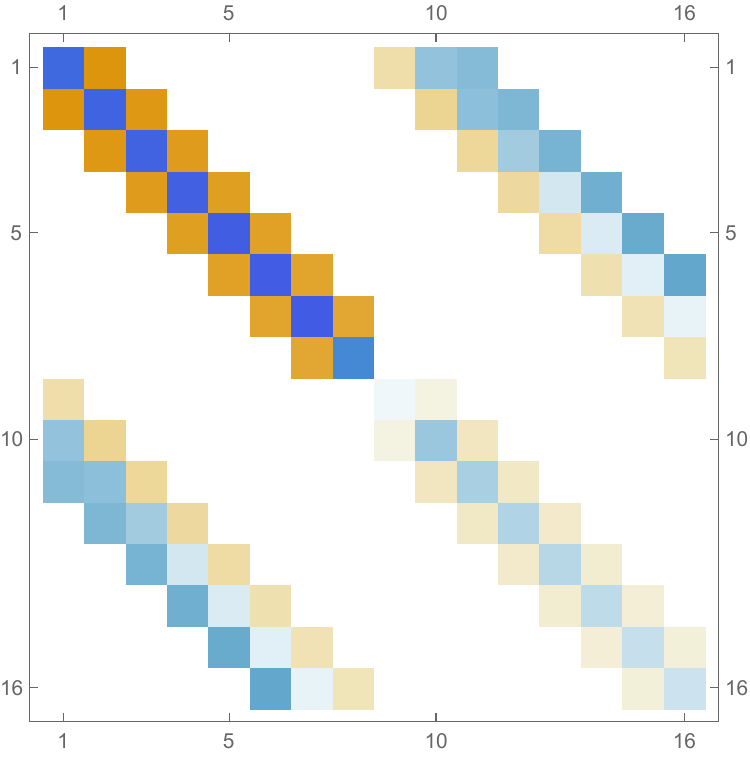}
    }
    \subfloat{
        \includegraphics[width=0.32\textwidth]{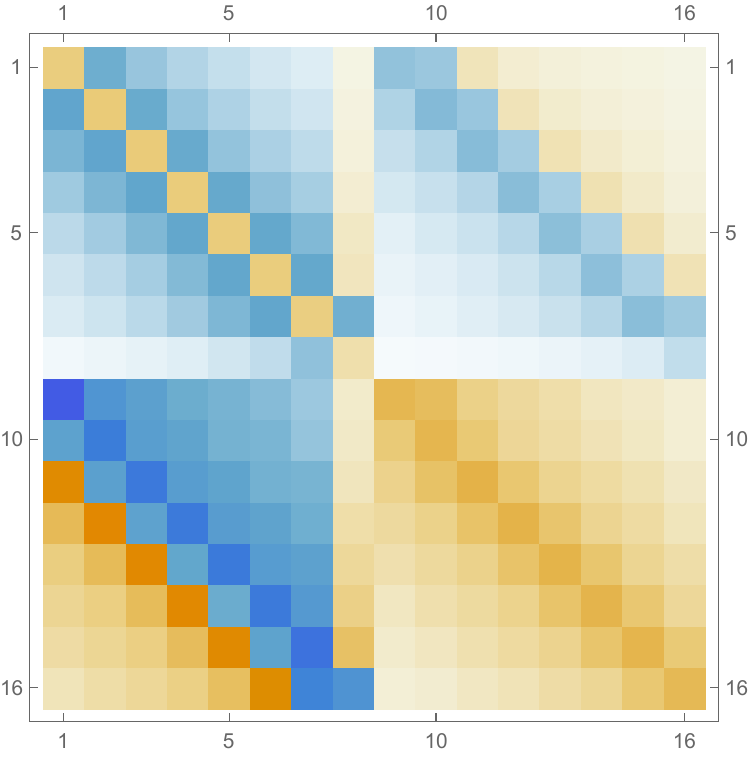}
    }
    \caption{{Coulomb} potential, $P$ ({\bf first column}), $Q$ ({\bf second column}), and $H$ matrices ({\bf last column}) for $N=2$ ({\bf first row}), $N=4$ ({\bf second row}), and $N=8$ ({\bf last row}) versions of first-order ContEvol, all with $r_N = Nh = 1$.}
    \label{fig:H_CE1_varN_mats}
\end{figure}

Figure~\ref{fig:H_CE1_varN_mats} shows $P$, $Q$, and $H$ matrices produced by $N=2$, $N=4$, and $N=8$ versions of first-order ContEvol.
Like in the case of infinite potential well (see Section~\ref{ss:IPW_CE1}, especially Figure~\ref{fig:IPW_mats_vecs}), each $P$ or $Q$ matrix has $2 \times 2$ tridiagonal blocks; because of the position-dependence of the Coulomb potential, elements on the same diagonal do not necessarily have the same value.
Most noticeable matrix elements are $P_{N,N}$ and $Q_{N,N}$, which are affected by the tail; the former are ``more positive'' in $P$ matrices, while the latter are ``less negative'' in $Q$ matrices. Consequently, the $N$th rows and $N$th columns of $H$ matrices do not follow the same pattern as other regions.

\begin{figure}[H]
    \subfloat{
        \includegraphics[width=0.48\textwidth]{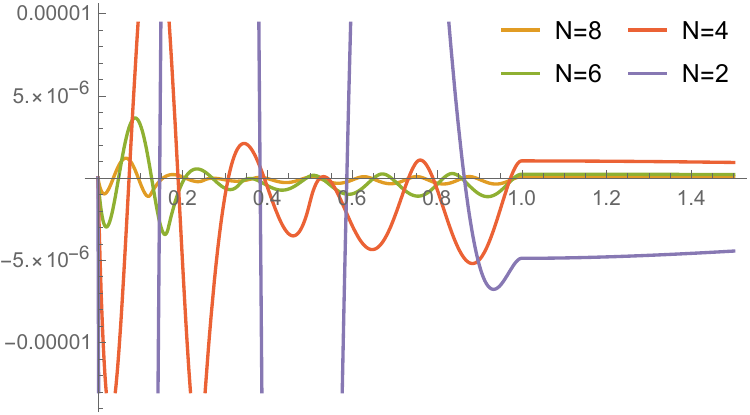}
    }
    \subfloat{
        \includegraphics[width=0.48\textwidth]{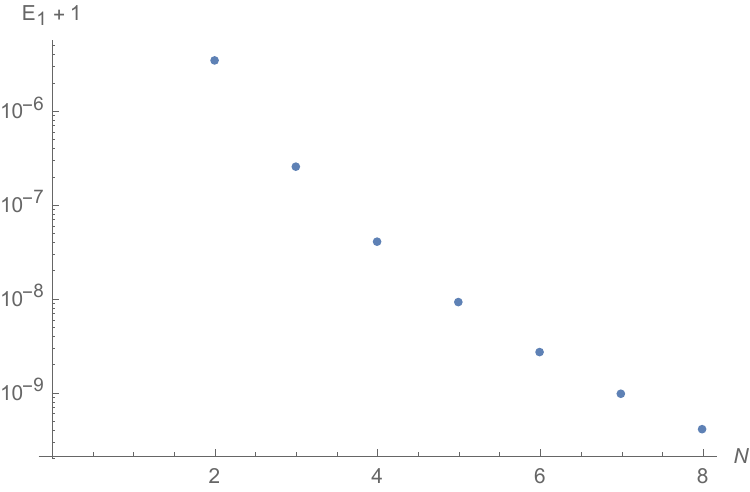}
    }
    \caption{{Coulomb} potential. ({\bf Left}): errors in rendered wavefunctions produced by $N=8$ (orange), $N=6$ (green), $N=4$ (red), and $N=2$ (purple) versions of first-order ContEvol, which are shown in different colors; all with $r_N = Nh = 1$. Note that peak of exact wavefunction is $2/e \approx 0.7358$.
    ({\bf Right}): errors in ground state energy produced by $N = 2, 3, \ldots, 8$ versions of first-order ContEvol, all with $r_N = Nh = 1$.}
    \label{fig:H_CE1_varN}
\end{figure}

In Figure~\ref{fig:H_CE1_varN}, the {left} 
 panel displays errors in rendered ground state wavefunctions of $N=2$, $N=4$, $N=6$ (not shown in Figure~\ref{fig:H_CE1_varN_mats}), and $N=8$ Hamiltonians, while the {right} panel plots errors in ground state energy predicted by first-order ContEvol with $N = 2, 3, \ldots, 8$.
Like in Section~\ref{ss:IPW_CE1}, the eigenpair is already remarkably accurate with $N=8$, which is arguably small.

\vspace{-12pt}

\vspace{-20pt}

\section{Discussion: Directions for Future Work}
\label{sec:discuss}

The ContEvol formalism has many potential applications inside and outside physics. For example, yearning for a ``smoother'' stellar evolution code has supplied the original motivation for this work.
As long as people want to represent continuous functions (of time, space, or both) with a finite sampling, ContEvol may help.
However, much work remains to be done to reveal its full potential. In this final section, we discuss some of the major directions for future development of ContEvol.

\subsection{{Mathematical Foundation}}

Although ContEvol appears to be successful, it lacks a solid mathematical foundation. Desirable justifications and auxiliary tools include but are not limited to:
\begin{itemize}
    \item Control over errors and non-symplecticity. With specific cases, this work seems to indicate that first-order ContEvol results have $\mathcal{O}(h^6)$ errors in values, $\mathcal{O}(h^5)$ errors in first derivatives, and $\mathcal{O}(h^5)$ error in deviation from equation(s) of motion---more specifically, the $\mathcal{O}(h^6)$ terms in values are usually just missing; see Equation~(\ref{eq:CHO_CE1_res}) for an example; second-order ContEvol does not improve order of errors in results, but does reduce deviation from EOM(s) to $\mathcal{O}(h^9)$; non-symplecticity (discrepancy between determinant of Jacobian and $1$) does not display a uniform pattern.
    Under what conditions do these statements hold? How do these quotes scale with the order of ContEvol? Such questions needs to be answered to solidify ContEvol results.

    \item Foundation for customized linear algebra. As hypothesized in Section~\ref{ss:IPW_CE1}, intuitively Hamiltonian $H = -P^{-1}Q$ based on Equation~(\ref{eq:IPW_CE1_PandQ}) should be a Hermitian operator, and the inner product defined in Equation~(\ref{eq:IPW_CE1_ortho}) is reasonable.
    Yet unless these statements are well justified, ContEvol does not guarantee an expected number of valid eigenpairs.

    \item Moments and transforms. This work has not included expressions for moments and transforms (e.g., Fourier and Laplace transforms) based on values and derivatives at nodes, yet such things are likely to be important for the analysis of ContEvol results.
    Do they reveal additional properties or limitations of ContEvol methods? The answer will inform choices for specific applications.
\end{itemize}

\subsection{{Higher Dimensions}}
This work has been focused on one-dimensional scenarios, either time or space; nevertheless, the combination of function representation with linear coefficients and cost function minimization can be generalized to high-dimensional cases.
In other words, the ContEvol formalism should be able to solve partial differential equations (PDEs) as well as ordinary differential equations (ODEs). Here we outline major directions of such extensions for first-order ContEvol.
\begin{itemize}
    \item Evolving one-dimensional functions. In this case, the full evolutionary history of the function $\psi(x, t)$, sampled at $N_t$ timestamps and $N_x$ nodes, can be fully characterized by $N_t \times N_x$ quadruples, $\{ \psi, \psi_{;x}, \psi_{;t}, \psi_{;x;t} \}$, where semicolons ``$;$'' in subscripts denote partial derivatives.
    Thus at each space-time location, the function can be rendered as the product of a cubic polynomial in $x$ and a cubic polynomial in $t$; such a representation has $16$ coefficients, corresponding to four quadruples at four corners of a space-time cell.

    \item Representing high-dimensional functions. Although there are no restrictions for use of curvilinear coordinates, the discussion here focuses on Cartesian coordinates.
    To fully characterize a spatial distribution, in principle one could use $\{ \psi, \psi_{;x}, \psi_{;y}, \psi_{;x;y} \}$ in two dimensions and $\{ \psi, \psi_{;x}, \psi_{;y}, \psi_{;x;y}, \psi_{;z}, \psi_{;x;z}, \psi_{;y;z}, \psi_{;x;y;z} \}$ in three dimensions. However, in $d$ dimensions, multiplying the $N^d$ growth of number of nodes and $2^d$ growth of number of features can easily make things computationally unaffordable.
    A less expensive version of the high-dimensional function representation would only use values and first derivatives, i.e., $\{ \psi, \psi_{;x}, \psi_{;y} \}$ in 2D and $\{ \psi, \psi_{;x}, \psi_{;y}, \psi_{;z} \}$ in 3D, so that the number of features only grows as $1+d$.
    A difficulty is that in 2D (3D), there are only $10$ (or $20$) zeroth- to third-order terms, but there are $2^2 \times (1+2) = 12$ (or $2^3 \times (1+3) = 32$) features to fit for each cell; to bypass inconsistency, it is recommended to add some higher-order terms (e.g., $x^2y^2$), but those involving fourth or higher order in a single variable should probably be avoided (e.g., $x^4$ or $y^4$).

    \item Evolving high-dimensional functions. Space and time coordinates could be viewed as equivalent from the perspective of special relativity, yet for most computational physics problems, time may play a different role than spatial coordinates.
    Thenceforth, for better representing the ``history'' of a dynamic system, $\{ \psi, \psi_{;x}, \psi_{;y}, \psi_{;t}, \psi_{;x;t}, \psi_{;y;t} \}$ in 2D and $\{ \psi, \psi_{;x}, \psi_{;y}, \psi_{;z}, \psi_{;t}, \psi_{;x;t}, \psi_{;y;t}, \psi_{;z;t} \}$ in 3D might be a more sensible choice.
\end{itemize}

{In} 
 addition to higher dimensions, we note that extension to multiple functions is also natural; vector and tensor functions can be decomposed into independent components, as we did in Section~\ref{sec:celestial}.

\subsection{{Technical Improvements}}
The last group of directions addresses some technical issues involved in the ContEvol formalism per se, which may lead to improvements in accuracy, precision, or performance.
\begin{itemize}
    \item Multistep version. This works has been focused on single-step ContEvol methods, regardless of the order, yet it is possible to extend ContEvol to multiple steps or intervals.
    For boundary value problems, if we want to study the function $f(x)$ for some interval $x_i \leq x \leq x_{i+1}$, while the combination of $\{ f_i, \dot f_i, f_{i+1}, \dot f_{i+1} \}$ can give us a cubic approximation, the combination of $\{ f_{i-1}, \dot f_{i-1}, f_i, \dot f_i, f_{i+1}, \dot f_{i+1}, f_{i+2}, \dot f_{i+2} \}$ (assuming sampling nodes $x_{i-1}$ and $x_{i+2}$ both exist or can be reasonably defined for convenience) can give us a septic approximation.
    For initial value problems, there are two basic strategies: backward, which for example approximates the evolution during the next interval as a quintic polynomial based on $\{ f_{-h}, \dot f_{-h}, f_0, \dot f_0, f_{h}, \dot f_{h} \}$; and forward, which for example approximates the evolution during the next two intervals as a pair of cubic polynomials or a unified quintic polynomial based on $\{ f_0, \dot f_0, f_{h}, \dot f_{h}, f_{2h}, \dot f_{2h} \}$. Of course one can include more steps or devise hybrid versions.
    Like higher orders (e.g., Section~\ref{ss:CHO_CE2}), inclusion of multiple steps complicates derivation and computation, but potentially improves accuracy or precision.

    \item Better sampling and evolving nodes. As mentioned in Section~\ref{ss:IPW_CE1}, the distribution of sampling nodes is by no means necessarily uniform; for some realistic applications, their distribution should not be fixed, for example in Section~\ref{ss:QHO_CE1}, when the potential function necessitates a flexible sampling.
    In short, the sampling is something ContEvol users are encouraged to fine-tune.
    In addition, when a field is evolved (see above for discussion on higher dimensions), drifting nodes (i.e., nodes with varying positions) and splitting or merging cells (i.e., adding or removing nodes) may be desirable.
    Because of the uniqueness of Hermite spline, splitting $[x_{\rm left}, x_{\rm right}]$ into $[x_{\rm left}, x_{\rm middle}]$ and $[x_{\rm middle}, x_{\rm right}]$ by inserting $f(x_{\rm middle})$ and $\dot f(x_{\rm middle})$ at an arbitrary location $x_{\rm middle}$ between $x_{\rm left}$ and $x_{\rm right}$ does not distort the ``current'' function representation at all; this fact should be applicable to higher dimensions as well.
    However, we note that such variations are preferably predefined (e.g., according to some strategy), not determined on-the-fly, as optimizing location of nodes often requires solving non-linear~equations.
    \item Computational efficiency. Let us consider arguably the most costly case of real-world physics problems, time evolution of a set of three-dimensional fields, e.g., cosmological simulations; we use single-step ContEvol with $N$ nodes in each dimension, and keep track of $N_q$ quantities, each with $N_f$ features (i.e., values or partial derivatives). Then the dimension of the matrix is $(N^3 N_q N_f) \times (N^3 N_q N_f)$, which can be overwhelmingly expensive.
    However, indexing each of the $(N_q N_f) \times (N_q N_f)$ blocks as $B_{\alpha \beta \gamma \alpha' \beta' \gamma'}$, where $\alpha('), \beta('), \gamma(') = 0, 1, \ldots, N-1$, the necessary condition for an element to be non-zero is $\max\{ |\alpha-\alpha'|, |\beta-\beta'|, |\gamma-\gamma'| \} \leq 1$. In other words, among the \linebreak  $(N^3)^2 = N^6$ elements of this block, only less than $3^3 N^3 = 27 N^3$ can possibly non-zero, i.e., such matrices are highly sparse when $N$ is large; a closer look would reveal many ``tridiagonal'' structures. Specialized data structures and algorithms could be designed to handle such matrices.
    Furthermore, when we compute the evolution of large-scale structures under gravitational interactions, information about specific chemical composition may not be particularly pertinent. In such cases, multi-tier strategy could be useful: at each step, we first evolve the ``dominating'' quantities, and then combine coarse-grained ``future'' and fine-grained ``present'' to evolve the ``dependent'' quantities.
\end{itemize}

\subsection{{Miscellany}}

\begin{figure}[H]
    \includegraphics[width=0.48\textwidth]{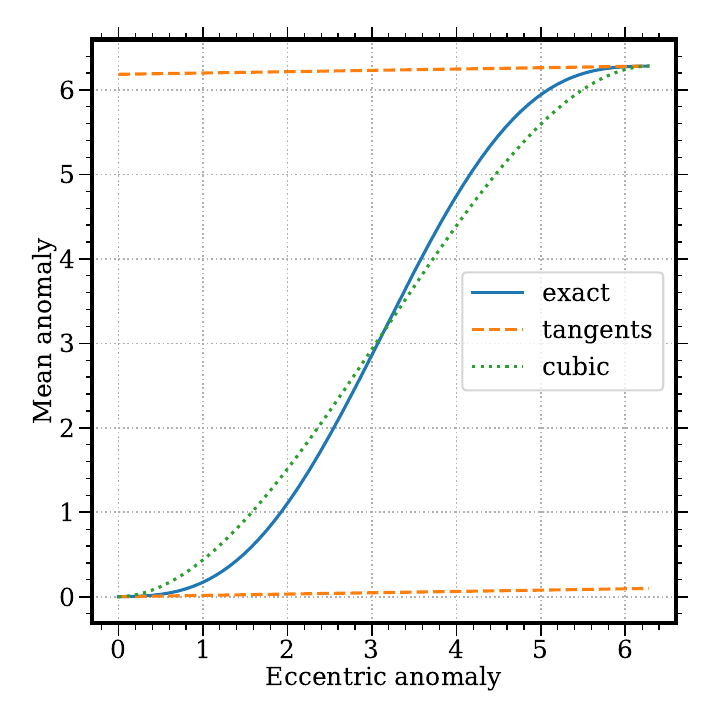}
    \caption{Mean anomaly versus eccentric anomaly based on Kepler's equation Equation~(\ref{eq:Kepler}) and eccentricity $e = 63/64$ (used in Section~\ref{ss:2BP_CE1N}).
    Exact solution, tangents at $(0, 0)$ and $(2\pi, 2\pi)$, and cubic approximation are shown as a ``tab:blue'' solid curve, a pair of ``tab:orange'' straight lines, and a ``tab:green'' dotted curve, respectively.}
    \label{fig:root_finding}
\end{figure}

In addition to the above directions, some miscellaneous topics are worth mentioning.
\begin{itemize}
    \item Root-finding. While this work has been focused on differential equations, the backbone function representation of ContEvol (Hermite spline) can be applied to algebraic equations as well: knowing both values and first derivatives at two sampling points, we can always find a cubic approximation of the function to help root-finding.
    For instance, Figure~\ref{fig:root_finding} displays Kepler's equation Equation~(\ref{eq:Kepler}) with $e = 63/64$; using Newton's method, one would have to carefully choose an initial guess to avoid divergence, while the cubic approximation is more robust.
    Admittedly, solution to a cubic equation is more complicated than that to a linear equation, yet cubic may work better in some cases; besides, one can use cubic for the first few steps, and then switch to linear for fine-tuning purposes.

    \item Numerical integration. Likewise, piece-wise cubic (or higher-order) polynomials may help numerical integration.
    As demonstrated in Section~\ref{sec:quantum}, using less sampling points, a ``compound'' sampling with both values and derivatives can outperform ``simple'' sampling with only values.
    Although fitting polynomials with multiple values (e.g., Simpson's rule) could effectively mitigate discreteness, usage of derivatives should rely less on a fine sampling.
    When the derivatives have to be evaluated numerically, in the first-order case, this technical is equivalent to a sampling like $\{ \ldots, x_i-\Delta/2, x_i+\Delta/2, x_{i+1}-\Delta/2, x_{i+1}+\Delta/2, \ldots \}$, where $\Delta \ll |x_{i+1} - x_i|$.

    \item Data structure of lookup tables. Due to the semi-analytic nature of the ContEvol formalism, its performance might be limited by lookup tables stored as hypercubes of values; fortunately, development of numerical methods may advance data structure of lookup tables as well.
    This section has already addressed how high-dimensional functions are supposed to be digitalized by combining values and derivatives; the three-dimensional plan can be naturally extended to higher dimensions.
    Even without ContEvol, ``continuous'' lookup tables have their own benefits, e.g., higher accuracy or less storage usage.
\end{itemize}

\subsection{{Limitations}} 
{It} should be noted that several limitations have been identified throughout the text.
\begin{itemize}
    \item Lack of strict symplecticity. As shown in Section~\ref{ss:CHO_CE1}, the ContEvol methods are not strictly symplectic, although the deviation is small. Therefore, caution is needed when studying long-term behavior of dynamic systems, for which geometric solvers \citep{Krantz2008book} may be a better choice. 
    \item Moderate benefits of higher-order methods. As shown in Section~\ref{ss:CHO_CE2}, compared to the first-order version, second-order ContEvol method only reduces, but does not eliminate, higher-order errors. Consequently, adopting higher-order ContEvol methods does not prevent the need for small step sizes.
    \item Challenges in handling infinite boundaries. As discussed in Section~\ref{ss:QHO_CE1}, for some boundary value problems, linearity of equations can only be achieved at the expense of limited flexibility while handling infinite boundaries. Therefore, for functions with significant higher-order moments, the ContEvol formalism may require a wider spread of sampling nodes.
\end{itemize}

{Some} of these limitations may be ameliorated or overcome in the future.

In conclusion, it is our hope that, with further developments, the ContEvol (continuous evolution) formalism can benefit some applications of computational physics.

 {The following} software is used on KC's personal computer (HP All-in-One 24-dp1xxx, Microsoft Windows 11 Home).
Most symbolic operations throughout this work are performed and figures in Section~\ref{sec:quantum} are made with Wolfram Mathematica 11.0 \citep{Mathematica11}.
Numerical tests in Section~\ref{sec:celestial} are conducted with Python 3.11 \citep{van2023python} codes developed using {\sc NumPy} {1.26.4} \citep{Harris2020Natur} and {\sc Numba} {0.59.1} \citep{Lam2015}, corresponding exact solution is derived with {\sc SciPy} {1.13.0} \citep{Virtanen2020NatMe}, while figures therein and that in Section~\ref{sec:discuss} are made with {\sc {Matplotlib}}~{3.8.3} \citep{Hunter2007CSE}.
Mathematica and Jupyter notebooks for this work {are} 
 available in the GitHub repository ContEvol\_formalism {\url{https://github.com/kailicao/ContEvol_formalism.git}}, {accessed on 8 May 2024}.  
This article is prepared with Overleaf, Online LaTeX Editor {\url{https://www.overleaf.com/}}, {accessed on 8 May 2024} 
 and Online LaTeX Equation Editor. {\url{https://latex.codecogs.com/eqneditor/editor.php}}, {accessed on 8 May 2024}.

\vspace{6pt}

\funding{{This research was funded by an internal funding source at The Ohio State University.}}

\dataavailability{{The original data presented in the study are openly available at} 
\url{https://github.com/kailicao/ContEvol_formalism.git}, {accessed on 8 May 2024}.}

\acknowledgments{{The author thanks his advisors, Christopher M. Hirata and Marc H. Pinsonneault, for inspirations through research projects in cosmological image processing and stellar evolution, respectively, as well as insights and encouragement during the preparation of this work. The author appreciates insightful feedback from (in chronological order) Anil K. Pradhan, Annika H.G. Peter, R.J. Furnstahl, and Todd A. Thompson. The author also thanks Li-Yong Zhou (Nanjing University, China) 
 and R.J. Furnstahl for introducing him to numerical methods in celestial mechanics and quantum mechanics, respectively.} {Finally, the author is grateful to the reviewers for their constructive comments on his original submission.}}

\conflictsofinterest{{The authors declare no conflict of interest.}}

\begin{adjustwidth}{-\extralength}{0cm}
\reftitle{References}

\PublishersNote{}

\end{adjustwidth}
\end{document}